\documentclass{jfm}

\usepackage{amsmath}
\usepackage{graphicx}
\usepackage{subcaption}
\usepackage[export]{adjustbox}
\usepackage{float}
\usepackage{graphicx}
\usepackage{epstopdf, epsfig}
\usepackage{flushend}
\usepackage[dvipsnames]{xcolor}
\usepackage{amsmath}
\usepackage{latexsym}
\usepackage{caption}
\usepackage{color}
\usepackage{soul}
\usepackage{physics}
\usepackage[normalem]{ulem}
\usepackage{physics}
\usepackage[colorlinks=true, allcolors=blue]{hyperref}
\usepackage{xcolor}
\usepackage{pgfplots}
\usepackage{bm}
\usepackage{comment}
\usepackage{amsmath,setspace}
\usepackage{graphicx}
\usepackage{epstopdf, epsfig}
\usepackage{tikz}
\usepackage{mathtools}
\usepackage{mathdots}
\usepackage{nicematrix}
\usepackage{multicol}

\usepackage{lipsum}
\usepackage{subcaption} 
\usepackage{algorithm}
\usepackage{algpseudocode}
\usepackage{amsfonts}
\usepackage{amssymb}
\usepackage{array}
\usepackage{xparse}
\usepackage{dsfont}
\usepackage[font=footnotesize]{caption}
\usepackage{bbm}

\usepackage{dsfont}
\DeclareMathAlphabet{\mathbbm}{U}{dsrom}{m}{n}
\DeclareMathAlphabet{\mathbbmsl}{U}{dsrom}{m}{sl}

\newcommand{\Wd}{\bm{\mathrm{W}}}
\newcommand{\Sd}{\bm{\mathrm{S}}}
\newcommand{\qd}{\bm{\mathrm{q}}}
\newcommand{\Qd}{\bm{\mathrm{Q}}}

\newcommand{\MCSd}{\bm{\mathsfb{S}}}
\newcommand{\QCd}{\bm{\mathsfb{Q}}}
\newcommand{\MCWd}{\bm{\mathsfb{W}}}
\newcommand{\Td}{\bm{\mathrm{T}}}
\newcommand{\Psibd}{\pmb{\rmPsi}}
\newcommand{\Thetabd}{\bm{\rmTheta}}
\newcommand{\Lambdad}{\bm{\rmLambda}}
\newcommand{\MC}{\mathsfb{M}} % Giant combined tensor
\newcommand{\etab}{\bm{\eta}}
\newcommand{\MCSb}{\mathsfbi{{S}}}
\newcommand{\QCb}{{\mathsfbi{Q}}}
\newcommand{\MCWb}{\mathsfbi{W}}
\newcommand{\TdCb}{\mathsfbi{T}}
\newcommand{\FCb}{\mathsfbi{F}}
\newcommand{\GCb}{\mathsfbi{G}}
\newcommand{\FC}{{F}}
\newcommand{\GC}{{G}}
\newcommand{\UC}{{U}}
\newcommand{\ICb}{\mathsfbi{I}}
\newcommand{\UCb}{\mathsfbi{U}}
\newcommand{\VCb}{\mathsfbi{V}}
\newcommand{\UCHb}{\hat{\mathsfbi{U}}}
\newcommand{\VCHb}{\hat{\mathsfbi{V}}}
\newcommand{\HCb}{\mathsfbi{H}}
\newcommand{\WCb}{\mathsfbi{W}}

\newcommand{\Wb}{\bm{{W}}}
\newcommand{\WVb}{\bm{{WV}}}
\newcommand{\Rb}{\bm{{R}}}
\newcommand{\Sb}{\bm{{S}}}
\newcommand{\qb}{\bm{{q}}}
\newcommand{\Qb}{\bm{{Q}}}
\newcommand{\gb}{\bm{g}}
\newcommand{\fb}{\bm{f}}
\newcommand{\phib}{\pmb{\phi}}
\newcommand{\psib}{\pmb{\psi}}
\newcommand{\Psib}{\pmb{\Psi}}
\newcommand{\Sigmab}{\bm{\rmSigma}}

\newcommand{\Rres}{\bm{R}}
\newcommand{\Ab}{\bm{A}}
\newcommand{\Gb}{\tilde{\bm{g}}}
\newcommand{\Hb}{\bm{H}}
\newcommand{\Ib}{\bm{I}}
\newcommand{\ucb}{\bm{u}}
\newcommand{\vcb}{\bm{v}}
\newcommand{\fscal}{q}

\newcommand{\pcoeff}{\xi}
\newcommand{\freq}{\gamma}
\newcommand{\fcik}{\gamma_{k}}
\newcommand{\fci}{\gamma}
\newcommand{\Fcik}{\Gamma_{k}}
\newcommand{\Fci}{\Gamma}
\newcommand{\dt}{\mathrm{d}t}
\newcommand{\df}{\mathrm{d}f}
\newcommand{\dx}{\mathrm{d}\bm{x}}
\newcommand{\f}{\mathbf{f}}  	% f vector
\newcommand{\si}[1]{\sum_{#1=-\infty}^\infty}
\newcommand{\ivar}{\bm{x}}
\newcommand{\ivarp}{\bm{x}^\prime}
\newcommand{\perindx}{k_\alpha}
\newcommand{\m}{\bm{m}}
\newcommand{\x}{\bm{x}}

\newcommand{\Fset}{\mathcal{\Omega}}

\makeatletter
\NewDocumentCommand{\LeftComment}{s m}{%
  \Statex \IfBooleanF{#1}{\hspace*{\ALG@thistlm}}\(\triangleright\) #2}
\makeatother
\makeatletter
\NewDocumentCommand{\LeftCommentE}{s m}{%
  \Statex \IfBooleanF{#1}{\hspace*{\ALG@thistlm}}\(\phantom{\triangleright}\) #2}
\makeatother
\pgfplotsset{compat=1.17}

\shorttitle{Cyclostationary spectral proper orthogonal decomposition}
\shortauthor{L. Heidt and T. Colonius}
\title{Spectral proper orthogonal decomposition of harmonically forced turbulent flows}

\author{Liam F. Heidt\aff{1}
  \corresp{\email{lheidt@caltech.edu}},
  \and Tim Colonius\aff{2}}

\affiliation{\aff{1}Graduate Aerospace Laboratories of the California Institute of Technology, California Institute of Technology, California, 91101, USA
\aff{2}Department of Mechanical and Civil Engineering, California Institute of Technology, California, 91101, USA
}

\begin{document}
\maketitle
\begin{abstract}
Many turbulent flows exhibit time-periodic statistics.  These include turbomachinery flows, flows with external harmonic forcing, and the wakes of bluff bodies.  Many existing techniques for identifying turbulent coherent structures, however, assume the statistics are statistically stationary. In this paper, we leverage cyclostationary analysis, an extension of the statistically stationary framework to processes with periodically varying statistics, to generalize the spectral proper orthogonal decomposition (SPOD) to the cyclostationary case. The resulting properties of the cyclostationary SPOD (CS-SPOD for short) are explored, a theoretical connection between CS-SPOD and the harmonic resolvent analysis is provided, simplifications for the low and high forcing frequency limits are discussed, and an efficient  algorithm to compute CS-SPOD with SPOD-like cost is presented. We illustrate the utility of CS-SPOD using two example problems: a modified complex linearized Ginzburg-Landau model and a high-Reynolds-number turbulent jet. 
\end{abstract}

\section{Introduction} 
Periodic and quasi-periodic forced turbulent flows are ubiquitous in engineering and nature.  Such flows include those in turbomachinery, weather and climate, and flow control with harmonic actuation. In cases where the forcing is slow compared to the turbulence time scales, the statistics may be modeled as quasi-stationary (comprising a series of stationary states). However, in many cases, the forcing is at frequencies commensurate with the turbulence, and the turbulence structure is not only modulated by, but also altered by, the forcing.  In such, a key goal is to identify coherent structures that can be compared and contrasted to their occurrence in similar but unforced flows but that are otherwise mutually uncorrelated.

The most commonly used technique to identify coherent structures in turbulence is proper orthogonal decomposition \citep{lumley1967,lumley1970,aubry1988dynamics,sirovich1989chaotic,aubry1991hidden},
which represents flow data as mutually orthogonal modes whose amplitudes optimally reconstruct the correlation tensor. When applied in its typical space-only form, the modes are not coherent in time, leading many researchers to apply DMD and its variants \citep{rowley2009spectral,schmid2010dynamic,schmid2011applications}.  However, for statistically stationary flows, spectral POD (SPOD) \citep{lumley1967,lumley1970,citriniti2000reconstruction,picard2000pressure,towne2018spectral} leads to an optimal reconstruction of the space-time statistics and results in modes that oscillate at a single frequency. 
A fundamental assumption required in both space-only POD and SPOD is statistical stationarity, meaning that the statistics are time-invariant. This assumption is appropriate for many unforced flows. However, when forced, this fundamental assumption is no longer valid as the flow, and its statistics, are now correlated to the forcing. 
Several works have developed extensions to SPOD to study forced turbulent flows. \citet{franceschini2022identification} studied flows where a high-frequency turbulent component develops on a low-frequency periodic motion. Subsequently, a quasi-steady assumption is made, and conditionally fixed coherent structures at each phase are determined. \citet{glezer1989development} developed an extended POD method for flows with periodic statistics by summing an ensemble of time series. However, since this method is based on POD, it still contains the shortcomings present in POD. \citet{heidt2021analysis} applied SPOD to the residual component of the triply decomposed fields \citep{hussain1970mechanics,hussain1972mechanics} to isolate the impact of the forcing on the turbulence but still required a stationary assumption. Clearly, SPOD and the aforementioned extensions are not sufficient to study forced turbulent flows. This motivates an extension of SPOD to these flows, which is the primary focus of this paper which we achieve by leveraging cyclostationary analysis. \par

Cyclostationary analysis is an extension to statistically stationary analysis to processes with periodic statistics that has been applied in a range of fields \citep{gardner2018statistically}, from economics to physics and mechanics. Initially developed by \citet{gudzenko1959periodic}, \citet{lebedev1959random}, and \citet{gladyshev1963periodically}, it was then extensively studied and popularized in \citet{hurd1969} and \citet{gardner1972representation}. The theory of second-order cyclostationary processes was further developed by \citet{boyles1983cycloergodic} and \citet{gardner1986introduction}, while \citet{brown1987theory} and \citet{gardner1986spectral} furthered the theory of complex-valued processes. Cyclostationary analysis provides a robust statistical theory to study these processes, and tools analogous to those used to study stationary processes (e.g. the mean, cross-correlation, cross-spectral density, etc) have been developed which naturally collapse back to their stationary counterparts when analyzing a stationary process. \par

\citet{kim1996eofs} developed cyclostationary empirical orthogonal-functions (CSEOFs) that essentially extends SPOD to cyclostationary processes for one-dimensional data. \citet{kim1997eofs} modified this technique to include multi-dimensional data by reducing the computational cost through several approximations. However, due to a lack of clarity in the literature regarding the derivation, properties, interpretation, and computation of these techniques, their use has been limited. Furthermore, despite the aforementioned approximations, both formulations are computationally intractable for high-dimensional data. In this paper, we extend SPOD to flows with time-periodic statistics through an extension to the exact form of CSEOFs \citep{kim1996eofs} to include large multi-dimensional data. We hereafter refer to this method as cyclostationary SPOD (CS-SPOD for short). \par

Methods used to model coherent structures are also considered.  Specifically, we consider resolvent analysis (also known as input/output analysis), where one seeks forcing modes that give rise to the most amplified response modes with respect to their energetic gain. When applied to turbulent fluid flows, the nonlinear modal interactions are regarded as forcing terms to the linearized time-averaged turbulent mean \citep{mckeon2010critical}. Resolvent analysis has been used to study a wide range of transitional and turbulent flows \citep{cossu2009optimal,mckeon2010critical, meliga2012sensitivity, sharma2013coherent, oberleithner2014impact,jeun2016input,schmidt2018spectral}, amounts others.  \citet{towne2018spectral} provided a theoretical connection between SPOD and resolvent, showing that resolvent output modes equal SPOD modes when the resolvent forcing modes are mutually uncorrelated. This provides a theoretical basis to use resolvent analysis to develop models of the space-time statistics of a turbulent flow \citep{moarref2013model,towne2020resolvent,amaral2021resolvent} and the development of various methods \citep{morra2019relevance,pickering2021optimal} to help whiten the forcing coefficients, thereby improving these models. Resolvent analysis was extended to flows with a time-periodic mean flow in \citet{padovan2020analysis} and \citet{padovan2022analysis} and is termed {\it harmonic resolvent analysis}.  This leads to a system of frequency-couple equations that provide the ability to study the first-order triadic interactions present in these time-periodic flows. Analogous to the relationship between SPOD and resolvent analysis, in the present paper, we establish a theoretical connection between CS-SPOD and harmonic resolvent analysis. \par

The remainder of the paper is organized as follows. Section \ref{sec:CStheory} introduces and outlines the theory of cyclostationary processes and reviews an algorithm to compute their statistics. In \S \ref{sec:CSSPOD}, CS-SPOD is derived, its properties explored, and an efficient computational algorithm is proposed.  After validating the method in \S \ref{sec:validation}, we demonstrate its utility of CS-SPOD in \S \ref{sec:examples}.  Finally, in \S \ref{sec:harres}, we explore the relationship between CS-SPOD and the harmonic resolvent analysis. Section \ref{sec:conclusion} concludes the manuscript and summarizes the main points. 

\section{Cyclostationary theory} \label{sec:CStheory}
This section provides an overview of the theory of cyclostationary analysis and the tools used to study them, with a focus on fluid dynamics. Comprehensive reviews can be found in \citet{gardner2006cyclostationarity}, \citet{antoni2009cyclostationarity}, and \citet{napolitano2019cyclostationary}. 

A complex-valued scalar process $\fscal(t)$ at time $t$ is cyclostationary in the wide sense if its mean and autocorrelation function are periodic with period $T_0$ \citep{gardner1986introduction}, giving  
\begin{subequations}
\begin{align}
    E\{\fscal(t)\} &= E\{\fscal(t + T_0)\}, \\
    R(t, \tau) &= R(t + T_0, \tau),
\end{align}
\end{subequations}
where $E\{\cdot\}$ is the expectation operator, $R$ is the autocorrelation function, and $\tau$ is a time-delay. Since the mean and autocorrelation are time-periodic, they can be expressed as a Fourier series
\begin{subequations}
\begin{align}
    E\{\fscal(t)\}  &= \si{\perindx} \hat{{\fscal}}_{\perindx \alpha_0} e^{i2\pi (\perindx \alpha_0) t}, \\
    R(t, \tau) &\equiv E\{\fscal(t + \tau/2) \fscal^*(t - \tau/2)\} = \si{\perindx} \hat{R}_{\perindx \alpha_0}(\tau) e^{i2\pi (\perindx \alpha_0)t} \nonumber,
\end{align}
\end{subequations}
where $\perindx \in \mathbb{Z}$ and the Fourier series coefficients are given by 
\begin{subequations}
\begin{align}
    \hat{\fscal}_{\perindx \alpha_0} & \equiv \frac{1}{T_0} \int_{-T_0/2}^{T_0/2} E\{\fscal(t)\} e^{-i2\pi (\perindx \alpha_0) t} dt, \\
    \hat{R}_{\perindx \alpha_0}(\tau) &\equiv \frac{1}{T_0} \int_{-T_0/2}^{T_0/2} R(t, \tau) e^{-i2\pi (\perindx \alpha_0) t} dt, 
\end{align}
\end{subequations}
where $\alpha_0 = 1/T_0$ is the fundamental cycle frequency. The Fourier coefficients $\hat{R}_{\perindx \alpha_0}(\tau)$ are known as the cyclic autocorrelation functions of $\fscal(t)$ at cycle frequency $\perindx \alpha_0$. If a process contains non-zero $\hat{{\fscal}}_{\perindx \alpha_0}$ and/or $\hat{R}_{\perindx \alpha_0}(\tau)$, it is said to exhibit first- and second-order cyclostationarity at cycle frequency $\perindx \alpha_0$, respectively. Wide-sense stationary processes are the special case for which $\hat{R}_{\perindx \alpha_0}(\tau) \ne 0$ for $k = 0$ only. \par

If the process $\fscal(t)$ contains a deterministic periodic component at cycle frequency $\perindx \alpha_0$,  it would exhibit both first-order and second-order (and any higher-order) cyclostationarity at cycle frequency $\perindx \alpha_0$. Thus, a deterministic component results in a pure first-order component and an impure (i.e. made up from components of a lower-order) second-order (or higher) component \citep{antoni2004cyclostationary}. \citet{antoni2004cyclostationary} and \citet{antoni2009cyclostationarity}  showed that in physical systems, it is crucial to analyze the first- and second-order components separately, where the second-order component $\fscal^{\prime\prime}(t)$ is defined as 
\begin{equation}
    \fscal^{\prime\prime}(t) \equiv \fscal(t) - E\{\fscal(t)\},
    \label{eqn:firstseconddecomp}
\end{equation}
such that $\fscal(t) = E\{\fscal(t)\} + \fscal^{\prime\prime}(t)$ and the mean $ E\{\fscal(t)\} =  E\{\fscal(t + T_0)\}$ is $T_0$ periodic. This approach makes physical sense considering that the first-order component is the deterministic tonal component that originates from the forcing, while the second-order component is a stochastic component that represents the underlying turbulence that is modified by the forcing. The sequential approach is analogous to the triple decomposition \citep{hussain1970mechanics,hussain1972mechanics} where the underlying flow is separated into the first-order (phase-averaged) and second-order (turbulent/residual) components. \par

In this manuscript, we assume that all processes analyzed using second-order analysis tools are zero-mean processes (or have had their first-order component removed). Thus, by stating that a process exhibits second-order cyclostationarity at cycle frequency $\perindx \alpha_0$, we mean that the process exhibits {\it pure} second-order cyclostationarity at $\perindx \alpha_0$. 

\subsection{Second-order cyclostationary analysis tools}
In fluid dynamics, we are frequently interested in the correlation between two quantities. Thus, we will now consider the complex-valued process $\qb(\ivar, t)$ at time $t$ and independent variables (or spatial locations) $\ivar$ instead of the scalar process $\fscal(t)$. Two processes are jointly cyclostationary if their cross-correlation function can be expressed as a Fourier series, such that
\begin{align}
    \Rb(\ivar, \ivarp, t, \tau) &\equiv E\{\qb(\ivar, t+\tau/2)\qb^*(\ivarp, t - \tau/2)\} = \si{\perindx} \hat{\Rb}_{\perindx \alpha_0}(\ivar, \ivarp,  \tau) e^{i2\pi(\perindx \alpha_0) t},
                     \label{eqn:cscorr}
\end{align}
where the Fourier series coefficients are given by 
\begin{align}
    \hat{\Rb}_{\perindx \alpha_0}(\ivar, \ivarp, \tau) &\equiv \frac{1}{T_0} \int_{-T_0/2}^{T_0/2} \Rb(\ivar, \ivarp, t, \tau) e^{-i2\pi (\perindx \alpha_0) t} dt,
\end{align}
and are known as the cyclic cross-correlation functions of between $\qb(\ivar)$ and $\qb(\ivarp)$ at cycle frequency $\perindx \alpha_0$ with $(\cdot)^{*}$ being the complex conjugate of $(\cdot)$. If the only non-zero cycle frequency is $\perindx \alpha_0 = 0$, then $\qb(\ivar)$ and $\qb(\ivarp)$ are jointly wide-sense stationary. Similar to the common assumption in stationary analysis, we assume that all processes are separately and jointly cyclostationary. \par

A cyclostationary process can be analyzed in the dual-frequency domain via the cyclic cross-spectral density (CCSD). The CCSD is the generalization of the cross-spectral density (CSD) for cyclostationary processes and  is related to the cyclic cross-correlation function via the cyclic Wiener-Khinchin relation \citep{gardner1989statistical} 
\begin{equation}
    {\Sb}_{\perindx \alpha_0}(\ivar, \ivarp, f) = \int_{-\infty}^\infty \hat{\Rb}_{\perindx \alpha_0}(\ivar, \ivarp, \tau) e^{-i2\pi f\tau} d\tau.
    \label{eqn:cwkt} 
\end{equation}
The CCSD can also be written as
\begin{align}
    & {\Sb}_{\perindx \alpha_0}(\ivar, \ivarp, f) \equiv     \label{eqn:CCSDcorr}\\ & \lim_{\Delta f \rightarrow 0} \lim_{T \rightarrow \infty} \frac{1}{T} \int_{-T/2}^{T/2} \Delta f E\left\{\hat{\qb}_{1/\Delta f} (\ivar, t, f + \frac{1}{2} \perindx \alpha_0) \hat{\qb}^*_{1/\Delta f} (\ivarp, t, f - \frac{1}{2} \perindx \alpha_0) \right\} dt, \nonumber
\end{align}

where $\hat{\qb}_{W} (\ivar, t, f) \equiv \int_{t - \frac{W}{2}}^{t + \frac{W}{2}} \qb(\ivar, t^\prime) e^{-i2\pi f t^\prime} dt^\prime$ is the short-time Fourier transform of $\qb(\ivar, t)$, $f$ is the spectral frequency, and $\perindx \alpha_0$ is the cycle frequency. This shows that the CCSD represents the time-averaged statistical correlation (with zero lag) of two spectral components at frequencies $f + \frac{1}{2} \perindx \alpha_0$ and $f - \frac{1}{2} \perindx \alpha_0$ as the bandwidth approaches zero \citep{napolitano2019cyclostationary}. For $\perindx = 0$, the CCSD naturally reduces to the CSD, i.e. $\Sb_0(\ivar, \ivarp, f)$. Correlation between spectral components in cyclostationary processes is critical in the derivation of CS-SPOD, and for stationary processes, the lack of correlation between spectral components is why SPOD can analyze each frequency independently. \par

The Wigner-Ville (WV) spectrum \citep{martin1982time,martin1985wigner,antoni2007cyclic} shows the spectral information of the process as a function of time (or phase) and, for a cyclostationary process, is given by 
\begin{equation}
    \WVb(\ivar, t, f) = \si{\perindx} \Sb_{\perindx \alpha_0}(\ivar, f)e^{i2\pi (\perindx \alpha_0) t}, 
    \label{eqn:wvdef}
\end{equation}
where $\Sb_{\perindx \alpha_0}(\ivar, f)$ is the cyclic power-spectral density (i.e.  $\Sb_{\perindx \alpha_0}(\ivar,\ivar, f)$). While nonphysical, the WV spectrum may contain negative energy densities due to the negative interaction terms in the WV spectrum \citep{antoni2007cyclic,flandrin1998time}. However, \citet{antoni2007cyclic} showed this could be arbitrarily reduced with increasing sampling time. The CCSD and WV spectrum can be integrated with respect to frequency \citep{gardner1994introduction,randall2001relationship}, which results in the instantaneous variance and the cyclic distribution of the instantaneous variance, respectively
\begin{subequations}
\begin{align}
    \m(\ivar, t) &= E\{\qb(\ivar, t)\qb^*(\ivar, t)\} = \int_{-\infty}^\infty \WVb(\ivar, t, f) df, \\
    \hat{\m}_{\perindx \alpha_0}(\ivar) &= \int_{-\infty}^\infty \Sb_{\perindx \alpha_0}(\ivar, f) df, 
    \label{eqn:cyclicvariance}
\end{align}
\end{subequations}
where ${\m}(\ivar, t)$ is the mean-variance of the process and  $\hat{\m}_{\perindx \alpha_0}(\ivar)$ quantifies the mean-variance contribution from each cycle frequency $\perindx \alpha_0$. \par 

So far, we have assumed that the cycle frequencies are known, but this may not always be the case. To determine the cycle frequencies present in the system, all possible cycle frequencies $\alpha$ are explored by rewriting the CCSD as
\begin{align}
    & {\Sb}(\ivar, \ivarp, \alpha, f) \equiv \nonumber \\
    & \lim_{\Delta f \rightarrow 0} \lim_{T \rightarrow \infty} \frac{1}{T} \int_{-T/2}^{T/2} \Delta f E\left\{\qb_{1/\Delta f} (\ivar, t, f + \frac{\alpha}{2}) \qb^*_{1/\Delta f} (\ivarp, t, f - \frac{\alpha}{2}) \right\} dt.
    \label{eqn:CCSDcorralpha}
\end{align}

A process exhibits cyclostationarity at cycle frequency $\alpha$ when ${\Sb}(\ivar, \ivarp, \alpha, f) \ne 0$.  The range of possible cycle frequencies is $\alpha = [-0.5/\Delta t\ \ 0.5/\Delta t]$, which must be searched over with a resolution $\Delta \alpha = 1/T$ \citep{gardner1986measurement} to ensure all cycle frequencies present are captured. For cyclostationary processes, because the cross-correlation function is periodic, the spectral correlation becomes discrete in $\alpha$ such that
\begin{equation}
    \Sb(\ivar, \ivarp, \alpha, f) = \si{\perindx} \Sb_{\perindx \alpha_0}(\ivar, \ivarp, f) \delta (\alpha - \perindx \alpha_0).
\end{equation}
The cyclic distribution of the instantaneous variance is rewritten as 
\begin{equation}
\hat{\m}(\ivar, \alpha) = \int_{-\infty}^\infty \Sb(\ivar, \alpha, f) df, 
\end{equation}
which similarly becomes discrete for a cyclostationary process. 

We must clarify one point of terminology. Considering stationary processes are a subset of cyclostationary processes, all stationary processes are also cyclostationary. We use the most restrictive description, i.e. stationary processes are referred to as stationary and not cyclostationary. By stating that a process exhibits cyclostationarity, we imply that at least one cycle frequency $\perindx \alpha_0, \perindx \ne 0$ exists. 
\subsection{Cycloergodicity} \label{sec:cye}
In fluid dynamics, it is laborious to require multiple realizations of a single process, and we often invoke ergodicity in stationary processes to equate the ensemble average with a long-time average of a single realization.  We can similarly leverage the concept of cycloergodicity as described in \citet{boyles1983cycloergodic}, allowing us to replace the expectation operator with a suitable time average, specifically, the cycle-averaging operator \citep{braun1975extraction} 
\begin{equation}
        \widetilde{\qb}(\ivar, t) = E\{\qb(\ivar, t)\} = \lim_{P \rightarrow \infty} \frac{1}{P} \sum_{p = 0}^P \qb(\ivar, t + pT_0), 
        \label{eqn:phaseaverage}
\end{equation}
where $\widetilde{\qb}(\ivar, t)$ is the mean. The cycle-averaging operator is used when the data is phase-locked to the forcing (i.e. sampled at an integer number of samples per cycle) and is identical to the phase-average used in the triple decomposition \citep{hussain1970mechanics,reynolds1972mechanics}. As the cycle average operator is periodic, it can be expressed as a Poisson sum
\begin{equation}
 \widetilde{\qb}(\ivar, t) = E\{\qb(\ivar, t)\} = \si{\perindx} e^{i2\pi (\perindx \alpha_0) t} \lim_{s\rightarrow \infty} \frac{1}{s}\int_{-s/2}^{s/2} \qb(\ivar, t)e^{-i2\pi (\perindx \alpha_0) t} dt.
\label{eqn:poissonsum}
\end{equation}
This definition is employed for non-phase-locked data or to filter out first-order components which are assumed to be statistical noise \citep{franceschini2022identification,sonnenberger2000fourier} and is identical to the harmonic-averaging procedure used by \citet{mezic2013analysis} and \citet{arbabi2017study} when restricted to a temporally periodic average.   

\subsection{Computing the CCSD} \label{sec:CompCS}
There are practical considerations and nuances to computing the CCSD from discrete data that we discuss in this section. Let the vector $\qd_k \in \mathbb{R}^N$ represent a flow snapshot, i.e. the instantaneous state of the process $\qb(\x, t)$ at time $t_k$ on a set of points in a spatial domain $\Omega$. The length of the vector $N$ is equal to the number of spatial points multiplied by the number of state variables. We assume that this data is available for $M$ equispaced snapshots, with $t_{k+1} = t_k + \Delta t$. In addition, we assume that this data is phase-locked, meaning that there are an integer number of time steps in the fundamental period, $T_0$, and define $N_\theta=T_0/\Delta t$ \footnote{This restriction simplifies and reduces the computational expense of the calculations but can in principle be relaxed by using the Poisson sum time-average as in \eqref{eqn:poissonsum} and the non-computationally-efficient form of CS-SPOD shown in algorithm \ref{algo:ccfdbase}. Alternatively, non-phased-locked data can be temporally interpolated to be phase-locked.}. Adopting similar notation to \citet{towne2018spectral}, we estimate the CCSD tensor $\Sb(\x, \x^\prime, \alpha, f)$, which represents the spectral correlation between $\qb(\x, t)$ and $\qb(\x^\prime, t)$ at cycle frequency $\alpha$ and spectral frequency $f$. For a cyclostationary process, $\Sb(\x, \x^\prime, \alpha, f)$ is non-zero for $\alpha = \perindx \alpha_0$ only, and therefore is written as $\Sb_{\perindx \alpha_0}(\x, \x^\prime, f)$ or equivalently $\Sb_{\perindx/T_0}(\x, \x^\prime, f)$. The space-time data can now be represented as the data matrix $\Qd$ and time vector $\Td$
\refstepcounter{equation}
$$
  \Qd = [\qd_1, \qd_2, \cdots, \qd_M] \in \mathbb{R}^{N\times M}, \quad
  \Td = [t_1, t_2, \cdots, t_M] \in \mathbb{R}^{M}.  \eqno{(\theequation, \refstepcounter{equation} \theequation)} 
$$

Although we have a formula for the CCSD as seen in (\ref{eqn:CCSDcorr} and \ref{eqn:CCSDcorralpha}), this does not result in a consistent estimator of the CCSD, as the variance of the \textit{estimate} of the CCSD does not tend to zero as the amount of available data becomes large \citep{jenkins1968spectral,antoni2007cyclic,napolitano2019cyclostationary}. Instead, this results in an estimate where the variance in the estimate is equal to the squared value of the estimate itself. A consistent estimate of the CCSD can be obtained by employing an appropriate averaging technique. The most common technique is the time-averaging Welch method \citep{welch1967use} due to its high computational efficiency. The Welch method averages a number of CCSDs to obtain a consistent \textit{estimate} of the CCSD.  From \eqref{eqn:CCSDcorralpha}, we see that to compute the CCSD the Welch procedure is performed on two frequency-shifted versions of the data, given by
\begin{equation}
    \Qd_{\pm \alpha/2}  = \Qd e^{-i 2\pi (\pm \alpha/2) \Td} = [{\qd}_{1, \pm \alpha/2},{\qd}_{2, \pm \alpha/2}, \cdots, {\qd}_{M, \pm \alpha/2} ],
\end{equation}
where ${\qd}_{k, \pm \alpha/2}$ are the $\pm \frac{1}{2}\alpha$ frequency-shifted data matrices corresponding to the $k^{th}$ snapshot, i.e. ${\qd}_{k, \pm \alpha/2} = {\qd}_{k} e^{-i 2\pi (\pm \alpha/2) t_k}$. Next, we split the two frequency-shifted data matrices into a number of, possibly overlapping, blocks. Each block is written as
\begin{equation}
    \Qd^{(n)}_{\pm \alpha/2} = [\qd_{1,\pm \alpha/2}^{(n)},\qd_{2,\pm \alpha/2}^{(n)}, \cdots, \qd_{N_f,\pm \alpha/2}^{(n)} ] \in \mathbb{C}^{N\times N_f}, 
\end{equation}
where $N_f$ is the number of snapshots in each block and the $k^{th}$ entry of the $n^{th}$ block is $\qd^{(n)}_{k,\pm \alpha/2} = \qd_{k+(n-1)(N_f - N_0), \pm \alpha/2}$.  The total number of blocks, $N_b$, is given by $N_b = \lfloor\frac{N - N_0}{N_f - N_0}\rfloor$, where $\lfloor \cdot \rfloor$ represents the floor operator and $N_0$ is the number of snapshots that each block overlaps. The cycloergodicity hypothesis states that each of these blocks is considered to be a single realization in an ensemble of realizations of this cyclostationary flow. Subsequently, the DFT of each block for both frequency-shifted matrices is computed using a window $w$, giving  
\begin{equation}
    \hat{\Qd}^{(n)}_{\pm \alpha/2} = [\hat{\qd}_{ 1, \pm \alpha/2}^{(n)},\hat{\qd}_{2, \pm \alpha/2}^{(n)}, \cdots, \hat{\qd}_{N_f, \pm \alpha/2}^{(n)} ],
\end{equation}
where
\begin{equation}
\hat{\qd}_{k, \pm \alpha/2}^{(n)} = \frac{1}{\sqrt{N_f}} \sum_{j = 1}^{N_f} {w_j}{\qd}_{j, \pm \alpha/2}^{(n)} e^{-i2\pi (k-1)[(j - 1)/N_f]},
\end{equation}
for $k = 1, \cdots, N_f$ and $n = 1, \cdots, N_b$ where $\hat{\qd}^{(n)}_{k, \pm \alpha/2}$ is the $k^{th}$ Fourier component of the $n^{th}$ block of the $\pm \alpha/2$ frequency-shifted data matrix, i.e. $f_{k, \pm \alpha_0/2}$. The nodal values $w_j$ of a window function are utilized to mitigate spectral and cyclic leakage arising from the non-periodicity of the data within each block. Due to the $\pm \alpha/2$ frequency-shifting applied, the $k^{th}$ discrete frequencies of the $\pm \alpha/2$ frequency-shifted data matrices represent a frequency of   
\begin{equation}
    f_{k, \pm \alpha/2} = f_{k}  \pm \alpha/2 =    \begin{dcases}
        \frac{k-1}{N_f\Delta t}  & \text{for } k \leq N_f/2, \\
        \frac{k-1 - N_f}{N_f \Delta t} & \text{for } k > N_f/2.
    \end{dcases} \pm \frac{\alpha}{2}
\end{equation}
This shows that the frequency components $f_k + \alpha/2$ and $f_k - \alpha/2$, as required by \eqref{eqn:CCSDcorralpha}, have the same index $k$ in the shifted frequency vectors $f_{k, \pm \alpha/2}$, respectively. The CCSD tensor $\Sb(\x, \x^\prime, \alpha, f)$ is then estimated at cycle frequency $\alpha$ and spectral frequency $f_k$ by
\begin{equation}
    \Sd_{f_k, \alpha}  = \frac{\Delta t}{s N_b} \sum_{n =1}^{N_b} \hat{\qd}^{(n)}_{k, \alpha/2} (\hat{\qd}^{(n)}_{k, -\alpha/2})^*,
\end{equation}
where $s = \sum_{j = 1}^{N_f}w_j^2$ is the normalization constant that accounts for the difference in power between the windowed and non-windowed signal. This is written compactly by arranging the Fourier coefficients at the same index $k$ into new frequency-data matrices 
\begin{align}
    \hat{\Qd}_{f_k, \pm \alpha/2} &= \sqrt{\kappa} [\hat{\qd}_{k, \pm \alpha/2}^{(1)},  \hat{\qd}_{k, \pm \alpha/2}^{(2)}, \cdots, \hat{\qd}_{k, \pm \alpha/2}^{(N_b-1)}, \hat{\qd}_{k, \pm \alpha/2}^{(N_b)}] \in \mathbb{C}^{N\times N_b}, 
\end{align}
where $\kappa = \frac{\Delta t}{s N_b}$. $\Sd_{f_k, \alpha}$ is then estimated by
\begin{equation}
     \Sd_{f_k, \alpha}  = \hat{\Qd}_{f_k, \alpha/2} (\hat{\Qd}_{f_k, -\alpha/2})^*.
\end{equation}
This estimate converges, i.e. the bias and variance become zero, as $N_b$ and $N_f$ are increased together \citep{welch1967use,bendat2011random,antoni2007cyclic}. The algorithm to compute the CCSD from data snapshots is outlined in algorithm \ref{algo:csdalgo}, from which all other second-order cyclostationary analysis tools can be computed. For efficient memory management, variables assigned with `$\leftarrow$' can be deleted after each iteration in their respective loop. Similar to the Welch estimate of the CSD, the estimate of the CCSD suffers from the standard bias-variance trade-off, and caution should be taken to ensure sufficiently converged statistics. In the CCSD, a phenomenon similar to spectral leakage is present and is called cyclic leakage \citep{gardner1986measurement} that results in erroneous cycle frequencies. Using  $67\%$ overlap when using a Hanning or Hamming window results in excellent cyclic leakage minimization and variance reduction \citep{antoni2007cyclic}. To reduce the variance sufficiently, $T\Delta f >> 1$ is required \citep{antoni2009cyclostationarity}.
\begin{algorithm}
\setstretch{1.25}
\caption{Algorithm to compute CCSD using frequency-shifted data matrices.}\label{alg:cap}
\begin{algorithmic}[1] 
\For{Each data block, $n = 1, 2, \cdots, N_b$} 
\Statex{\hspace{4.5mm} $\triangleright$ Compute the frequency-shifted block data matrices}
\State \scalebox{0.95}{${\Qd}^{(n)}_{\pm \alpha/2} \gets [\qd_{1+(n-1)(N_f - N_0),\pm \alpha/2}, \qd_{2+(n-1)(N_f - N_0), \pm \alpha/2}, \cdots, \qd_{N_f+(n-1)(N_f - N_0),\pm \alpha/2}]$} 
\Statex{\hspace{4.5mm} $\triangleright$ Using a (windowed) fast Fourier transform, calculate and store the row-wise}
\Statex{\hspace{4.5mm} DFT for each frequency-shifted block data matrix}
\State $\hat{\Qd}^{(n)}_{\pm \alpha/2} = \text{FFT}({\Qd}^{(n)}_{\pm \alpha/2}) = [\hat{\qd}_{1, \pm \alpha/2}^{(n)},\hat{\qd}_{2, \pm \alpha/2}^{(n)}, \cdots, \hat{\qd}_{N_f, \pm \alpha/2}^{(n)} ]$
\Statex{\hspace{4.5mm} $\triangleright$ The column $\hat{\qd}_{k, \pm \alpha/2}^{(n)}$ contains the $n^{th}$ realization of the Fourier mode}
\Statex{\hspace{4.5mm} at the $k^{th}$ discrete frequency $f_{k, \pm \alpha/2}$ }
\EndFor
\For{Each frequency $k = 1, 2, \cdots, N_f$ (or some subset of interest)} 
\Statex{\hspace{4.5mm} $\triangleright$ Assemble the matrices of Fourier realizations from the $k^{th}$ column of each $\hat{\Qd}^{(n)}_{\pm \alpha/2}$}
\State $\hat{\Qd}_{f_k, \pm \alpha/2} \gets \sqrt{\kappa} [\hat{\qd}_{k, \pm \alpha/2}^{(1)}, \hat{\qd}_{k, \pm \alpha/2}^{(2)}, \cdots, \hat{\qd}_{k, \pm \alpha/2}^{(N_b-1)}, \hat{\qd}_{k, \pm \alpha/2}^{(N_b)}]$ 
\Statex{\hspace{4.5mm} $\triangleright$ Compute the CCSD at spectral frequency $f_k$ and cycle frequency $\alpha$}
\State  $ \Sd_{f_k, \alpha}  = \hat{\Qd}_{f_k, \alpha/2} (\hat{\Qd}_{f_k, -\alpha/2})^*.$
\EndFor
\end{algorithmic}
\label{algo:csdalgo}
\end{algorithm}
\section{Cyclostationary spectral proper orthogonal decomposition} \label{sec:CSSPOD}

\subsection{Derivation}
The objective of CS-SPOD is to find deterministic functions that best approximate, on average, a zero-mean stochastic process.  For clarity, we derive CS-SPOD using an approach and notation analogous to the SPOD derivation presented in \citet{towne2018spectral} and refer the reader to \citet{berkooz1993proper}, \citet{towne2018spectral}, and \citet{schmidt2020guide} for detailed discussions on POD and SPOD. Like SPOD, we seek deterministic modes that depend on both space and time such that we can optimally decompose the space-time statistics of the flow. Thus, we assume that each realization of the stochastic process belongs to a Hilbert space with an inner product 
\begin{equation}
    \langle \qb_1, \qb_2 \rangle_{x,t} = \int_{-\infty}^\infty \int_{\Omega} \qb_2^*(\x, t) \Wb(\x) \qb_1(\x, t) \dx \dt,
    \label{eqn:spacetimenorm}
\end{equation}
where $\qb_1(\x, t),\ \qb_2(\x, t)$ are two realizations of the flow, $\Wb(\x)$ is a positive-definite weighting tensor, and $\Omega$ denotes the spatial domain of interest. We then seek to maximize
\begin{equation}
    \lambda = \frac{E\{|\langle \qb(\x, t), \phib(\x, t)\rangle_{x,t}|^2\}}{\langle \phib(\x, t), \phib(\x, t)\rangle_{x,t}}, 
\end{equation}
which leads to 
\begin{align}
\int_{-\infty}^\infty \int_{\Omega} \Rb(\x, \x^\prime, t, t^\prime) \Wb(\x^\prime) \phib(\x^\prime, t^\prime)  \dx^\prime \dt^\prime  &= \lambda \phib(\x, t), 
\label{eqn:KLEquation}
\end{align}
where $\Rb(\x, \x^\prime, t, t^\prime) \equiv E\{\qb(\x, t)\qb^*(\x^\prime, t^\prime)\}$ is the two-point space-time correlation tensor. Until this stage, no assumptions about the flow has been made and is therefore identical to the derivation of SPOD \citep{lumley1967,lumley1970,towne2018spectral}. \par

Since cyclostationary flows persist indefinitely, they have infinite energy in the space-time norm, as shown in \eqref{eqn:spacetimenorm}. Consequently, the eigenmodes of \eqref{eqn:KLEquation} do not possess any of the useful quantities relied upon in POD or SPOD. To solve this, a new eigenvalue decomposition is obtained in the spectral domain from which modes with the desired properties are determined. We employ a solution ansatz of 
\begin{equation}
    \phib(\x, t) = \sum_{m \in \mathcal{A}_m} \psib(\x, \fci + m\alpha_0) e^{i2\pi (\fci + m\alpha_0)t}.
    \label{eqn:ansatz1}
\end{equation}
The set of frequencies present in the solution ansatz $\phib(\x, t)$, is called the $\fci$ set of solution frequencies $\Fset_{\fci} = \{\ \cdots,\ \fci- 2\alpha_0,\ \fci- \alpha_0,\ \fci,\ \fci+ \alpha_0,\ \fci + 2\alpha_0,\ \cdots\ \}$. \par

In appendix \ref{appen:csspodderivation}, we then use theory from \S \ref{sec:CStheory} to derive the infinite-dimensional CS-SPOD eigenvalue problem, written compactly as  
\begin{equation}
     \int_{\Omega} \MCSb(\x, \x^\prime, \fci) \MCWd(\x^\prime) \Psib(\x^\prime, \fci) \dx^\prime = \lambda \Psib(\x, \fci),
      \label{eqn:finalcsspod}
\end{equation}
where

\refstepcounter{equation}
\begin{align}
&\MCSb(\x, \x^\prime, \fci) =  \label{eqn:infiniteeig} \tag{\theequation{a}} \\
&\begin{bmatrix}
\ddots & \ddots & \ddots & \ddots & \ddots \\
\ddots & \Sb_0(\x, \x^\prime, \fci - \alpha_0) & \Sb_{-\alpha_0}(\x, \x^\prime, \fci - \frac{\alpha_0}{2} ) & \Sb_{-2\alpha_0}(\x, \x^\prime,\fci) & \ddots \\
\ddots & \Sb_{\alpha_0}(\x, \x^\prime,\fci- \frac{\alpha_0}{2}) & \Sb_0(\x, \x^\prime,\fci) & \Sb_{-\alpha_0}(\x, \x^\prime,\fci+ \frac{\alpha_0}{2} )  & \ddots \\
\ddots & \Sb_{2\alpha_0}(\x, \x^\prime,\fci) & \Sb_{\alpha_0}(\x, \x^\prime,\fci+ \frac{\alpha_0}{2} )  & \Sb_0(\x, \x^\prime,\fci+ \alpha)  & \ddots \\
\ddots & \ddots & \ddots & \ddots & \ddots 
\end{bmatrix}, \nonumber
\end{align}
\hspace{-5mm}\noindent\begin{minipage}{.6\linewidth}
\begin{equation}
\MCWb(\x) = 
\begin{bmatrix}
\ddots &  &  &  & \\ 
 & \hspace{-2mm}\Wb(\x) &  &  & \\ 
 &  & \hspace{-2mm}\Wb(\x) &  & \\ 
 &  &  & \hspace{-2mm}\Wb(\x) & \\ 
 &  &  &  & \hspace{-2mm}\ddots
\end{bmatrix},
\label{eqn:weightmatrixblock} \tag{\theequation{b}}
\end{equation} 
\end{minipage}
\begin{minipage}{.42\linewidth}
\begin{equation}\Psib(\x, \fci) = 
\begin{bmatrix}
\vdots\\ 
\psib(\x,\fci- \alpha_{0})\\ 
\psib(\x,\fci)\\ 
\psib(\x,\fci+ \alpha_{0})\\ 
\vdots
\end{bmatrix}. \tag{\theequation{c}}
\end{equation}
\end{minipage}
$\MCSb(\x, \x^\prime, \fci)$ is the CS-SPOD decomposition tensor, $\MCWb(\x)$ is the concatenated weight tensor,  and $\Psib(\x, \fci)$ are the CS-SPOD eigenvectors. The CS-SPOD eigenvectors $\phib(\x, t)$ have Fourier series coefficients, at each $f \in \Fset_{\fci}$, of $\psib(\x, f)$. \par

This coupling of frequencies in CS-SPOD occurs because frequency components separated by $n\alpha_0$ are correlated to each other, as shown in \eqref{eqn:CCSDcorr}.  In contrast, stationary processes do not exhibit correlation between frequencies, and thus each frequency can be solved independently via SPOD. Due to this coupling, CS-SPOD performed at $\fci$ and $\fci + \alpha_0$ solve the same problem, i.e. giving $\Fset_{\fci} = \Fset_{\fci+z \alpha_0}$, where $z \in \mathbb{Z}$,  meaning that CS-SPOD only contains unique solutions for the frequency sets corresponding to $\fci \in \Fci$, where $\Fci  = (-\alpha_0/2,\ \alpha_0/2]$. \par

In practice, the infinite-dimensional problem is not solved, and we restrict our solution frequencies by limiting $\mathcal{A}_m$ to $a_1$ harmonics, giving
$\mathcal{A}_m = \{-a_{1}, -a_1+1, \cdots, 0, \cdots, a_1-1, a_1\}$ and $\Fset_{\fci} = \{-a_{1}\alpha_0 + \fci,\ (-a_1+1)\alpha_0 + \fci,\ \cdots,  \fci,\  \cdots, (a_1-1)\alpha_0+ \fci,\ a_1\alpha_0+ \fci\}$. In addition, the flow may only exhibit cyclostationarity at $a_2$ harmonics of the fundamental cycle frequency giving $\mathcal{A}_n = \{ -a_2, -a_2 + 1, \cdots,  0,  \cdots, a_2 - 1, a_2\}$. We employ identical notation to restrict the harmonics used to compute various second-order tools, such as the Wigner-Ville spectrum. These limits result in $2a_1+1$ coupled equations, resulting in a $2a_1+1 \times 2a_1+1$ block eigensystem that is $2a_2+1$ banded-block-diagonal. In practice, $a_1$ should be chosen such that $\Fset_{\fci}$ encompasses all frequencies of interest, $a_2$ should be chosen to encompass all the cycle frequencies present in the flow, and $a_2 < a_1$. An example for $a_1 = 2, a_2 = 1$ is (for compactness, we have dropped the explicit dependence on $\x$ in this equation)
\begin{align}
&\MCSb(\fci)  = \label{eqn:finiteeprob}\\
&\resizebox{1\linewidth}{!}{$\begin{bmatrix}
\Sb_{0}(\fci - 2\alpha_0) & \Sb_{-\alpha_0}(\fci - \frac{3}{2}\alpha_0) &  0 & 0 & 0\\ 
\Sb_{\alpha_0}(\fci - \frac{3}{2} \alpha_0) & \Sb_{0}(\fci - \alpha_0) &  \Sb_{-\alpha_0}(\fci - \frac{1}{2} \alpha_0) &  0  & 0 \\ 
 0 & \Sb_{\alpha_0}(\fci - \frac{1}{2} \alpha_0)  & \Sb_{0}(\fci ) & \Sb_{-\alpha_0}( \fci + \frac{1}{2} \alpha_0) & 0 \\ 
 0 & 0 & \Sb_{\alpha_0}(\fci + \frac{1}{2} \alpha_0) & \Sb_{0}(\fci + \frac{1}{2} \alpha_0) & \Sb_{-\alpha_0}(\fci + \frac{3}{2} \alpha_0)\\ 
 0 & 0 & 0 & \Sb_{\alpha_0}(\fci + \frac{3}{2}\alpha_0) & \Sb_{0}(\fci + 2\alpha_{0})
\end{bmatrix}.$} \nonumber
\end{align}
In the limiting case that $a_2 = 0$, we obtain a block-diagonal CS-SPOD decomposition matrix where each diagonal block is the standard SPOD eigenvalue problem. 

\subsection{CS-SPOD properties}
Since $\MCSb(\x, \x^\prime, \fci)$ is compact and finite, Hilbert–Schmidt theory guarantees a number of properties analogous to those for POD and SPOD \citep{lumley1967,lumley1970,towne2018spectral}.  There are a countably infinite set of eigenfunctions $\Psib_{j}(\x, \fci)$ at each unique frequency set $\Fset_{\fci}$ that are orthogonal to all other modes at the same frequency set $\Fset_{\fci}$ in the spatial inner norm $\langle \qb_1, \qb_2 \rangle_x = \int_{\Omega} \qb_2^*(\x, t) \Wb(\x) \qb_1(\x, t) \dx$, i.e. $\langle \Psib_{j}(\x, \fci), \Psib_{k}(\x, \fci) \rangle_x = \delta_{j,k}$. The following concatenated vector of each flow realization at the solution frequencies is optimally expanded as
\refstepcounter{equation}
$$ \hat{\Qb}(\x, \fci) =  \begin{bmatrix}
\vdots\\ 
\hat{\qb}(\x, \fci - \alpha_0)\\ 
\hat{\qb}(\x, \fci)\\ 
\hat{\qb}(\x, \fci + \alpha_0)\\ 
\vdots
\end{bmatrix}, \quad \hspace{5mm}
  \hat{\Qb}(\x, \fci) = \sum_{j = 1}^\infty a_j(\fci) \Psib_{j}(\x, \fci),  \eqno{(\theequation{\mathit{a},\mathit{b}})}  $$
where $\hat{\qb}(\x, f)$ is the temporal Fourier decomposition of each flow realization $\qb(\x, t)$ at frequency $f$ and $a_j(\fci) = \langle \hat{\Qb}(\x, \fci), \Psib_{j}(\x, \fci)\rangle_{x}$ are the expansion coefficients, which are uncorrelated i.e. $E\{a_j(\fci) a^*_k(\fci)\} = \lambda_j(\fci)\delta_{j, k}$. \par

$\MCSb(\x, \x^\prime, \fci)$ is positive semi-definite meaning that $\MCSb(\x, \x^\prime, \fci)$ has the following unique diagonal representation
\begin{equation}
    \MCSb(\x, \x^\prime, \fci) = \sum_{j = 1}^\infty \lambda_j(\fci) \Psib_{j}(\x, \fci) \Psib^*_{j}(\x^\prime, \fci), 
    \label{eqn:MSCDR}
\end{equation}
in which the CS-SPOD modes are its principal components. This shows that CS-SPOD determines the modes that optimally reconstruct the second-order statistics, one frequency set $\Fset_{\fci}$ at a time. \par

CS-SPOD modes are optimal in terms of their total energy reconstruction of $\MCSb(\x, \x^\prime, \fci)$ \textit{only}. Thus, although each of the CCSDs present in $\MCSb(\x, \x^\prime, \fci)$ have a diagonal representation, the individual components of $\Psib_{j}(\x, \fci)$ are, in general, \textit{not orthogonal} in the space norm, i.e. $\langle \psib_j(\x, f), \psib_k(\x, f) \rangle_x \ne \delta_{j,k}$. One exception is for stationary processes where the correlation between different frequency components is zero, resulting in a block-diagonal matrix where $\Psib_{j}(\x, \fci)$ contains just a single non-zero $\psib_j(\x, \fci)$ component, with $\langle \psib_j(\x, \fci), \psib_k(\x, \fci) \rangle_x = \delta_{j,k}$. \par

Transforming the eigenvectors $\Psib_{j}(\x, \fci)$ back into the time domain, noting the ansatz defined in \eqref{eqn:ansatz1}, gives $\phib_{\fci, j}(\x, t) = \sum_{m \in \mathcal{A}_m} \psib_j(\x, \fci + m\alpha_0) e^{i2\pi (\fci + m\alpha_0) t}$, which are orthogonal in the space-time inner product integrated over a complete period. Thus, every mode occurring at each frequency set $\Fset_{\fci}$ can be viewed as a unique space-time mode. \par

The two-point space-time correlation tensor can be written as 
\begin{equation}
    \Rb(\x, \x^\prime, t, t^\prime) = \int_{-\alpha_0/2}^{\alpha_0/2} \sum_{j = 1}^\infty \lambda_j(\fci) \phib_{\fci, j}(\x, t) \phib^*_{\fci, j}(\x^\prime, t^\prime) d\fci.
\end{equation}
Substituting in the frequency expansion of $\phib_{\fci, j}(\x, t)$ and applying $t^\prime = t - \tau$ gives
\begin{align}
    \Rb(\x, \x^\prime, t, \tau) = \int_{-\alpha_0/2}^{\alpha_0/2} &\sum_{j = 1}^\infty \lambda_j(\fci) \sum_{m\in \mathcal{A}_m} \sum_{m^\prime\in \mathcal{A}_m} \\ \nonumber
    &\psib_{j}(\x, \fci + m\alpha_0) \psib^*_{j}(\x^\prime, \fci + m^\prime\alpha_0) e^{i2\pi (m - m^\prime) \alpha_0 t} e^{i2\pi (\fci + m^\prime \alpha_0) \tau} d\fci,
\end{align}
resulting in a reconstruction that is time-periodic due to $e^{i2\pi (m - m^\prime) \alpha_0 t}$, which is why the ansatz defined by \eqref{eqn:ansatz1} was chosen. \par

In summary, for cyclostationary flows, CS-SPOD leads to modes that oscillate at a set of frequencies ($\Fset_{\fci}$) and optimally represent the second-order space-time flow statistics. 

\subsection{Computing CS-SPOD modes in practice} \label{sec:csspodcomp}
We now detail how to compute CS-SPOD modes from data along with a technique to reduce the cost and memory requirements to levels similar to those of SPOD.  Since the dimension of the CCSD is $N\times N$, the overall eigensystem $\MCSd_{\fcik}$ (which is the discrete approximation of $\MCSb(\x, \x^\prime, \fci)$) becomes $(2a_1+1)N \times (2a_1+1)N$ in size. For common fluid dynamics problems, this can become a dense matrix $\textit{O}(10^6-10^9) \times \textit{O}(10^6-10^9)$ in size, which is computationally intractable to store in memory, let alone compute its eigendecomposition. This is also the dimension of the inversion required in the CSEOF methods by \citet{kim1996eofs} and \citet{kim1997eofs}. Thus, we derive a method-of-snapshots approach similar to the technique employed in POD \citep{sirovich1987turbulence} and SPOD \citep{citriniti2000reconstruction,towne2018spectral} that reduces the size of the eigenvalue problem from $(2a_1+1)N \times (2a_1+1)N$ to  $ (2a_1+1)N_b \times  (2a_1+1)N_b$. Since $N_b << N$, the method-of-snapshots technique makes the eigenvalue problem computationally tractable. 

\par
To determine CS-SPOD with a finite amount of discrete data, we substitute in the Welch computational procedure for the CCSD into each term of the frequency-limited version of \eqref{eqn:infiniteeig}. We numerically evaluate this as
\refstepcounter{equation}
$$ \MCSd_{\fcik} = \widetilde{\QCd}_{\fcik} \widetilde{\QCd}^*_{\fcik}, 
    \label{eqn:shortform2} \quad \hspace{5mm}
  \widetilde{\QCd}_{\fcik} =   \begin{bmatrix}
\hat{\Qd}_{\fcik, - a_1\alpha_0}\\ 
\vdots \\
\hat{\Qd}_{\fcik}\\ 
\vdots \\
\hat{\Qd}_{\fcik, a_1\alpha_0}
\end{bmatrix},  \eqno{(\theequation{\mathit{a},\mathit{b}})} \label{eqn:CompactCSSPOD} $$
where 
\begin{align}
    \hat{\Qd}_{\fcik, m\alpha_0} &= \sqrt{\kappa} [\hat{\qd}_{k, m\alpha_0}^{(1)},  \hat{\qd}_{k, m\alpha_0}^{(2)}, \cdots, \hat{\qd}_{k, m\alpha_0}^{(N_b-1)}, \hat{\qd}_{k, m\alpha_0}^{(N_b)}] \in \mathbb{C}^{N\times N_b}.
\end{align}
$\widetilde{\QCd}_{\fcik}$ is called the concatenated frequency-data matrix at the discrete $\Fset_{\fcik}$ set of solution frequencies and $\hat{\qd}_{k, m\alpha_0}^{(n)}$ is the $k^{th}$ DFT component of the $n^{th}$ block of the $m\alpha_0$ frequency-shifted data matrix.  As stated previously, the solution frequency sets are only unique for $\fci \in \Fci$, thus the corresponding DFT frequencies are
\begin{equation}
    \fcik =    \begin{dcases}
        \frac{k-1}{N_f\Delta t}  & \text{for } k \leq \lfloor\frac{\alpha_0 N_f \Delta t}{2}\rfloor + 1, \\
        \frac{k-1 - N_f}{N_f \Delta t} & \text{for } N_f - \lceil\frac{\alpha_0 N_f \Delta t}{2}\rceil + 1 < k \le N_f, \\
    \end{dcases}
\end{equation}
 \par
which forms the elements $\fcik \in \Fcik$. Expanding \eqref{eqn:CompactCSSPOD} gives 
\begin{align}
\MCSd_{\fcik}  &=  \label{eqn:finiteeprobpart} \\ \nonumber &\begin{bmatrix} 
\hat{\Qd}_{\fcik, - a_1\alpha_0}\hat{\Qd}_{\fcik, - a_1\alpha_0}^* & \cdots & \hat{\Qd}_{\fcik, - a_1\alpha_0}\hat{\Qd}_{\fcik}^* & \cdots & \hat{\Qd}_{\fcik, - a_1\alpha_0}\hat{\Qd}_{\fcik, a_1\alpha_0}^* \\
\vdots & \vdots & \vdots & \vdots & \vdots \\
\hat{\Qd}_{\fcik}\hat{\Qd}_{\fcik, - a_1\alpha_0}^* & \cdots & \hat{\Qd}_{\fcik}\hat{\Qd}_{\fcik}^* & \cdots & \hat{\Qd}_{\fcik}\hat{\Qd}_{\fcik, a_1\alpha_0}^* \\
\vdots & \vdots & \vdots & \vdots & \vdots \\
\hat{\Qd}_{\fcik, a_1\alpha_0}\hat{\Qd}_{\fcik, - a_1\alpha_0}^* & \cdots & \hat{\Qd}_{\fcik, a_1\alpha_0}\hat{\Qd}_{\fcik}^* & \cdots & \hat{\Qd}_{\fcik, a_1\alpha_0}\hat{\Qd}_{\fcik, a_1\alpha_0}^* 
\end{bmatrix}.
\end{align}

This expression shows that $\MCSd_{\fcik}$ contains off-diagonal terms that represent spectral correlations that are not present in the process (i.e. not present in $\mathcal{A}_n$).  However, as $N_b$ and $N$ are increased together, this system converges and becomes a consistent estimate of  \eqref{eqn:infiniteeig}. Thus, all terms that represent spectral correlations not present in $\mathcal{A}_n$ converge to zero. Furthermore, the estimate is numerically positive semi-definite resulting in CS-SPOD modes that will inherit the desired properties. We note the restriction of cycle frequencies to $\mathcal{A}_n$ is not required for the numerical computation, and only $a_1$ is chosen. \par

 Equation \eqref{eqn:shortform2} shows that the final eigenvalue problem can be compactly written as 
\begin{subeqnarray}
    \MCSd_{\fcik} \MCWd \Psibd_{\fcik} &=& \Lambdad_{\fcik} \Psibd_{\fcik}, \\
    \widetilde{\QCd}_{\fcik} \widetilde{\QCd}_{\fcik}^* \MCWd \Psibd_{\fcik} &=& \Lambdad_{\fcik} \Psibd_{\fcik}.
    \label{eqn:eigproblemd}
\end{subeqnarray}
The spatial inner weight
\begin{equation}
    \langle \qb_1, \qb_2 \rangle_x = \int_{\Omega} \qb_1^*(\x, t) \Wb(\x) \qb_2(\x, t) \dx
    \label{eqn:spacenorm2}
\end{equation}
is approximated as $\langle \qb_1, \qb_2 \rangle_x = \qd_1^* \Wd \qd_2$ where $\Wd \in \mathbb{C}^{N\times N}$ is a positive-definite Hermitian matrix that accounts for both the weight and the numerical quadrature of the integral on the discrete grid and $\MCWd \in \mathbb{C}^{(2a_1+1)N \times (2a_1+1)N}$  is the block-diagonal matrix of $\Wd$ (similar to \ref{eqn:weightmatrixblock}). The CS-SPOD modes are then given by the columns of $\Psibd_{\fcik}$ and are ranked by their corresponding eigenvalues given by the diagonal matrix $\Lambdad_{\fcik}$. These discrete CS-SPOD modes hold analogous properties to all those previously discussed, including that they are discretely orthogonal $\Psibd^*_{\fcik} \Wd \Psibd_{\fcik} = \ICb$ and optimally decompose the estimated CS-SPOD decomposition matrix $\MCSd_{\fcik} = \Psibd_{\fcik} \Lambdad_{\fcik} \Psibd^*_{\fcik}$ (i.e. the second-order statistics).

At most, $\text{min}(N, N_b )$ number of non-zero eigenvalues can be obtained. Thus, it is possible to show that the following $N_b  \times N_b $ eigenvalue problem
\begin{equation}
     \widetilde{\QCd}_{\fcik}^* \MCWd \widetilde{\QCd}_{\fcik} \Thetabd_{\fcik} = \widetilde{\Lambdad}_{\fcik} \Thetabd_{\fcik},
\end{equation}
contains the same non-zero eigenvalues as \eqref{eqn:eigproblemd}. This approach is known as the method-of-snapshots \citep{sirovich1987turbulence}. The corresponding eigenvectors are exactly recovered as 
\begin{equation}
\widetilde{\Psibd}_{\fcik} = \widetilde{\QCd}_{\fcik} \Thetabd_{\fcik} \widetilde{\Lambdad}_{\fcik}^{-1/2}.
\end{equation}

 Other than the simple weighting matrix $\MCWd$, only the concatenated data matrix $\widetilde{\QCd}_{\fcik}$ must be determined, which is easily achieved by computing each term ($\hat{\Qd}_{\fcik, m\alpha_0}$) in $\widetilde{\QCd}_{\fcik}$ using algorithm \ref{algo:csdalgo}. Once $\widetilde{\QCd}_{\fcik}$ is determined, one computes $\widetilde{\QCd}_{\fcik}^* \MCWd \widetilde{\QCd}_{\fcik} $ and then performs the eigenvalue decomposition.  Typically, only the first few modes are of physical interest, which allows us to employ a truncated decomposition where we determine a limited number of the most energetic CS-SPOD modes using randomized linear algebra methods \citep{martinsson2020randomized}. The total energy can be efficiently evaluated by taking the trace of $\widetilde{\QCd}_{\fcik}^* \MCWd \widetilde{\QCd}_{\fcik}$.  \par
 
 Algorithm \ref{algo:ccfdbase} implements the CS-SPOD in a practical, but computationally inefficient, manner.  The algorithm requires computing $2a_1+1$ CCSDs, and those the cost is approximately $2a_1 + 1$ times that of the SPOD. The memory requirement scales similarly. This can be prohibitive when analyzing large data sets. 
 
 \begin{algorithm}
\setstretch{1.25}
\caption{Naive algorithm to compute CS-SPOD. } \label{algo:ccfdmbase}
\begin{algorithmic}[1]
\For{Each data block, $n = 1, 2, \cdots, N_b$} 
\Statex{\hspace{4.5mm} $\triangleright$ Construct the block data matrix}
\State ${\Qd}^{(n)} = [\qd_{1+(n-1)(N_f - N_0)}, \qd_{2+(n-1)(N_f - N_0)}, \cdots, \qd_{N_f+(n-1)(N_f - N_0)}]$  
\Statex{\hspace{4.5mm} $\triangleright$ Construct the block time matrix}
\State $\Td^{(n)} = [t_{1+(n-1)(N_f - N_0)}, t_{2+(n-1)(N_f - N_0)}, \cdots, t_{N_f+(n-1)(N_f - N_0)}]$  
\EndFor
\For{Each $m \in \mathcal{A}_m$, where $\mathcal{A}_m = \{-a_1, -a_1+1, \cdots, 0,  \cdots, a_1-1, a_1\}$ } 
\For{Each data block, $n = 1, 2, \cdots, N_b$} 
\Statex{\hspace{9mm} $\triangleright$ Compute the frequency-shifted block data matrices}
\State ${\Qd}^{(n)}_{ m\alpha_0} \gets \Qd^{(n)} e^{-i 2\pi (m\alpha_0) \Td^{(n)}} $  
\Statex{\hspace{9mm} $\triangleright$ Using a (windowed) fast Fourier transform, calculate and store the row-wise}
\Statex{\hspace{9mm} DFT for each frequency-shifted block data matrix }
\State $\hat{\Qd}^{(n)}_{ m\alpha_0} = \text{FFT}({\Qd}^{(n)}_{ m\alpha_0})  = [\hat{\qd}^{(n)}_{1, m\alpha_0}, \hat{\qd}^{(n)}_{2, m\alpha_0}, \cdots, \hat{\qd}^{(n)}_{N_w, m\alpha_0} ]$ 
\Statex{\hspace{9mm} where, the column $\hat{\qd}_{k, m\alpha_0}^{(n)}$ contains the $n^{th}$ realization of the Fourier mode at }
\Statex{\hspace{9mm} the $k^{th}$ discrete frequency of the $m\alpha_0$ frequency-shifted block-data matrix}
\EndFor
\EndFor
\For{Each $\fcik \in \Fcik $ (or some subset of interest)} 
\Statex{\hspace{9mm} $\triangleright$ Assemble the concatenated frequency-data matrix for frequency set $\Fset_{\fcik}$ } 
\State $\widetilde{\QCd}_{\fcik} \gets \begin{bmatrix}
\hat{\Qd}_{\fcik, -a_1\alpha_0} \\ 
\vdots \\
\hat{\Qd}_{\fcik, 0} \\
\vdots \\
\hat{\Qd}_{\fcik, a_1\alpha_0}
\end{bmatrix},$
\Statex{\hspace{4.5mm} where $\hat{\Qd}_{\fcik, m\alpha_0} \gets \sqrt{\kappa} [\hat{\qd}_{k, m\alpha_0}^{(1)},  \hat{\qd}_{k, m\alpha_0}^{(2)}, \cdots, \hat{\qd}_{k, m\alpha_0}^{(N_b-1)}, \hat{\qd}_{k, m\alpha_0}^{(N_b)}]$ is the matrix of }
\Statex{\hspace{4.5mm} Fourier realizations corresponding to the ${k}^{th}$ column of the  $m\alpha_0$ frequency-shifted}
\Statex{\hspace{4.5mm}  block-data matrix $\hat{\Qd}^{(n)}_{ m\alpha_0}$}
\State Compute the matrix $\MC_{\fcik} \gets \widetilde{\QCd}_{\fcik}^* \MCWd \widetilde{\QCd}_{\fcik}$
\State Compute the eigenvalue decomposition $\MC_{\fcik} = \Thetabd_{\fcik} \widetilde{\Lambdad}_{\fcik} \Thetabd^*_{\fcik} $
\State Compute and save the CS-SPOD modes $\widetilde{\Psibd}_{\fcik} = \widetilde{\QCd}_{\fcik} \Thetabd_{\fcik} \widetilde{\Lambdad}_{\fcik}^{-1/2}$ 
\Statex{\hspace{4.5mm} and energies $\widetilde{\Lambdad}_{\fcik}$ for the $\fcik$ frequency set $\Fset_{\fcik}$}
\EndFor
\end{algorithmic}
\label{algo:ccfdbase}
\end{algorithm} 

However, significant savings are realized since all the terms in $ \widetilde{\QCd}_{\fcik}$ are in the form of $\hat{\Qd}_{\fcik, m\alpha_0}$, which represent the $k^{th}$ frequency component of the temporal Fourier transform of the $m\alpha_0$ frequency-shifted data matrix. The temporal Fourier transform of the $n^{th}$ realization of the $m\alpha_0$ frequency-shifted data is given by
\begin{subeqnarray}
    \hat{\qd}_{k, m\alpha_0}^{(n)} &=&  \frac{1}{\sqrt{N_f}} \sum_{j = 1}^{N_f}w_j{\qd}_{j, m\alpha_0}^{(n)}e^{-i2\pi (k-1)[\frac{j - 1}{N_f}]}, \\
    &=&  \frac{1}{\sqrt{N_f}} \sum_{j = 1}^{N_f}w_j{\qd}_{j}^{(n)} e^{-i 2\pi (m\alpha_0\Delta t)[{(j-1) + (n-1)(N_f - N_0)}]} e^{-i2\pi (k-1)[\frac{j - 1}{N_f}]} ,
\end{subeqnarray}
where $e^{-i 2\pi (m\alpha_0 \Delta t)[{(j-1) + (n-1)(N_f - N_0)}]}$ is the frequency-shifting operation. We separate these components into a phase-shifting component and a  zero-phase-shift frequency-shifting component, by 
\begin{subeqnarray}
    &&\hat{\qd}_{k, m\alpha_0}^{(n)} =  e^{-i 2\pi (m\alpha_0 \Delta t)[{(n-1)(N_f - N_0)}]} \frac{1}{\sqrt{N_f}} \sum_{j = 1}^{N_f}w_j{\qd}_{j}^{(n)} e^{-i2\pi (m\alpha_0\Delta t N_f + k-1)[\frac{j - 1}{N_f}]}, \phantom{asdsad}\\
    &&\hat{\qd}_{k, m\alpha_0}^{(n)} =  e^{-i 2\pi (m\alpha_0 \Delta t)[{(n-1)(N_f - N_0)}]} \hat{\qd}_{\ell(k, m)}^{(n)}, 
\end{subeqnarray}
where $\ell(k, m)$ is the $\ell^{th}$ frequency that is a function of $k, m$. This shows that the $f_k$ discrete frequency of the $m\alpha_0$-frequency-shifted data matrix ($f_{k, m\alpha_0}$) can be exactly computed as a phase-shifted version of the $f_{\ell(k, m)}$ discrete frequency component of the non-frequency-shifted data matrix. To employ this method, $m\alpha_0\Delta t N_f \in \mathbb{Z}$. This ensures that the change in frequency due to the applied frequency-shifting operator is equal to an integer change in the index of the frequency vector. Since $\alpha_0 \Delta T = 1/N_\theta$, this gives $\frac{m N_f}{N_\theta} \in \mathbb{Z}$, which requires $N_f = N_{osc} N_\theta,\ N_{osc} \in \mathbb{Z}$. With this restriction, the frequency spectrum of the DFT of a $N_f$ length record is
\begin{equation}
    f_{k} =    \begin{dcases}
        \frac{(k-1)\alpha_0}{N_{osc}}  & \text{for } k \leq \frac{N_{osc} N_\theta}{2}, \\
        \frac{(k-1 - N_{osc}N_\theta)\alpha_0}{N_{osc}} & \text{for } k > \frac{N_{osc} N_\theta}{2}, \\
    \end{dcases}
\end{equation}
and the unique frequency sets become
\begin{equation}
    \fcik =    \begin{dcases}
        \frac{(k-1)\alpha_0}{N_{osc}}  & \text{for } k \leq \lfloor\frac{N_{osc}}{2}\rfloor + 1, \\
        \frac{(k-1 - N_{osc}N_\theta)\alpha_0}{N_{osc}} & \text{for } N_f - \lceil\frac{N_{osc}}{2} \rceil + 1 < k \le N_f. \\
    \end{dcases}
\end{equation}
This demonstrates that a frequency shift of $m\alpha_0$ corresponds to an integer change in the frequency index, i.e. the $k^{th}$ frequency component of the $m\alpha_0$-frequency-shifted data matrix corresponds to the phase-shifted version of the $\ell(k, m)^{th}$ frequency component ($f_{\ell(k, m)}$) of the non-frequency-shifted data matrix, i.e. $f_{k, m\alpha_0} = f_{\ell(k, m)}$, where
\begin{equation}
    \ell(k, m) =    \begin{dcases}
        \begin{dcases}
        k + m N_{osc}  & \text{for } m \ge 0 \\
        k + m N_{osc} + N_{f}  & \text{for } m < 0
        \end{dcases} & 
         \text{\ \ for } k \leq \lfloor\frac{N_{osc}}{2}\rfloor + 1, \\
        \begin{dcases}
        k + m N_{osc} - N_{f}    & \text{for } m \ge 0 \\
        k + m N_{osc} & \text{for } m < 0
        \end{dcases}
          & \text{\ \ for } N_f - \lceil\frac{N_{osc}}{2} \rceil + 1 < k \le N_f.  \\
    \end{dcases}
    \label{eqn:k2val}
\end{equation}
This means that all the data required for CS-SPOD (for all frequency sets $\Fset_{\fcik}$) is contained within the Fourier transform of the original data matrix. \par

Algorithm \ref{algo:ccfdeff} incorporates these savings and requires only a single DFT of the data matrix, making it similar in computational cost and memory requirement to SPOD. The memory usage to compute CS-SPOD for complex input data is $\approx (\frac{1}{1 - N_0/N_f} + 1)\times \text{mem}(\Qd)$, which is the memory required to store the, possibly overlapping, block data matrix and the original data matrix. Additional memory is required to store the temporary matrix $\widetilde{\QCd}_{\fcik}$, although the size of this matrix is minimal as typically $2a_1 + 1 << N_f$. In extreme cases where only a single snapshot can be loaded at a time, a streaming CS-SPOD algorithm could be developed analogous to the streaming SPOD method by \citet{schmidt2019efficient}. 

\begin{algorithm}
\setstretch{1.25}
\caption{Efficient algorithm to compute CS-SPOD. }\label{algo:ccfdm}
\begin{algorithmic}[1]
\For{Each data block, $n = 1, 2, \cdots, N_b$} 
\Statex{\hspace{4.5mm} $\triangleright$ Construct the block data matrix}
\State ${\Qd}^{(n)} \gets [\qd_{1+(n-1)(N_f - N_0)}, \qd_{2+(n-1)(N_f - N_0)}, \cdots, \qd_{N_f+(n-1)(N_f - N_0)}]$  
\Statex{\hspace{4.5mm} $\triangleright$ Using a (windowed) fast Fourier transform, calculate and store the row-wise}
\Statex{\hspace{4.5mm} DFT for each frequency-shifted block data matrix }
\State $\hat{\Qd}^{(n)} = \text{FFT}({\Qd}^{(n)}) $  \Comment{Discard any frequency components that are not required}
\LeftCommentE{to compute $\widetilde{\QCd}_{\fcik}$ (if one is not computing $\widetilde{\QCd}_{\fcik}$ over all $\fcik \in \Fcik$)}
\EndFor
\For{Each $\fcik \in \Fcik$ (or some subset of interest)} 
\Statex{\hspace{4.5mm} $\triangleright$ Assemble the concatenated frequency-data matrix for frequency set $\Fset_{\fcik}$ } 
\State $\widetilde{\QCd}_{\fcik} \gets   \begin{bmatrix}
\hat{\Qd}_{\fcik, -a_1\alpha_0} \\ 
\vdots \\
\hat{\Qd}_{\fcik, 0} \\
\vdots \\
\hat{\Qd}_{\fcik, a_1\alpha_0}
\end{bmatrix}$ 

\Statex{\hspace{4.5mm} where $\hat{\Qd}_{\fcik, m\alpha_0} \gets \sqrt{\kappa} [\hat{\qd}_{k, m\alpha_0}^{(1)},  \hat{\qd}_{k, m\alpha_0}^{(2)}, \cdots, \hat{\qd}_{k, m\alpha_0}^{(N_b-1)}, \hat{\qd}_{k, m\alpha_0}^{(N_b)}]$ is the matrix of }
\Statex{\hspace{4.5mm} Fourier realizations corresponding to the ${k}^{th}$ column of the  $m\alpha_0$ frequency-shifted}
\Statex{\hspace{4.5mm}  block-data matrix $\hat{\Qd}^{(n)}_{ m\alpha_0}$, evaluated efficiently by $\hat{\qd}_{k, m\alpha_0}^{(n)} =  $}
\Statex{\hspace{4.5mm} $e^{-i 2\pi (m\alpha_0 \Delta t)[{(n-1)(N_f - N_0)}]} \hat{\qd}_{\ell(k, m)}^{(n)}$,  where the index $l(k, m)$ is given by \eqref{eqn:k2val}}
\State Compute the matrix $\MC_{\fcik} \gets \widetilde{\QCd}_{\fcik}^* \MCWd \widetilde{\QCd}_{\fcik}$
\State Compute the eigenvalue decomposition $\MC_{\fcik} = \Thetabd_{\fcik} \widetilde{\Lambdad}_{\fcik} \Thetabd^*_{\fcik} $
\State Compute and save the CS-SPOD modes $\widetilde{\Psibd}_{\fcik} = \widetilde{\QCd}_{\fcik} \Thetabd_{\fcik} \widetilde{\Lambdad}_{\fcik}^{-1/2}$ 
\Statex{\hspace{4.5mm} and energies $\widetilde{\Lambdad}_{\fcik}$ for the $\fcik$ frequency set $\Fset_{\fcik}$}
\EndFor
\end{algorithmic}
\label{algo:ccfdeff}
\end{algorithm}

\section{Validation of our CCSD and CS-SPOD algorithms} \label{sec:validation}

We validate our implementation of the CCSD and CS-SPOD using a model problem that has an analytical solution. Let $n(x, t)$ be a zero-mean, complex-valued, stationary random process with uniformly distributed phase (between $0$ and $2\pi$), normally distributed unit variance, and a covariance kernel $c(x, x^\prime) = E\{n(x, t)n^*(x^\prime, t)\}$, of
\begin{equation}
c(x, x^\prime) = \frac{1}{\sqrt{2\pi}\sigma_\eta} \text{exp}\left[ -\frac{1}{2} \left(\frac{x - x^\prime}{\sigma_\eta} \right)^2\right] \text{exp}\left[ -i2\pi \frac{x - x^\prime}{\lambda_\eta}\right],
\end{equation}
where $\sigma_\eta = 4$ is the standard deviation of the envelope, $\lambda_\eta = 20$ is the wavelength of the filter, and $x_0 = 1.5$ is the center off-set distance. This covariance kernel is identical to the one used by \citet{towne2018spectral} as its structure is qualitatively similar to statistics present in real flows (e.g. a turbulent jet).  The filtered process $\widetilde{n}(x, t)$ is defined as the convolution between a filter $f_\ell(x, t)$ and $n(x, t)$, given by
\begin{equation}
\widetilde{n}(x, t)  = f_\ell(x, t) \circledast n(x, t).
\end{equation}
We sinusoidally modulate $\widetilde{n}(x, t)$ to create a cyclostationary process
\begin{equation}
    g(x, t) = \widetilde{n}(x, t)\text{cos}(2\pi f_0 t + \phi_0), 
    \label{eqn:modulatingdummy}
\end{equation}
where $f_0 = 0.5$ is the modulation frequency and $\phi_0 = \frac{1}{3}2\pi$ is a phase offset. Using the theory developed in \S \ref{sec:CStheory}, the CCSD of $g(x, t)$ is analytically determined as
\begin{equation}
S_{g}(x, x^\prime, \alpha, f)=
    \begin{cases}
        \frac{1}{4}e^{\pm i2\theta}S_{\widetilde{n}}(x, x^\prime, 0
        , f ) & \text{for } \alpha = \pm 2f_0 \\
        \frac{1}{4} S_{\widetilde{n}}(x, x^\prime, 0, f + f_0) + \frac{1}{4} S_{\widetilde{n}}(x, x^\prime, 0, f - f_0) & \text{for } \alpha = 0 \\
        0 & \text{otherwise }
    \end{cases}
    \label{eqn:CCSDanalytical}
\end{equation}
where $S_{\widetilde{n}}(x, x^\prime, 0, f)$ is the CCSD of $\widetilde{n}(x, t)$ at cycle frequency $\alpha = 0$ (thus equaling the CSD). The fundamental and only non-zero cycle frequency present is $\alpha_0= \pm 2f_0$, indicating that this process exhibits cyclostationarity. The CSD of $\widetilde{n}(x, t)$ is given by  
\begin{equation}
    S_{\widetilde{n}}(x, x^\prime, 0, f) = c(x, x^\prime) F_\ell(x, f)  F_\ell^*(x^\prime, f), 
\end{equation}
where $F_\ell(x, f)$ is the temporal Fourier transform of the filter $f_\ell(x, t)$. The filter employed is a $5^{th}$-order finite-impulse-response filter with a cutoff frequency $f_{co}$, that varies as a function of the spatial location $f_{co} = 0.2|x - x_0|/\text{max}(x) + 0.2$. This results in a filter exhibiting a more rapid spectral decay at $x_0$ and a flatter spectrum moving away from this location.  A domain $x \in [-10, 10]$ is employed and is discretized using 2001 equispaced grid points resulting in a grid spacing of $\Delta x = 0.01$. All estimates of the CCSD and CS-SPOD are performed using a Hamming window with $L_w = 10N_\theta$ and an overlap of $67\%$. Snapshots are saved in time with $\Delta t  = 0.04$, resulting in $N_\theta = 25$ time steps per period of the fundamental cycle frequency, $T_0 = 1/\alpha_0 = 1/(2f_0)$. Data is saved for $t_{end} = 2000T_0$, resulting in 50000 snapshots and 593 blocks (realizations) of the process.

\begin{figure}
    \centering
    \begin{minipage}[t]{.45\textwidth}
        \centering
        \includegraphics[height= 0.7 \textwidth]{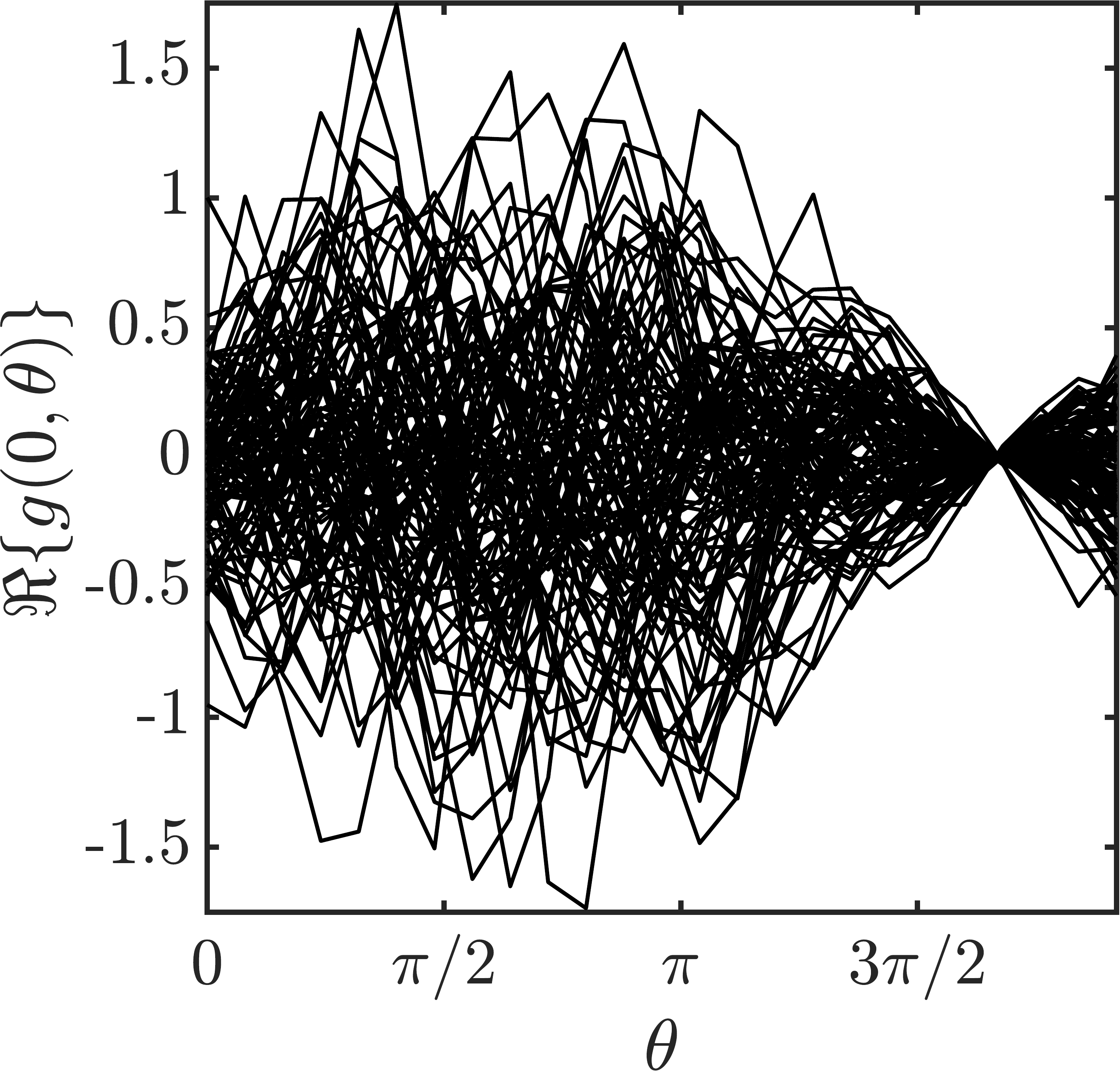}
        \caption{Model process sample paths at $x = 0$.}
        \label{fig:Dummy:example}
    \end{minipage} \hfill
    \begin{minipage}[t]{0.45\textwidth}
        \centering
        \includegraphics[height=0.718 \textwidth]{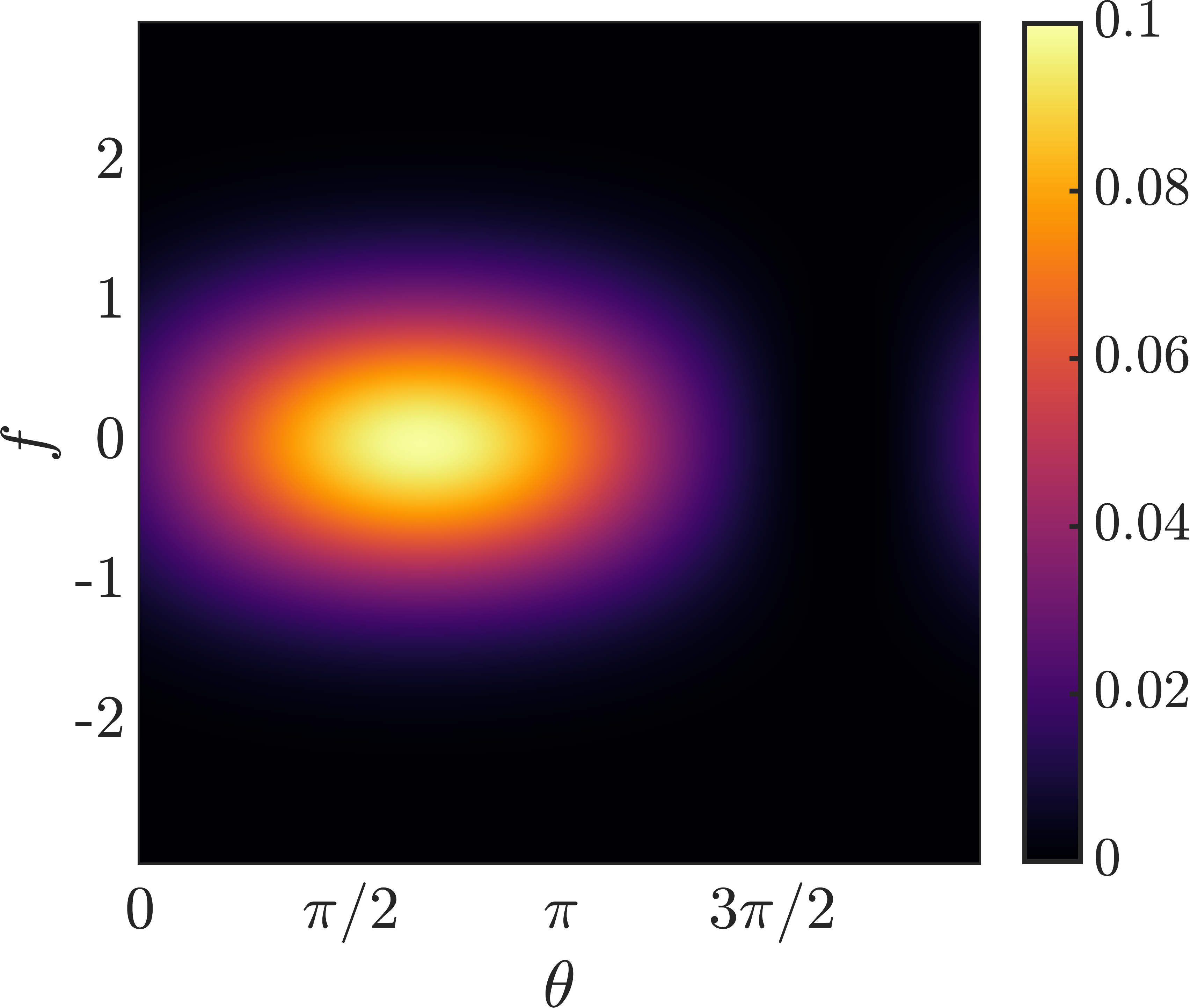}
        \caption{Analytical WV spectrum at $x = x^\prime = 0$ for the model process.}
        \label{fig:Dummy:WV}
    \end{minipage}
\end{figure}

\begin{figure}
\centering
    \begin{subfigure}[b]{0.2\textwidth}
        \centering
        \includegraphics[height=0.9425\textwidth]{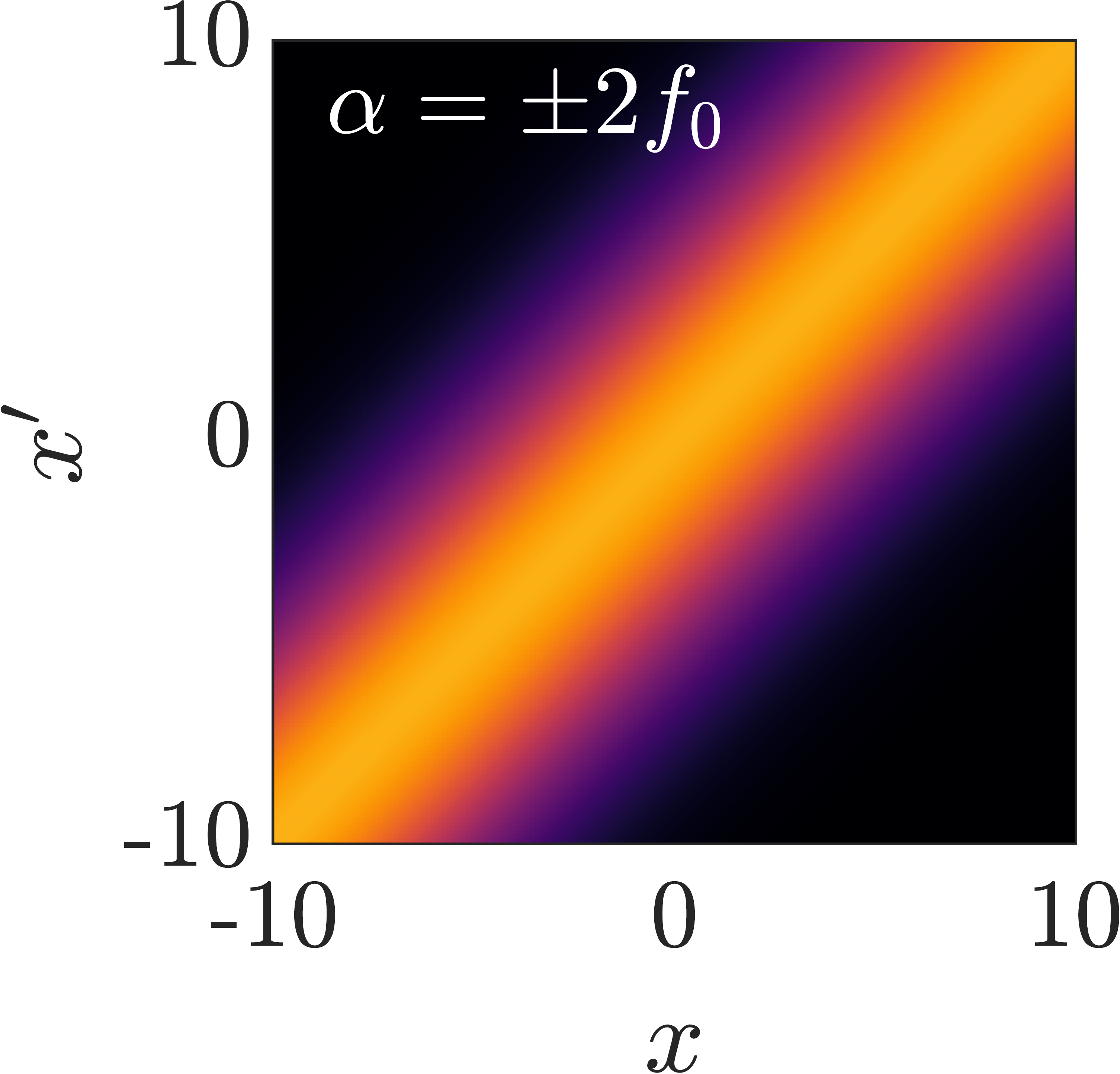}
        \caption{Analytical \hfill}  
    \end{subfigure} %
    \begin{subfigure}[b]{0.29\textwidth}  
        \centering 
       \includegraphics[height=0.65\textwidth]{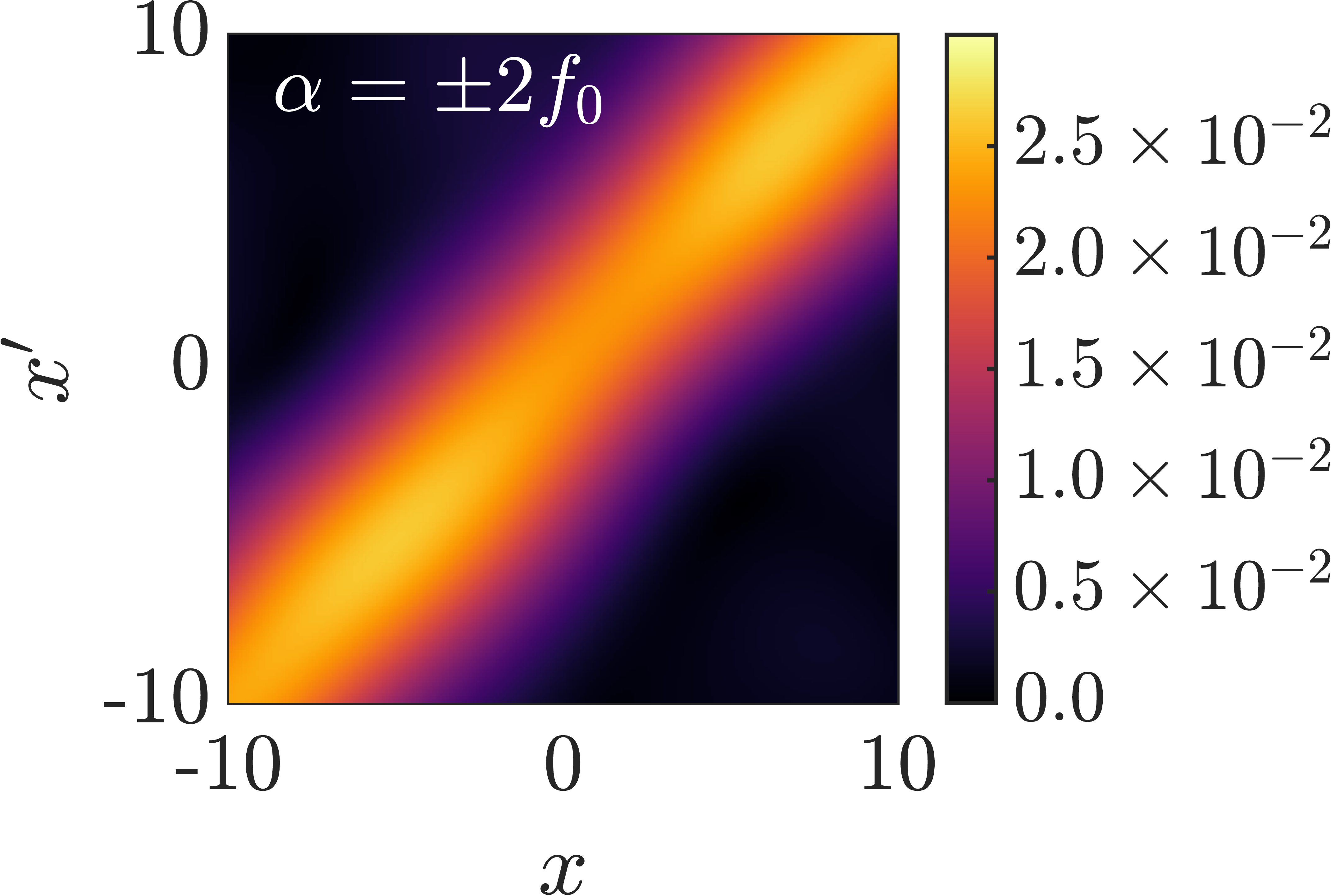}
        \caption{Numerical \hfill}   
    \end{subfigure}%
    \begin{subfigure}[b]{0.2\textwidth}   
        \centering 
       \includegraphics[height=0.9425\textwidth]{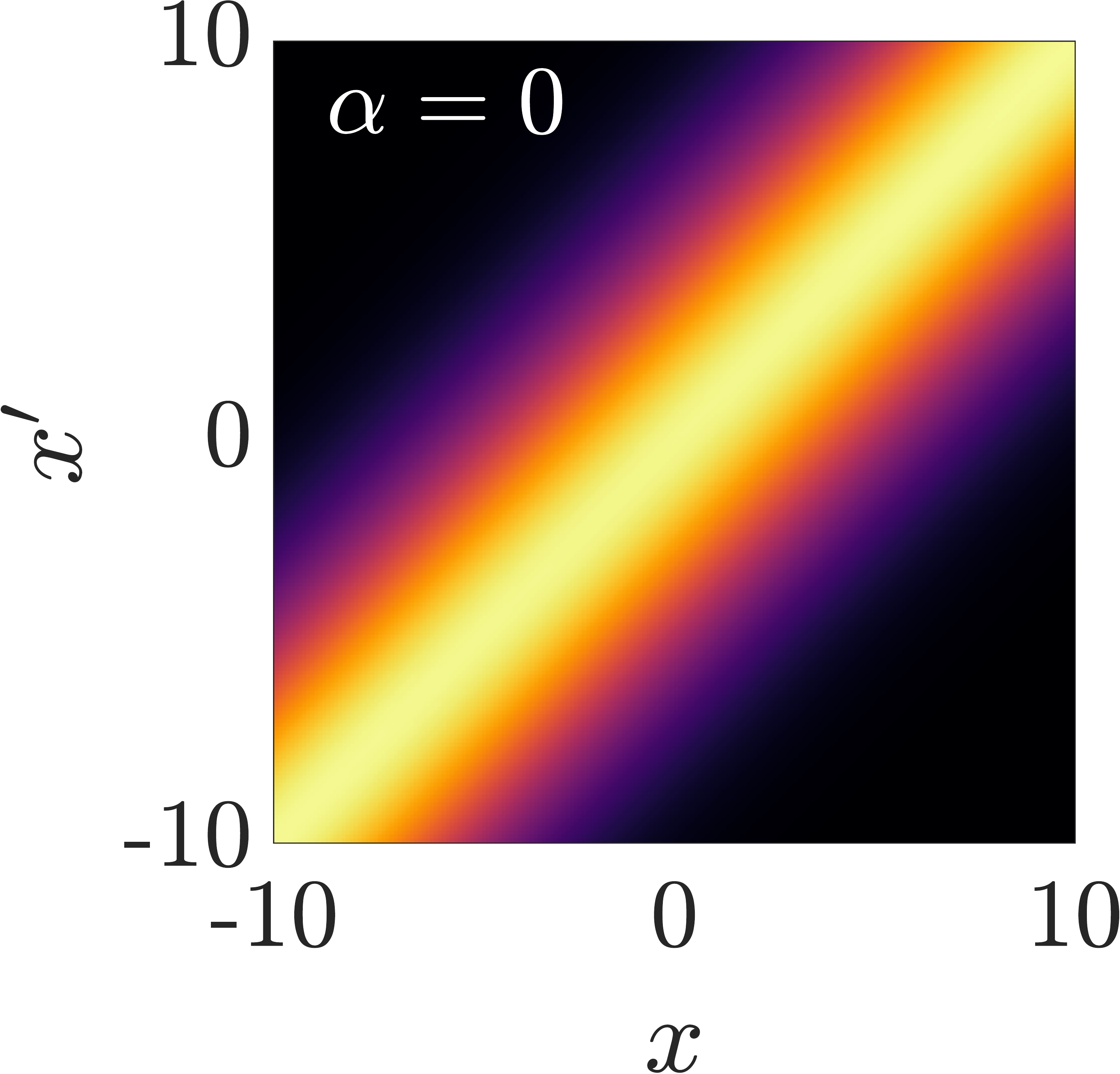}
        \caption{Analytical \hfill}   
    \end{subfigure}%
    \begin{subfigure}[b]{0.29\textwidth}   
        \centering 
        \includegraphics[height=0.65\textwidth]{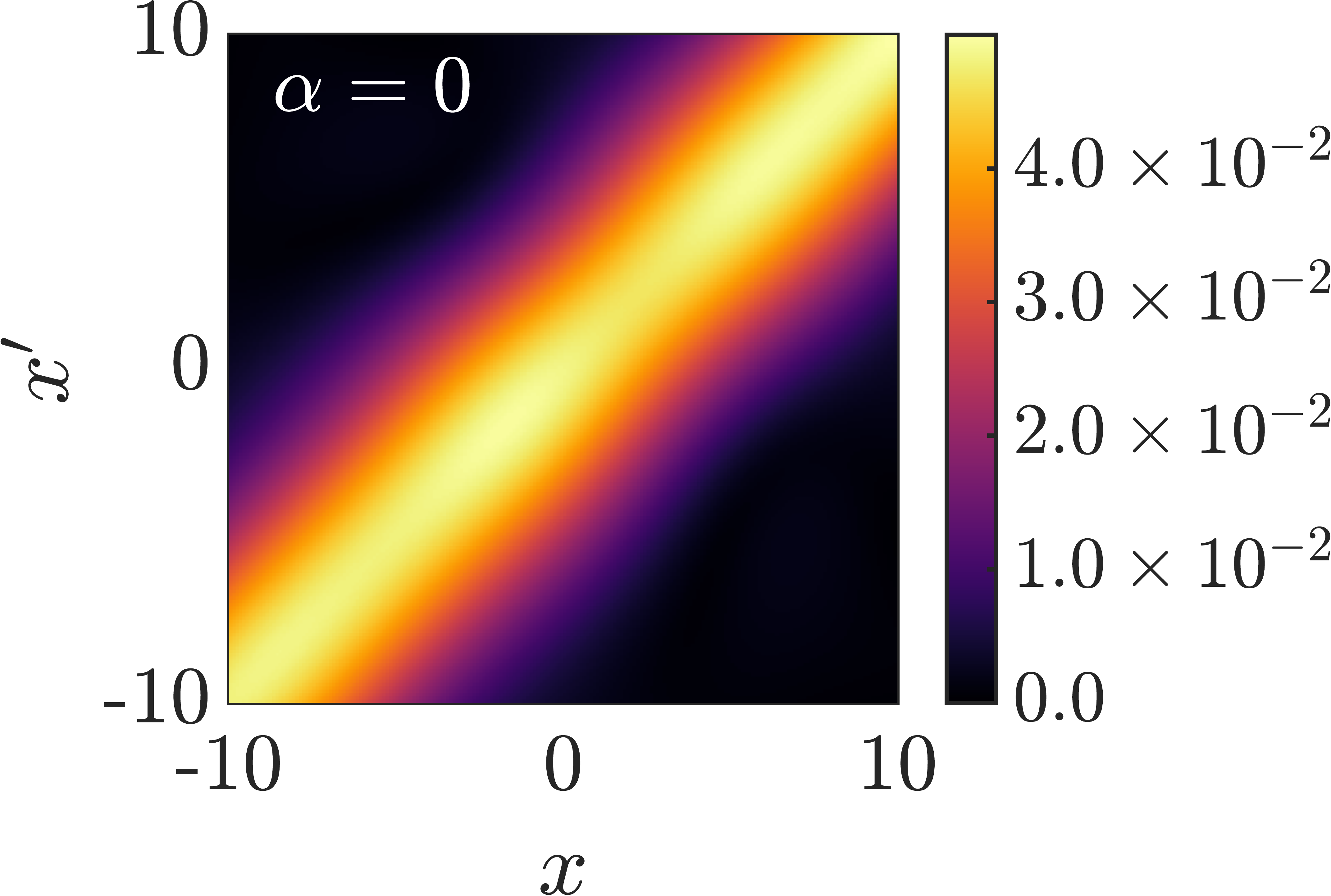}
         \caption{Numerical \hfill}  
    \end{subfigure}%
        \caption{Plot of the magnitude of the CCSD for $f = 0.1$ of the dummy process for the analytical and numerically generated results.  }
    \label{fig:CCSDan}
\end{figure}

Sample paths of the process at $x = 0$, as a function of the phase of the fundamental cycle frequency, are shown in figure \ref{fig:Dummy:example}.  As theoretically predicted, we observe a modulation in the amplitude of the process as a function of the phase. Since $\alpha_0 = 2f_0$, the phase offset $\phi_0 = \frac{1}{3} 2\pi$ applied to the sinusoidal modulation results in a phase offset of $\frac{1}{6}2\pi$ in the sample paths. This modulation is observed in figure \ref{fig:Dummy:WV}, where we plot the analytical WV spectrum computed using (\ref{eqn:wvdef} and \ref{eqn:CCSDanalytical}) at $x = x^\prime = 0$. This shows the sinusoidal modulation of the PSD as a function of the phase and a decay in the amplitude of the spectrum with increasing $|f|$ due to the applied filter. In figure \ref{fig:CCSDan}, we compare the magnitude of the analytical and numerical CCSD at $f = 0.1$ and $\alpha =0, \pm 2f_0$. Here, we observe the aforementioned key structures of the covariance kernel along with the excellent agreement between the numerical and analytical CCDSs, which would further improve with an increasing number of realizations, thereby validating our CCSD implementation (algorithm \ref{alg:cap}).  
\noindent

\begin{figure}
\centering
    \begin{minipage}[t]{0.46\textwidth}
        \centering
        \includegraphics[height=0.8\textwidth,valign=t]{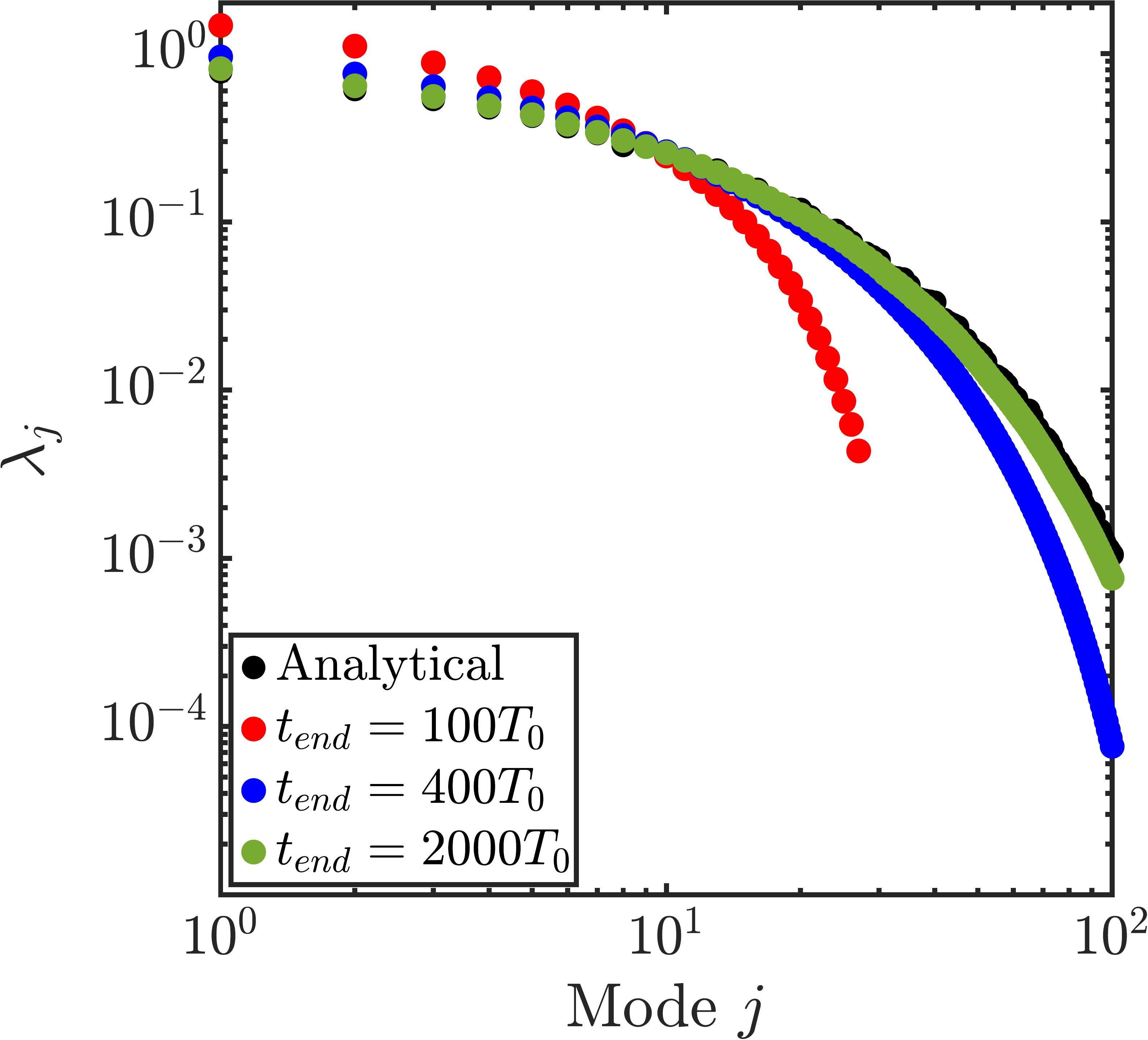}
        \caption{Plot of the analytical and numerical CS-SPOD eigenspectrum at $\fci = 0.2$ for the dummy problem at multiple signal durations. }
        \label{fig:evalall}
    \end{minipage}\hfill
    \begin{minipage}[t]{0.46\textwidth}  
        \centering \includegraphics[height=0.8\textwidth,valign=t]{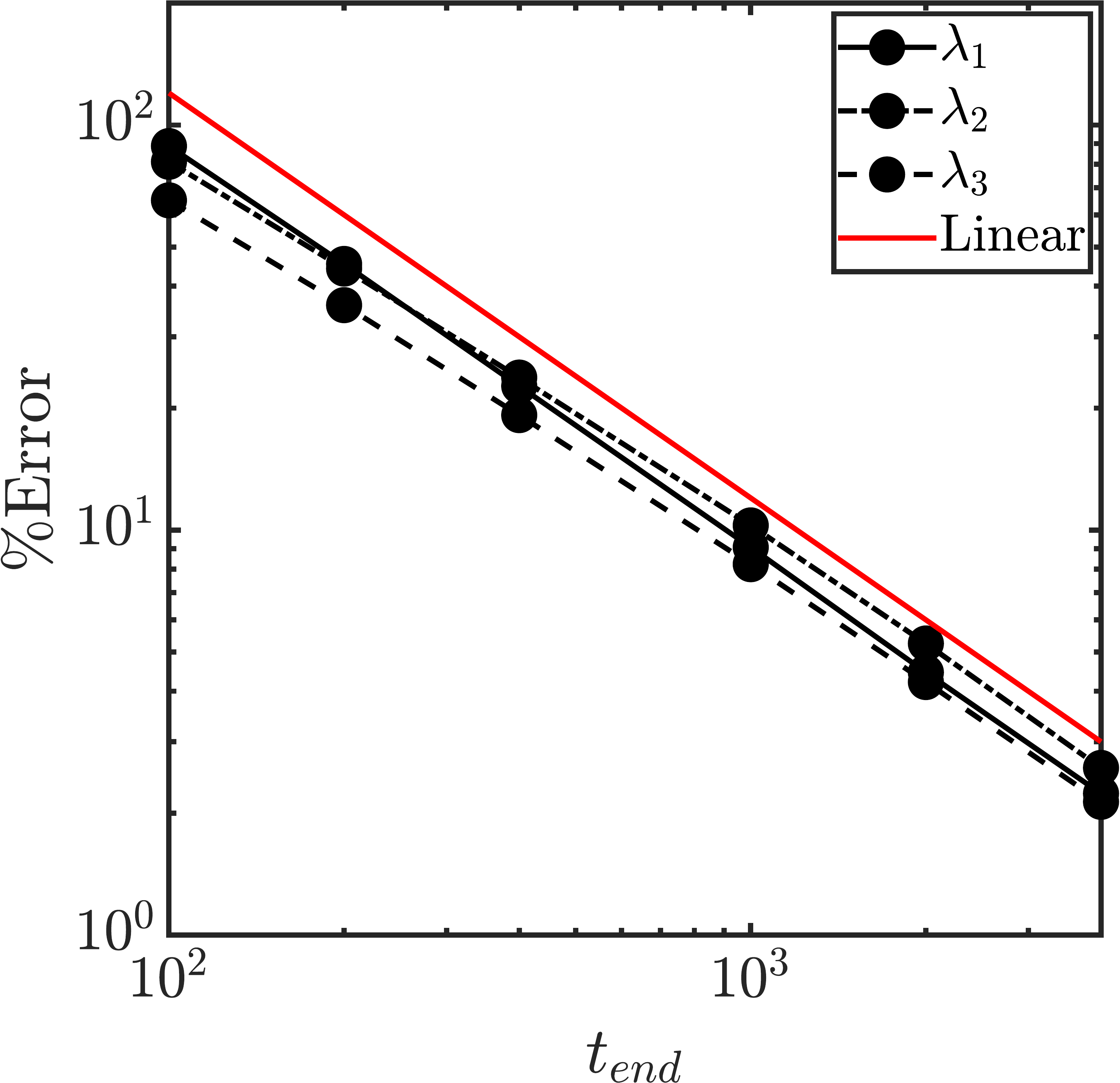}
       \caption{Convergence of CS-SPOD eigenvalues as a function of the total signal duration of the dummy problem.}
       \label{fig:Dummy:errorovertime}
    \end{minipage}%
\end{figure}

Next, we validate our efficient algorithm to compute CS-SPOD (algorithm \ref{algo:ccfdeff}) and determine its convergence with increasing data by comparing the numerical results to the analytical results. The analytical solution is determined by forming the CS-SPOD eigensystem defined via \eqref{eqn:infiniteeig} through evaluating the analytical CCSDs (given by \eqref{eqn:CCSDanalytical}) and then numerically evaluating the final eigenvalue problem. To encompass the range of relevant frequencies we use $a_1 = 10$ to construct $\mathcal{A}_m$, resulting in $\Fset_{\fci} = [-10, 10] + \fci$. Figure \ref{fig:evalall} shows a comparison of the analytical and numerical CS-SPOD eigenspectrums (averaged over 10000 realizations of the process), at $\fci = 0.2$ for  $t_{end} = 100T_0, 400T_0$, and $ 2000T_0$, which corresponds to 27, 117, and 593 blocks, respectively. As the duration of the process increases, we observe an increasingly converged estimate of the eigenspectrum. This is reflected in the percentage error between the averaged numerical eigenvalues and the analytical eigenvalues for the three most dominant CS-SPOD modes, which we show in figure \ref{fig:Dummy:errorovertime}. We see that these eigenvalues linearly converge to the true value as the duration of the process increases, which is theoretically expected due to the linear reduction in the variance of the Welch estimate of the CCSD with increasing realizations \citep{antoni2007cyclic}. Overall, we obtain a consistent estimate of the CS-SPOD eigenvalues and conclude that our implementation of CS-SPOD is correct. 

\section{Example problems} \label{sec:examples}
\subsection{Application to a modified linearized complex Ginzburg-Landau equation}  \label{sec:FGL}
Our first example is the simple and well-understood linearized complex Ginzburg-Landau equation, which has been used as a model for a convectively unstable flow that exhibits non-modal growth \citep{chomaz1988bifurcations,cossu1997global,hunt1991instability}. It can be written in the form of a generic linear forced system  
\begin{equation}
    \frac{\partial q(x, t)}{\partial t} - L(x, t) q(x, t) = f(x, t), 
    \label{eqn:GLEqn}
\end{equation}
where $q(x, t)$ and $f(x, t)$ represent the state and forcing, respectively, with $|q(x\rightarrow\pm \infty, t)| \rightarrow 0$, and $L(x, t)$ is the linear operator
\begin{equation}
    L(x, t) = -\nu_1\frac{\partial}{\partial x} + \nu_2 \frac{\partial^2}{\partial x^2} - \mu(x, t).
    \label{eqn:GLEqnmu}
\end{equation}
We use the commonly used form $\mu(x) = \mu_0 - c_{\mu}^2 + \mu_2 \frac{x^2}{2}$ \citep{hunt1991instability,bagheri2009input,chen2011h2,towne2018spectral}, resulting in time-invariant dynamics. All constants in (\ref{eqn:GLEqn}, \ref{eqn:GLEqnmu}), except for $\mu_0$, use the values in \citet{bagheri2009input}. Similar to \citet{franceschini2022identification}, we construct periodic dynamics by  using $\mu_0 = \overline{\mu}_0 + A_{\mu_0} \text{sin}(2\pi f_0 t)$, where $\overline{\mu}_0$ is the average value of $\mu_0$, $A_{\mu_0}$ is the amplitude of the periodic modulation of $\mu_0$, and $f_0$ is the frequency of the periodic modulation. For $A_{\mu_0} = 0$ the system has time-invariant dynamics, while for $|A_{\mu_0}| > 0$ the system has time-periodic dynamics, resulting in a stationary and cyclostationary response, respectively. By varying $A_{\mu_0}$, we modify the degree to which the system is cyclostationary.  We choose $f_0 = 0.1$, which is substantial compared to the frequencies of interest ($\approx [-0.5, 0.5]$), meaning that the quasi-steady approach of \citet{franceschini2022identification} can not be employed. Like \citet{towne2018spectral}, we use $\overline{\mu}_0 = 0.23$, which for $A_{\mu_0} = 0$ strongly amplifies external noise due to the non-normality of $L(x, t)$ and results in a degree of low-rankness typically present in turbulent flows. As per \citet{franceschini2022identification}, we confirm the stability of the system using Floquet analysis (results not shown). To demonstrate the utility of CS-SPOD and to facilitate its interpretation, we compare CS-SPOD performed at several levels of cyclostationarity $A_{\mu_0} = 0.0, 0.2,\text{and}\ 0.4$. \par

A pseudo-spectral approach utilizing Hermite polynomials is employed to discretize the equations \citep{bagheri2009input,chen2011h2}, where the collocation points $[x_1, x_2,\cdots,x_{N_H}]$ correspond to the first $N_H$ Hermite polynomials with scaling factor $\mathfrak{R}\{(-\mu_2/(2\nu_2))^\frac{1}{4}\}$. Following \citet{bagheri2009input} and \citet{towne2018spectral}, we use $N_H = 221$, leading to a computational domain $x \in [-85.19, 85.19]$, which is large enough to mimic an infinite domain. The boundary conditions are implicitly satisfied through the use of Hermite polynomials \citep{bagheri2009input}. For CS-SPOD, the value of the weighting matrix at $x_i$ is determined as the distance between the midpoints of the neighbouring grid points. Temporal integration is performed using the embedded $5^{th}$ order Dormand–Prince Runge-Kutta method \citep{dormand1980family,shampine1997matlab}. After the initial transients have decayed, a total of 40000 solution snapshots are saved with $\Delta t = 0.5$, giving a Nyquist frequency of $f_{\text{Nyquist}} = 1$. \par

To mimic a turbulent system, similar to \citet{towne2018spectral}, we force our system using spatially correlated band-limited noise. This is performed by constructing spatially correlated noise with the following covariance kernel 
\begin{equation}
    g(x, x^\prime) = \frac{1}{\sqrt{2\pi}\sigma_\eta} \text{exp}\left[ -\frac{1}{2} \left(\frac{x - x^\prime}{\sigma_\eta} \right)^2\right] \text{exp}\left[ -i2\pi \frac{x - x^\prime}{\lambda_\eta}\right],
    \label{eqn:spatialkernel}
\end{equation}
where $\sigma_\eta$ is the standard deviation of the envelope and $\lambda_\eta$ is the wavelength of the filter. Spatial correlation is introduced by multiplying white noise by the Cholesky decomposition of the covariance kernel. The white noise has a uniform phase, normally distributed amplitude with unit variance, and is generated as in \citet{towne2018spectral}. The forcing is spatially restricted to an interior portion of the domain via the window $\text{exp}[-(x/L)^p]$, where $L = 60, p = 10$. The spatially correlated noise is low-pass filtered using a $10^{th}$-order finite-impulse-response filter with a cutoff frequency equal to $0.6f_{\text{Nyquist}}$. This results in a stationary forcing that is approximately constant in amplitude up to the cutoff frequency ($-6\text{dB}$ in amplitude at the cutoff frequency) but has non-zero spatial correlation as defined by \eqref{eqn:spatialkernel}. The forcing is then linearly interpolated to the temporal locations required by the temporal integration. To compute the WV spectrum, SPOD, and CS-SPOD, we employed a window length $N_w = 10N_\theta$ and an overlap $67\%$, resulting in $N_b = 595$ (realizations) of the process and a frequency discretization of $\Delta f= 0.01$.

\begin{figure}
\centering
    \begin{subfigure}[b]{0.33\textwidth}
        \centering
        \includegraphics[height=1.1\textwidth]{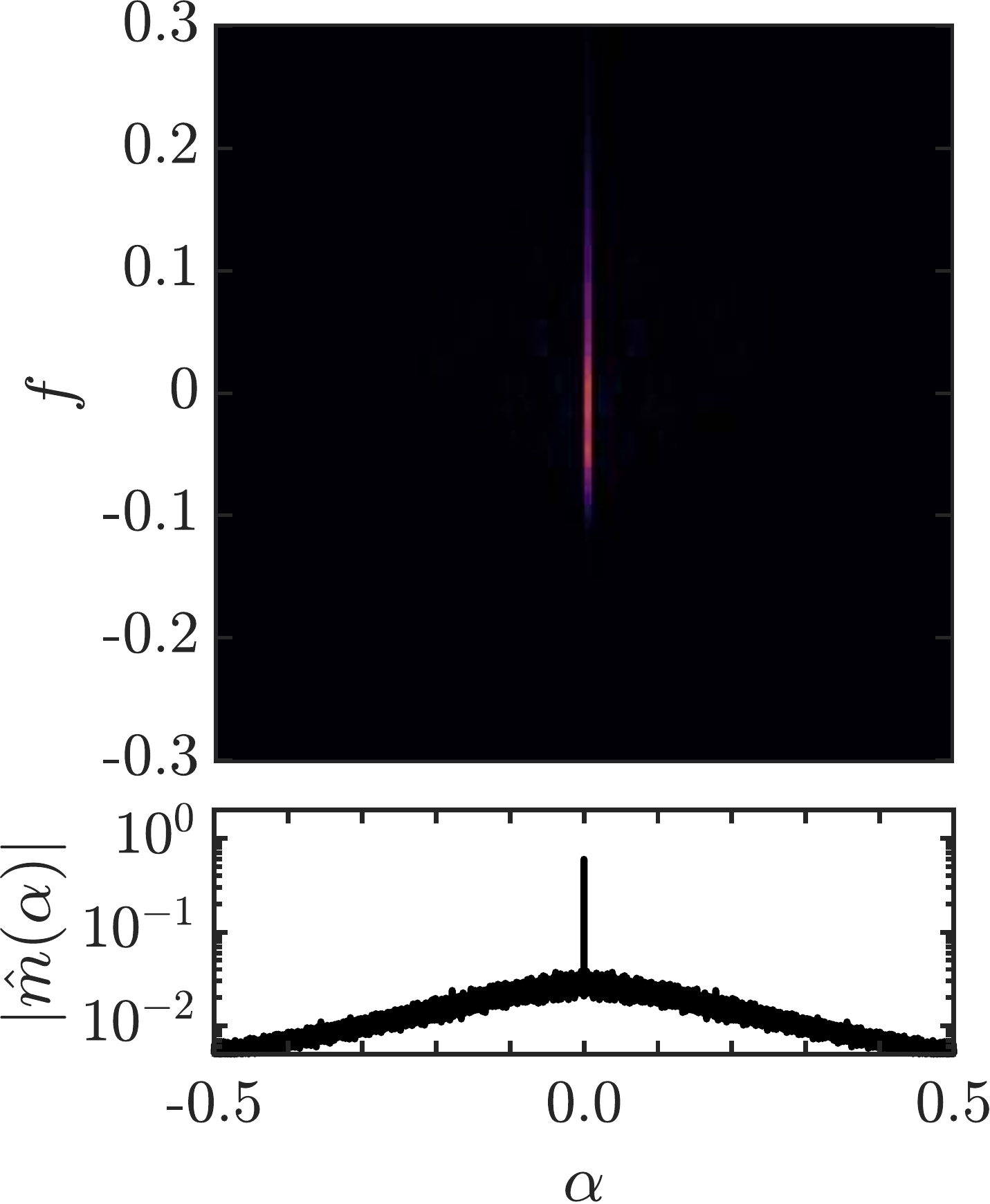}
        \caption{$A_{\mu_0} = 0$}
    \end{subfigure}% 
    \begin{subfigure}[b]{0.33\textwidth}  
        \centering 
       \includegraphics[height=1.1\textwidth]{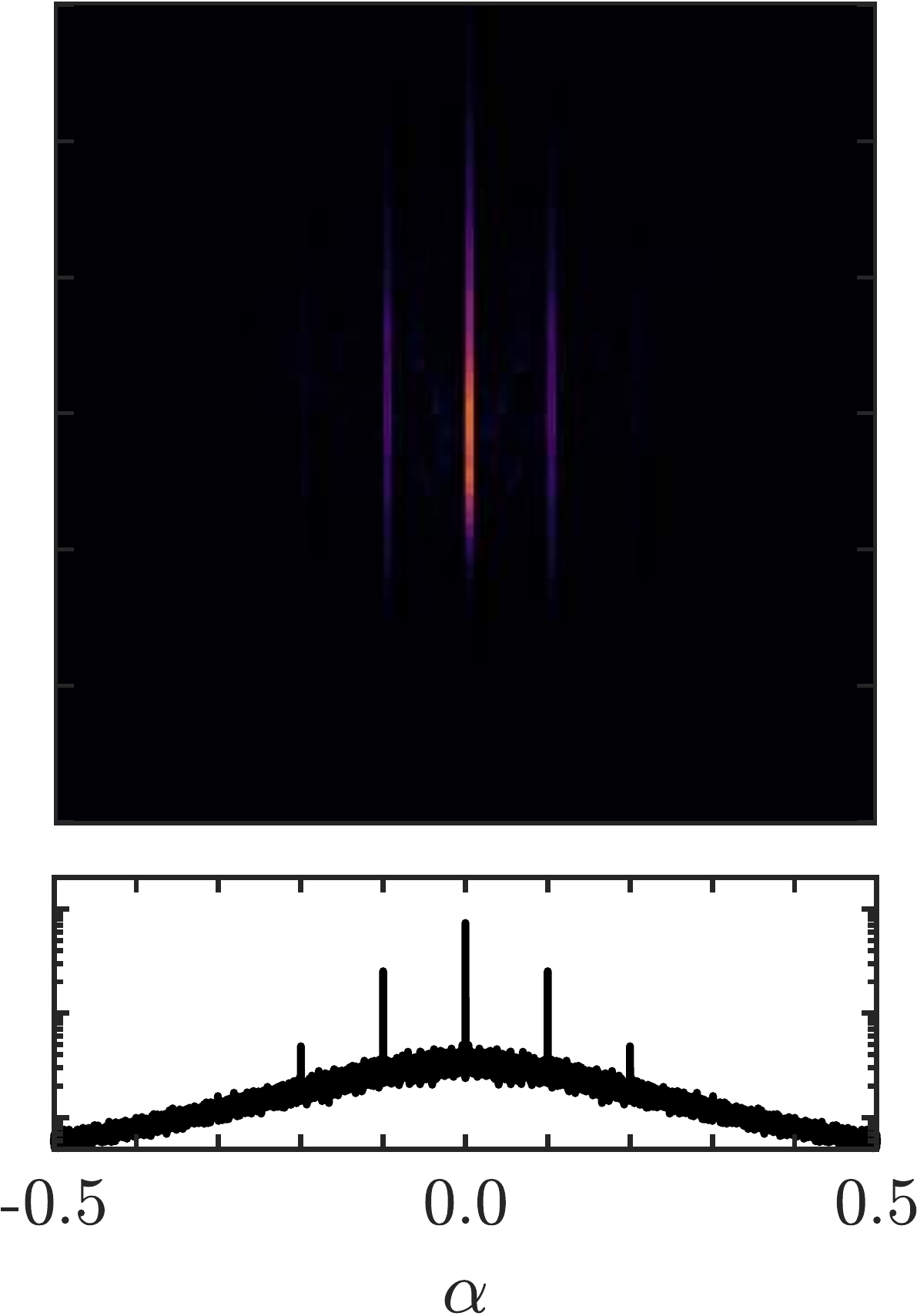}
       \caption{$A_{\mu_0} = 0.2$}
    \end{subfigure}%
    \begin{subfigure}[b]{0.33\textwidth}  
        \centering 
        \includegraphics[height=1.1226\textwidth]{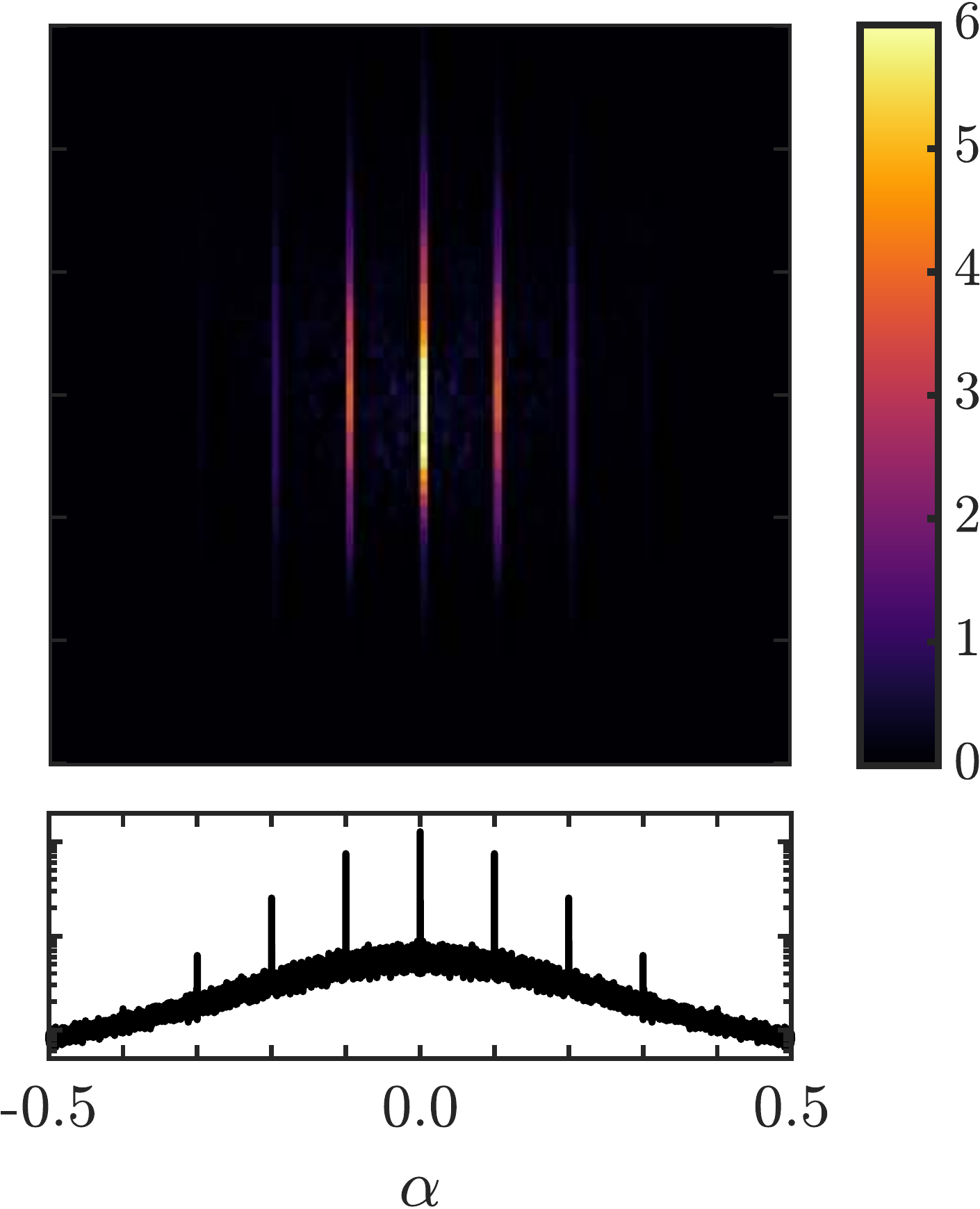}
        \caption{$A_{\mu_0} = 0.4$}
    \end{subfigure}
       \caption{CCSD (top) and integrated CCSD (bottom) for the Ginzburg-Landau system at $x = 0$.}
    \label{fig:CCSDint}
\end{figure}

In analyzing the fabricated data, we must first determine those frequencies, if any, where the system exhibits cyclostationarity. To do this, we compute the CCSD and search over all possible values of $\alpha$ in the range of possible cycle frequencies $\alpha \in [-1, 1]$, noting the $\alpha$ discretization required as discussed in \S \ref{sec:CStheory} to ensure no possible cycle frequencies are missed. Figure \ref{fig:CCSDint} shows the CCSD and integrated CCSD for the three values of $A_{\mu_0}$ at $x = 0$, and confirms that the system is cyclostationary when $A_{\mu_0} > 0$ as high values of the CCSD and the integrated CCSD are seen at $\alpha = 0$, the modulation frequency ($f_0$), and an increasing number of harmonics as $A_{\mu_0}$ is further increased. 

\begin{figure}
\centering
    \begin{subfigure}[b]{0.33\textwidth}
        \raggedright
        \hspace{2mm}
        \includegraphics[height=0.648\textwidth]{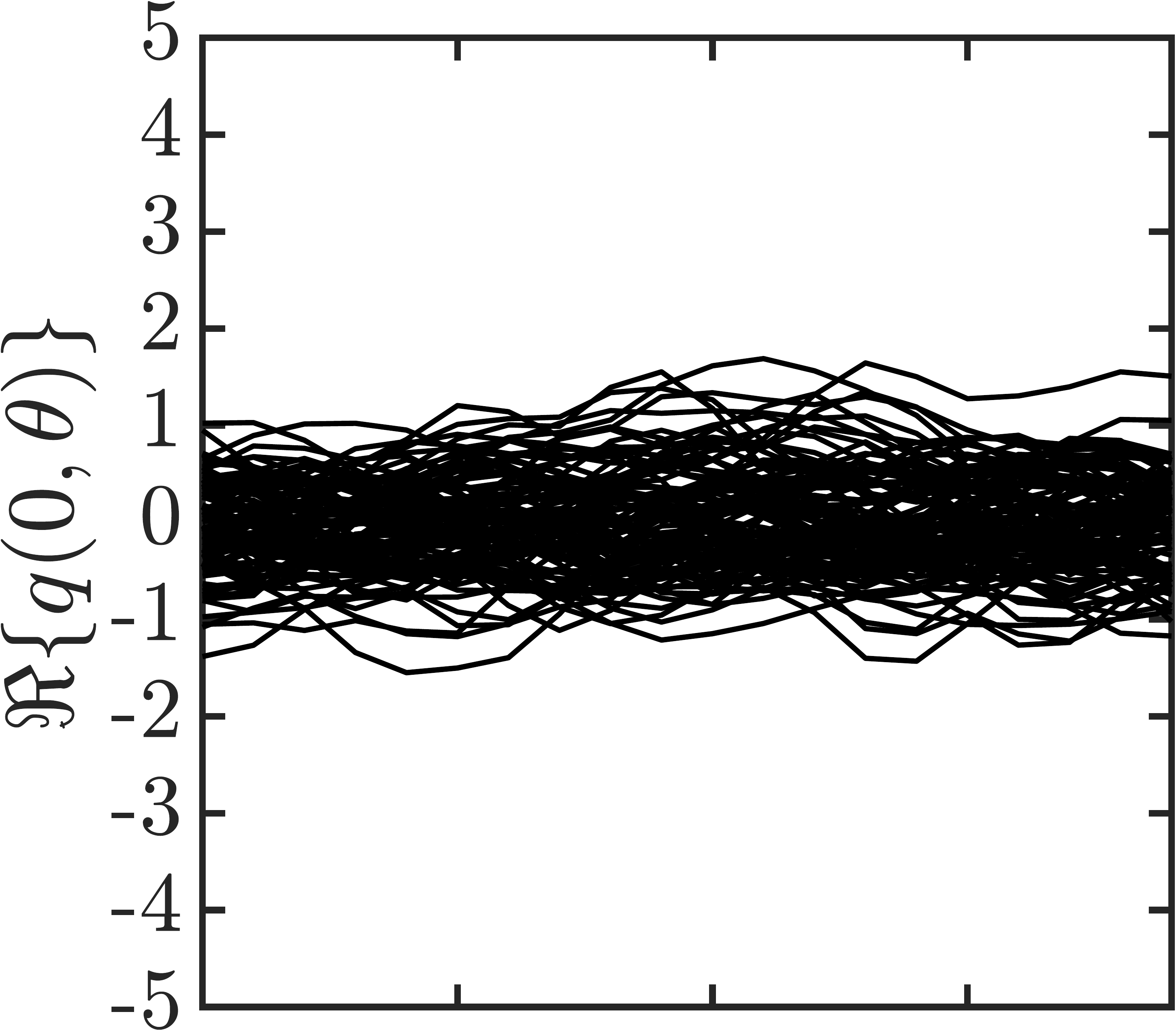}
    \end{subfigure}% 
    \begin{subfigure}[b]{0.33\textwidth}  
        \raggedright 
        \hspace{7mm}
       \includegraphics[height=0.62\textwidth]{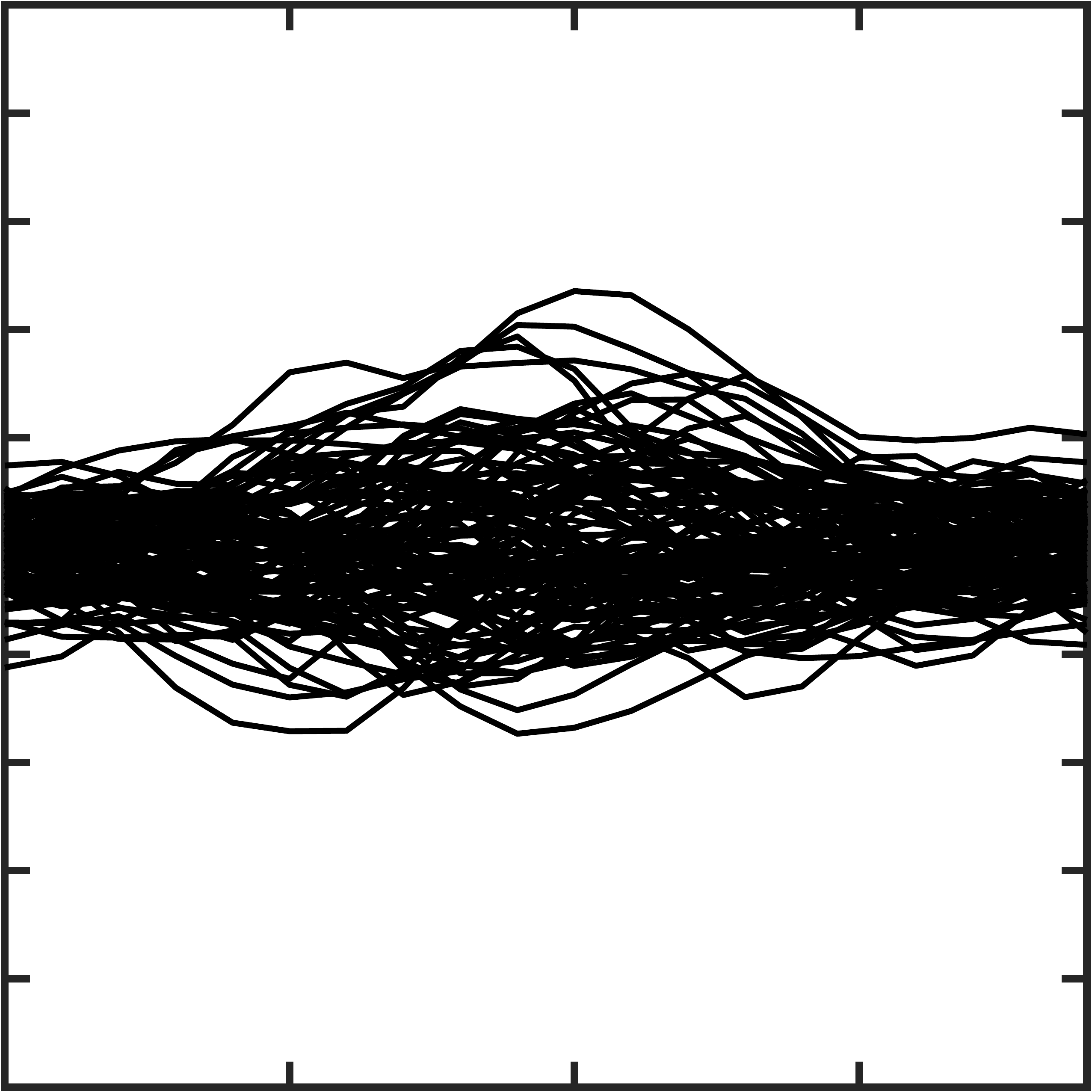}
    \end{subfigure}%
    \begin{subfigure}[b]{0.33\textwidth}  
        \raggedright 
        \hspace{7mm}
        \includegraphics[height=0.62\textwidth]{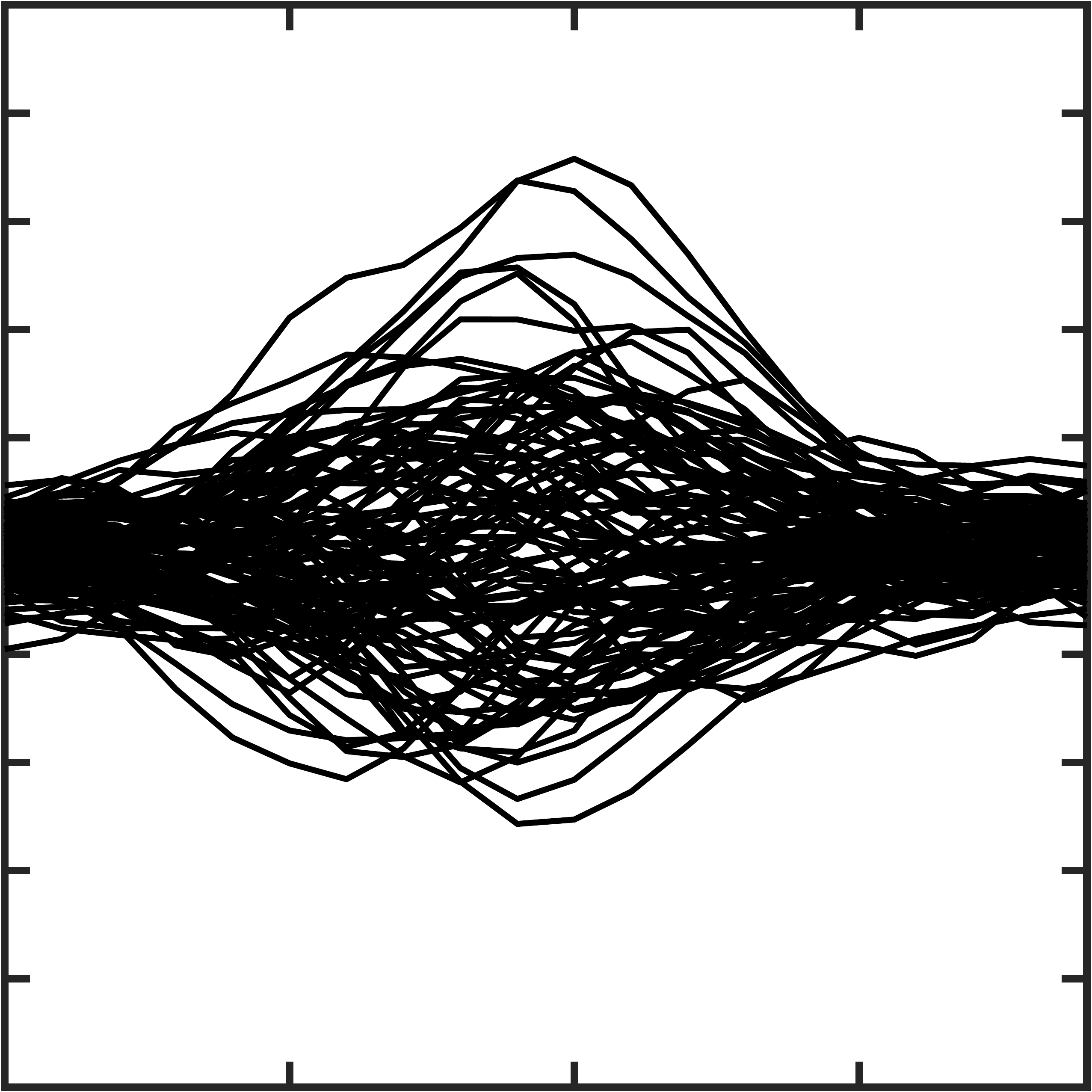}
    \end{subfigure} \\ \vspace{0.5mm} 
    \begin{subfigure}[b]{0.33\textwidth}
        \raggedright
        \includegraphics[height=0.75\textwidth]{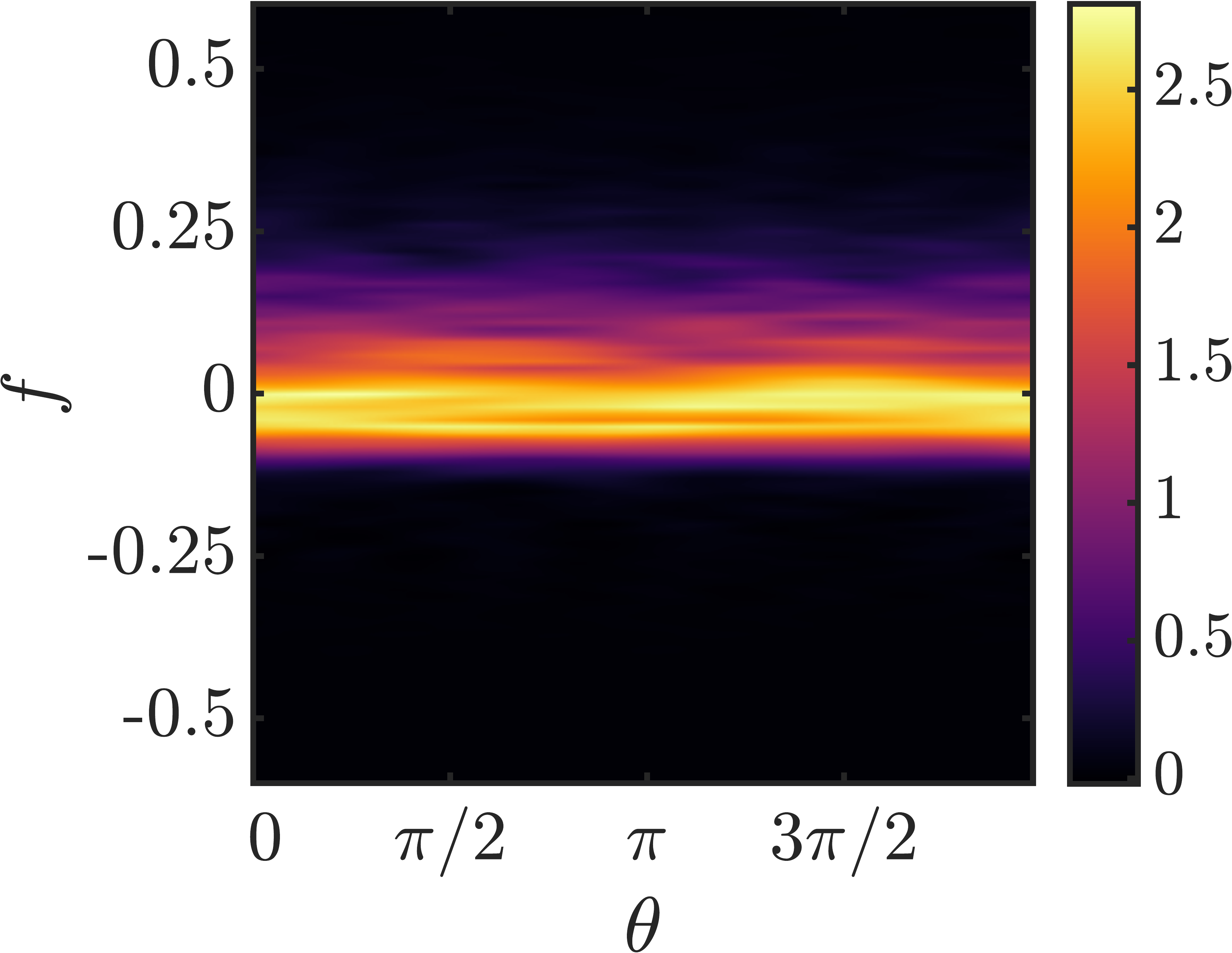}
        \caption{$A_{\mu_0} = 0$} 
    \end{subfigure}% 
    \begin{subfigure}[b]{0.33\textwidth}  
        \raggedright 
        \hspace{7mm}
       \includegraphics[height=0.75\textwidth]{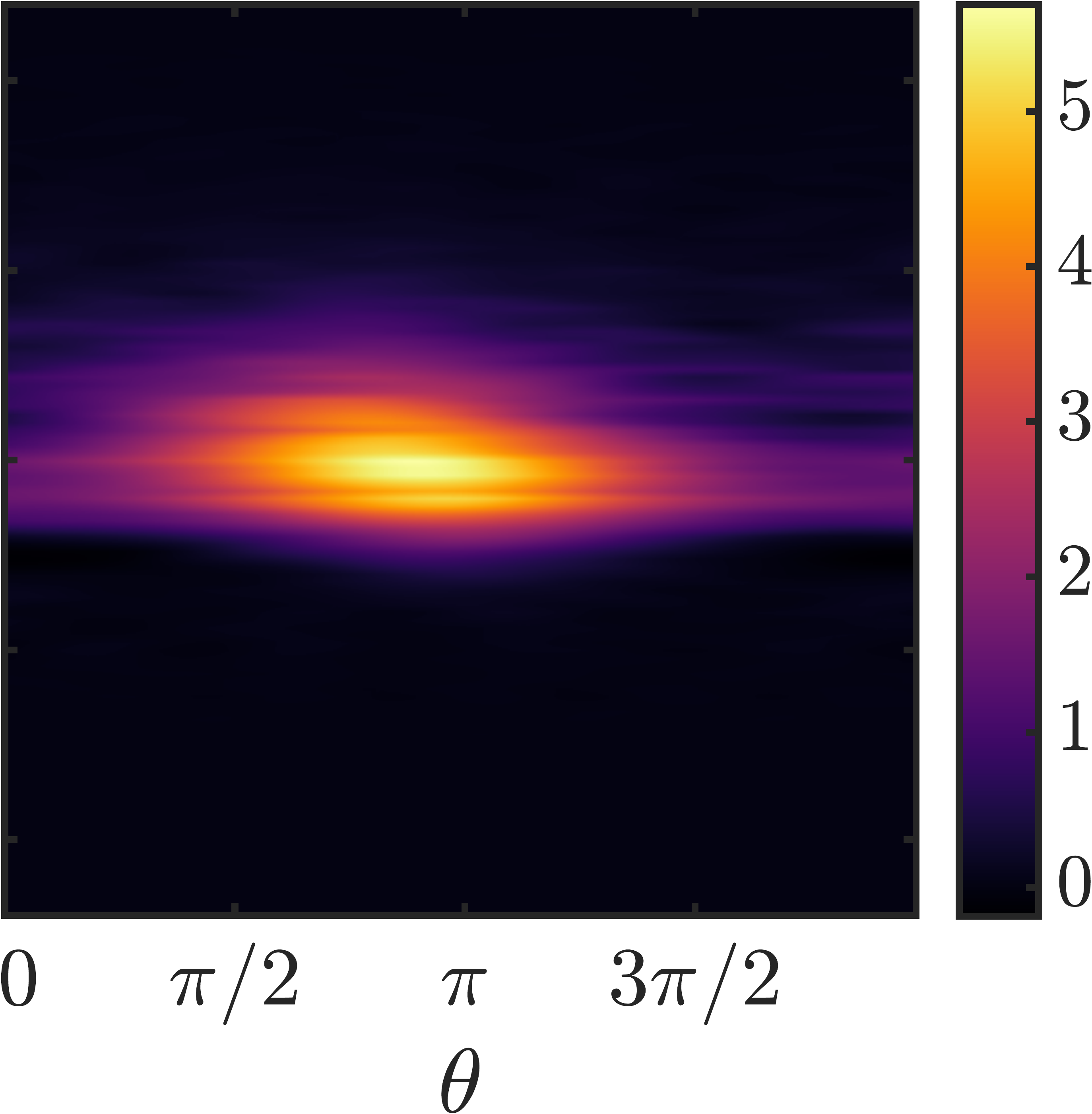}
       \caption{$A_{\mu_0} = 0.2$}
    \end{subfigure}%
    \begin{subfigure}[b]{0.33\textwidth}  
        \raggedright 
        \hspace{7mm}
        \includegraphics[height=0.75\textwidth]{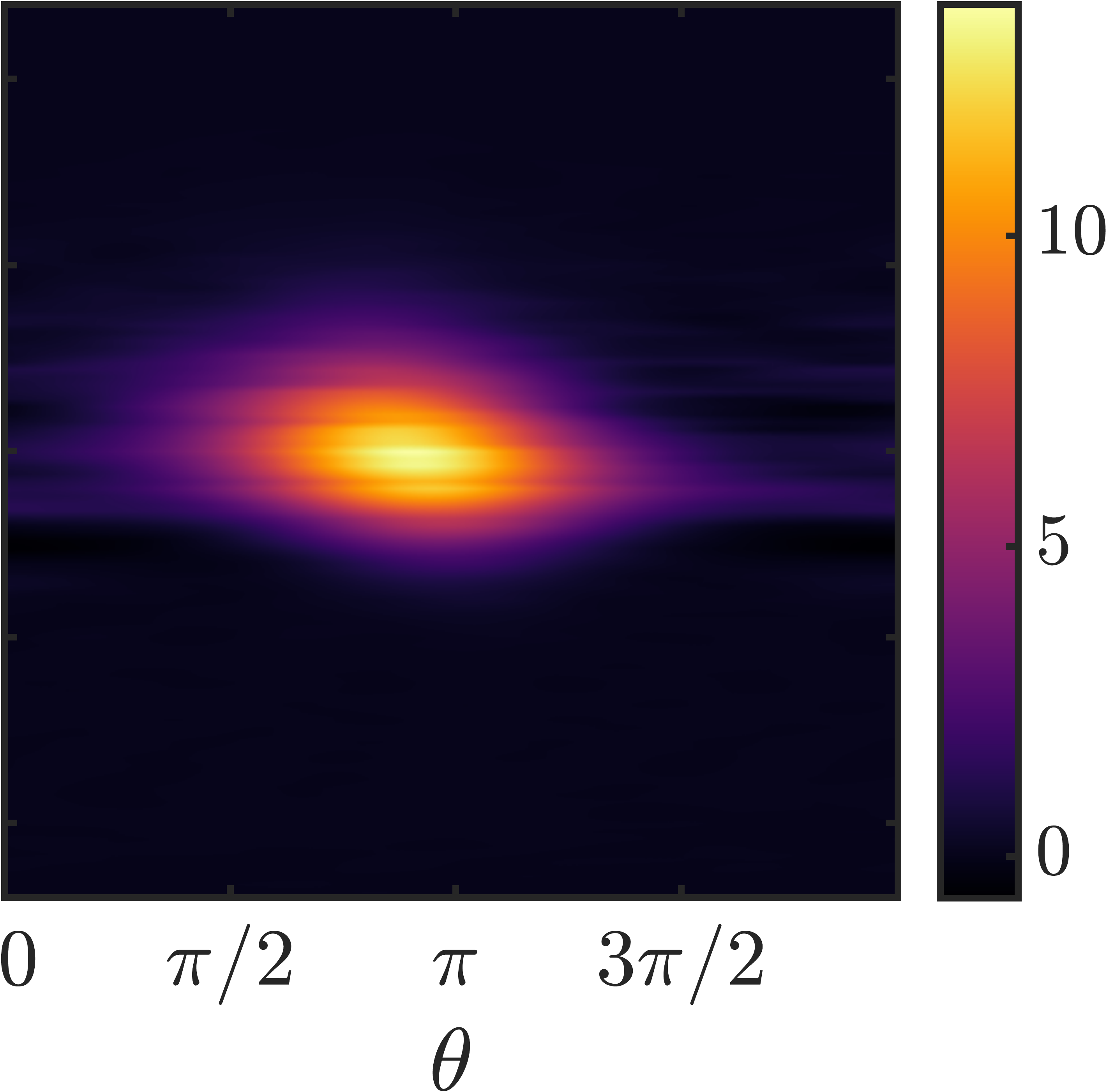}
       \caption{$A_{\mu_0} = 0.4$}
    \end{subfigure}
       \caption{Example Ginzburg-Landau sample paths (top) and WV spectrum at $x = 0$ (bottom).}
    \label{fig:GLpaths2}
\end{figure}

We show 100 realizations of the process for each $A_{\mu_0}$ along with the WV spectrum at $x = x^\prime = 0$ as a function of the phase of $\alpha_0$ in figure \ref{fig:GLpaths2}. The WV spectrum is computed using $a_2 = 5$ to encompass all cycle frequencies present. Figure \ref{fig:GLpaths2} (a) shows that the statistics are almost constant as a function of phase for $A_{\mu_0} = 0$, which is expected given the time-invariant dynamics. The small degree of modulation observed is due to statistical error. In figures \ref{fig:GLpaths2} (b, c), we observe increasing levels of modulation in the statistics as $A_{\mu_0}$ increases. Furthermore, the peak value of the spectrum also increases due to the increasing non-normality of the system with increasing $\mu_0$. Given that the largest value of $\mu_0$ occurs at $\theta = 0.5\pi$ and the peak of the WV spectrum occurs at $\theta \approx 0.95\pi$, there is a phase delay of $\approx 0.45\pi$ between when the dynamics of the system are the least stable and when the perturbations are, on average, the largest.

Based on the preceding analysis and to ensure we encompass all frequencies of interest, we compute CS-SPOD using $a_1 = 5$, resulting in a frequency range of $\Fset_{\fci} = [-0.5, 0.5] + \fci$.  We first consider the stationary process with $A_{\mu_0} = 0.0$. Although CS-SPOD modes are theoretically equivalent to SPOD for the stationary case, finite data length leads to differences. 

\begin{figure}
        \centering
        \includegraphics[width=0.65 \textwidth]{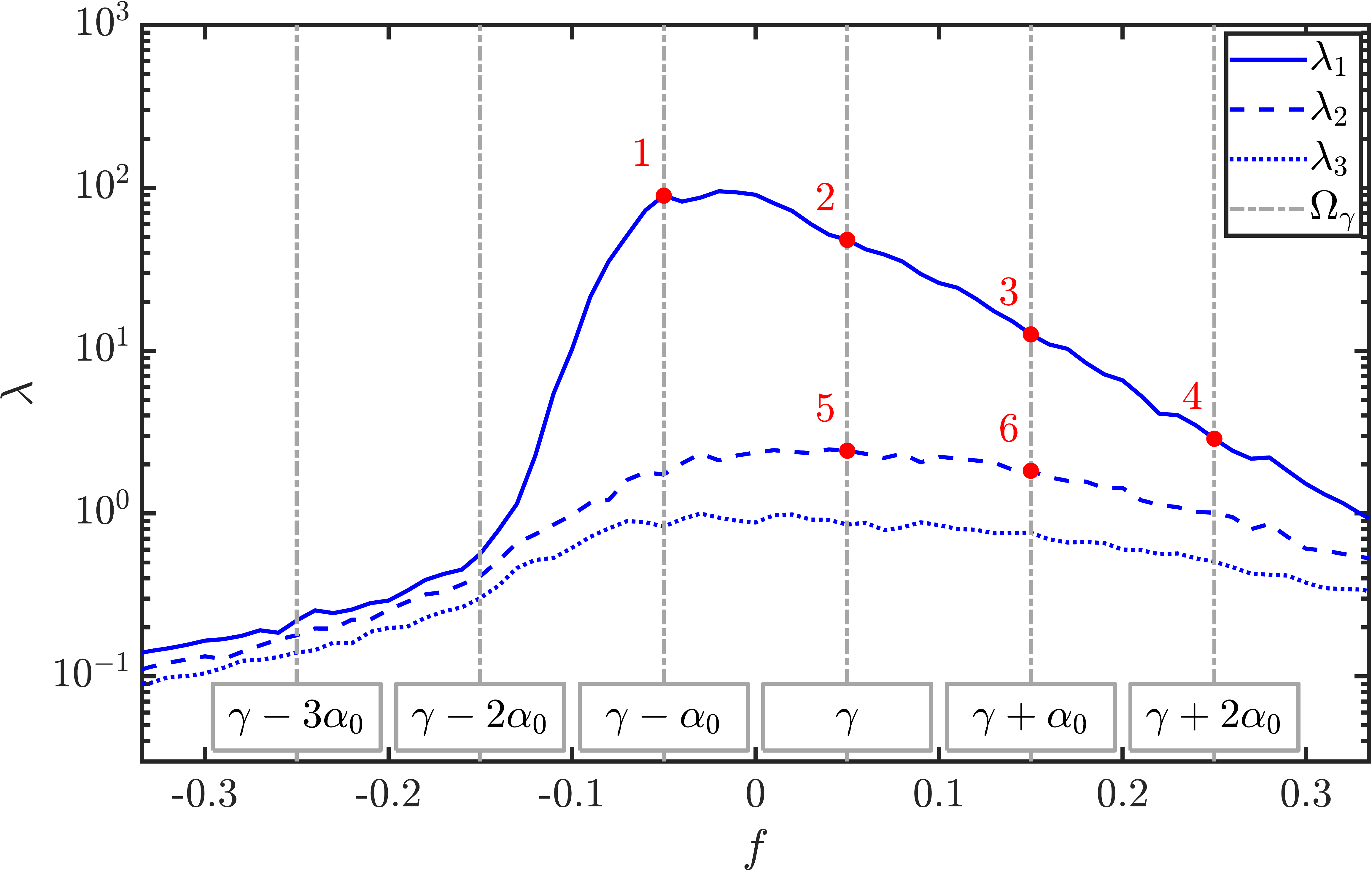}
        \caption{SPOD eigenspectrum for the Ginzburg-Landau system at $A_{\mu_0} = 0.0$ showing the three most energetic modes at each discrete frequency $f$. The 6 highest-energy modes occurring at the frequencies present in the CS-SPOD solution frequencies, i.e. $f \in \Fset_{\fci}$, are depicted with the red dots.}
        \label{fig:GLSPOD}
\end{figure}

\begin{figure}
\centering
\includegraphics[height=0.078\textwidth]{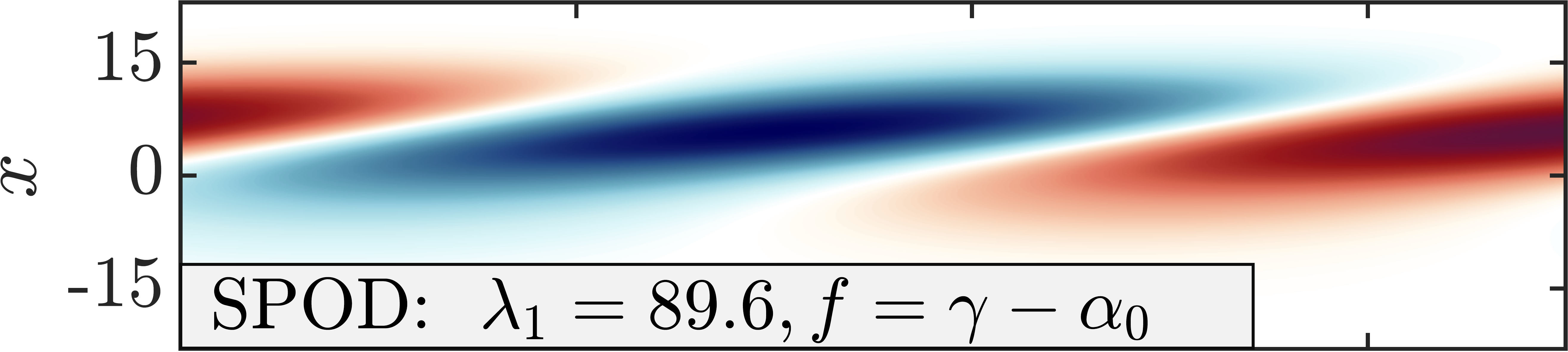}
\includegraphics[height=0.078\textwidth]{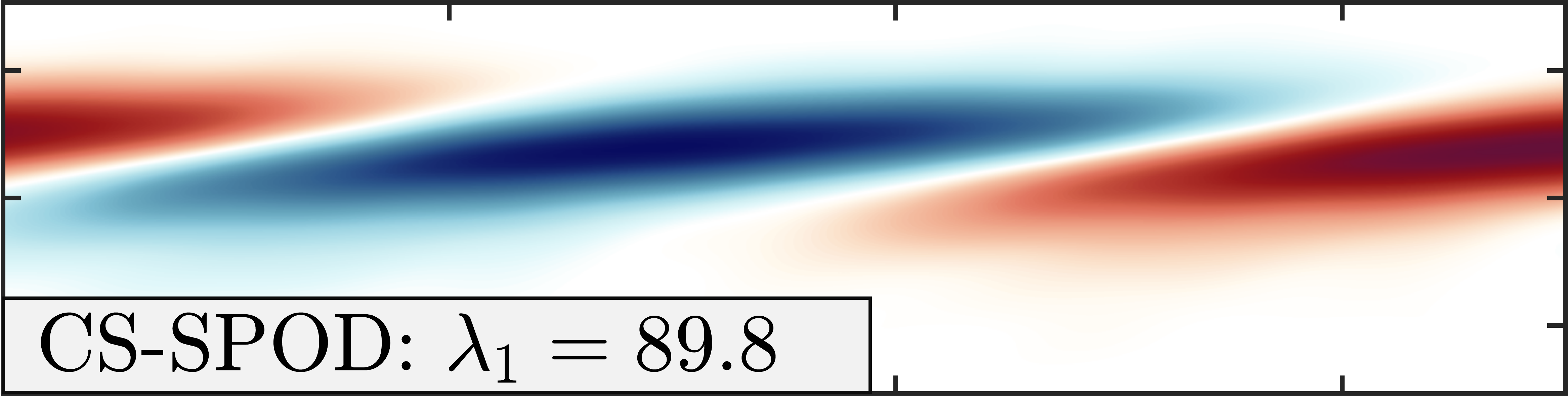} \vspace{0.8mm} \\
\includegraphics[height=0.078\textwidth]{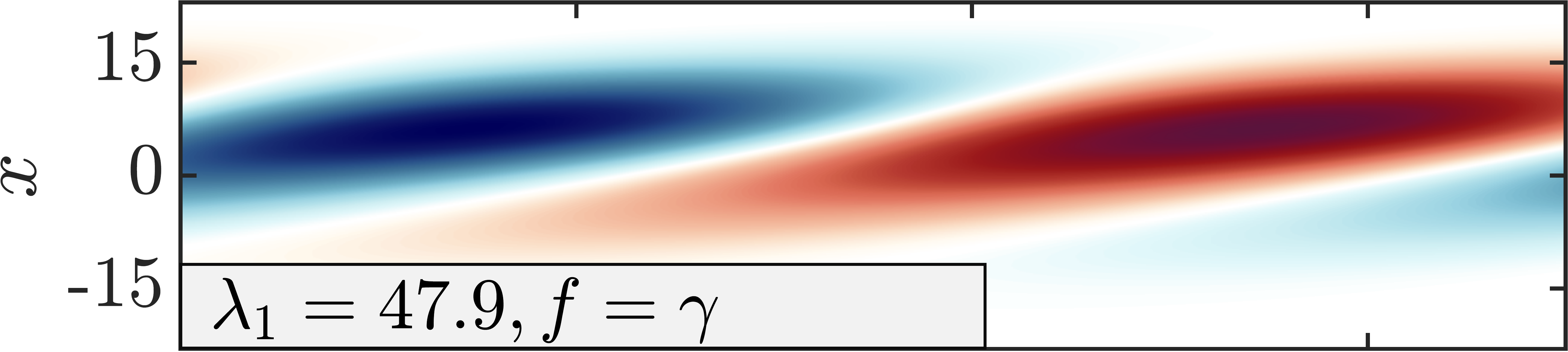}
\includegraphics[height=0.078\textwidth]{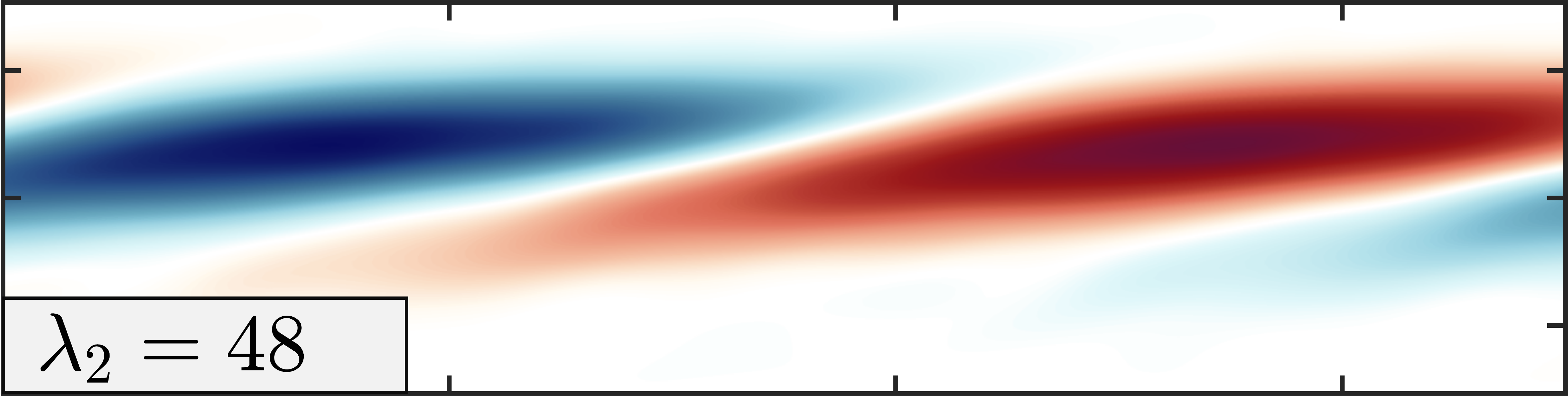} \vspace{0.8mm} \\
\includegraphics[height=0.078\textwidth]{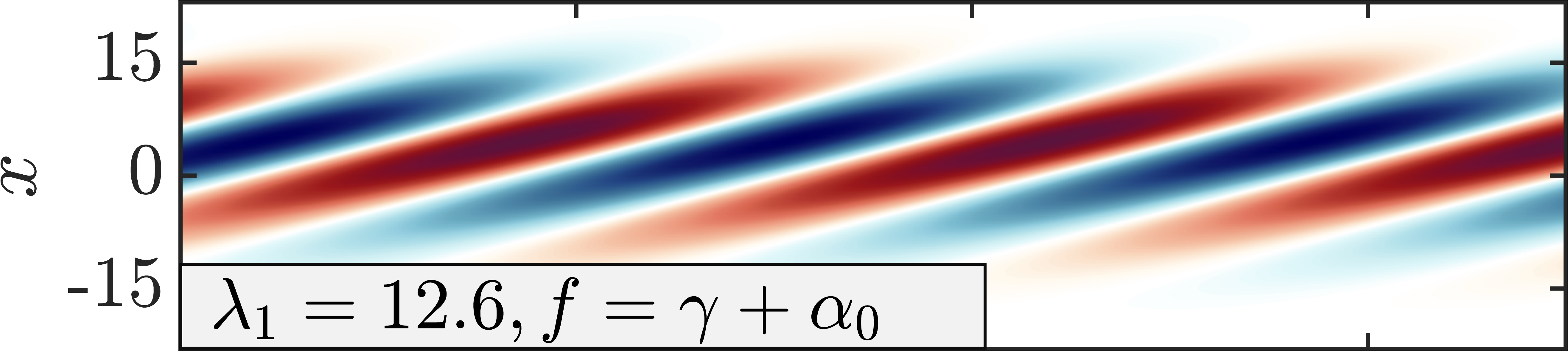}
\includegraphics[height=0.078\textwidth]{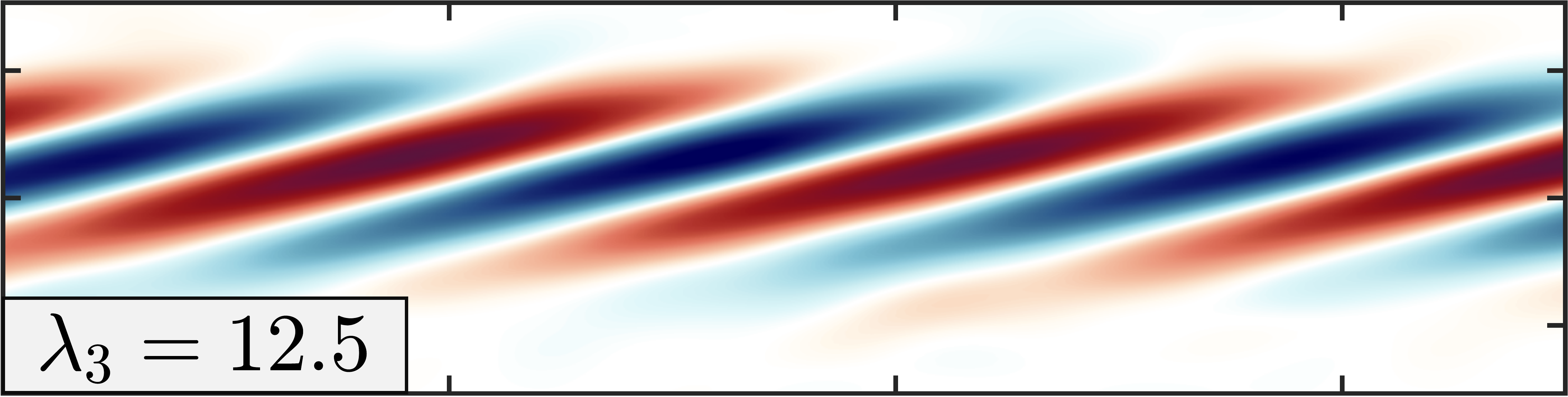} \vspace{0.8mm} \\
\includegraphics[height=0.078\textwidth]{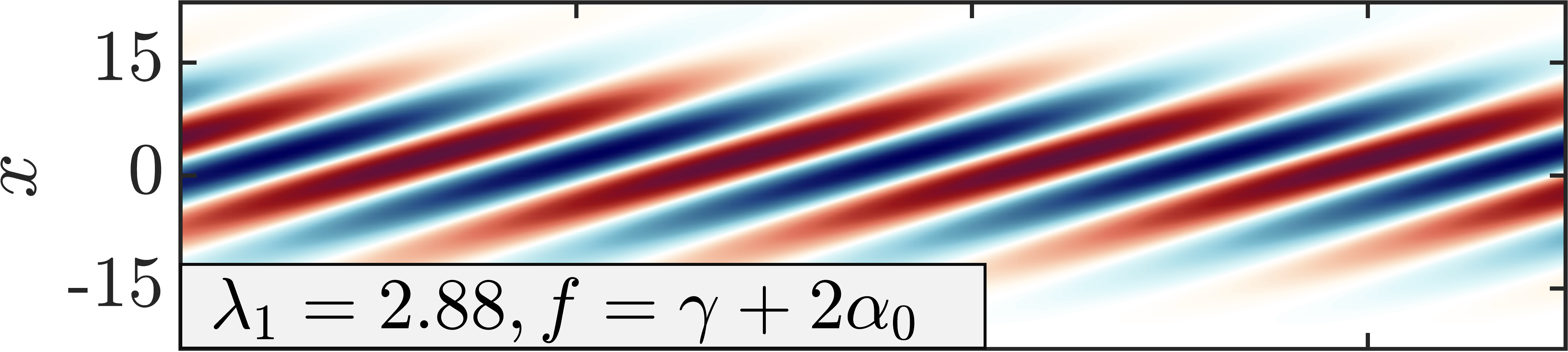}
\includegraphics[height=0.078\textwidth]{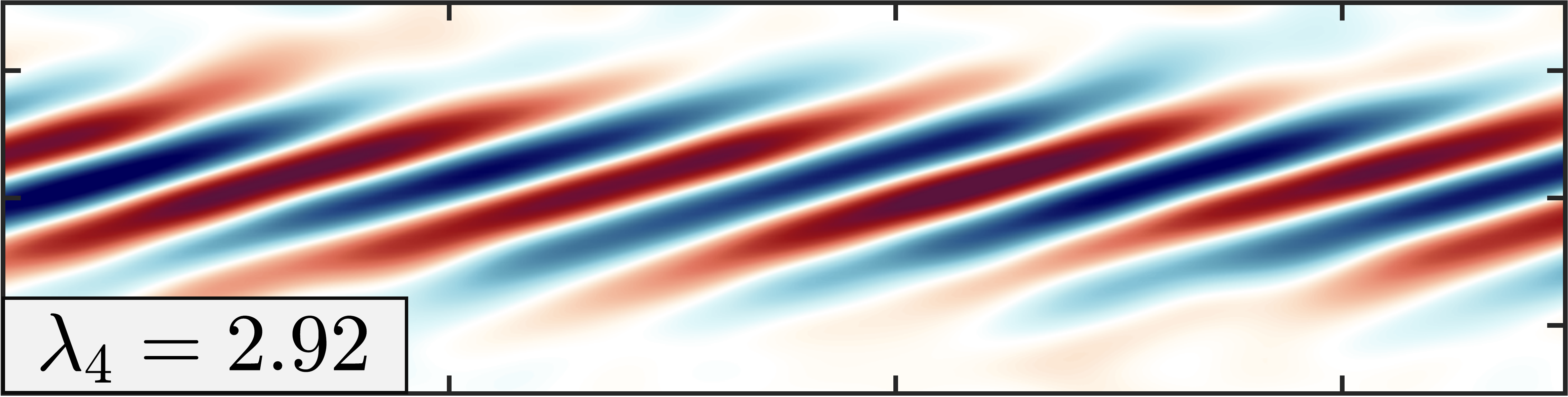} \vspace{0.8mm} \\
\includegraphics[height=0.078\textwidth]{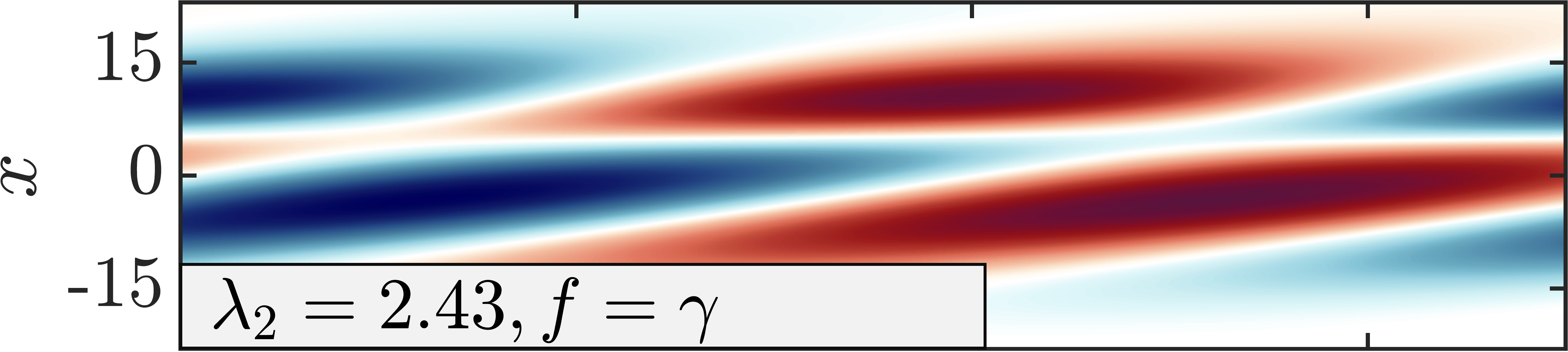}
\includegraphics[height=0.078\textwidth]{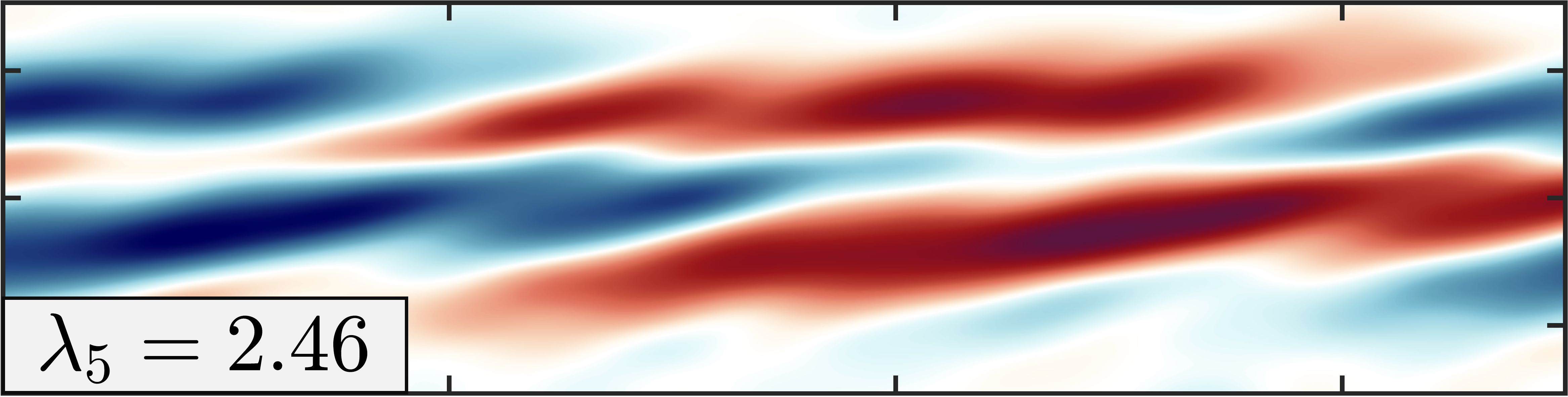} \vspace{0.8mm} \\
\includegraphics[height=0.115636364\textwidth]{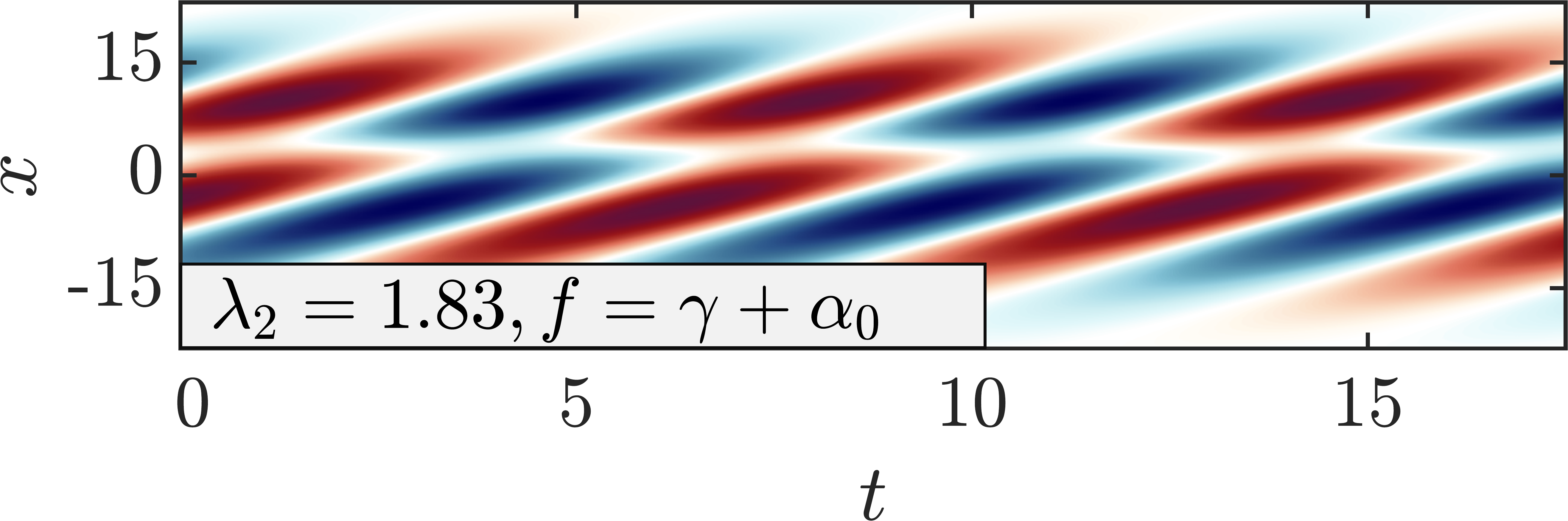}
\includegraphics[height=0.115636364\textwidth]{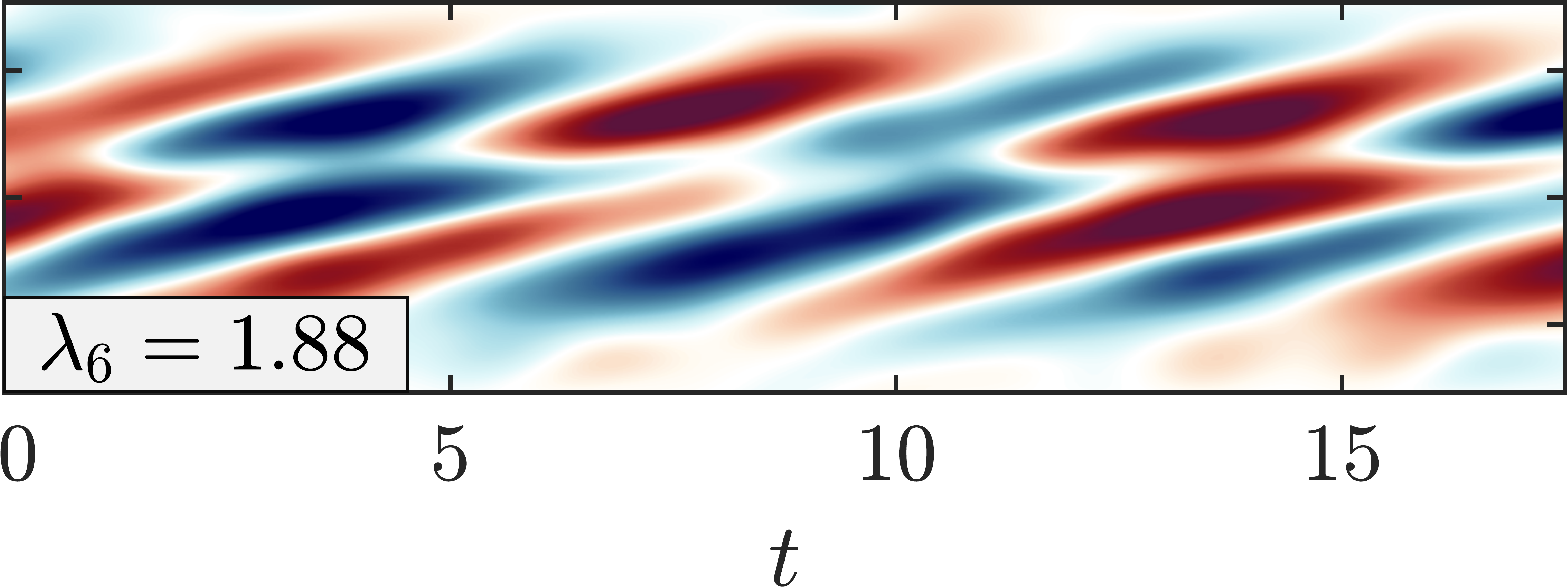} \hfill
        \caption{Comparison of SPOD (left) and CS-SPOD modes (right) for the Ginzburg-Landau system at $A_\mu = 0$.   From top to bottom are the six most dominant CS-SPOD modes and the six points identified in figure~\ref{fig:GLSPOD}. The contour limits for the CS-SPOD eigenfunctions are set equal to corresponding SPOD mode $\pm ||\psib_j(\x, t)||_\infty$.}
    \label{fig:CSSPODModes}
\end{figure}

Figure \ref{fig:GLSPOD} shows the SPOD eigenspectrum for $A_{\mu_0} = 0.0$.  Note that the spectrum is not symmetric in $f$ because the Ginzburg-Landau system is complex. We superpose on the SPOD spectra the set of frequencies $f \in \Fset_{\fci}$ for $\fci = 0.05$, and mark and rank the 6 intersections with the highest energy.  Based on the plot, we should find that the 4 most dominant CS-SPOD modes correspond to the dominant SPOD mode at a frequency of $\fci - \alpha_0, \fci, \fci + \alpha_0$, and $\fci + 2\alpha_0$, respectively. Similarly, the $5^{th}$ and $6^{th}$ CS-SPOD modes should correspond to the first subdominant SPOD modes at a frequency of $\fci$ and $ \fci + \alpha_0$, respectively. Figure \ref{fig:CSSPODModes} makes comparisons between SPOD and CS-SPOD (performed assuming a fundamental cycle frequency of $\alpha_0 = f_0$) for the energy and eigenfunctions for each of these six modes. While the results are quite similar in each case, there are differences associated with statistics convergence, and this, as expected, occurs when there is a small energy separation between two distinct modes (e.g. modes 5 and 6).  
 % Script is SPOD_GL_NoInterp_Plotter

\begin{figure}
\centering
    \begin{minipage}[t]{0.44\textwidth}
            \centering
            \vskip 0pt
            \includegraphics[height=48mm,valign=t]{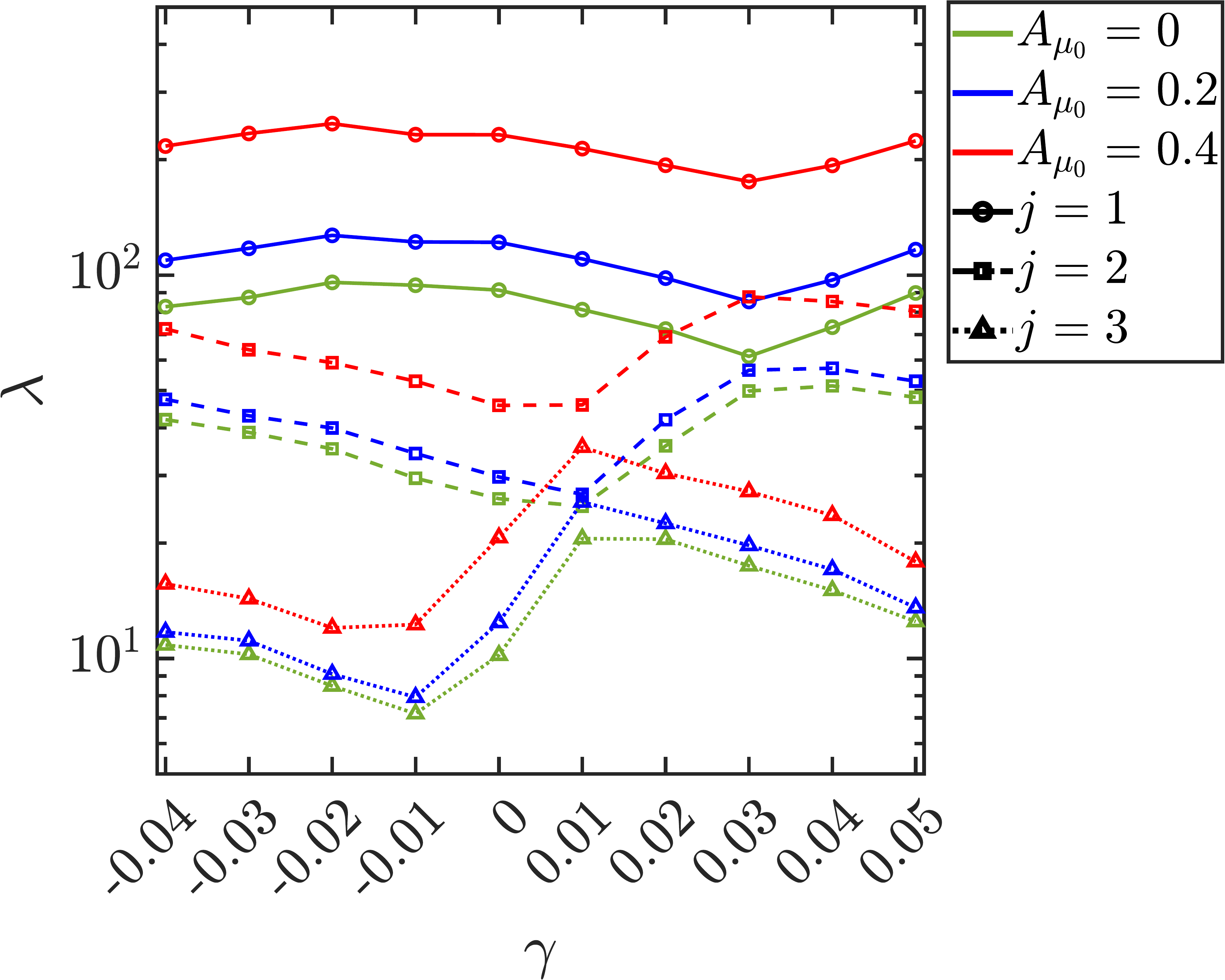}
            \caption{CS-SPOD energy spectrum of the three Ginzburg-Landau systems. }
            \label{fig:csespecgl}
    \end{minipage}\hfill
    \begin{minipage}[t]{0.53\textwidth}
            \centering
            \vskip 0pt
            \includegraphics[height=48mm,valign=t]{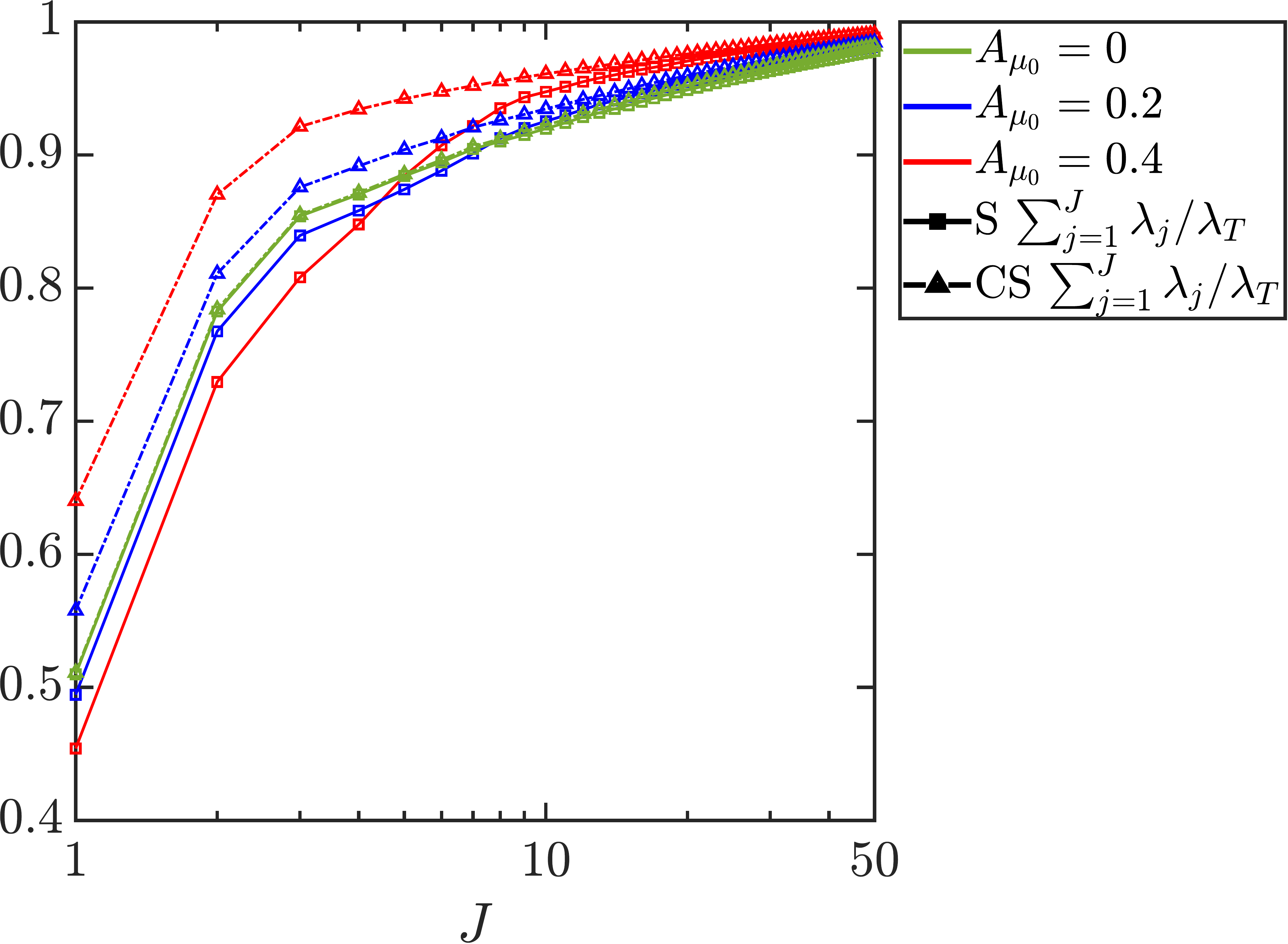}
            \caption{Total fractional energy captured by a truncated set of CS-SPOD (CS) and SPOD (S) modes for the three Ginzburg-Landau systems at $\fci = 0.05$.}
            \label{fig:scstotalenergygl}
    \end{minipage}
\end{figure}

In figure \ref{fig:csespecgl}, we now compare the CS-SPOD eigenspectrum for all $\fcik \in \Fset_\fci$ for the three different values of $A_{\mu_0}$. As $A_{\mu_0}$ increases, so does the energy, as the disturbances are increasingly amplified by the increasing non-normality of the linear operator at phases corresponding to positive $A_{\mu_0} \text{sin}(2\pi f_0t)$, consistent with the trend shown previously in figure \ref{fig:GLpaths2}.  A large energy separation between the dominant and sub-dominant CS-SPOD modes is observed, which increases for greater $A_{\mu_0}$, indicating that the process is increasingly low rank. In figure \ref{fig:scstotalenergygl}, for $\fci = 0.05$, we show the fraction of the total energy ($\lambda_T = \sum_{j} \lambda_j$) that the first $J$ CS-SPOD or SPOD modes recover. As theoretically expected for $A_{\mu_0} =0$, CS-SPOD and SPOD result in almost identical energy distribution. In contrast, with increasing $A_{\mu_0}$, CS-SPOD captures an increasingly greater amount of energy than SPOD. For example, at $A_{\mu_0} = 0.4$, the first CS-SPOD mode captures $64\%$ of the total energy, while the first SPOD mode captures just $45\%$. Furthermore, the first three CS-SPOD modes capture $92\%$ of the total energy, while seven SPOD modes are required to capture a similar amount of energy. As theoretically expected, the energy captured by SPOD does not exceed the energy captured by CS-SPOD (since SPOD modes are a subset of CS-SPOD modes). Thus, as the statistics become increasingly cyclostationary (i.e. more phase-dependent), CS-SPOD is able to capture an increasingly larger fraction of the phase-dependent statistics present in the process, which SPOD, due to the fundamentally flawed assumption of statistical stationarity, is unable to achieve. \par

\begin{figure}
        \includegraphics[height=0.078\textwidth]{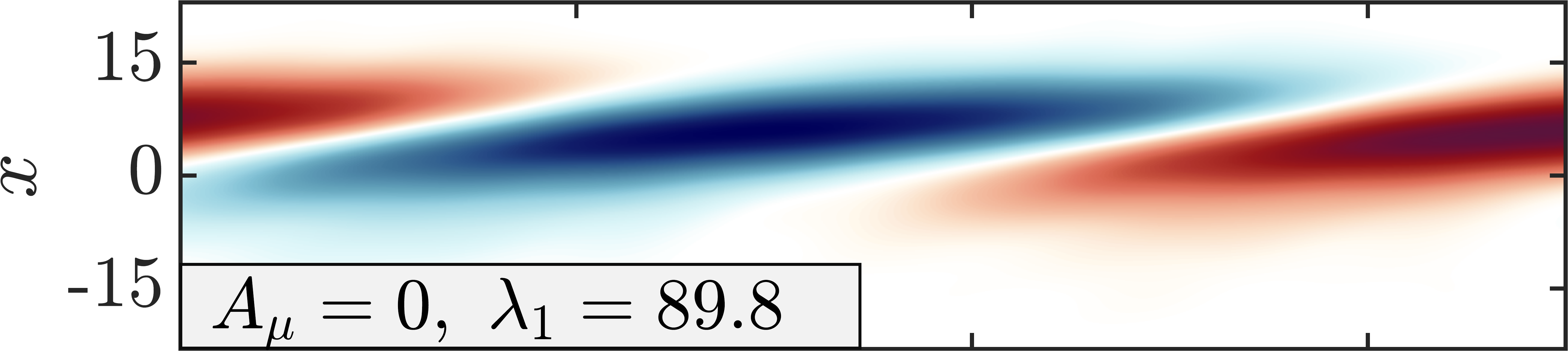}
        \includegraphics[height=0.078\textwidth]{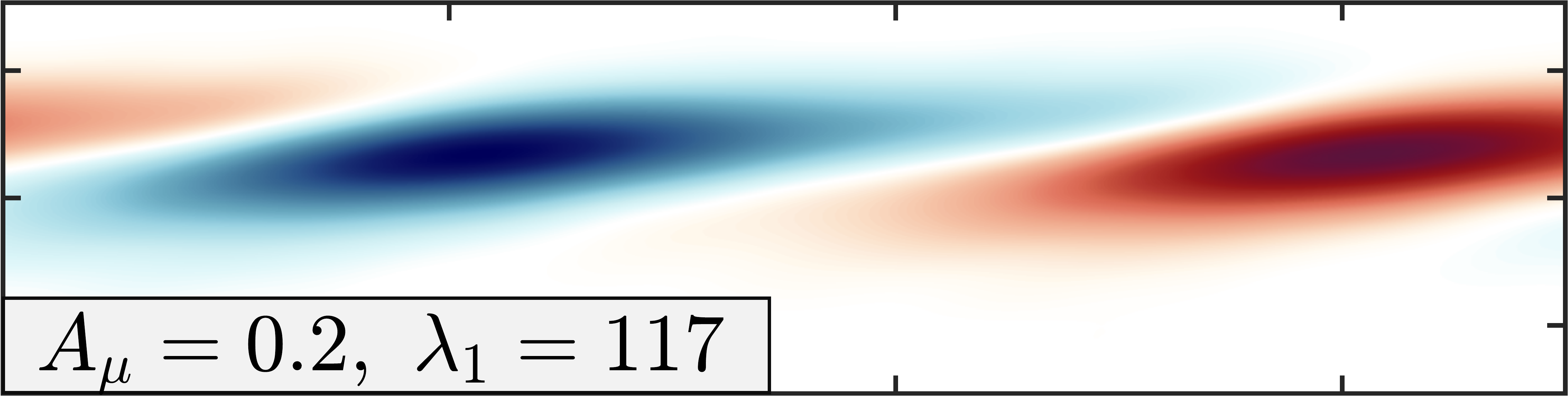}
        \includegraphics[height=0.078\textwidth]{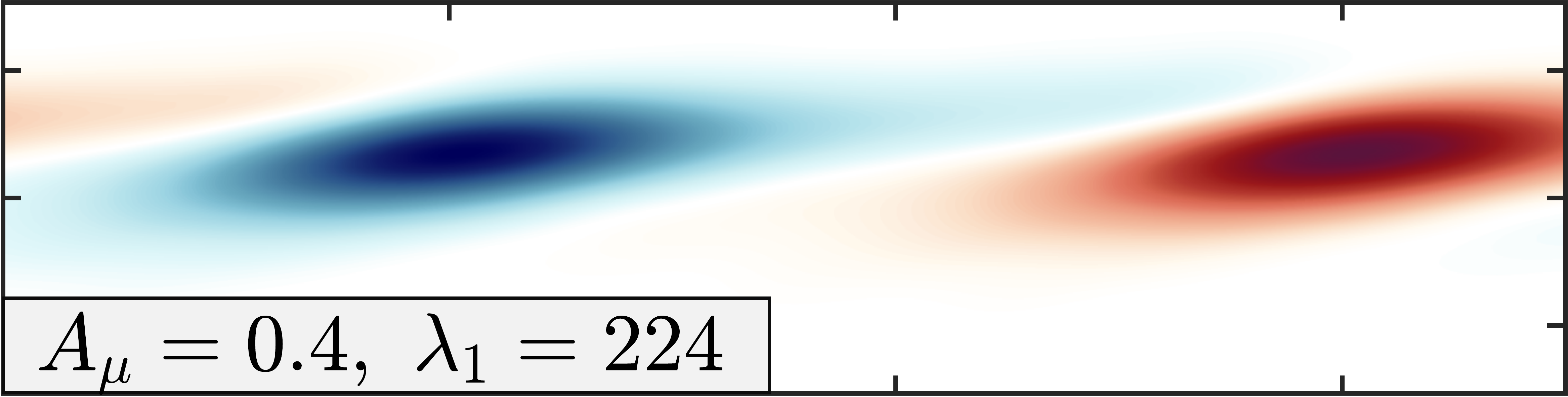}\vspace{1mm}\\
        \includegraphics[height=0.078\textwidth]{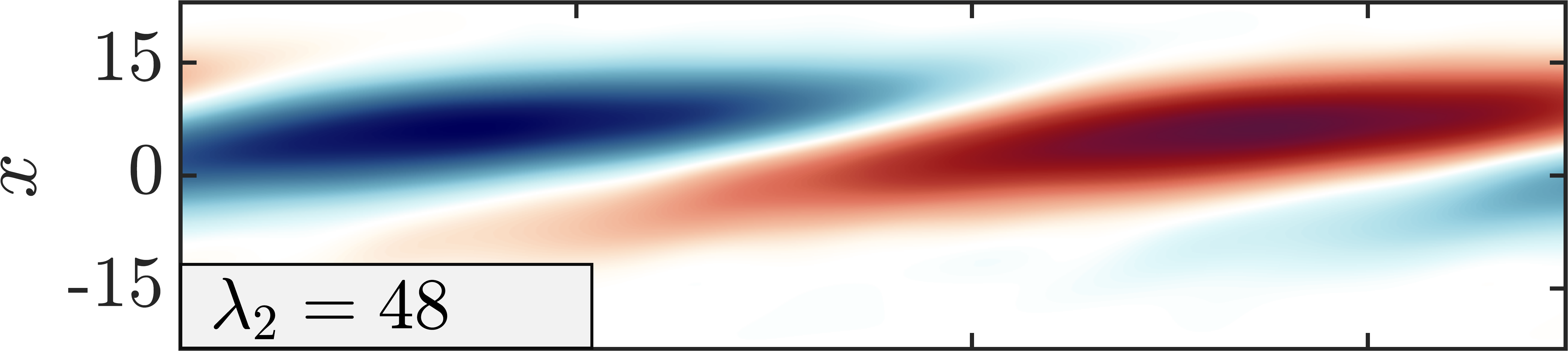}
        \includegraphics[height=0.078\textwidth]{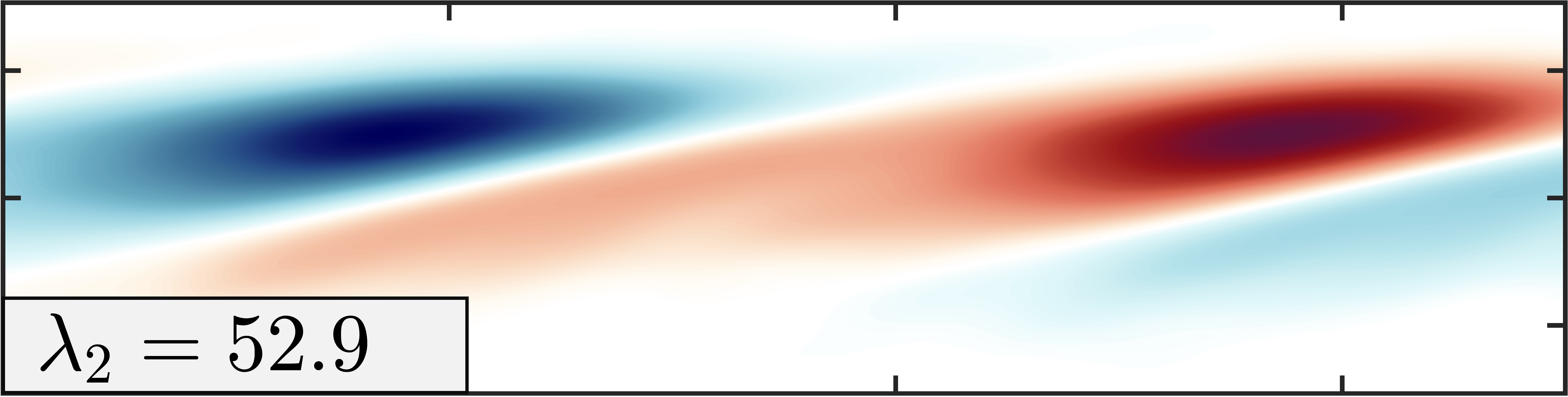}
        \includegraphics[height=0.078\textwidth]{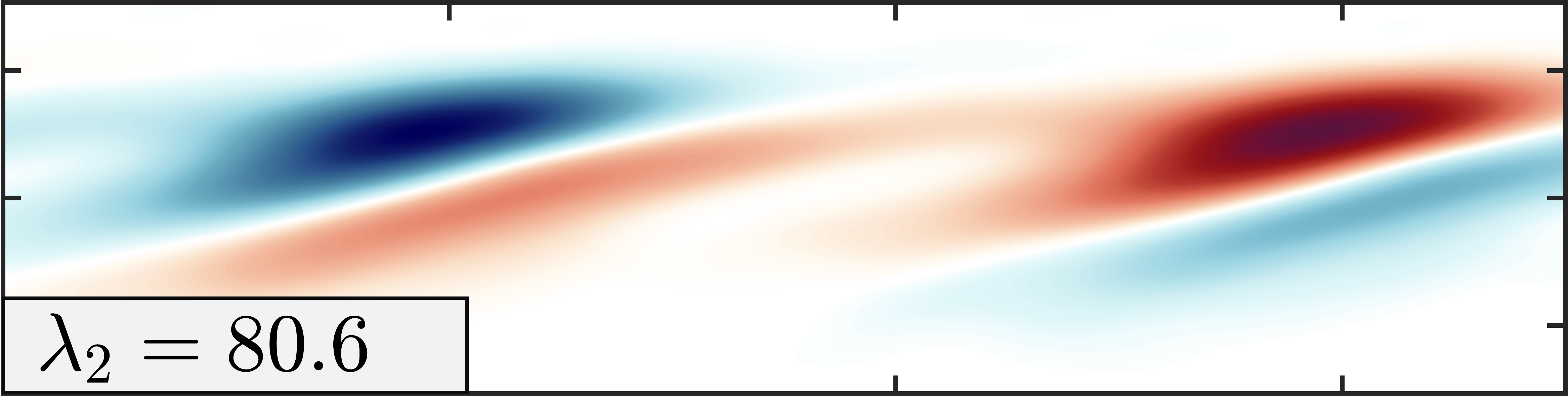}\vspace{1mm}\\
        \includegraphics[height=0.1157\textwidth]{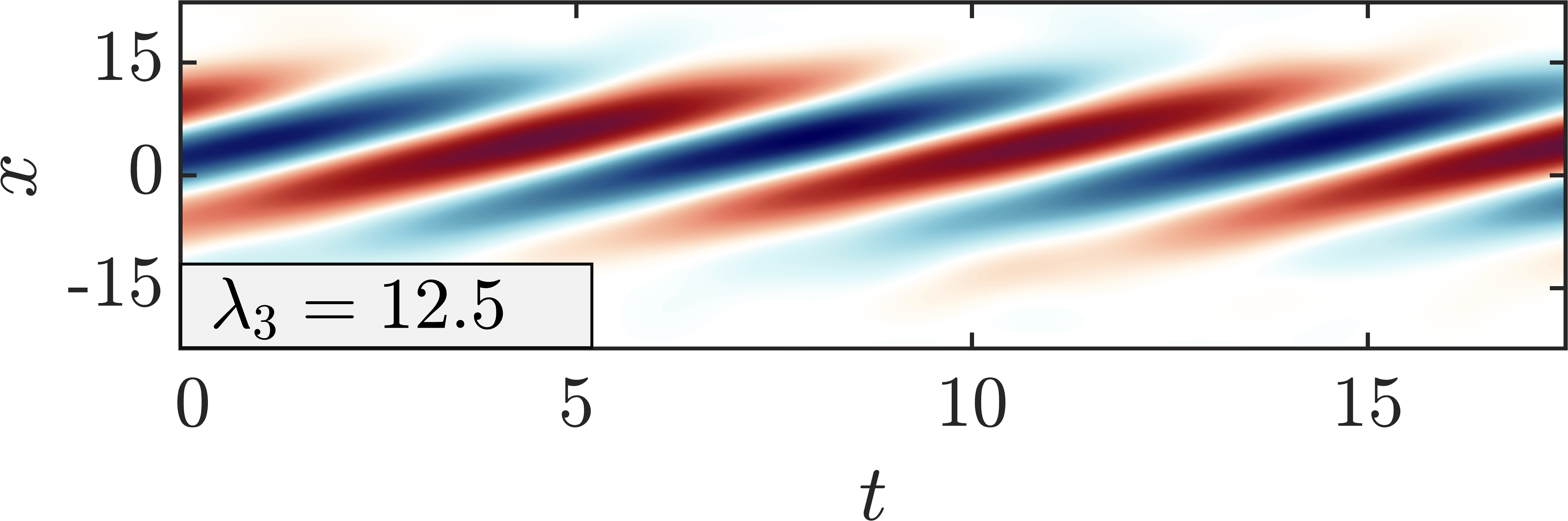}
        \includegraphics[height=0.1157\textwidth]
        {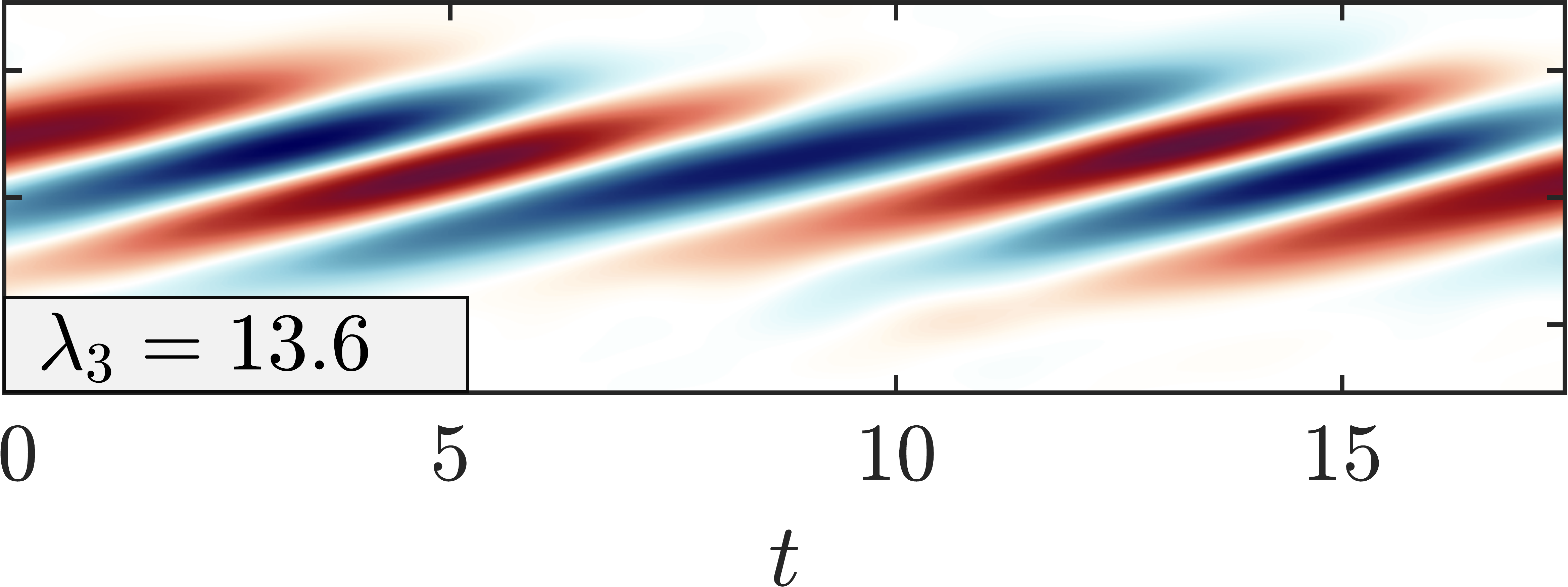}
        \includegraphics[height=0.1157\textwidth]{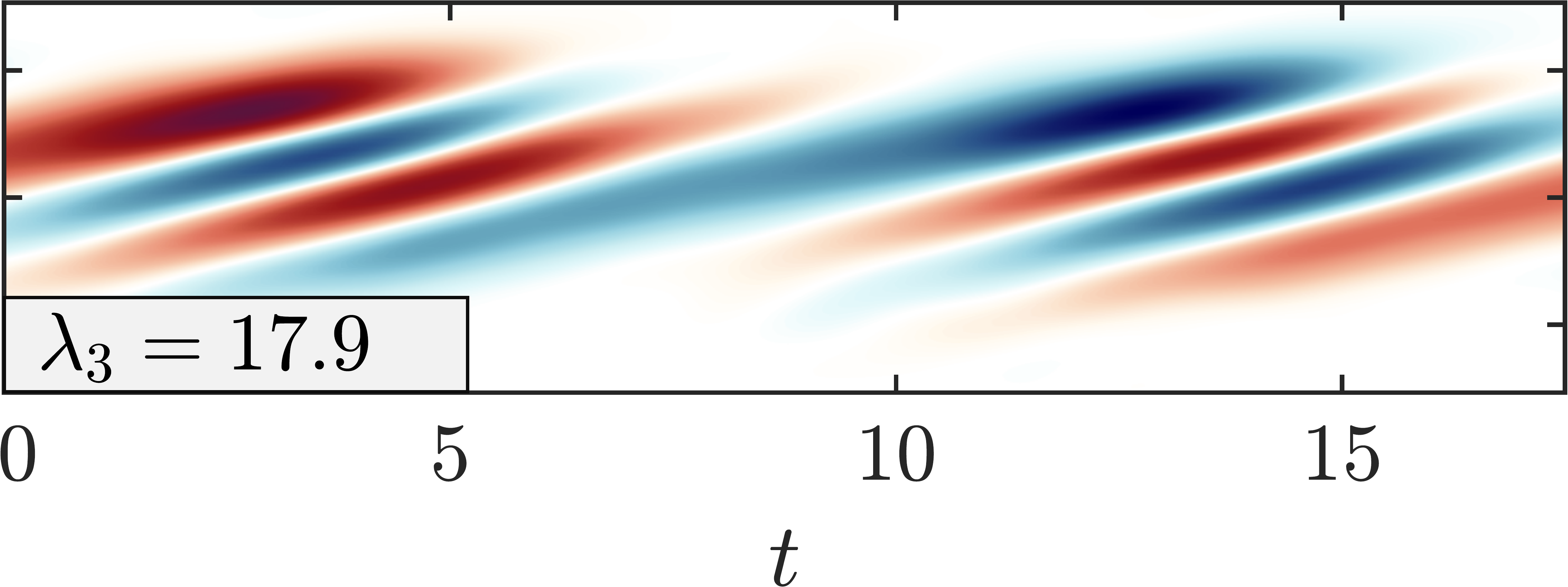}
        \caption{Real component of the three dominant CS-SPOD modes at $\fci = 0.05$ for the three Ginzburg-Landau systems. The contour limits for each CS-SPOD mode is $\pm |\mathfrak{R}\{\psib_j(\x, t)\}|_\infty$.}
    \label{fig:realmodesalf}
\end{figure}

\begin{figure}
\centering 
        \includegraphics[height=0.078\textwidth]
        {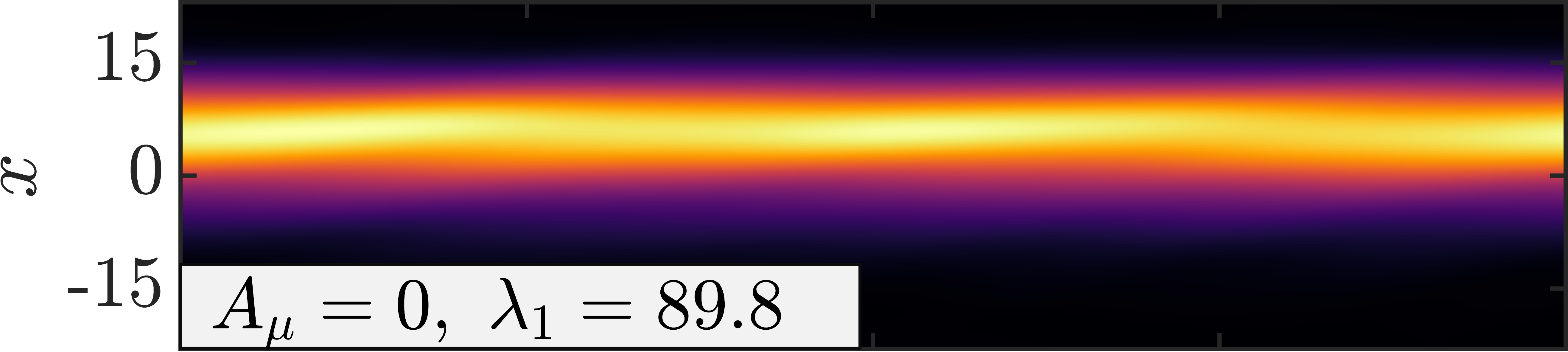}
        \includegraphics[height=0.078\textwidth]{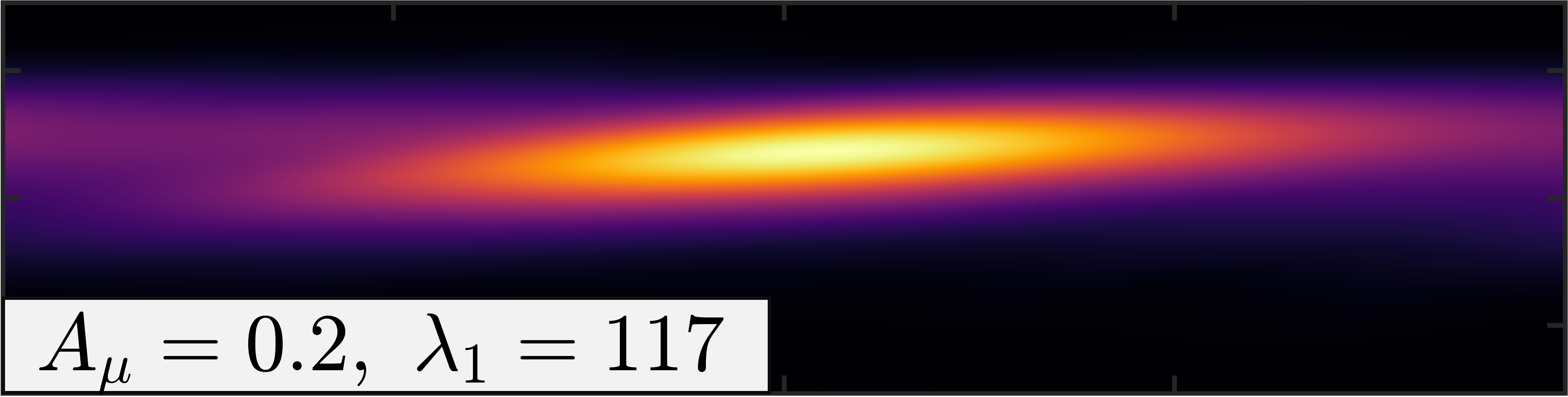}
        \includegraphics[height=0.078\textwidth]{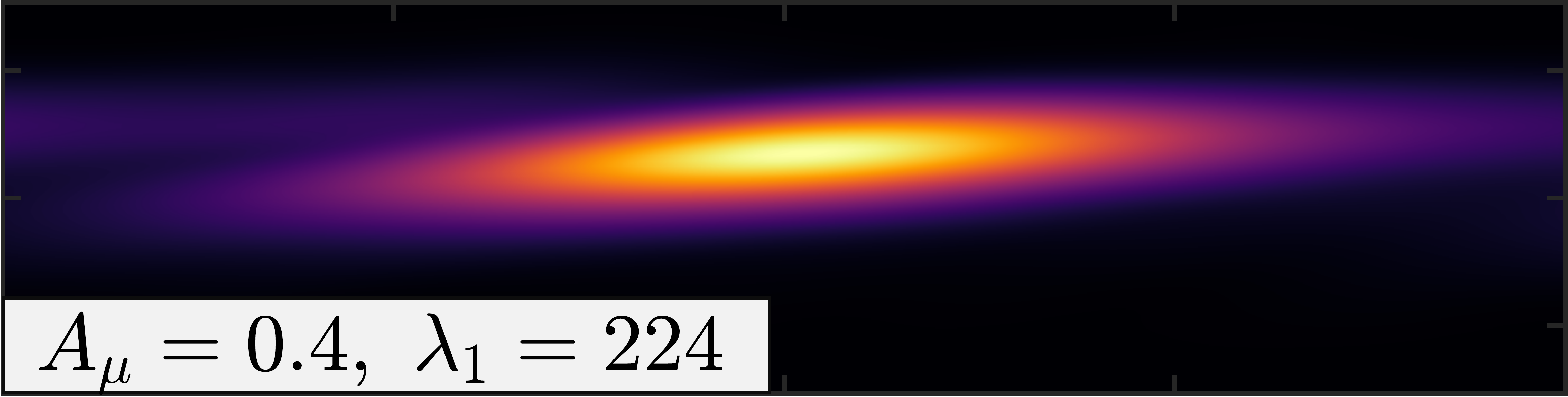}\vspace{1mm}\\
        \includegraphics[height=0.078\textwidth]{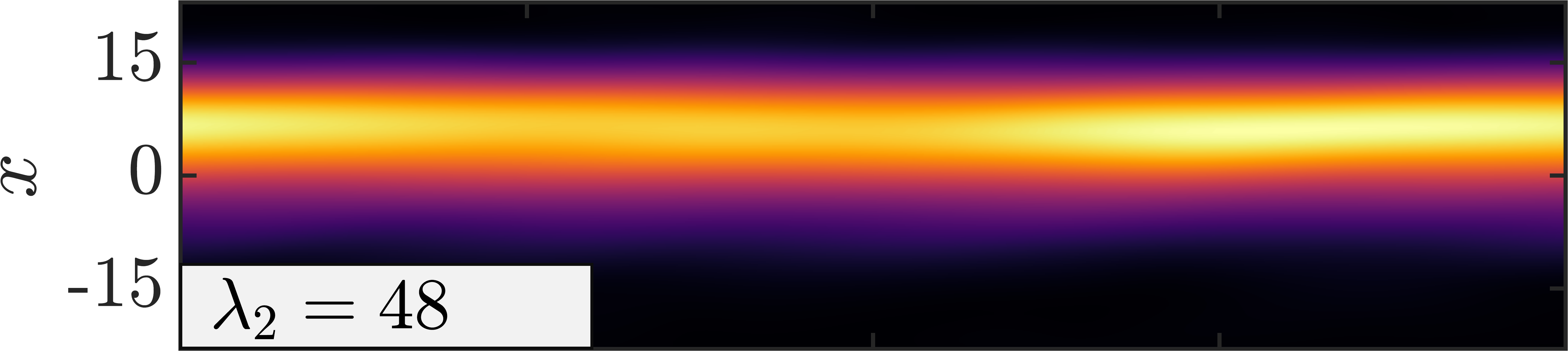}
        \includegraphics[height=0.078\textwidth]{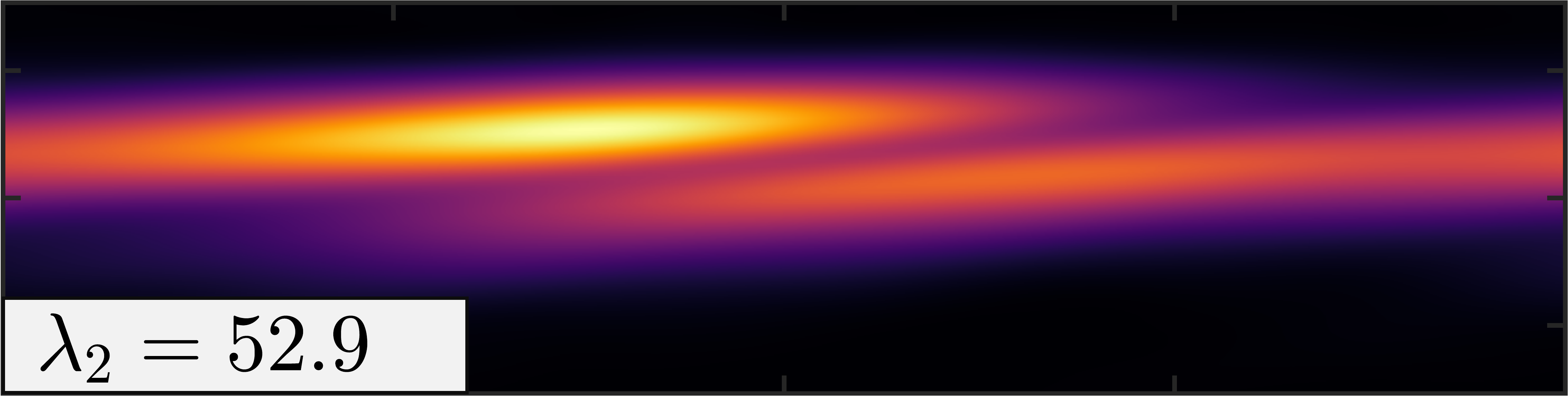}
        \includegraphics[height=0.078\textwidth]{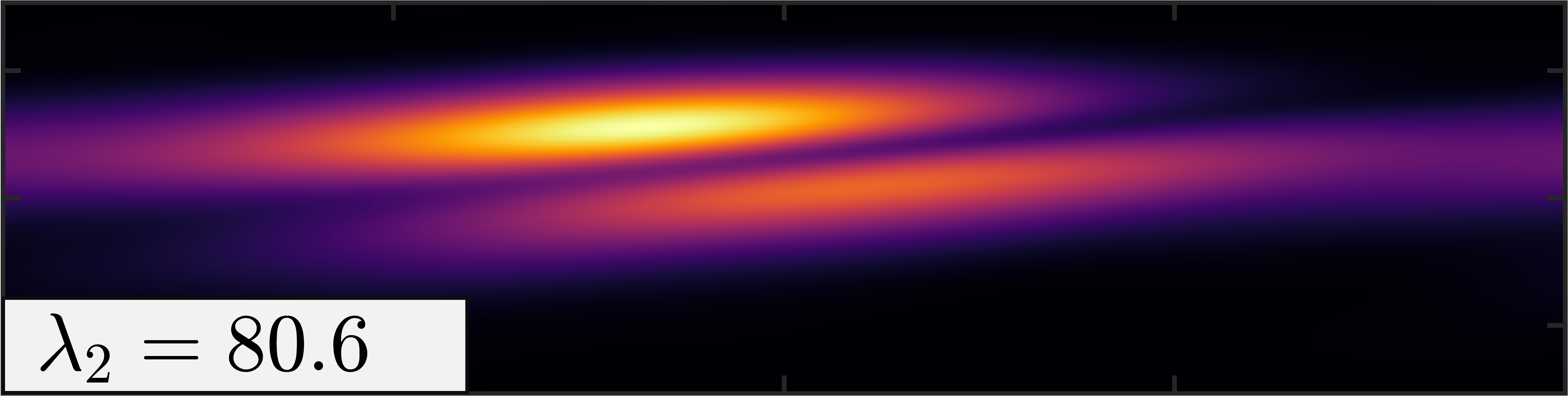}\vspace{1mm}\\
        \includegraphics[height=0.1157\textwidth]{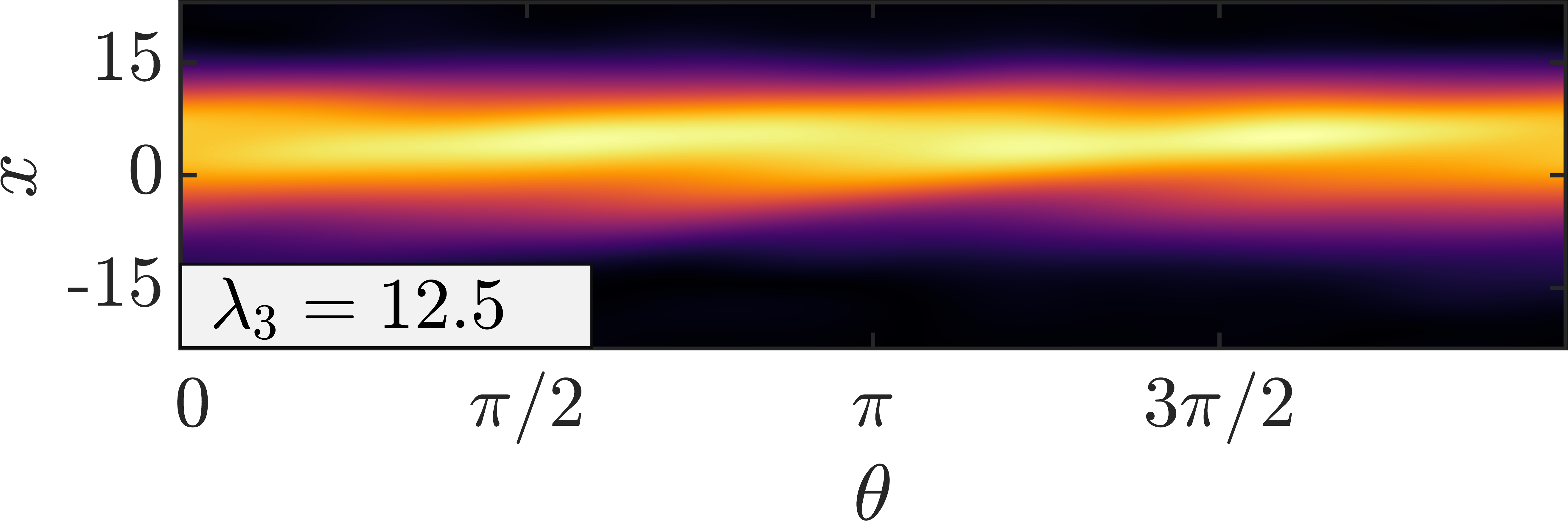}
        \includegraphics[height=0.1157\textwidth]{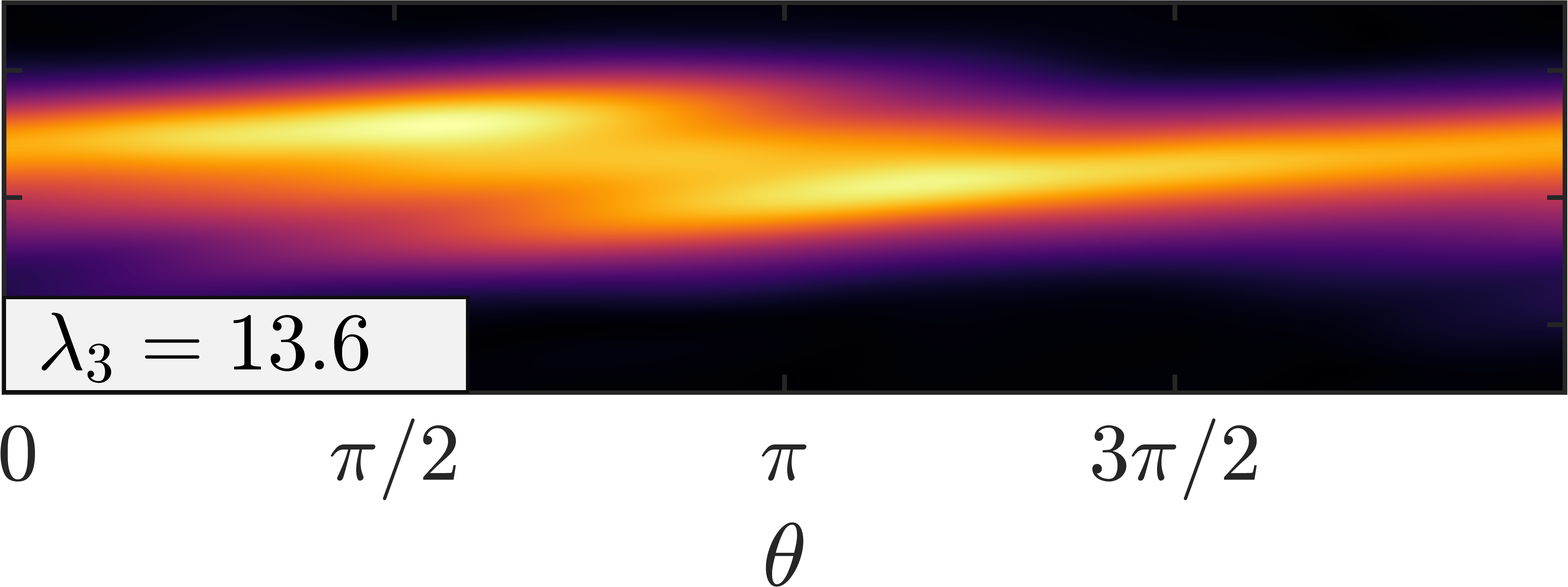}
        \includegraphics[height=0.1157\textwidth]{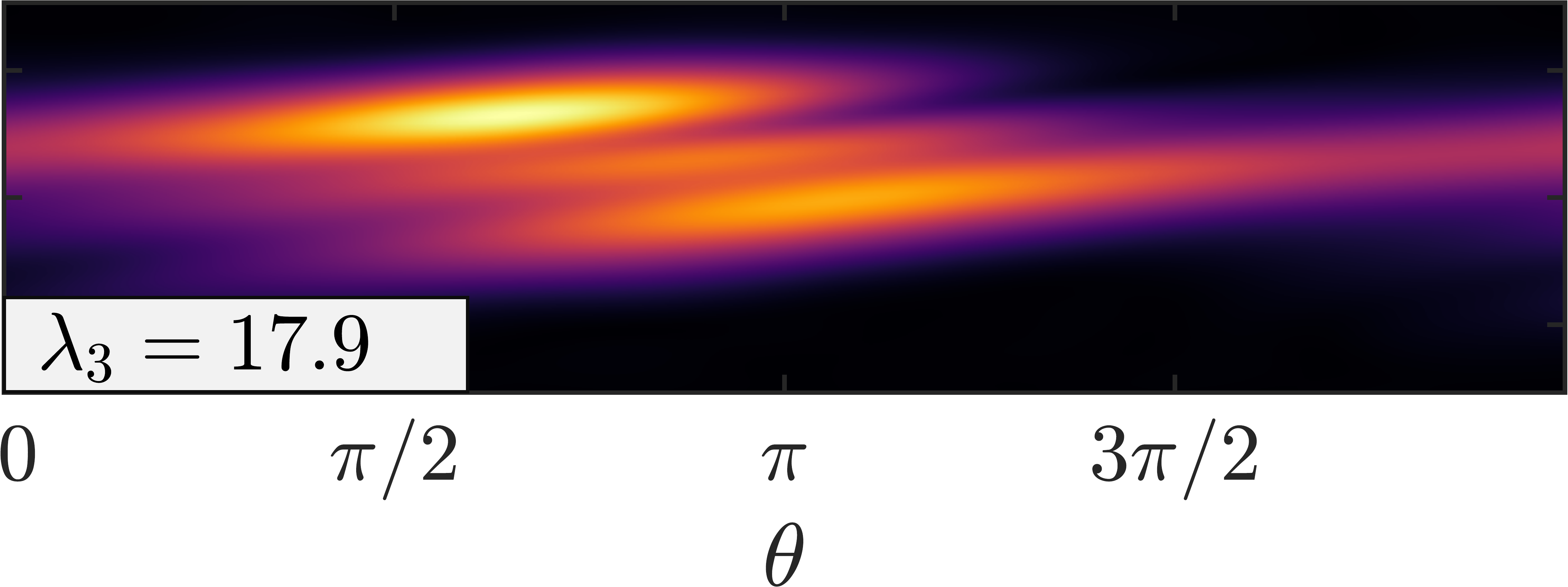} 
        \caption{Magnitude of the three dominant CS-SPOD modes at $\fci = 0.05$ for the three Ginzburg-Landau systems. The contour limits for each CS-SPOD mode is $[0, ||\psib_j(\x, t)||_\infty]$.}
    \label{fig:absmodesalf}
\end{figure}

We now investigate how $A_{\mu_0}$ modifies the dominant CS-SPOD modes, at $\fci = 0.05$, by showing the real component and the magnitude of the temporal evolution of the modes $\phib_j(\x, t)$ in figures \ref{fig:realmodesalf} and \ref{fig:absmodesalf}, respectively. We note that due to the multiple frequency components ($\Fset_{\fci}$) present in $\phib_j(\x, t)$, $\phib_j(\x, t)$ can, unlike SPOD, no longer be completely represented by a single snapshot and instead must be displayed as a function of time. Similarly, the amplitude of the mode is periodic in time with period $T_0 = 1/\alpha_0$, unlike SPOD where the amplitude is constant in time. Thus, the amplitude is displayed as a function of phase $\theta$.  Similar results are observed for other values of $\fci$ not shown here. Overall, across all values of $A_{\mu_0}$, the real component of the CS-SPOD modes shows a similar structure. However, as $A_{\mu_0}$ is increased, an additional modulation is seen that results in increasingly time/phase-dependent magnitudes. 
 
\begin{figure}
\centering
    \begin{subfigure}[b]{1\textwidth}
        \includegraphics[height=0.1485\textwidth]{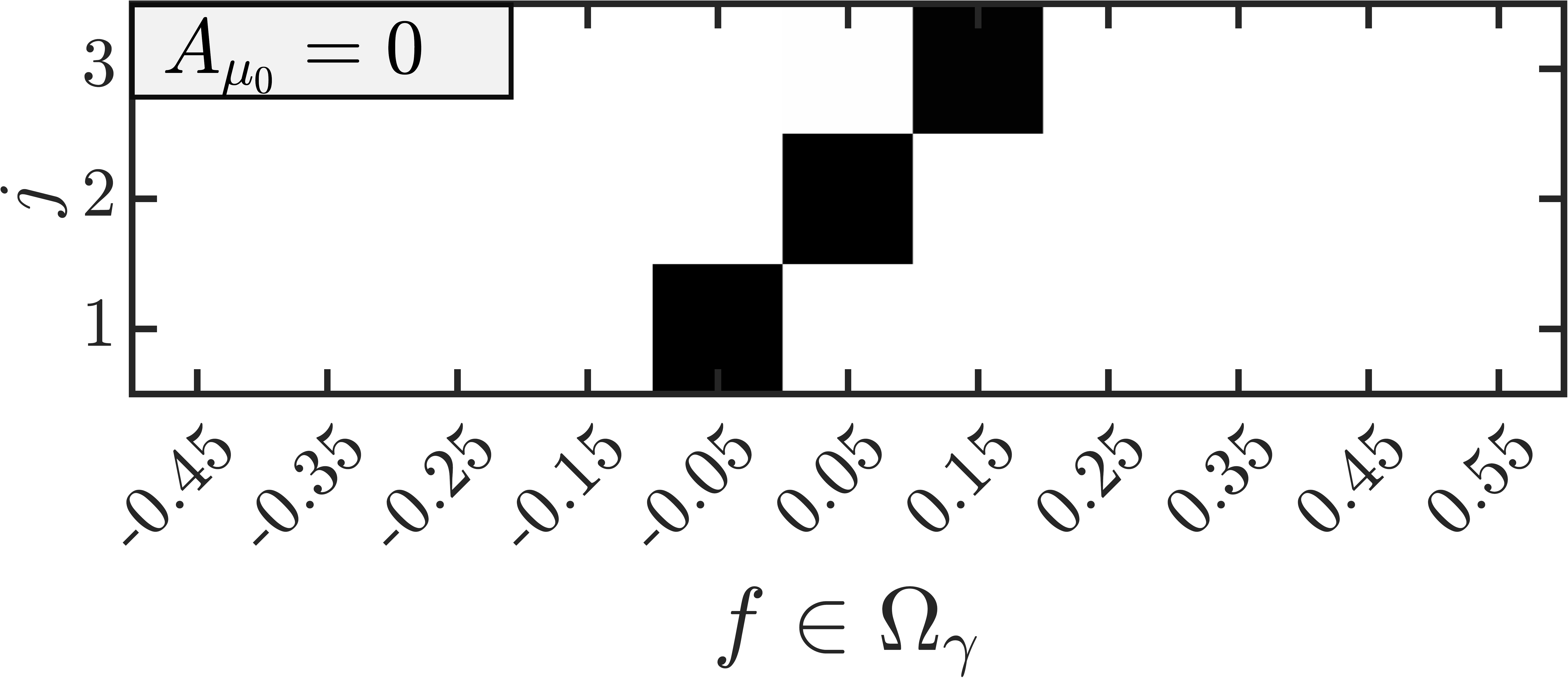} 
        \includegraphics[height=0.1485\textwidth]{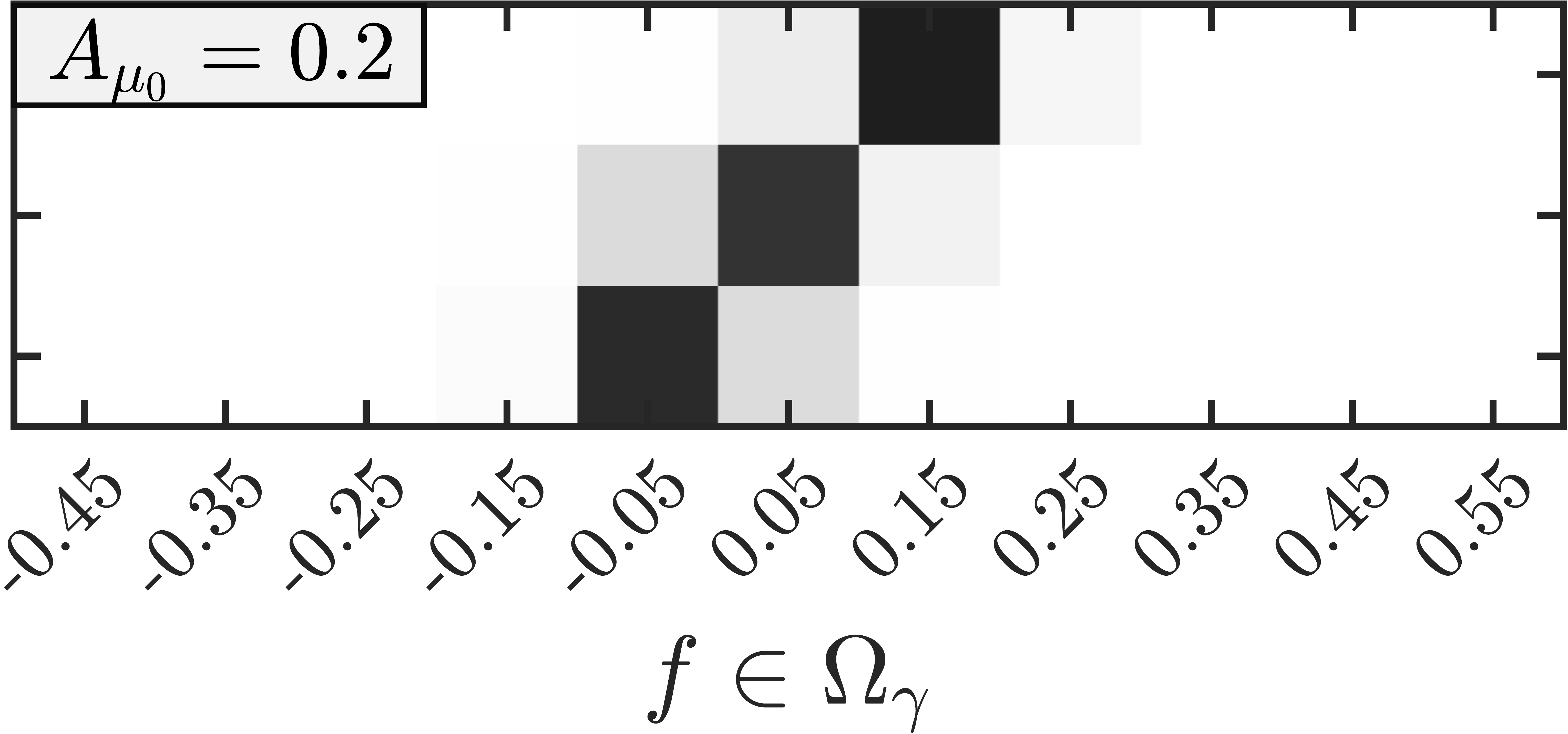}
        \includegraphics[height=0.15004\textwidth]{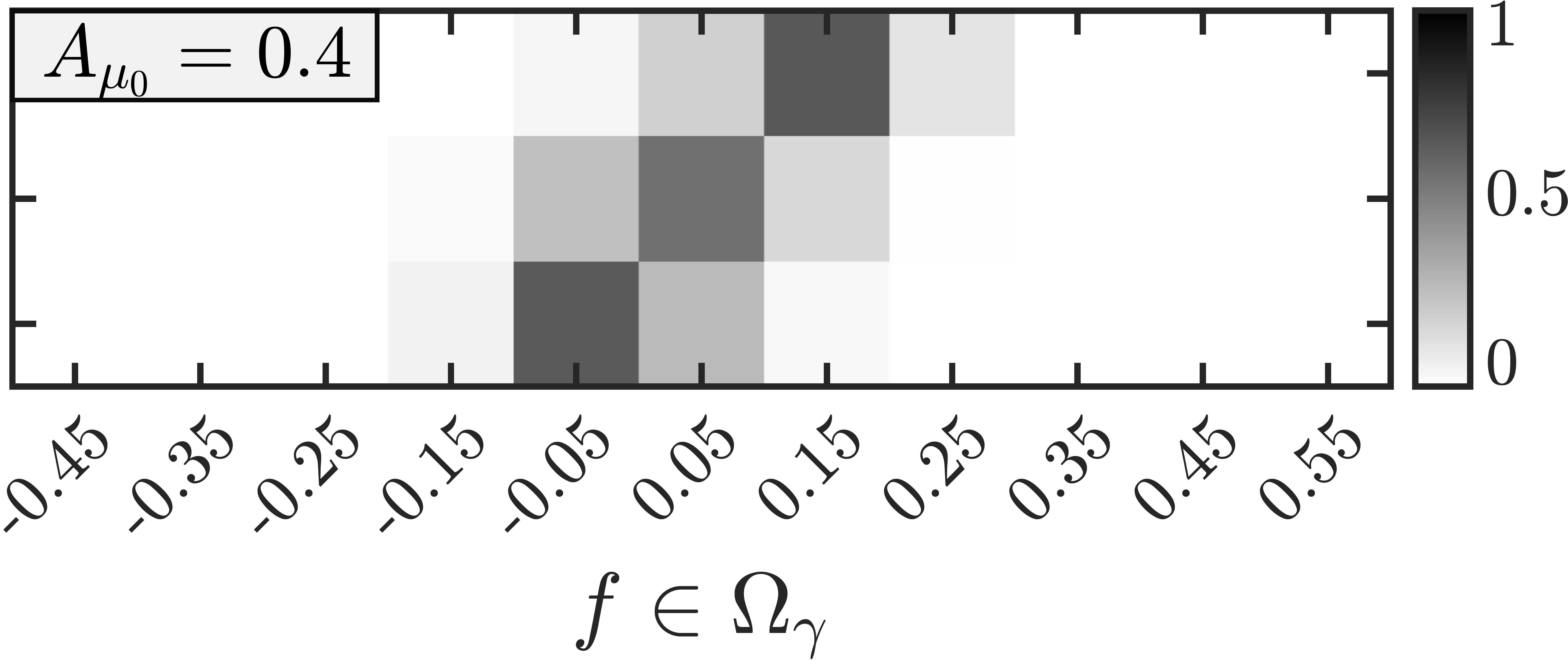}
    \end{subfigure}%
    \caption{Fractional CS-SPOD modal energy, $E_{f, j}$, at $\fci = 0.05$ for the Ginzburg-Landau systems.}
    \label{fig:fracenergy}
\end{figure}

Finally, in figure \ref{fig:fracenergy}, we investigate which frequency components are most energetic via the fractional energy of each frequency component $f \in \Fset_{\fci}$ for each CS-SPOD mode, defined as $E_{f, j} \equiv \psib_j(\x, f)^*  \Wb(\x) \psib_j(\x, f)$, where $\sum_{f \in \Fset_{\fci}} E_{f, j} = 1$.  As $A_{\mu_0}$ increases, the CS-SPOD modes are constructed from an increased number of non-zero-energy frequency components and at higher energy levels. For example, at $\fci = 0.05$, the dominant frequency component, $f = 0.05$, contains $\approx 100\%,\ 83\%$, and $64\%$ of the total energy of the corresponding CS-SPOD mode for $A_{\mu_0} = 0, 0.2$, and $0.4$, respectively. This occurs because of the increasing amount of correlation present between different frequency components as $A_{\mu_0}$ increases. Alternatively, this phenomenon can be understood as the following; as $A_{\mu_0}$ increases, the statistics become more time-dependent, and thus, the amount of interaction between frequency components in $\Fset_{\fci}$ increases such that the summation of these frequency components result in CS-SPOD modes that capture the time-periodic modulation experienced by the flow. 

\subsection{Forced turbulent jet}  \label{sec:CSForcedTurb}

We now consider a forced turbulent, isothermal, subsonic jet for which data is available from a previous study \cite{heidt2021analysis}. The LES was computed using the Charles solver by Cascade Technologies using a setup similar to previous, experimentally validated simulations of turbulent jets \citep{bres2017unstructured,bres2018importance}. The jet has a Mach number of $M_j = U_j/c_j =0.4$ and a Reynolds number of $Re_j = {\rho_j U_j D}/{\mu_j} = 4.5\times10^5$, where $\rho$ is the density, $\mu$ is the viscosity, $U$ is the velocity, $c$ is the speed of sound, $D$ is the nozzle diameter, and the subscripts j and $\infty$ represent the jet and free-stream conditions, respectively. Frequencies are reported with respect to the Strouhal number $St = fD/U_j$, where f is the frequency. 

\begin{figure}
\centering

  \begin{subfigure}[t]{0.5318784\textwidth}
    \includegraphics[height=0.7983\textwidth]{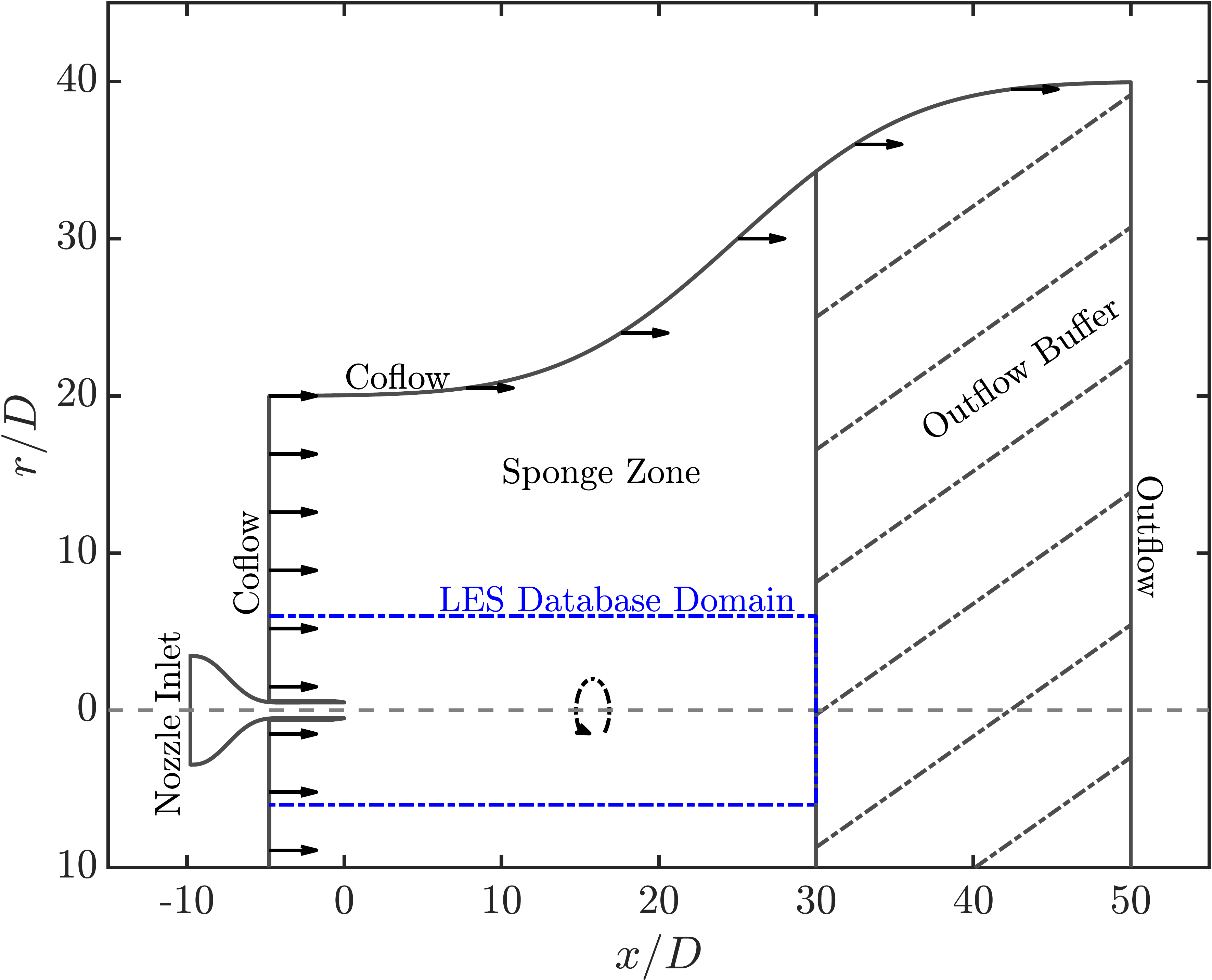}
    \caption{Overall domain}
    \label{fig:nozzlezoomedout}
  \end{subfigure} \hfill
  \begin{subfigure}[t]{0.350892\textwidth}
    \includegraphics[height=1.21\textwidth]{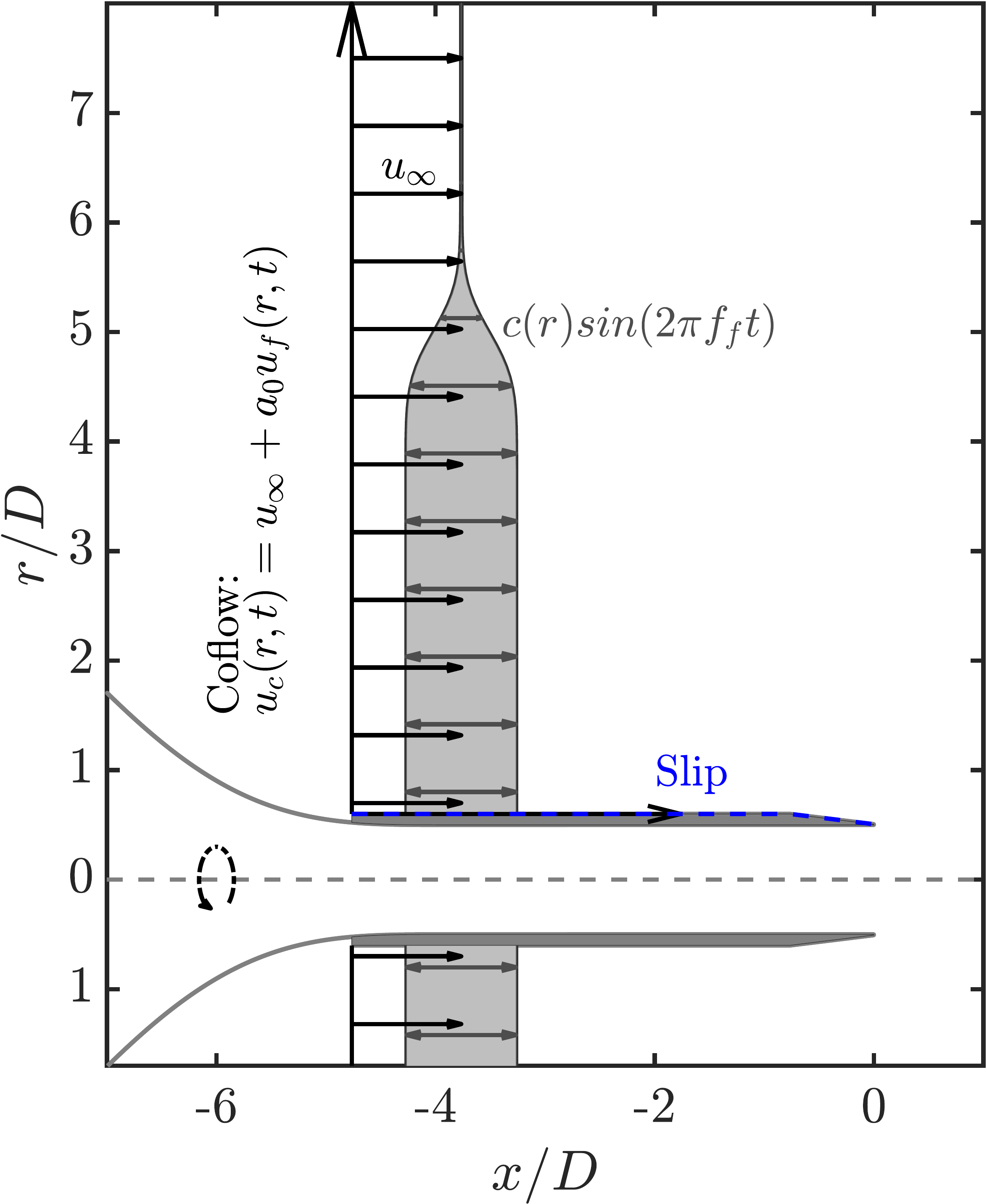}
    \caption{Nozzle region}
    \label{fig:nozzlezoomedin}
  \end{subfigure} \hspace{5mm}
  \caption{Schematic of the forced Mach 0.4 turbulent jet, adapted from \citet{bres2018importance}.}
  \label{fig:nozzlemain}
\end{figure}

A schematic of the simulation setup is shown in figure \ref{fig:nozzlemain}. An acoustic forcing is applied at a frequency $St_f = f_fD/U_j=0.3$ and amplitude $a_0/U_j = 0.1$. This forcing was chosen to roughly model the forced jet experiments of \citet{crow_champagne_1971}, and we chose $St_f=0.3$ to match what they observed as the frequency that led to the largest amplification by the flow (i.e. the {\it jet preferred mode}).  We intentionally used a high amplitude of forcing as we wanted to clearly establish cyclostationarity in the resulting turbulence. The forcing is applied in an annular region surrounding the jet up to $r/D = 5$. The acoustic forcing inlet co-flow is defined by: 
\begin{subeqnarray}
c(r)&=& 0.5 \left[1 - \text{erf}\left(2 (r - 5)\right) \right], \\
u_f(r, t) &=& c(r)  \text{sin}(2 \pi f_f t), \\
u_x(r, t) &=& u_{\infty} + a_0 u_f(r, t),\\
u_r(r, t) &=& u_\theta(r, t) = 0, \\
\rho(r, t) &=& \rho_\infty + \rho_\infty (u_x(r, t) - u_\infty)/a_\infty,\\
p(r, t) &=& p_\infty + a_\infty\rho_\infty (u_x(r, t) - u_\infty).    
\end{subeqnarray} 
The simulation was run, post-transient, with a time-step of $\Delta t D/c_\infty = 0.001$, for $480$ periods of the forcing frequency (or a total time of $t_{sim}D/c_\infty \approx 4000$), during which $N_\theta = 48$ snapshots were saved over each cycle of the forcing.  The unstructured LES data were interpolated onto a structured cylindrical grid ($n_x \times n_r \times n_\theta = 656 \times 138 \times 128)$ spanning $x/D \in [0, 30]$, $r/D \in [0, 6]$, and $\theta \in [0, 2\pi]$, which was employed in the subsequent analyses. For the stochastic estimates, we use a window length $N_w = 6N_\theta$ and an overlap of $67\%$, resulting in $N_b = 237$ blocks and a non-dimensional frequency discretization of $\Delta St \approx 0.05$.

\begin{figure}
\centering
  \begin{subfigure}[t]{0.49\textwidth}
    \includegraphics[width=0.99\textwidth]{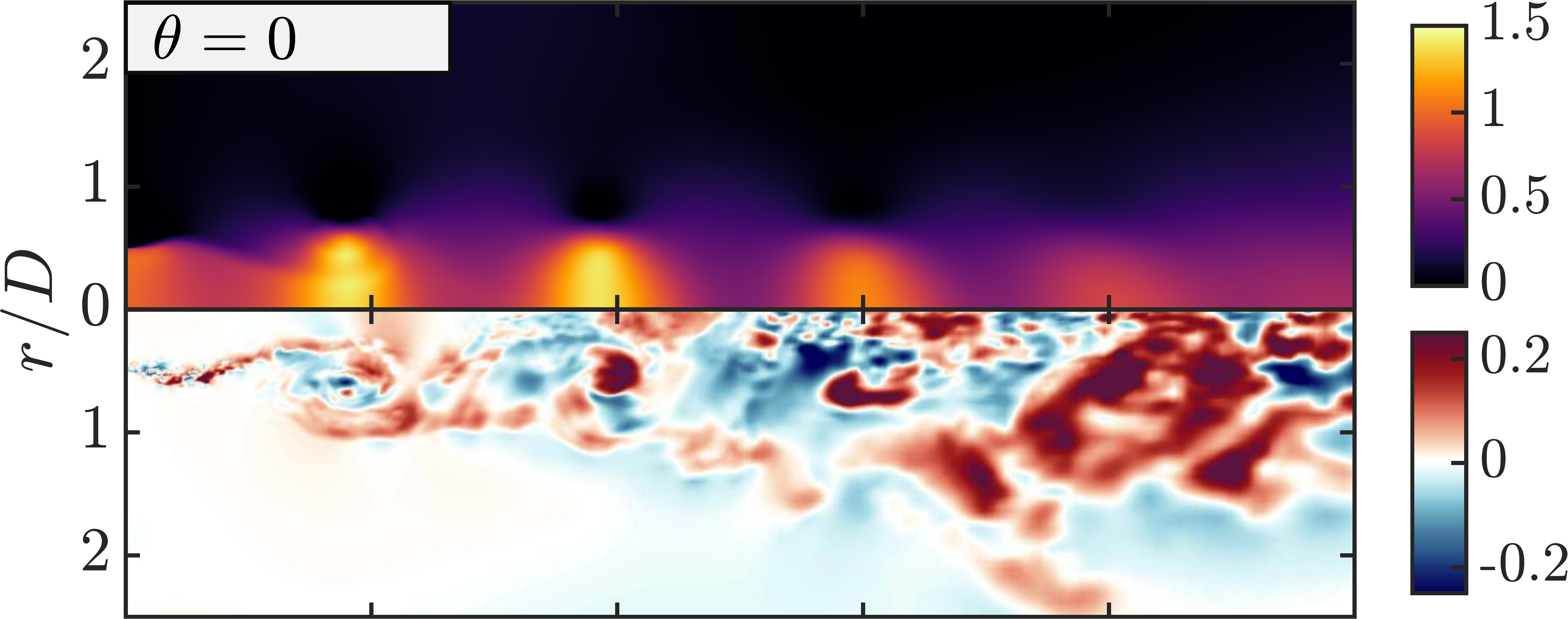}
  \end{subfigure}  \hfill
  \begin{subfigure}[t]{0.49\textwidth}
    \includegraphics[width=0.99\textwidth]{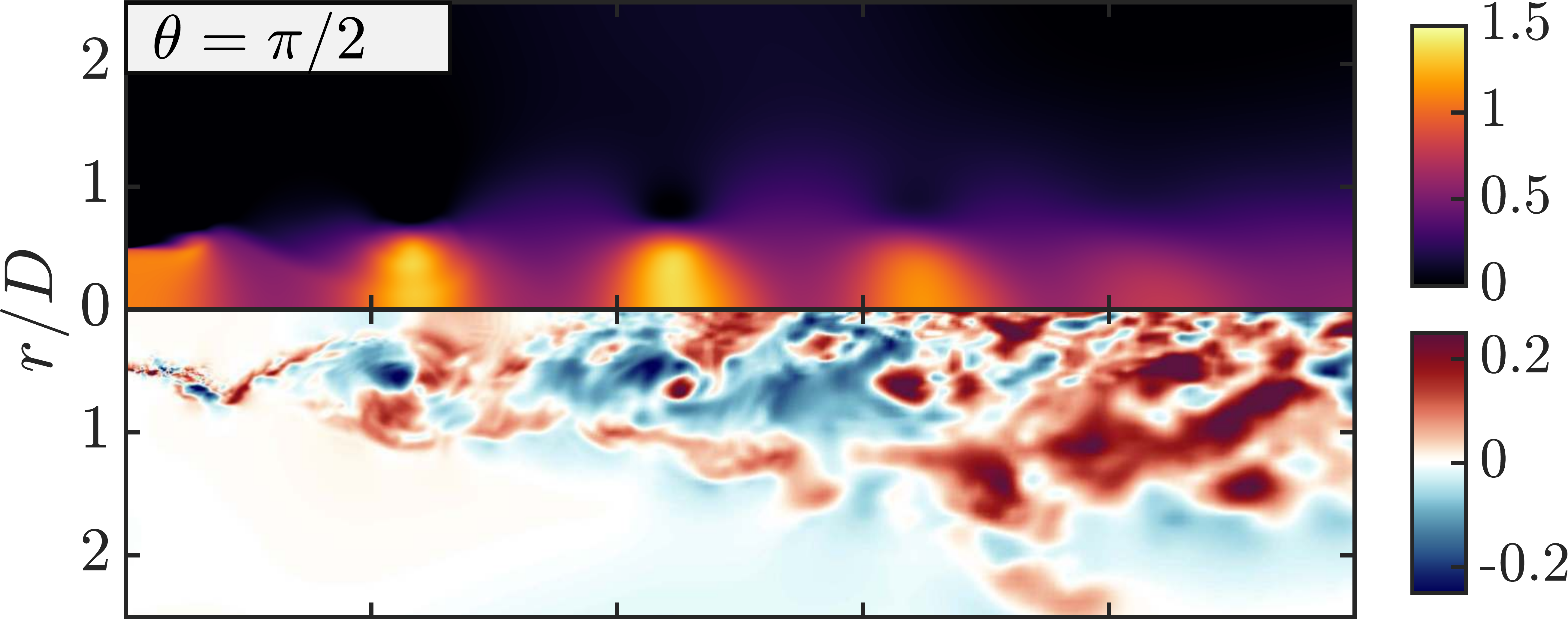}
  \end{subfigure} \\ \vspace{2.2mm}
  \begin{subfigure}[t]{0.49\textwidth}
    \includegraphics[width=0.99\textwidth]{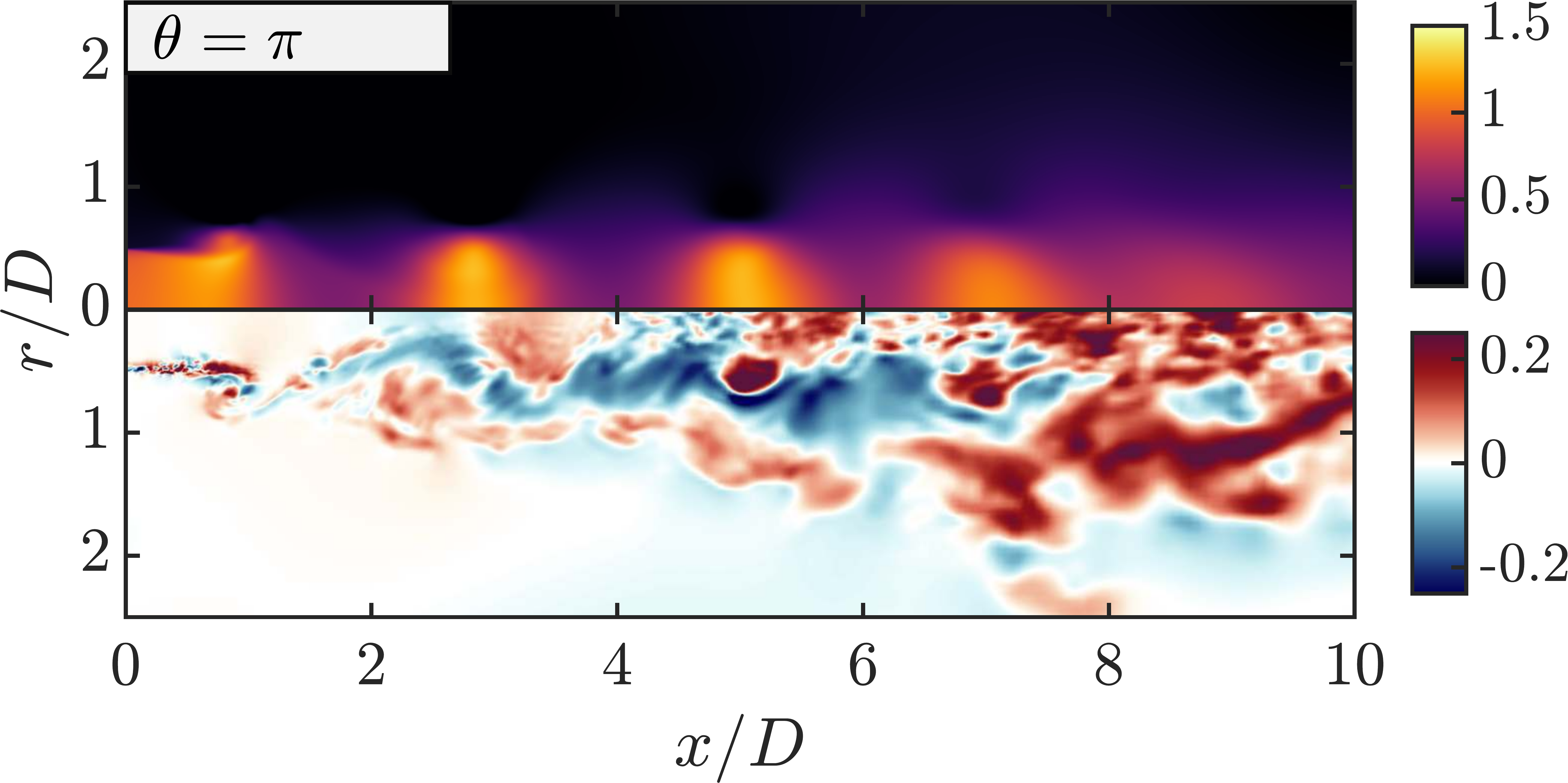}
  \end{subfigure}   \hfill
  \begin{subfigure}[t]{0.49\textwidth}
    \includegraphics[width=0.99\textwidth]{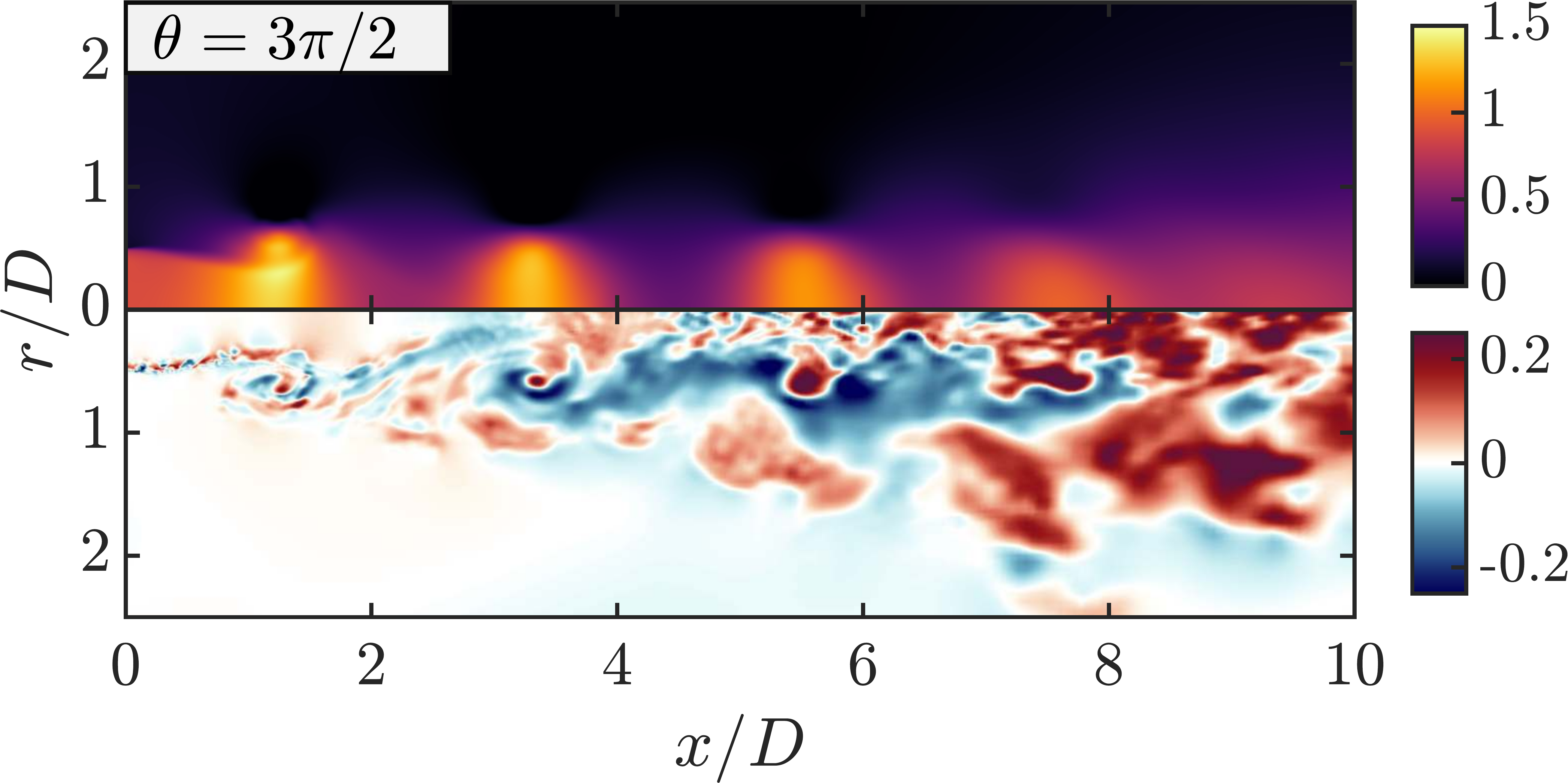}
  \end{subfigure}
\caption{Top of each pair of images is $\widetilde{u}/U_j$ at $\theta = 0,\ \pi/2,\ \pi,\ 3\pi/2$ for the forced Mach 0.4 turbulent jet. Bottom of each pair of images is ${u_x^{\prime\prime}}/U_j$ at a time instant corresponding to a forcing phase of $\theta = 0,\ \pi/2,\ \pi,\ 3\pi/2$.}
\label{fig:jetphase_new}
\end{figure}

\begin{figure}
\centering
  \begin{subfigure}[b]{0.49\textwidth}
    \includegraphics[width=0.95\textwidth]{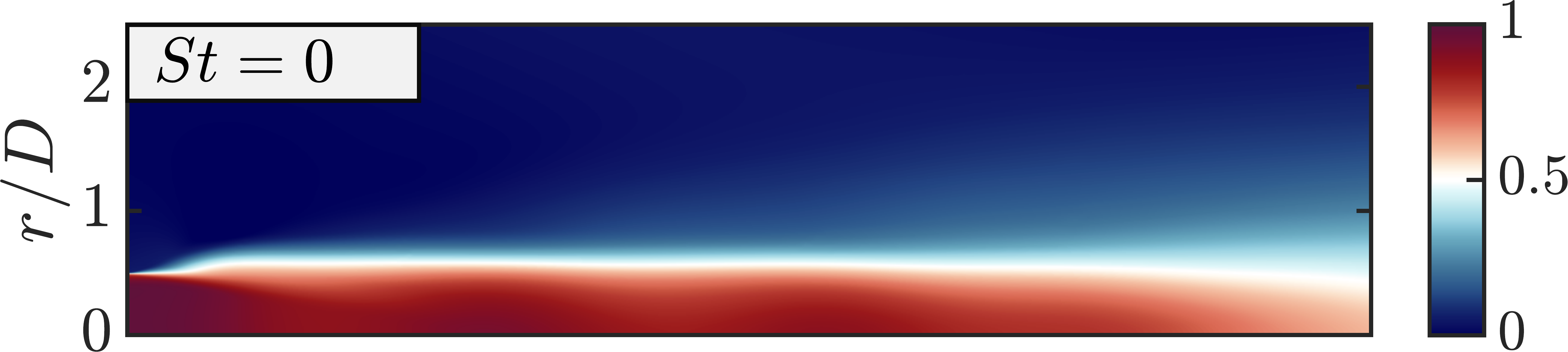} \vspace{0.25mm}
  \end{subfigure} \hfill
  \begin{subfigure}[b]{0.49\textwidth}
    \includegraphics[width=0.95\textwidth]{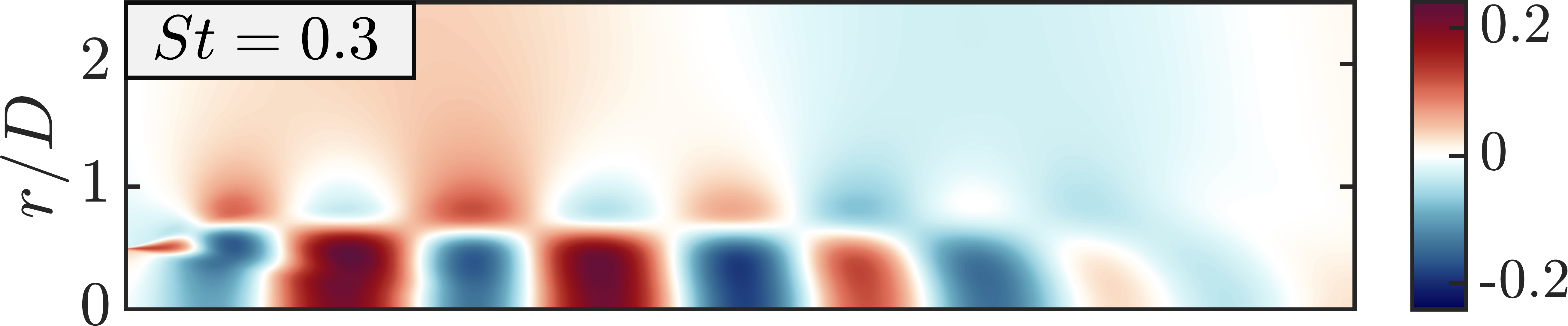}
  \end{subfigure} \\ \vspace{1mm}
  \begin{subfigure}[b]{0.49\textwidth}
    \includegraphics[width=0.98\textwidth]{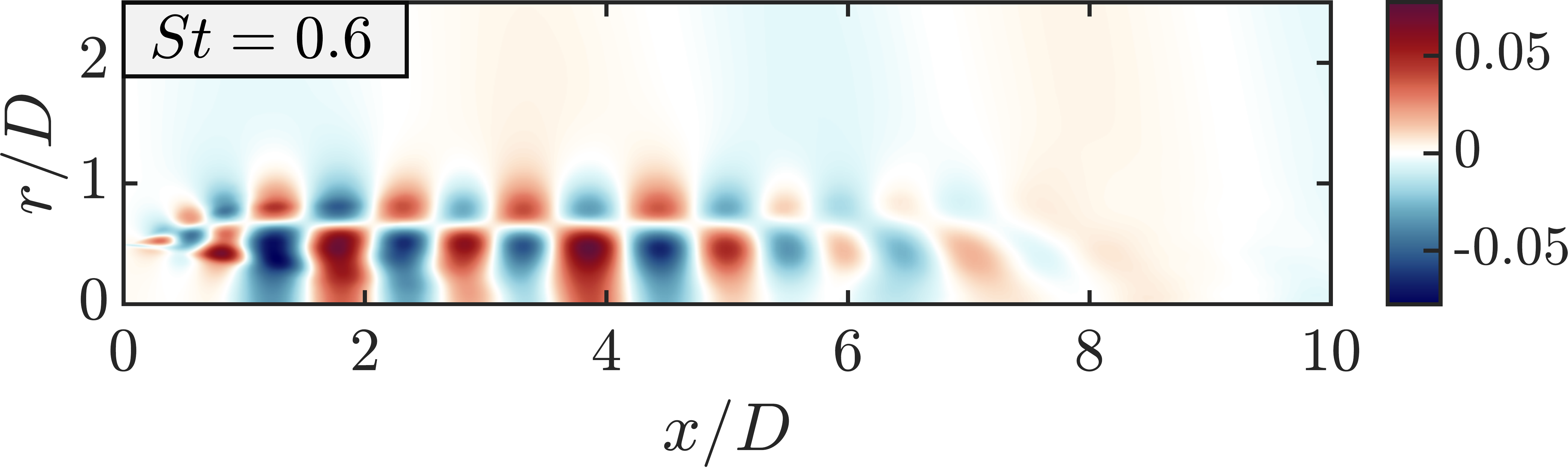}
  \end{subfigure}  \hfill
  \begin{subfigure}[b]{0.49\textwidth}
    \includegraphics[width=0.967888\textwidth]{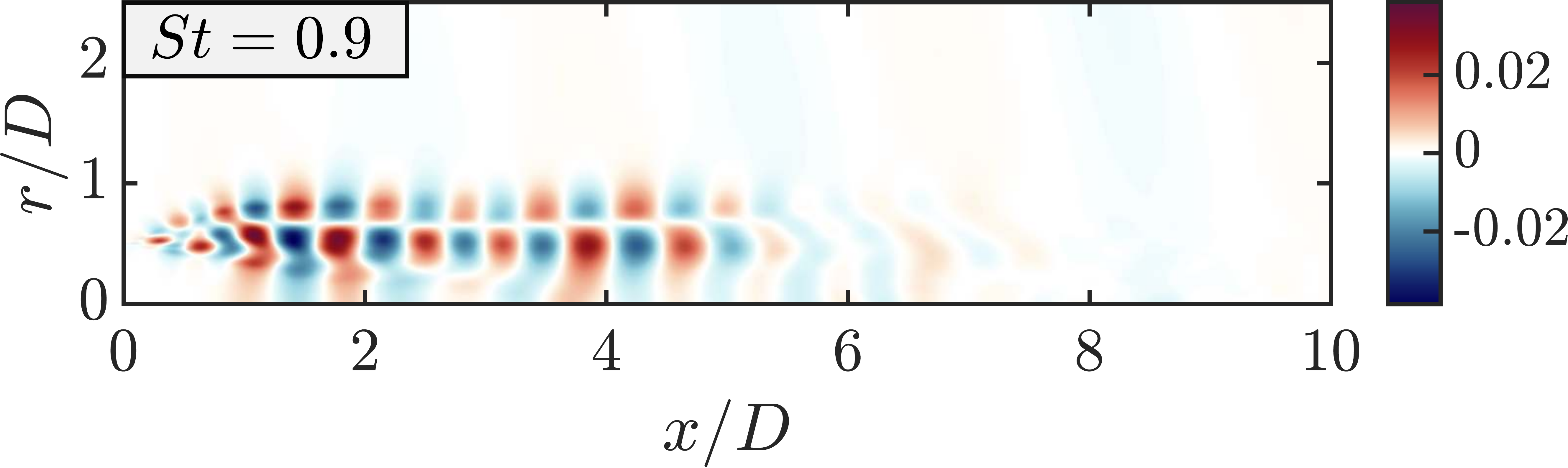}
  \end{subfigure}
\caption{$\mathfrak{R}\{\hat{{u}}_{x, St}/U_j \}$ at $St = 0, 0.3, 0.6$, and $0.9$ for the forced Mach 0.4 turbulent jet.}
\label{fig:firstorderjet}
\end{figure}

In figure \ref{fig:jetphase_new}, we plot the instantaneous and phase-averaged \eqref{eqn:phaseaverage} velocity at four phases of one forcing cycle.  Though not shown, we verified that the phase-averaged field is axisymmetric, consistent with the axisymmetric jet forcing.  In the phase-averaged field, a large modulation in the axial velocity of the jet is observed with a vortex roll-up occurring around $x/D = 2.0$. The fundamental frequency fluctuation is primarily located in the potential core region and drives the large-scale periodic modulation.  In figure~\ref{fig:firstorderjet}, we extract the first four frequency components ($f = 0, 0.3, 0.6, 0.9$) of the phase-averaged field.  The total fluctuation level, i.e. $2\times \mathfrak{R}\{\hat{u}_{x, \alpha}/U_j \}$, for each non-zero frequency is $\approx 40\%, 15\%$, and $8\%$ thereby indicating that a substantial, nonlinear periodic modulation of the mean occurs. Harmonic generation similarly peaks near $x=2$ where the strong roll-up is occurring.


Next, we analyze the second-order stochastic component to determine the cycle frequencies present in order to apply CS-SPOD. Similar to the previous example, to determine what cycle frequencies are present in the flow, we interrogate the CCSD and integrated CCSD for $\alpha = [-3, 3]$ (not shown), again noting the $\alpha$ discretization as discussed in \S \ref{sec:CStheory}. We confirm that only the cycle frequencies present are harmonics of the forcing frequency (i.e. $\mathbb{Z} f_f$). \par
\begin{figure}
    \centering
    \begin{minipage}[t]{.45\textwidth}
        \centering
        \includegraphics[width = 0.9 \textwidth]{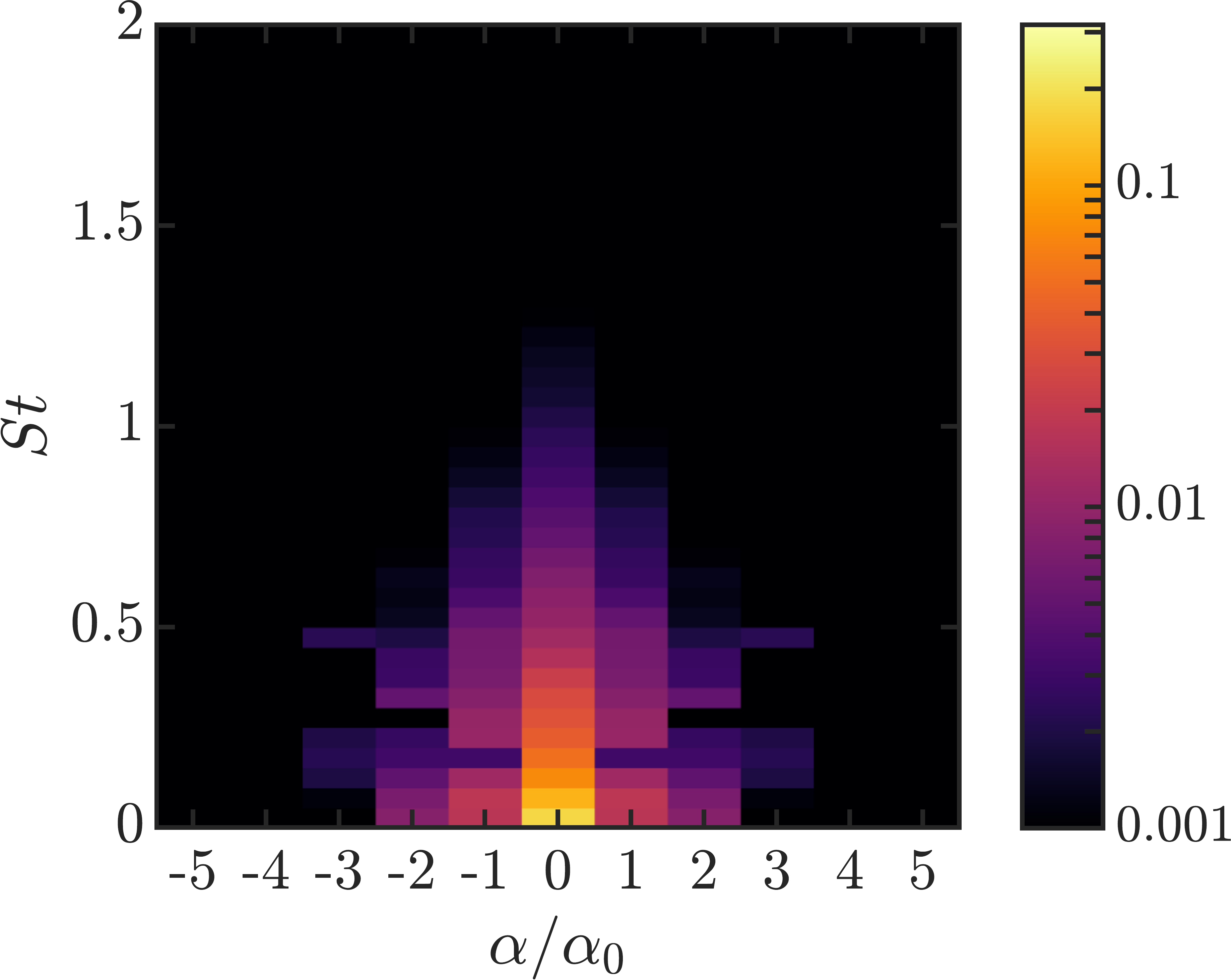}
        \caption{Absolute value of the CCSD density of $u_x^{\prime\prime}/U_j$ at $x/D = 7, r/D = 0.75$ for the forced Mach 0.4 turbulent jet.}
        \label{fig:jet:CCSD}
    \end{minipage}\hfill
    \begin{minipage}[t]{0.45\textwidth}
        \centering
        \includegraphics[width= 0.9 \textwidth]{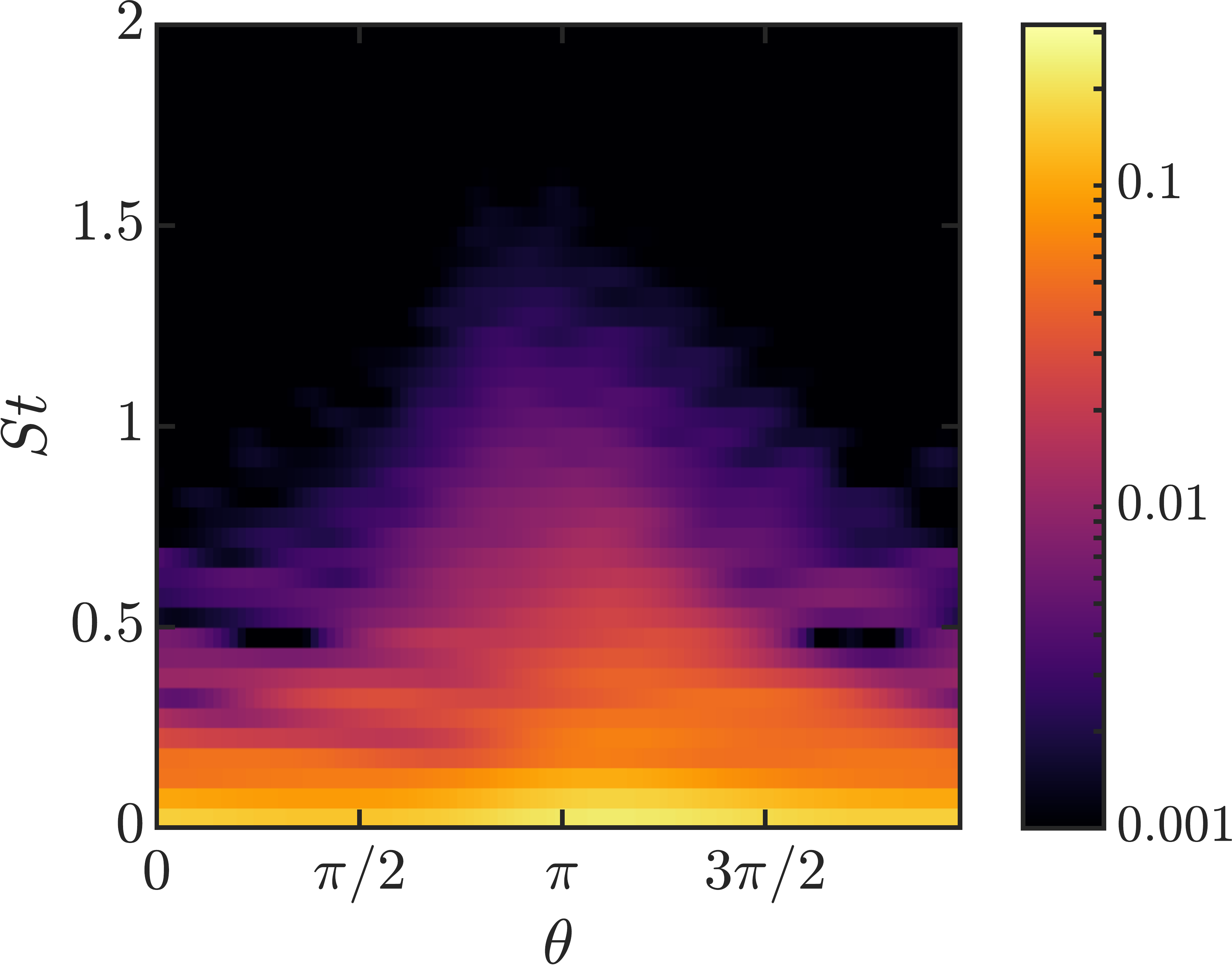}
        \caption{WV spectrum of $u_x^{\prime\prime}/U_j$ at $x/D = 7, r/D = 0.75$ for the forced Mach 0.4 turbulent jet.}
        \label{fig:jet:WV}
    \end{minipage}
\end{figure}

Figures \ref{fig:jet:CCSD} and  \ref{fig:jet:WV} show the CCSD and corresponding WV spectrum, respectively, of the axisymmetric component of the axial velocity at $x/D = 5, r/D = 0.75$. For clarity, the CCSD is only shown for $\alpha/\alpha_0 \in \mathbb{Z}$ since all other values of $\alpha$ are $\approx 0$ (to within statistical convergence). A large modulation occurs for $\alpha/\alpha_0 = 0, \pm 1, \pm 2$.  The WV spectrum shows this large modulation of the statistics, where the phase of the high-energy regions corresponds to when the high-velocity regions pass. Overall, it is clear that the forced turbulent jet exhibits cyclostationarity at frequencies equal to the harmonics of the forcing frequency. 
Finally, we demonstrate the utility of CS-SPOD on a forced turbulent jet.  Recalling that both SPOD and CS-SPOD modes are decoupled amongst the azimuthal modes of the jet (owing to the statistical axisymmetry of the flow), we focus for brevity only on the axisymmetric $m=0$ component of the fluctuations. We seek modes that are orthogonal in the Chu-compressible energy norm \citep{chu1965energy} that has been applied in previous SPOD studies \citep{schmidt2018spectral}
\begin{equation} \langle\mathbf{q}_{j}, \mathbf{q}_{k}\rangle_{E}=\iiint \mathbf{q}_{1}^{H} \operatorname{diag}\left(\frac{\overline{T}}{\gamma_g \overline{\rho} M^{2}}, \overline{\rho}, \overline{\rho}, \overline{\rho}, \frac{\overline{\rho}}{\gamma_g(\gamma_g-1) \overline{T} M^{2}}\right) \mathbf{q}_{k} r \mathrm{d} x \mathrm{d} r \mathrm{d} \theta = \mathbf{q}_j^* \Wd \mathbf{q}_2,
\end{equation}
where $M$ is the Mach number, $\gamma_g$ is the ratio of specific heats, and the matrix $\mathbf{W}$ takes into account the energy and domain quadrature weights. To compute CS-SPOD, we choose $a_1 = 10$, resulting in a non-dimensional frequency range of $\Fset_{\fci_{St}} = [-3, 3] + \fci_{St}$, which encompasses all frequencies of interest. 

\begin{figure}
\centering
    \begin{minipage}[t]{0.48\textwidth}
            \centering
            \vskip 0pt
            \includegraphics[height=45mm]{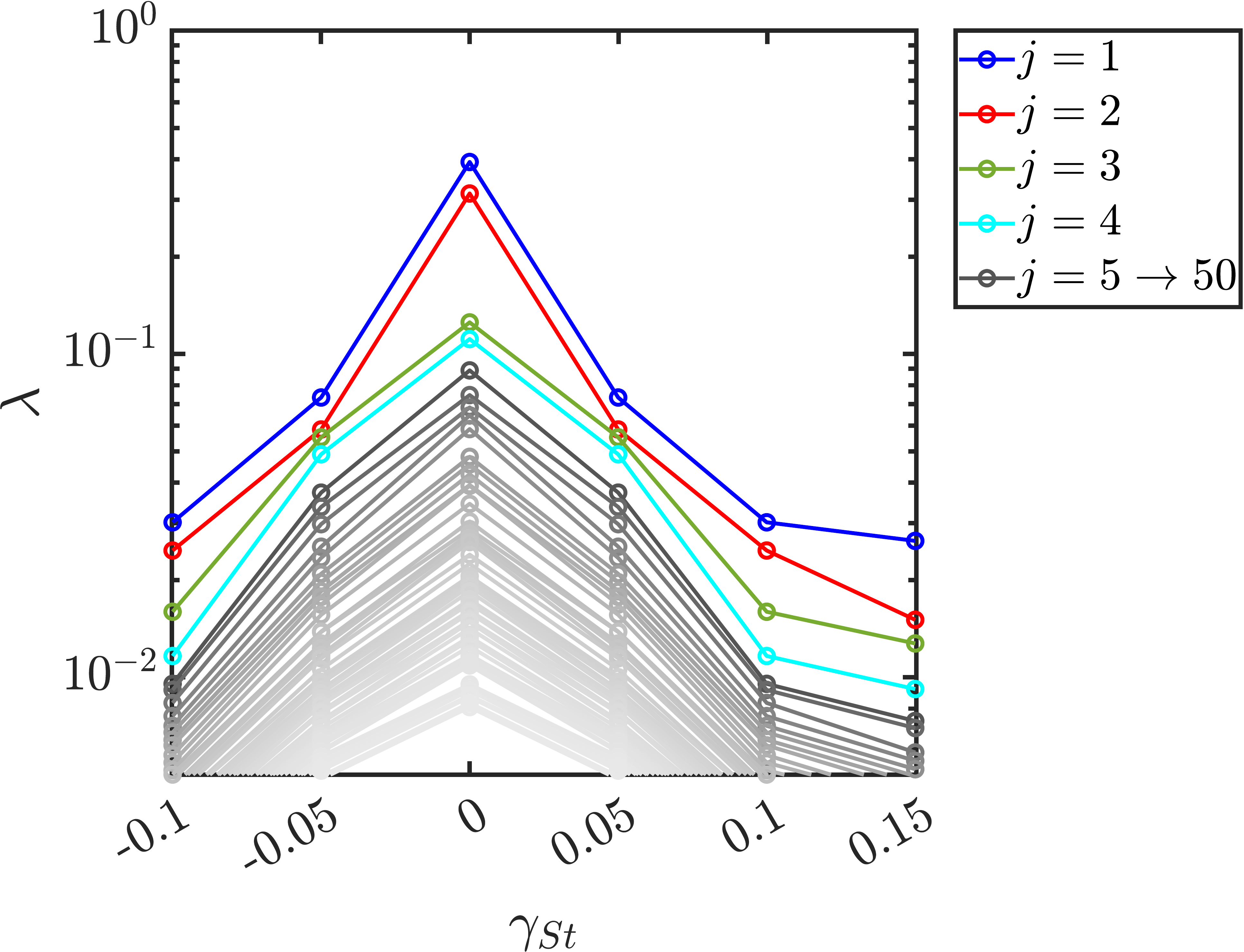}
            \caption{CS-SPOD Energy Spectrum for the forced Mach 0.4 turbulent jet.}
            \label{fig:jet:csespec}
    \end{minipage}\hfill %
    \begin{minipage}[t]{0.48\textwidth}
            \centering
            \vskip 0pt
            \includegraphics[height=45mm]{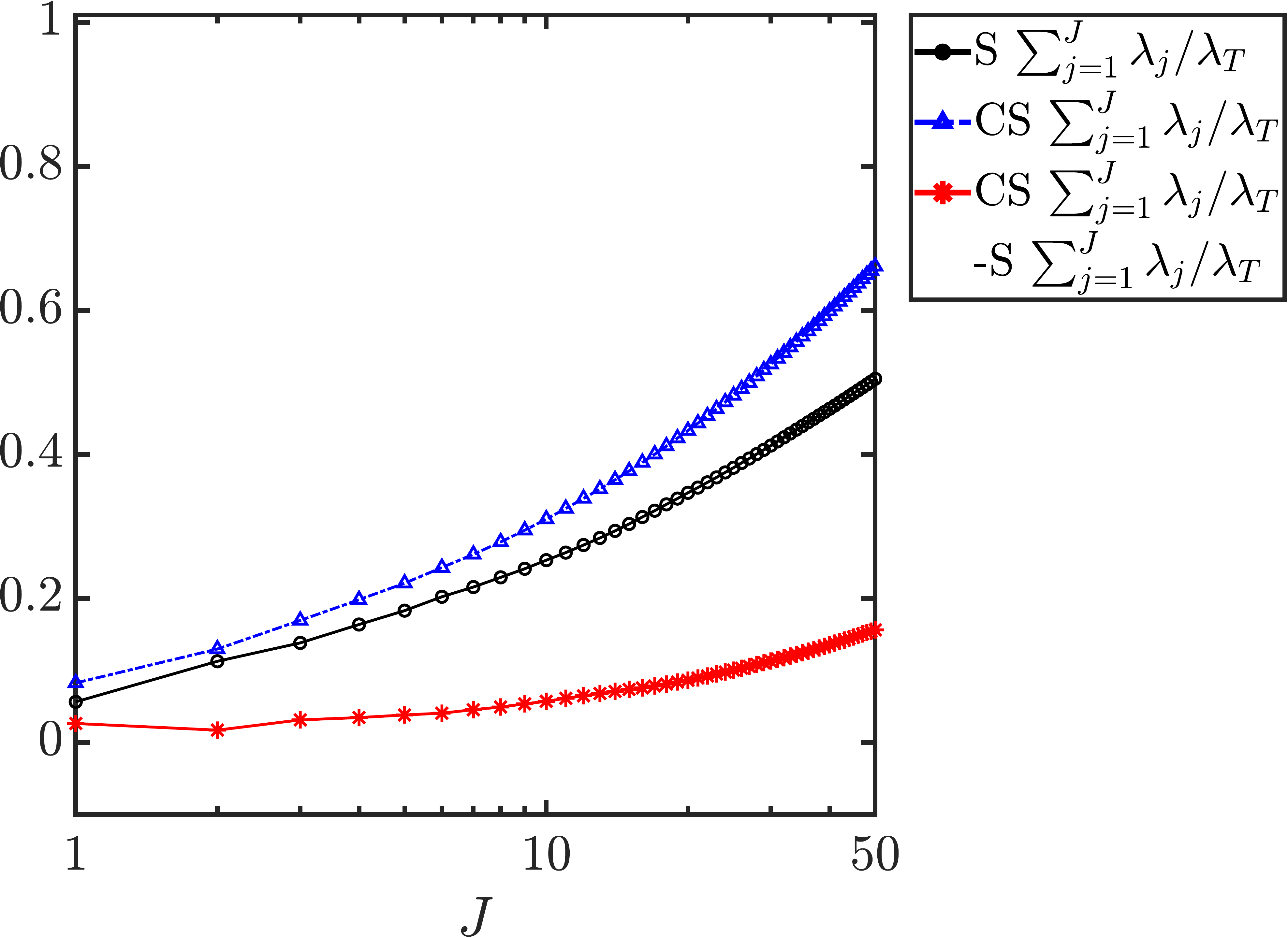}
            \caption{Total energy captured by a truncated set of CS-SPOD (CS) and SPOD (S) modes for the forced Mach 0.4 turbulent jet at $\fci_{St} = 0.15$.}
            \label{fig:jet:scstotalenergy}
    \end{minipage}
\end{figure}

We show the CS-SPOD eigenspectrum for the turbulent jet in figure \ref{fig:jet:csespec}. A large energy separation between the first three CS-SPOD modes is observed. Since CS-SPOD solves for multiple frequencies at a time, the energy separation will be smaller than with SPOD, in particular, with a flatter spectrum. The spectrum peaks at $\fci_{St} = 0$ and decays as $|\fci_{St}| \rightarrow 0.15$ which, because the smallest $|St| \in F$ occurs at $|\fci_{St}|$, occurs due to the decaying energy spectrum typically present in a turbulent jet. This low-rank behaviour, which is expected based on previous literature on natural turbulent jets (e.g. \citet{schmidt2018spectral}), is observed in figure \ref{fig:jet:scstotalenergy} where we show the fraction of the total energy captured by the first $J$ SPOD and CS-SPOD modes. The first CS-SPOD mode captures $8\%$ of the total energy present in the flow at the set of frequencies $\Fset_{\fci}$, 2 modes capture $13\%$, 10 modes capture $31\%$, and 50 modes capture $66\%$.  At $\fci_{St} = 0$ this increases to $16\%$, 29\%, $56\%$, and $87\%$ for 1, 2, 10, and 50 modes, respectively. Surprisingly, in contrast to the Ginsburg-Landau model, the energy separation between the most energetic CS-SPOD and SPOD modes is not large despite the high level of modulation present. However, despite this small difference, a large variation in the structure and temporal evolution of the most energetic SPOD and CS-SPOD modes is seen, which we explore next.

\begin{figure}
        \includegraphics[height=0.097326294\textwidth,valign=t]{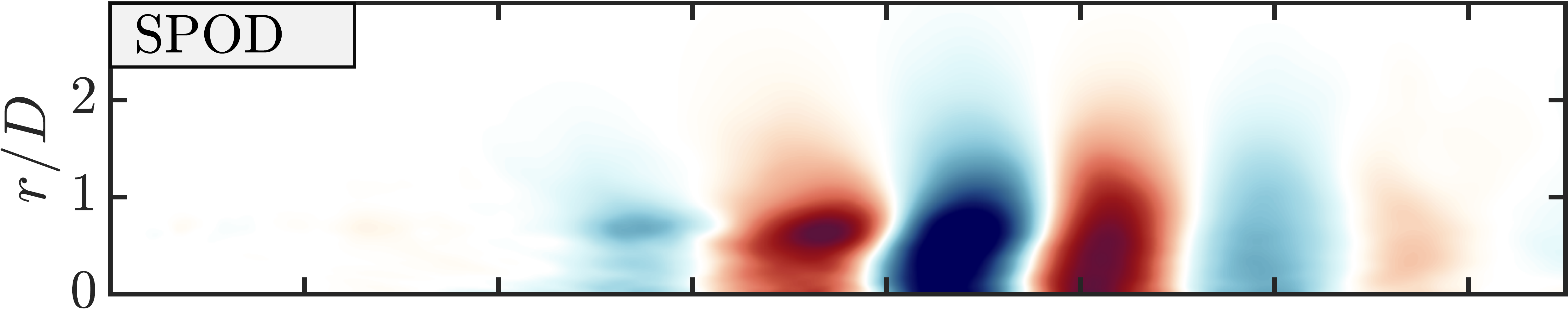} \hspace{0.2mm} 
        \includegraphics[height=0.09375\textwidth,valign=t]{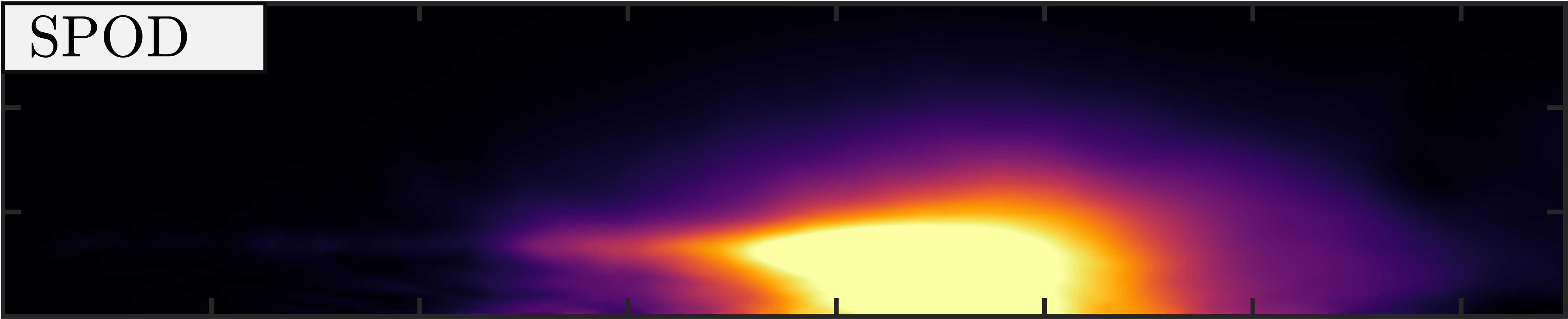}  \vspace{2mm} \hrule \vspace{1mm} \hrule \vspace{2mm} 
        \includegraphics[height=0.097326294\textwidth,valign=t]{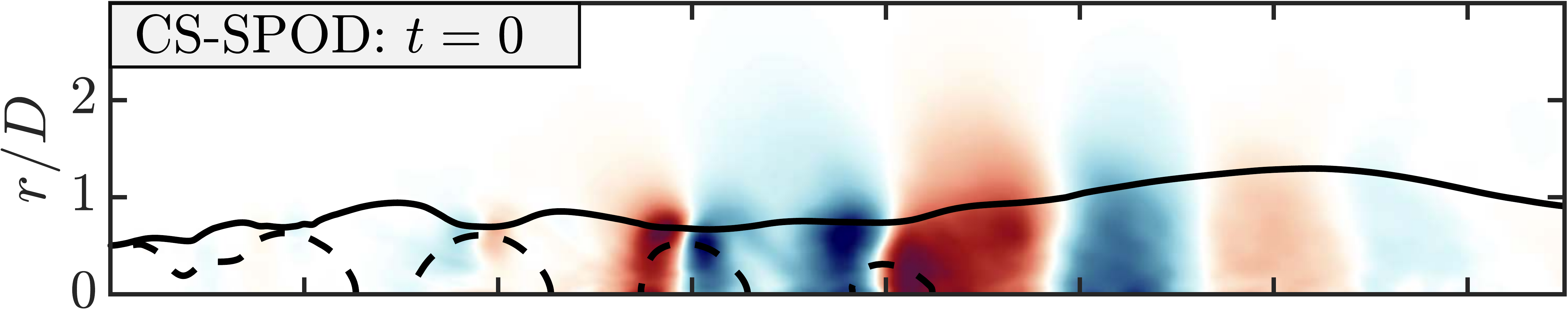} \hspace{0.2mm}
        \includegraphics[height=0.09375\textwidth,valign=t]{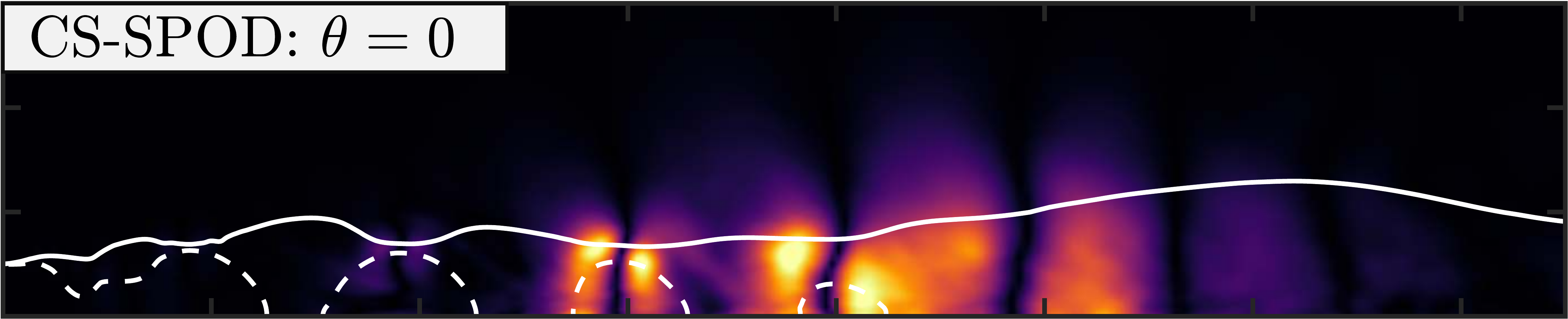}\vspace{1mm}\\ 
                \includegraphics[height=0.097326294\textwidth,valign=t]{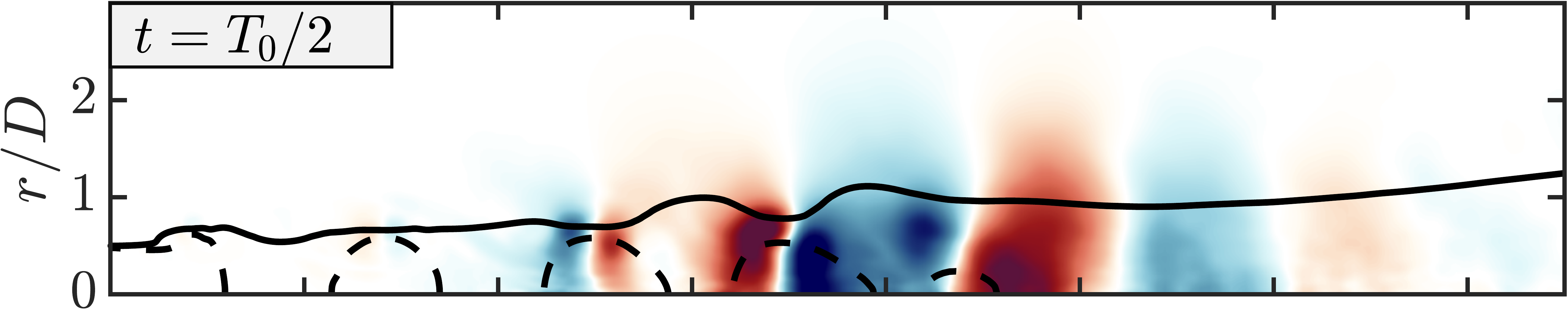} \hspace{0.2mm}
        \includegraphics[height=0.09375\textwidth,valign=t]{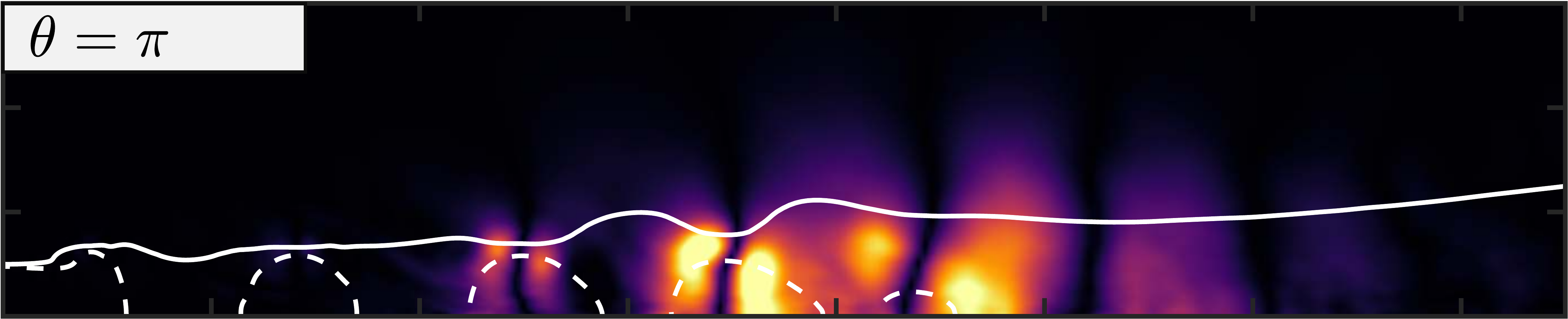}\vspace{1mm}\\ 
                \includegraphics[height=0.097326294\textwidth,valign=t]{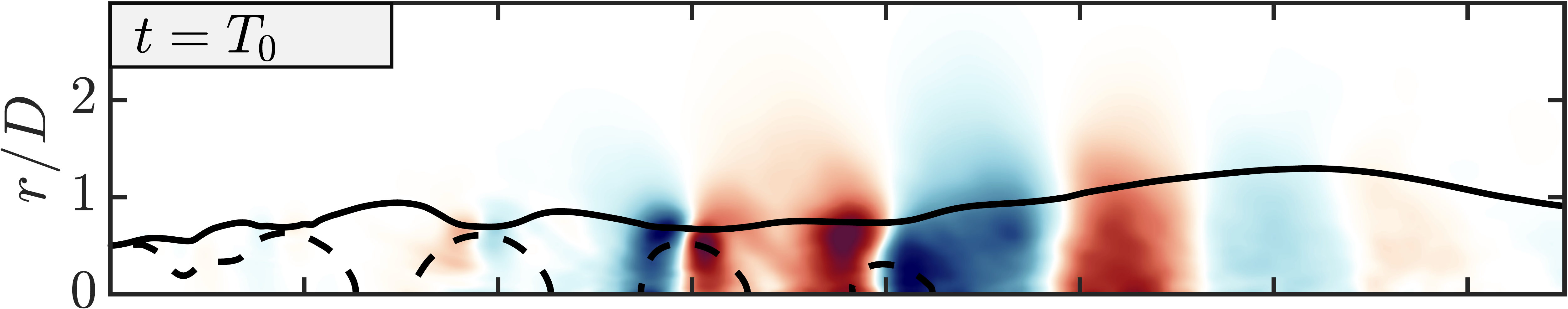} \hspace{0.2mm}
        \includegraphics[height=0.09375\textwidth,valign=t]{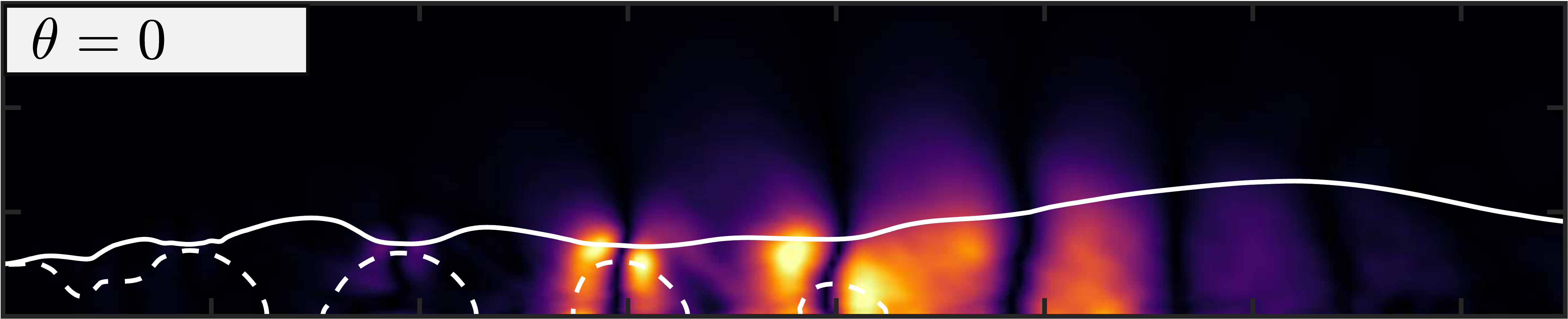}\vspace{1mm}\\ 
                \includegraphics[height=0.138208763\textwidth,valign=t]{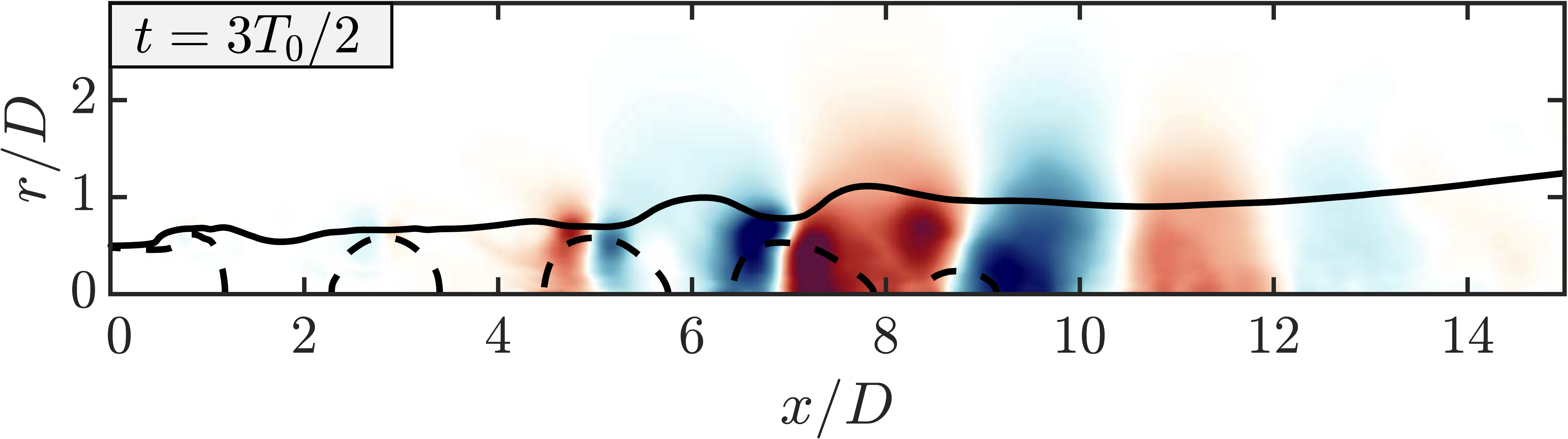} \hspace{0.2mm}
        \includegraphics[height=0.138208763\textwidth,valign=t]{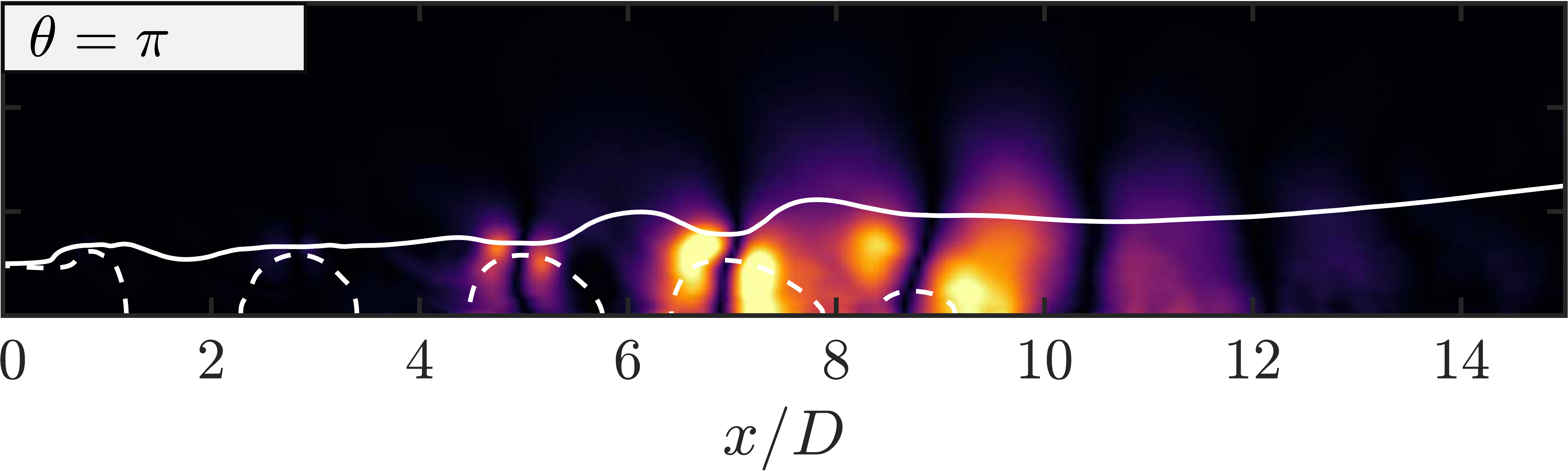}
        \caption{Comparison of the real and magnitude component of the dominant CS-SPOD mode to the dominant SPOD mode for $\fci_{St} = 0.15$ of the forced Mach 0.4 turbulent jet. All contours are set to $\pm 0.75|\mathfrak{R}\{\phi_{p,1}(x, r, t)\}|_\infty$ and $[0, 0.75|\phi_{p, 1}(x, r, t)|_\infty]$ for the real and magnitude contours, respectively.}
    \label{fig:jet:jetmodesboth}
\end{figure}

We show the real and absolute value of the pressure component of the most energetic SPOD and CS-SPOD mode at $\fci_{St} = 0.15$ in figure \ref{fig:jet:jetmodesboth}. The solid and dashed lines in these figures correspond to the contour lines of $\tilde{u}_x/U_j = 0.25, 0.75$. SPOD modes are only shown at a single time instance due to their time-invariant evolution, while CS-SPOD modes are shown at several time instances to show their temporal evolution. The most dominant SPOD mode is focused downstream at $x/D \approx 6-12$, has a frequency $St = 0.15$, and has a structure typical of the so-termed ``Orr modes'' previously observed in unforced turbulent jets \citep{schmidt2018spectral,pickering2020liftup}. By construction, the amplitude of the SPOD mode remains constant over time, and the region of maximum amplitude corresponds to $x/D \approx 6-12$ and $r/D \approx 0-1$. The real component of the most energetic CS-SPOD mode has a structure similar to the respective SPOD modes but with an additional modulation localized to the shear layer in regions of high velocity.  This is also observed in amplitude contours, where the amplitude of the mode substantially varies as a function of phase in a region similar to the amplitude profile of SPOD, but the high-amplitude regions always follow the high-velocity regions of the jet. The CS-SPOD modes follow this region since it is where the greatest amount of shear occurs along with the vortex roll-up (as seen in figures \ref{fig:firstorderjet} and \ref{fig:jetphase_new}). 

\begin{figure}
\centering
        \includegraphics[height=0.275\textwidth,valign=t]{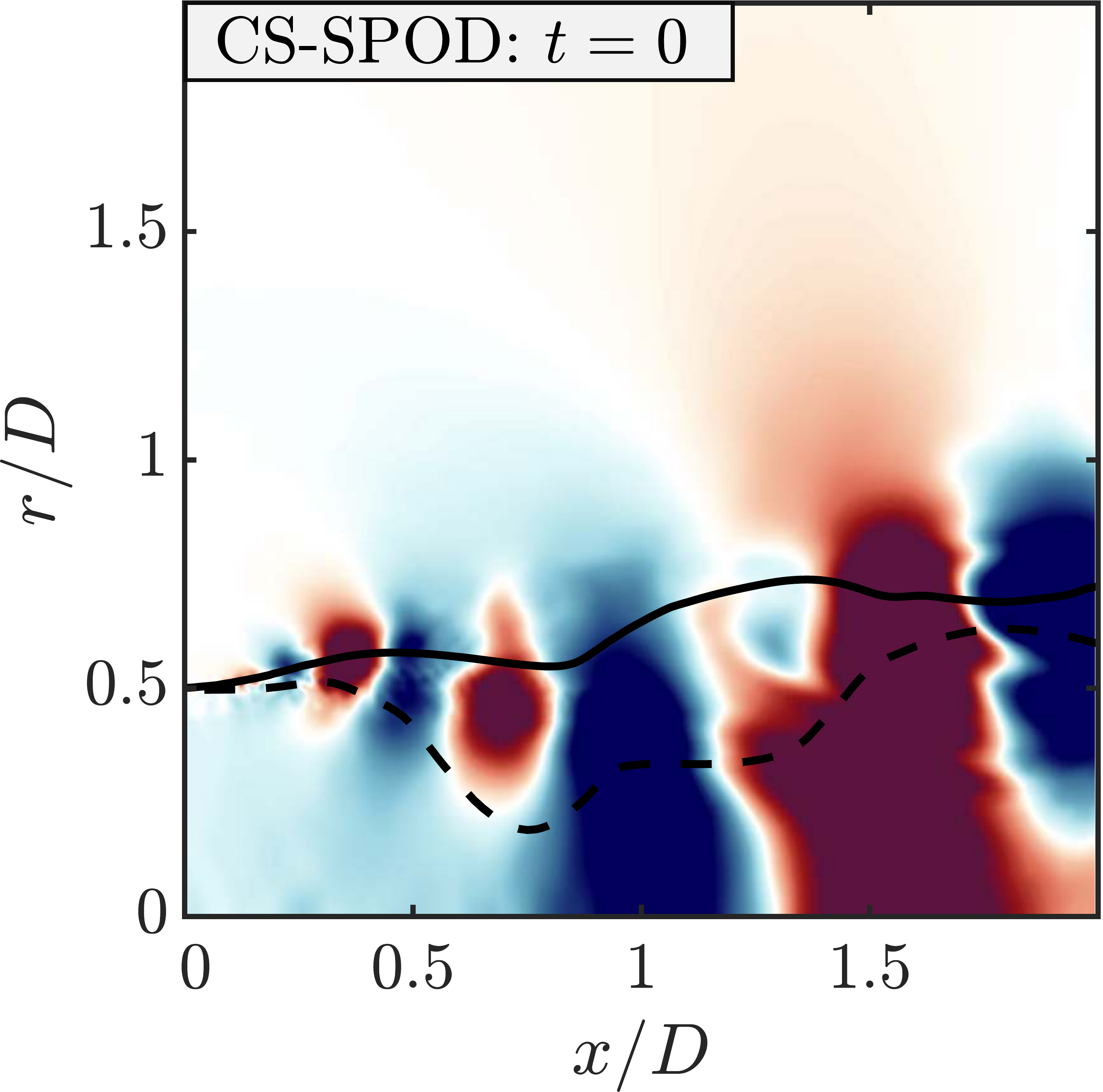}
        \includegraphics[height=0.275\textwidth,valign=t]{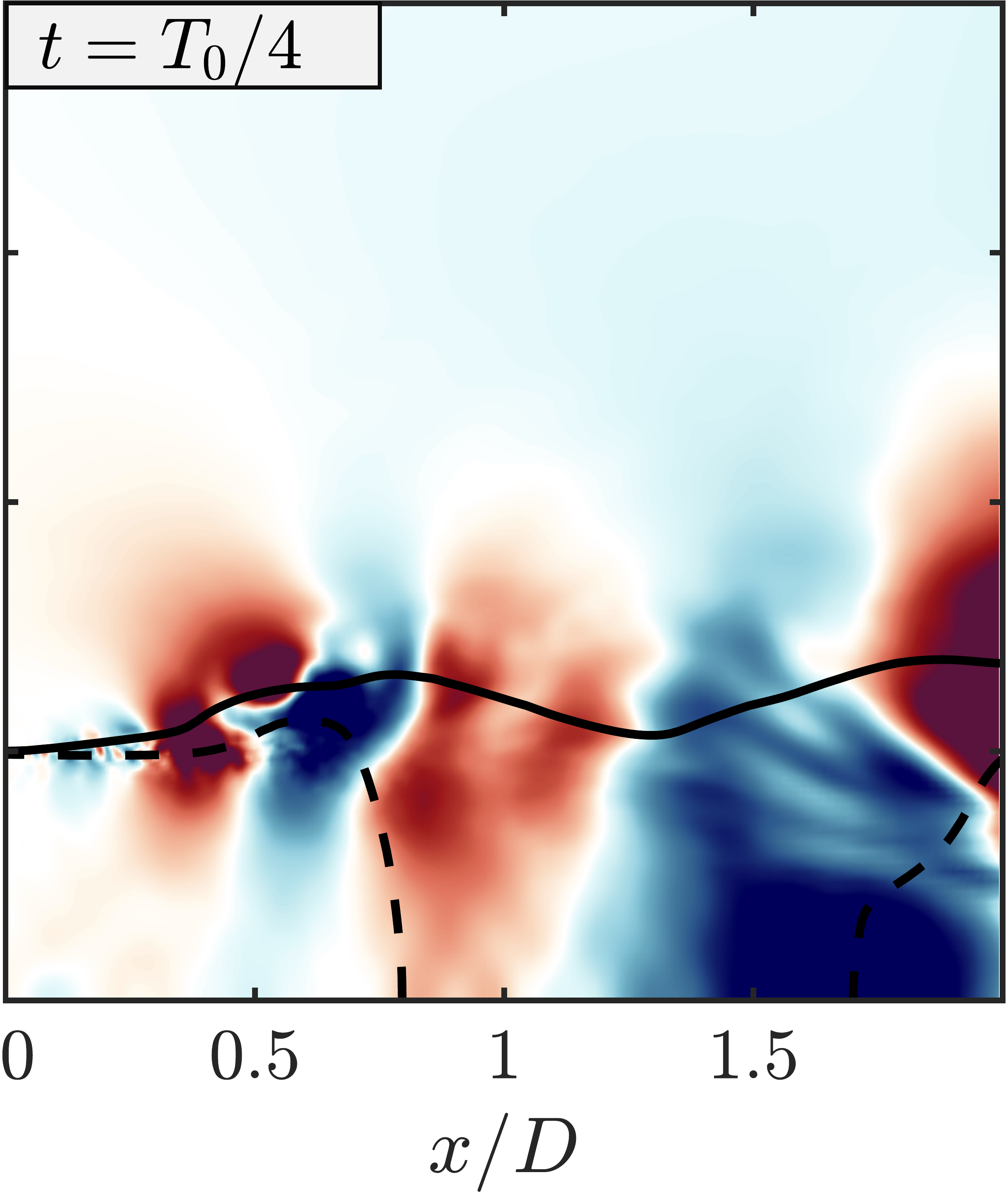}
        \includegraphics[height=0.275\textwidth,valign=t]{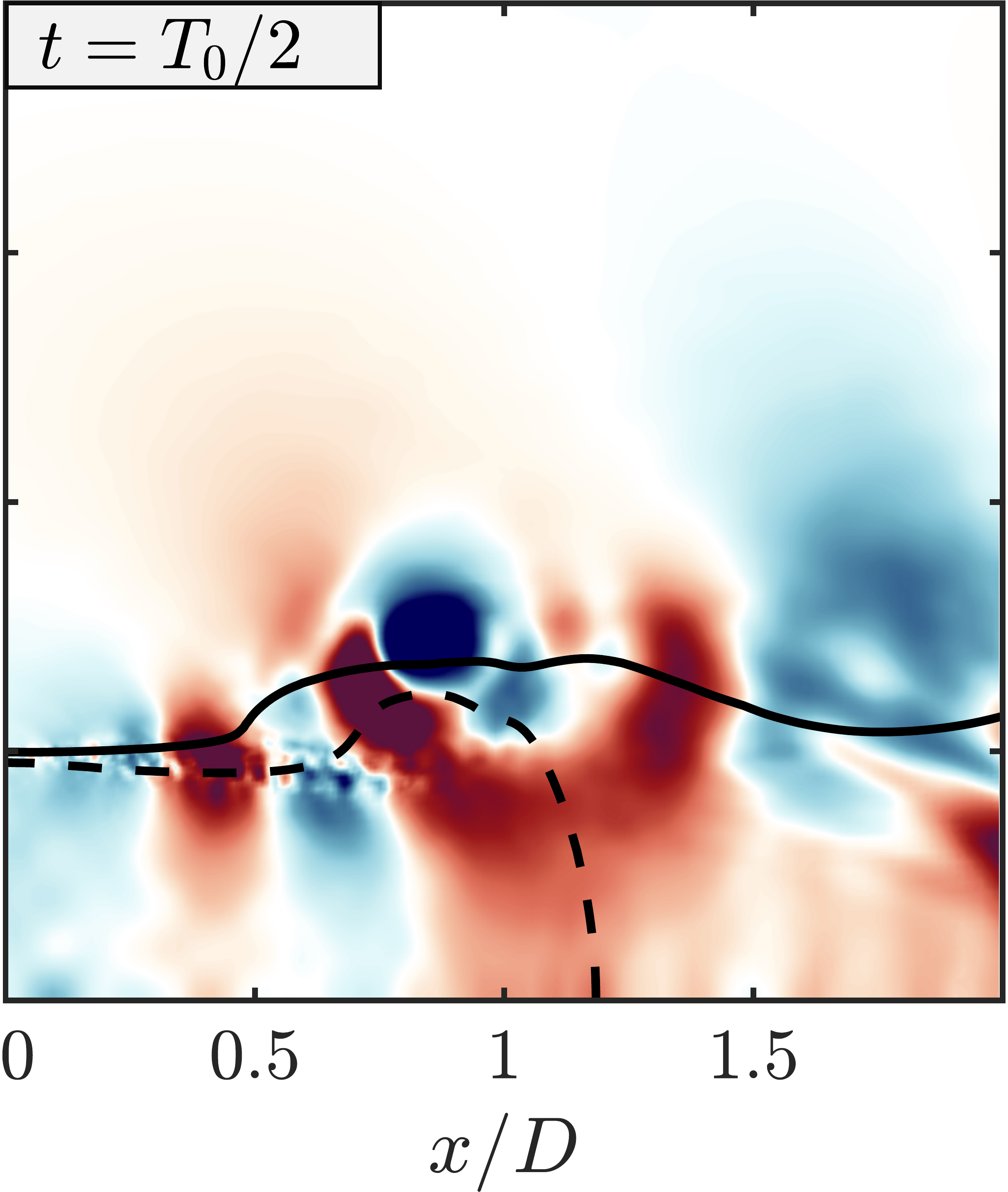}
        \includegraphics[height=0.275\textwidth,valign=t]{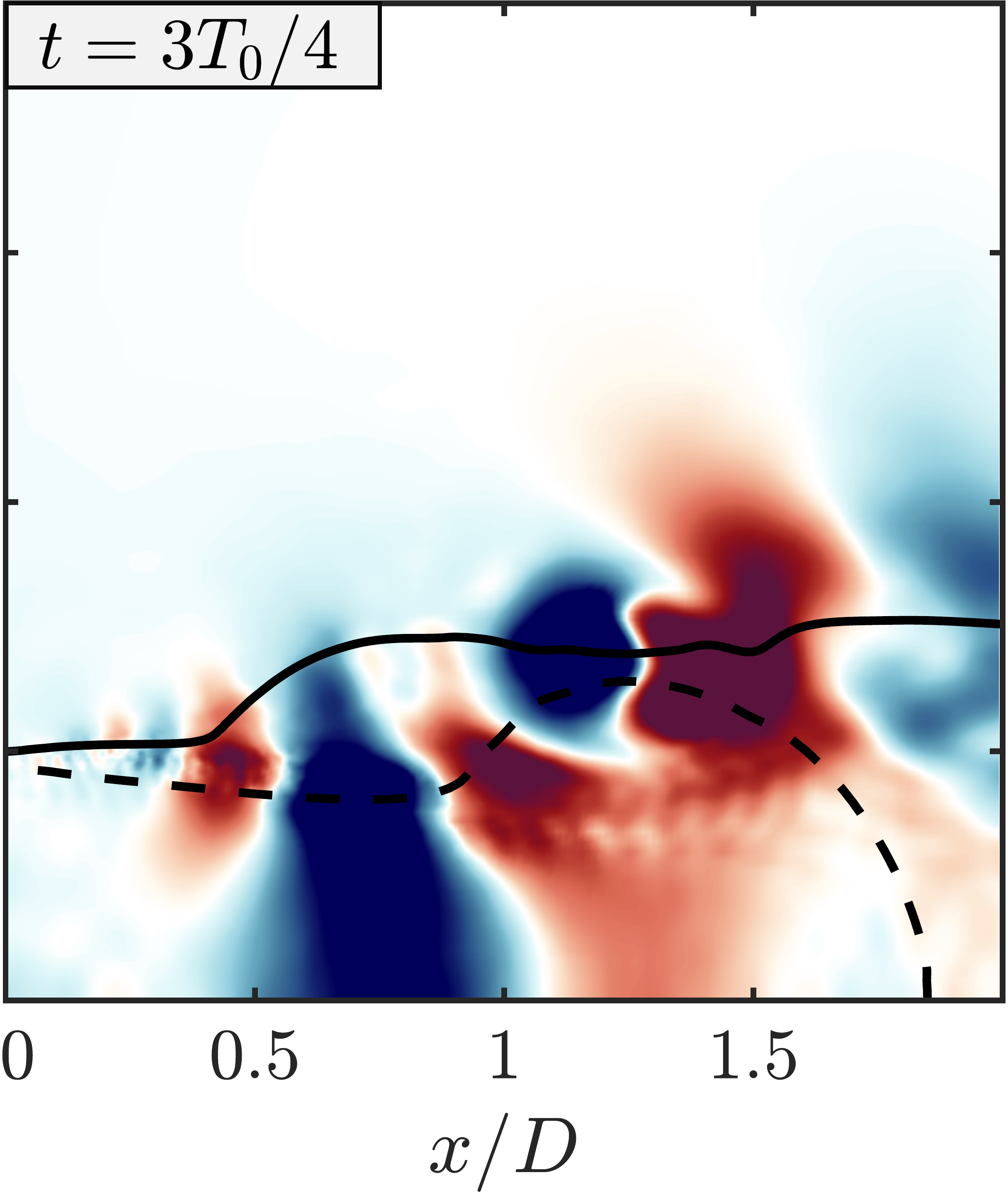} \\ 
        \caption{Real component of the dominant CS-SPOD mode for $\fci_{St} = 0.15$ for the forced Mach 0.4 turbulent jet (zoomed into $x/D = [0,\ 2], r/D = [0,\ 2])$. All contours are set to $\pm 0.25|\mathfrak{R}\{\phi_{p, 1}(x= [0,\ 2], r = [0,\ 2], t) \}|_\infty$.}
    \label{fig:jet:jetmodesrealzoomed}
\end{figure}

Figure \ref{fig:jet:jetmodesrealzoomed} shows the same CS-SPOD mode in a zoomed-in region near the nozzle exit, plotted with lower contour levels since the fluctuation levels are smaller there.  At $t = 0$ (i.e. $\theta = 0$), a short wavelength Kelvin-Helmholtz (KH) mode that is located between the $25\%$ and $75\%$ velocity lines in the $x/D = [0,\ 1]$ region is seen. The KH mode is angled towards the centerline due to the modulation of the mean flow.  Next, at  $t = T_0/4$ the KH mode has propagated slightly downstream and has become significantly weaker due to the much thinner shear layer at this phase of the motion. From $t = T_0/4$ to $t = 3T_0/4$, the KH mode increases in strength as it continues to propagate downstream due to the increasing thickness of the boundary layer. The KH mode also rotates due to the roll-up induced by the forcing, as seen in figure \ref{fig:jetphase_new}. At $t = 3T_0/4$, the KH mode is substantially stronger than at $t = T_0/4$ and is a lower-frequency structure located around $x/D = [0.6,\ 1]$ region and is angled away from the centerline. A corresponding interrogation of the SPOD mode shows no near-nozzle Kelvin-Helmholtz activity at this frequency, highlighting the ability of CS-SPOD to reveal potentially important dynamical effects that are slaved to the forcing frequency. 

\begin{figure}
\centering
    \begin{minipage}[b]{0.48\textwidth}
        \centering
       \includegraphics[height=0.7\textwidth,valign=t]{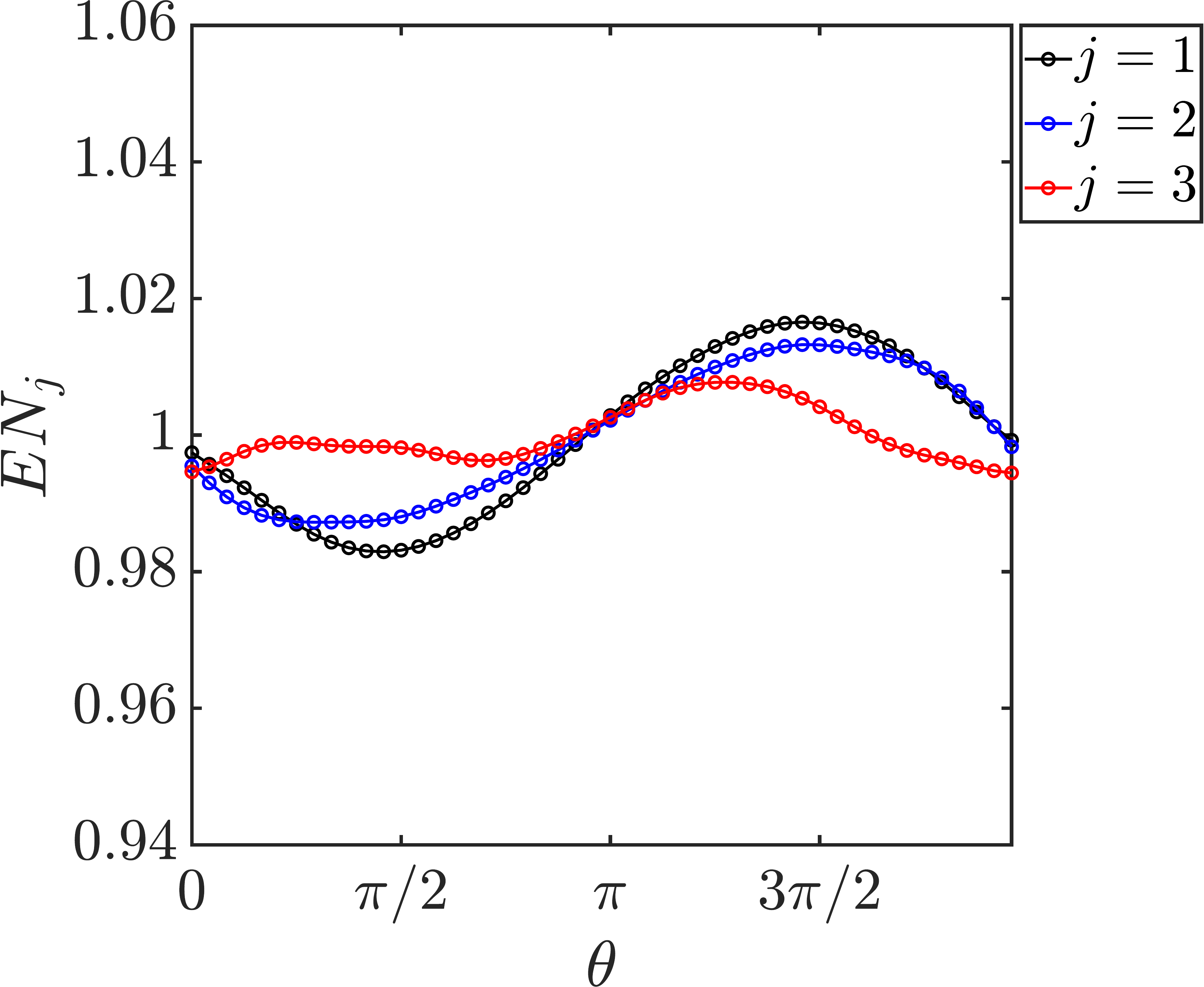}
        \caption{Energy of the dominant CS-SPOD modes over the phase of the external forcing for $\fci_{St} = 0.15$ for the forced Mach 0.4 turbulent jet.}
        \label{fig:jet:jeteoverphase}
    \end{minipage}\hfill %
    \begin{minipage}[b]{0.48\textwidth}  
        \centering 
       \includegraphics[height=0.7\textwidth,valign=t]{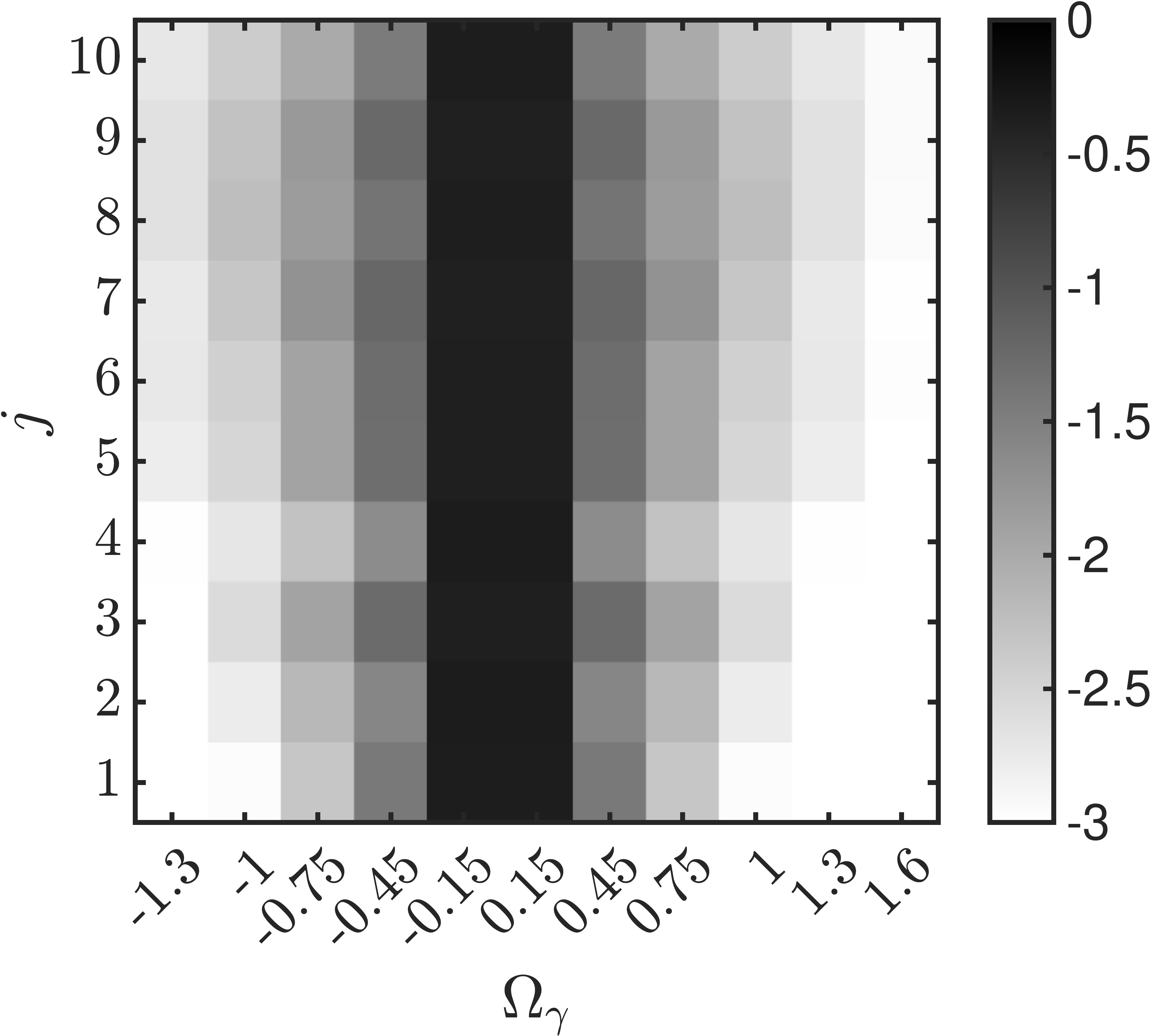}
       \caption{Fractional CS-SPOD eigenmode energy, by frequency, for $\fci_{St} = 0.15$ shown in $\text{log}_{10}$ scale for the forced Mach 0.4 turbulent jet.}
       \label{fig:jet:jetfracenergy}
    \end{minipage}% 
\end{figure}

Figure \ref{fig:jet:jeteoverphase} shows the normalized energy as a function of phase for the three dominant modes at $\fci_{St} = 0.15$. The energy, despite the large phase-dependent modulation seen in figure \ref{fig:jet:jetmodesboth}, varies by just $\pm 2\%$ as a function of phase. This demonstrates that, despite the strong phase-dependent structure of the mode and of the statistics present in the jet, on average, over the flow, the total energy contained within these modes is not strongly phase-dependent. Finally, in figure \ref{fig:jet:jetfracenergy}, we show the fractional energy of each frequency component $f \in \Fset_{\fci}$ for the CS-SPOD modes. The large amount of frequency interaction previously observed is visible, where for $j = 1$, the 8 highest energy frequency components are $\pm 0.15, \pm 0.45, \pm 0.75, \pm 1.05$ which contain $45.6\%, 3.7\%, 0.47\%, 0.11\%$ of the energy, respectively. Thus, a large amount of interaction occurs between the frequency components in $\Fset_{\fci}$, which results in the large periodicity observed. It is important to note that although a frequency component may only contain a small fraction of the total energy in a CS-SPOD mode, in many cases, it is still a physically important feature, such as the modulated KH mode discussed previously, and thus should be carefully studied.  \par

Overall, we see that the forcing clearly results in a large modulation of the KH and Orr modes present, an effect that SPOD is unable to capture. Thus, the utility of CS-SPOD to describe the coherent structures in a forced turbulent jet is demonstrated. 

\section{Harmonic resolvent analysis and its relationship to CS-SPOD} \label{sec:harres}

Harmonic resolvent analysis \citet{padovan2022analysis} extends resolvent analysis to time-periodic mean flows.  Starting with the nonlinear governing equations
\begin{equation}
    \frac{\partial \gb(t)}{\partial t} = \Hb(\gb(t)), 
    \label{eqn:harr:goveqn}
\end{equation}
where $\Hb$ is the time-independent continuity, momentum, and energy equations and $\gb(t) \in \mathbb{C}^N$ is the state vector of flow variables, we decompose the state as $\gb(\x, t) = \Gb(\x, t) + \gb^{\prime\prime}(\x, t)$, where $\Gb(t) = \Gb(t + T_0)$ is the $T_0$ periodic mean flow component (first-order component) and $ \gb^{\prime\prime}(t)$ is the turbulent component (second-order component). Since $\Gb(t)$ is periodic, it can be expressed as a Fourier series, giving $\Gb(t) = \sum_{n \in \mathcal{A}_n} \hat{\Gb}_{n} e^{i2\pi (n\alpha_0) t}$, where $\hat{\Gb}_{n}$ are harmonics of the fundamental frequency $\alpha_0 = 1/T_0$ of the mean flow (i.e. the Fourier series components), $T_0$ is the period of oscillation of the mean flow, and $n \in \mathcal{A}_n$ is defined as previous. The cycle frequencies, which in the context of linear analysis must be the frequencies present in the mean flow, are $n\alpha_0$. By substituting this decomposition into \eqref{eqn:harr:goveqn}, we obtain 
\begin{equation}
    \frac{\partial\gb^{\prime\prime}(t)}{\partial t} = {D}_{\gb} (\Hb(\Gb(t)) \gb^{\prime\prime}(t)+ \f^{\prime\prime}(t), 
    \label{eqn:harr:goveqndecomp}
\end{equation}
where $\f^{\prime\prime}(t)$ contains higher-order terms in $\gb^{\prime\prime}(t)$. The Jacobian $\Ab(t) = {D}_{\gb} (\Hb(\Gb(t))$ is also a periodic function in time, which, following the discussion in \citet{padovan2022analysis}, we assume is a differentiable function of time thereby guaranteeing a unique solution of \eqref{eqn:harr:goveqndecomp}. Subsequently, it is also expanded as a Fourier series $    \Ab(t) = \sum_{n \in \mathcal{A}_n} \hat{\Ab}_{n} e^{i2\pi (n\alpha_0) t}.$
Inserting this expansion into \eqref{eqn:harr:goveqndecomp}, gives
\begin{equation}
    \frac{\partial\gb^{\prime\prime}(t)}{\partial t} = \sum_{n \in \mathcal{A}_n} \hat{\Ab}_{n} e^{i2\pi (n \alpha_0) t} \gb^{\prime\prime}(t)+ \f^{\prime\prime}(t), 
    \label{eqn:harr:goveqndecompsub}
\end{equation}
which we Fourier transform in time and then separate by frequency, obtaining
\begin{equation}
    i2\pi \freq\hat{\gb}_{\freq} = \sum_{n \in \mathcal{A}_n} \hat{\Ab}_{n} e^{i2\pi (n \alpha_0) t} \hat{\gb}_{\freq- n \alpha_0}+ \hat{\f}_{\freq}, 
    \label{eqn:harr:goveqndecompsubfourier}
\end{equation}
where $\hat{\gb}_{\freq}$ and $\hat{\f}_{\freq}$ are the $\freq$-frequency components of  $\gb^{\prime\prime}(t)$ and $\f^{\prime\prime}(t)$, respectively. Equation \eqref{eqn:harr:goveqndecompsub} represents a system of coupled equations where perturbations at frequency $\freq$ are coupled to perturbations at frequency $\freq- n \alpha_0$ through the $n \alpha_0$ frequency component of the mean flow. In general, this results in an infinite-dimensional problem similar to the infinite-dimensional CS-SPOD eigenvalue problem. In practice, identically to CS-SPOD, we restrict the perturbation frequencies to $[\freq- a_1\alpha_0,\ \freq+ a_1\alpha_0]$ and thus, we seek time-periodic perturbations of  $\gb^{\prime\prime}(t) = \sum_{m\in \mathcal{A}_m} \hat{\gb}_{\freq+ m\alpha_0} e^{i2\pi (\freq+ m\alpha_0) t}$, where $\mathcal{A}_m = \{-a_1, \cdots, -1, 0, 1, \cdots, a_1\}$. This results in a solution frequency set of  $\Fset_{\fci} = \{-a_{1}\alpha_0 + \fci,\ (-a_1+1)\alpha_0 + \fci,\ \cdots,  \fci,\  \cdots, (a_1-1)\alpha_0+ \fci,\ a_1\alpha_0+ \fci\}$.

We also limit the mean flow frequencies to $[-a_2,\ a_2]$ with $a_2 \le a_1$. The final problem is compactly written as 
\refstepcounter{equation}
\begin{equation}
    (i2\pi\freq\ICb - \hat{\TdCb})\hat{\GCb} = \hat{\FCb}, 
    \label{eqn:HRsystem}
\end{equation} 
where 
\refstepcounter{equation}
\begin{equation}
\hat{\TdCb} = 
\renewcommand\arraystretch{1.7}
\begin{bmatrix}
\ddots&\ddots&\ddots&\ddots& \\
\ddots&  \hat{\Rres}_{- \alpha_0} &  \hat{\Ab}_{-\alpha_0} & \hat{\Ab}_{-2\alpha_0} & \ddots & \\
\ddots&  \hat{\Ab}_{\alpha_0} &  \hat{\Rres}_{0} & \hat{\Ab}_{-\alpha_0}   & \ddots\\
\ddots& \hat{\Ab}_{2\alpha_0} & \hat{\Ab}_{\alpha_0}  &  \hat{\Rres}_{\alpha_0} &\ddots \\
& \ddots&  \ddots & \ddots & \ddots
\end{bmatrix}, \quad \hspace{-3mm}
\hat{\GCb}= \begin{bmatrix}
\vdots\\ 
\hat{\gb}_{\freq- \alpha_{0}}\\ 
\hat{\gb}_{\freq}\\ 
\hat{\gb}_{\freq+ \alpha_{0}}\\ 
\vdots
\end{bmatrix}, \quad \hspace{-3mm}
  \hat{\FCb}= \begin{bmatrix}
\vdots\\ 
\hat{\fb}_{\freq- \alpha_{0}}\\ 
\hat{\fb}_{\freq}\\ 
\hat{\fb}_{\freq+ \alpha_{0}}\\ 
\vdots
\end{bmatrix},  \quad \hspace{-3mm} 
\tag{\theequation{$\mathit{a},\mathit{b}, \mathit{c}$}}
\end{equation}
$\hat{\Rres}_{k \alpha_0} = (-i k\alpha_0 \Ib + \hat{\Ab}_{0}) \in \mathbb{C}^{N\times N}$, and $\Ib \in \mathbb{R}^{(2a_1 + 1)N \times (2a_1 + 1)N}$ is the identity operator. The harmonic resolvent operator is then defined as $\hat{\HCb} = (i2\pi\freq\ICb - \hat{\TdCb})^{-1} \in \mathbb{C}^{(2a_1 + 1)N \times (2a_1 + 1)N}$ and has $(2a_1+1)$ coupled equations and is $(2a_2 + 1)$ banded-block-diagonal due to the periodicity of the mean flow. If the flow is time-invariant, then all off-diagonal blocks are zero, i.e. there is no cross-frequency coupling, and the system becomes block-diagonal where each diagonal block is the standard resolvent problem at frequency $\freq + k\alpha_{0}, k \in \mathbb{Z}$. As detailed by \citet{padovan2022analysis}, the singularity in the harmonic resolvent operator must be removed to avoid numerical difficulties. \par

Similar to CS-SPOD, harmonic resolvent analysis is periodic in $\freq$, and thus we must only solve over the range $\fci \in \Fset_\fci = (-\alpha_0/2,\ \alpha_0/2]$. We then seek to solve the forcing mode $\hat{\FCb}$ that results in the most energetic response $\hat{\GCb}$, expressed as the following optimization problem 
\begin{equation}
\sigma^2=\frac{\langle{\hat{\GCb}}, {\hat{\GCb}}\rangle_{\GC}}{\langle\hat{{\FCb}}, \hat{{\FCb}}\rangle_{\FC}},
\end{equation} 
where $\langle{\hat{\GCb}}_{j}, {\hat{\GCb}_{k}}\rangle_{\GC}$ and $\langle\hat{{\FCb}}_{j}, \hat{{\FCb}}_{k}\rangle_{\FC}$ are inner products on the output and input spaces, respectively, and are given by
\begin{subequations}
\begin{align}
    \langle{\hat{\GCb}}_{j}, {\hat{\GCb}_{k}}\rangle_{\GC} = \int_{\Omega} \hat{\GCb}^*_{k}(\x, f) \WCb_{\GC}(\x) \hat{\GCb}_{j}(\x, f) \dx, \\
    \langle{\hat{\FCb}}_{j}, {\hat{\FCb}_{k}}\rangle_{\FC} = \int_{\Omega} \hat{\FCb}^*_{k}(\x, f) \WCb_{\FC}(\x) \hat{\FCb}_{j}(\x, f) \dx.
\end{align}
\label{eqn:normdef}
\end{subequations}
The solution to this optimization problem is given by the singular value decomposition of the weighted harmonic resolvent operator
\begin{equation}
\widetilde{\HCb}    = \WCb_{\GC}^{1/2} \hat{\HCb} \WCb_{\FC}^{-1/2} = \widetilde{\UCb} \Sigmab \widetilde{\VCb}^*, 
\end{equation}
where the diagonal matrix $\Sigmab = \text{diag}[\sigma_1^2, \sigma_2^2, \cdots]$ contains the ranked gains and the columns of $\VCHb = \WCb_{\FC}^{-1/2} \widetilde{\VCb}$ and  $\UCHb= \WCb_{\GC}^{1/2} \widetilde{\UCb}$ contain the forcing and response modes, respectively. These modes have an to analogous structure to $\hat{\FCb}$ or $\hat{\GCb}$, and the $j^{th}$ forcing and response modes ($\UCHb_{j}$,  $\VCHb_{j}$) can be reconstructed in the time-domain as
\begin{subequations}
\begin{align}
    \UCb_j = \UCb_j(\x, t) &= \sum_{m\in \mathcal{A}_m}  \hat{\ucb}_{j, \freq+ m\alpha_0} e^{i2\pi (\freq+ m\alpha_0) t} \\
    \VCb_j = \VCb_j(\x, t) &= \sum_{m\in \mathcal{A}_m}  \hat{\vcb}_{j, \freq+ m\alpha_0} e^{i2\pi (\freq+ m\alpha_0) t}.
\end{align}
\end{subequations}
respectively. These modes are orthonormal in their respective spatial norms \newline $\langle{\UCHb_{j}}, {\UCHb_{k}}\rangle_{\GC} = \langle \VCHb_{j},\VCHb_{k}\rangle_{\FC} = \delta_{j,k}$ and the temporal modes are orthogonal in their respective space-time norms $\langle{{\UCb}}_{j}, {{\UCb}_{k}}\rangle_{\GC(x, t)},  \langle{{\VCb}}_{j}, {{\VCb}_{k}}\rangle_{\FC(x, t)}$, where
\begin{subequations}
\begin{align}
    \langle{{\UCb}}_{j}, {{\UCb}_{k}}\rangle_{\UC(x, t)} &= \int_{\Omega} {\UCb}^*_{k}(\x, t) \WCb_{\GC}(\x) {\UCb}_{j}(\x, t)\ d\x dt,  \\
    \langle{{\VCb}}_{j}, {{\VCb}_{k}}\rangle_{\GC(x, t)} &= \int_{\Omega} {\VCb}^*_{k}(\x, t) \WCb_{\FC}(\x) {\VCb}_{j}(\x, t)
    \ d\x dt, 
\end{align}
\end{subequations}
The decomposition is complete, allowing the output to be expanded as
\begin{equation}
    \hat{\GCb}(\x, \freq)= \sum_{j = 1}^\infty  \UCHb_j(\x, \freq) \sigma_j(\freq) \beta_j(\freq),
\end{equation}
where
\begin{equation}
    \beta_j(\freq) = \langle \hat{\FCb}(\x, \freq), \UCHb_j(\x, \freq)\rangle_{\FC}.
\end{equation}
A connection between harmonic resolvent analysis and CS-SPOD is obtained using an approach analogous to that of \citet{towne2018spectral} and is similar to relationship between resolvent analysis and SPOD. In \S \ref{sec:CStheory}, it was shown that $ \Sb(\x, \x^\prime, \alpha, f)$ can be compactly written as
\begin{equation}
    \Sb(\x, \x^\prime, \alpha, f) = E\{ \hat{\qb}(\x, f - \alpha/2) \hat{\qb}^*(\x^\prime, f + \alpha/2)\}, 
\end{equation}
where $\hat{\qb}(\x, f)$ is the short-time Fourier transform of $\qb(\x, t)$. Similarly, the CS-SPOD decomposition tensor for the process $\qb(\x, t)$ can be written as
\begin{equation}
    \MCSb(\x, \x^\prime, \fci) =  E\{ \hat{\QCb}(\x, \fci) \hat{\QCb}^*(\x^\prime, \fci)\}.
\end{equation}
To develop a relationship between CS-SPOD and harmonic resolvent analysis, we equate the CS-SPOD and harmonic resolvent expansions of the CS-SPOD decomposition matrix and set all norms to be equal, i.e. $\langle\ \cdot\ \rangle = \langle\ \cdot\ \rangle_{\GC} = \langle\ \cdot\ \rangle_{\FC} = \langle\ \cdot\ \rangle_{x}$, giving
\begin{subequations}
\begin{align}
\MCSb(\x, \x^\prime, \fci) &= \sum_{j = 1}^\infty \lambda_j(\fci) \Psib_{j}(\x, \fci) \Psib^*_{j}(\x^\prime, \fci)\\
& =\sum_{j=1}^{\infty} \sum_{k=1}^{\infty}\UCHb_j(\x, \fci)\UCHb_k^*(\x^{\prime}, \fci) \sigma_j(\fci) \sigma_k(\fci) S_{\beta_j \beta_k}(\fci), 
\end{align}
\end{subequations}
where $S_{\beta_j \beta_k}(\fci) = E\{\beta_j(\fci) \beta^*_k(\fci) \}$ is the scalar CSD between the $j^{th}$ and $k^{th}$ expansion coefficients. Identical to \citet{towne2018spectral}, the output harmonic resolvent modes and singular values were moved outside of the expectation operator since they are deterministic quantities. Conversely, the expansion coefficients depend on the forcing $\hat{\FCb}(\x, \freq)$, which is stochastic due to the random nature of turbulent flows and thus is described by the CSD. In the case of a stationary process, $\MCSb(\x, \x^\prime, \fci)$ is block-diagonal, meaning that $\Psib_{j}(\x, \fci)$ and $\UCHb_j(\x, \fci)$ contain only a single non-zero frequency component per mode, and this relationship simplifies to that in \citet{towne2018spectral}. For uncorrelated expansion coefficients $S_{\beta_j \beta_k}(\fci) = \mu_j(\fci)\delta_{jk}$, the relationship simplifies to
\begin{subequations}
\begin{align}
\MCSb(\x, \x^\prime, \fci) &= \sum_{j = 1}^\infty \lambda_j(f) \Psib_{j}(\x, \fci) \Psib^*_{j}(\x^\prime, \fci), \\
& =\sum_{j=1}^{\infty}\UCHb_j(\x, \fci)\UCHb_j^*(\x^{\prime}, \fci) \sigma_j^2(f) \mu_j(\fci). 
\end{align}
\end{subequations}
Since orthogonal diagonalizations are unique, this shows that CS-SPOD modes and harmonic resolvent modes are identical, and the $k^{th}$ most energetic CS-SPOD mode corresponds to the resolvent mode with the $k^{th}$ greatest $\sigma_j^2(\fci) \mu_j(\fci)$. If $\mu_j = 1$ for all $j$, then  $\sigma_j^2(\fci) = \lambda_j(\fci)$ and $\Psib_{j}(\x, \fci) = \UCHb_j(\x, \fci)$ showing that the ranked CS-SPOD eigenvalues equal the ranked harmonic resolvent gains. To determine the conditions when the expansion coefficients are uncorrelated, we perform identical manipulation to \citet{towne2018spectral}, and show that 
\begin{equation}
     S_{\beta_j \beta_k}(\fci) = \langle \langle  \MCSb_{\FC \FC}(\x, \x^\prime, \fci), \UCHb_j(\x, \fci)\rangle^* , \UCHb_k(\x, \fci)\rangle^*, 
     \label{eqn:sbetabeta}
\end{equation}
where $\MCSb_{\FC \FC}(\x, \x^\prime, \fci) =  E\{ \hat{\FCb}(\x, \fci) \hat{\FCb}^*(\x^\prime, \fci)\}$ is the CS-SPOD decomposition tensor of  $\hat{\FCb}(\x, \fci)$. Since harmonic resolvent modes are orthogonal, if $\langle  \MCSb_{\FC \FC}(\x, \x^\prime, \fci), \UCHb_j(\x, \fci)\rangle^* = \mu_j(\fci) \UCHb_j(\x, \fci)$ then $S_{\beta_j \beta_k}(\fci) = \mu_j(\fci) \delta_{jk}$. This can be written as
\begin{equation}
     \int_{\Omega} \MCSb_{\eta \eta}(\x, \x^\prime, \fci) \MCWb_\eta(\x^\prime) \VCHb_j(\x^\prime, \fci) \dx^\prime = \mu_j(\fci) \VCHb_j(\x, \fci),
\end{equation}
which is identical to the CS-SPOD of the input. One can then show that the expansion coefficients are uncorrelated if and only if the harmonic resolvent input modes correspond exactly with the CS-SPOD modes of the input.  Thus, we conclude that the relationship between CS-SPOD and harmonic resolvent analysis is identical to that of SPOD and resolvent analysis. \par

We can then specialize for $\mu_j = 1$, giving
\begin{equation}
    \MCSb_{\eta \eta}(\x, \x^\prime, \fci) \MCWb_\eta(\x^\prime) = \ICb \delta(\x - \x^\prime), 
\end{equation}
which for $\MCWb_\eta(\x^\prime) = \ICb$ results in $\MCSb_{\eta \eta}(\x, \x^\prime, \fci) = \ICb \delta(\x - \x^\prime)$, i.e. the forcing is unit-amplitude white noise. This results in identical harmonic resolvent and CS-SPOD modes along with equal identical energies/gains, i.e. $\sigma_j^2 = \lambda_j$. 

We demonstrate this result by comparing the CS-SPOD and harmonic resolvent analysis results for the modified forced Ginzburg-Landau for $A_{\mu} = 0.4$. For both CS-SPOD and harmonic resolvent analysis, we employ $a_1 = 5$ resulting in a frequency range of $\Fset_{\fci} = [-0.5, 0.5] + \fci$. To compute CS-SPOD, we force the system with unit variance band-limited white noise. This is constructed similarly to the spatially correlated case previously considered in \S \ref{sec:FGL} without the step to introduce the spatial correlation. We employ identical computational parameters to those used in  \S \ref{sec:FGL}. \par

\begin{figure}
        \centering
        \includegraphics[width=0.55 \textwidth]{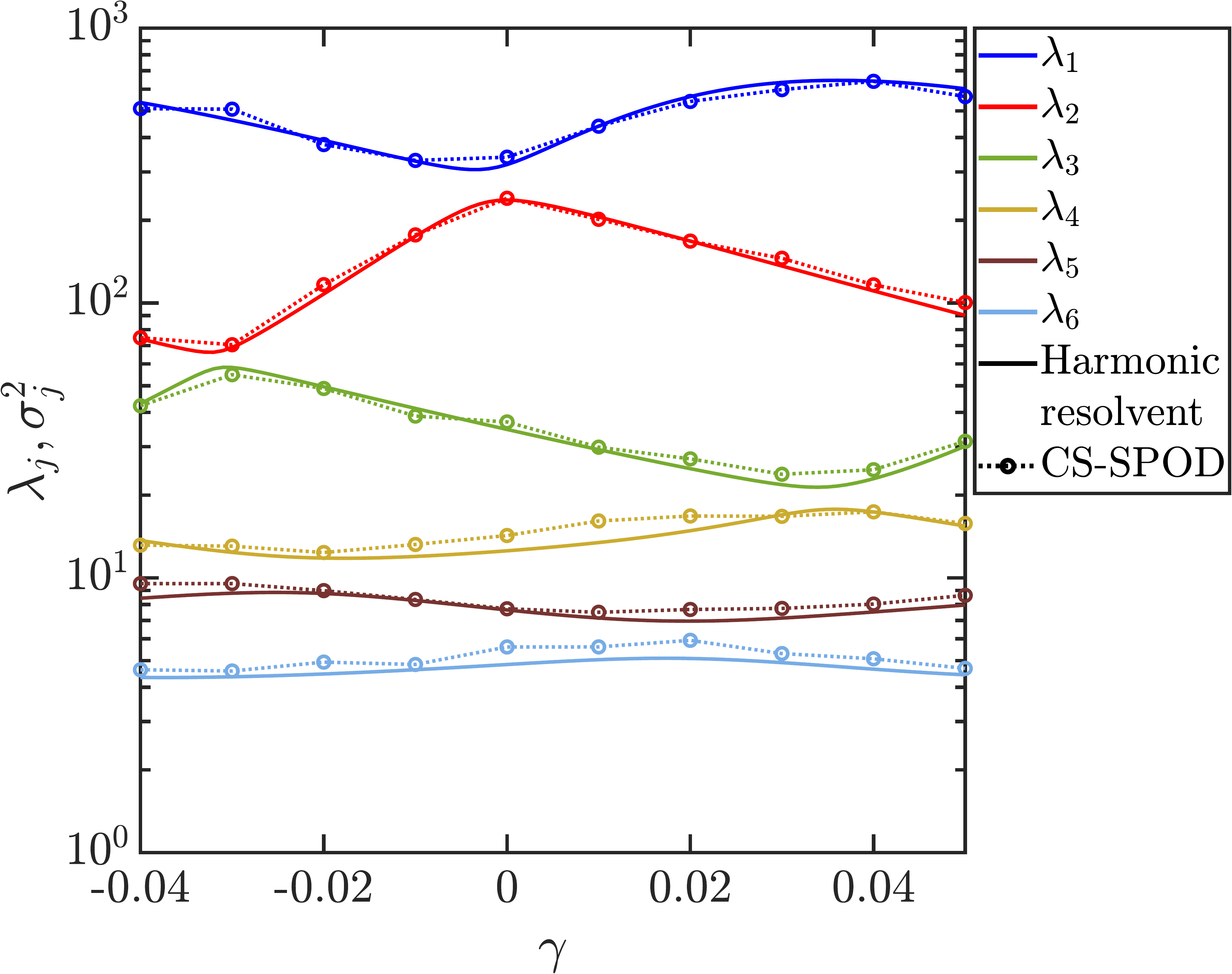}
        \caption{Comparison of the first six harmonic resolvent gains $\sigma_j^2$ and CS-SPOD eigenvalues $\lambda_j$ as a function of $\fci$ for the white noise forced Ginzburg-Landau system with $A_{\mu} = 0.4$.}
        \label{fig:CSHAREig}
\end{figure}

As demonstrated in \S \ref{sec:harres}, because the forcing is white CS-SPOD modes and harmonic resolvent analysis modes are theoretically identical. Furthermore, since the inner product has unit weight, the CS-SPOD eigenvalues equal the harmonic resolvent analysis gains. Figure \ref{fig:CSHAREig} shows the first six CS-SPOD eigenvalues and harmonic resolvent gains. Overall, excellent agreement is observed between the CS-SPOD eigenvalues and harmonic resolvent gains. The small amount of jitter present in the CS-SPOD eigenvalues is due to statistical convergence.  The minor overshoot or undershoot is associated with spectral and cycle leakage, which can be reduced by increasing the frequency resolution of the estimate. As with any spectral estimate, increasing the length of the blocks reduces the number of blocks leading to the well-known bias-variance tradeoff. Improved control over the bias-variance tradeoff in SPOD was achieved using multi-taper methods \citep{schmidt2022spectral} and could similarly be used for CS-SPOD. \par

\begin{figure}
\centering
\includegraphics[height=0.09438\textwidth]{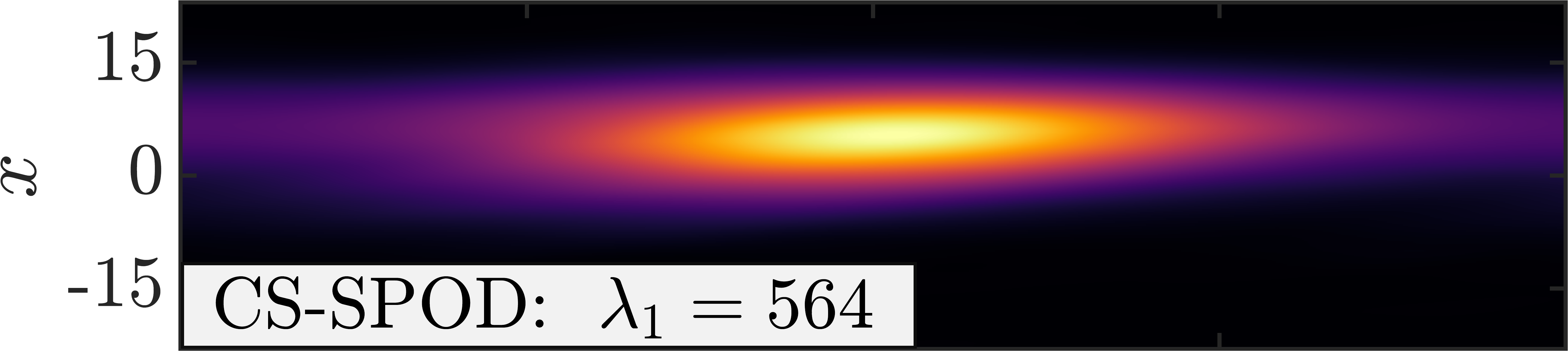}
\includegraphics[height=0.09438\textwidth] {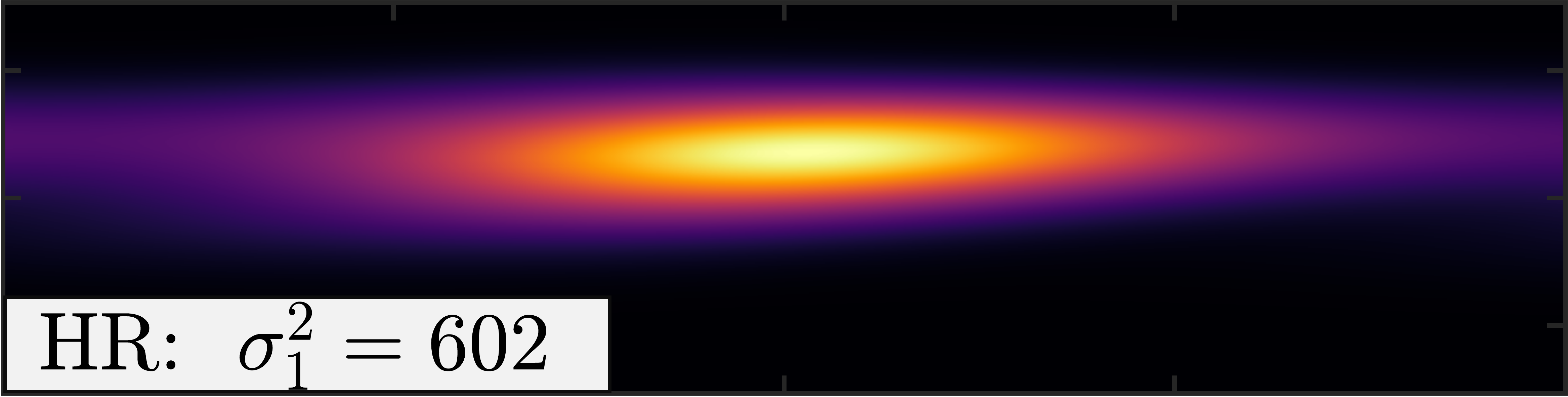} \vspace{0.8mm} \hfill  \\
\includegraphics[height=0.09438\textwidth]{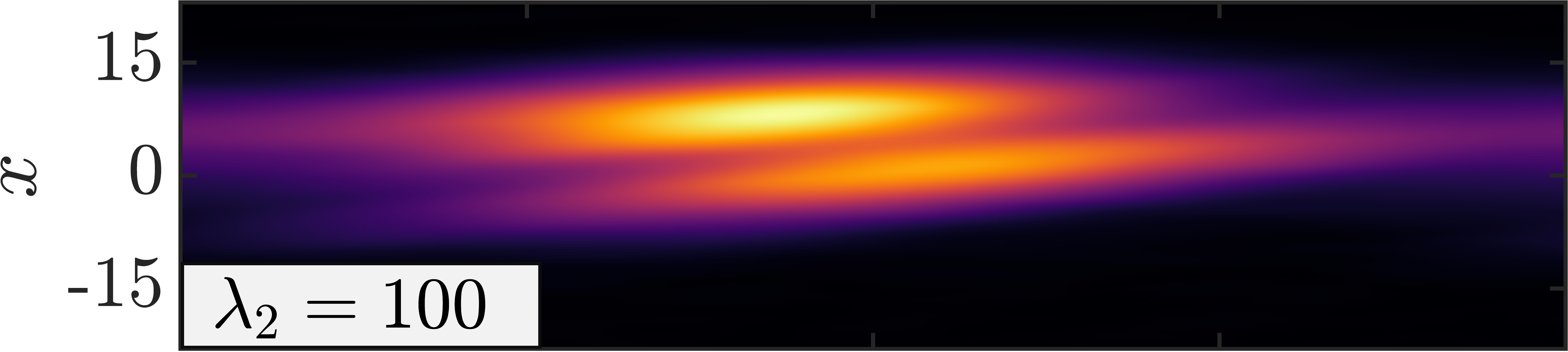}
\includegraphics[height=0.09438\textwidth]{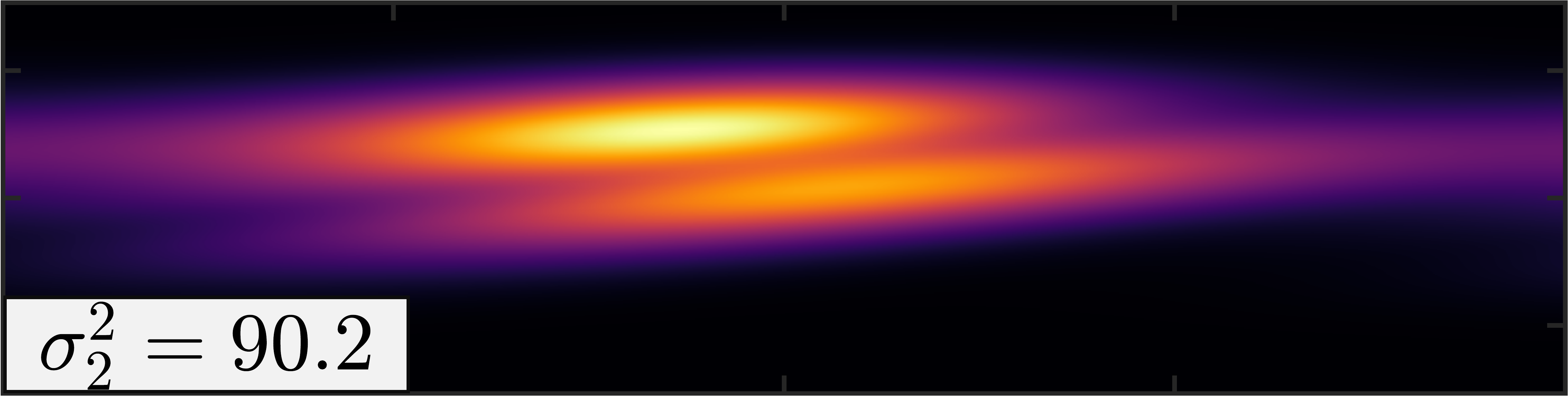} \vspace{0.8mm} \hfill  \\
\includegraphics[height=0.14003\textwidth]{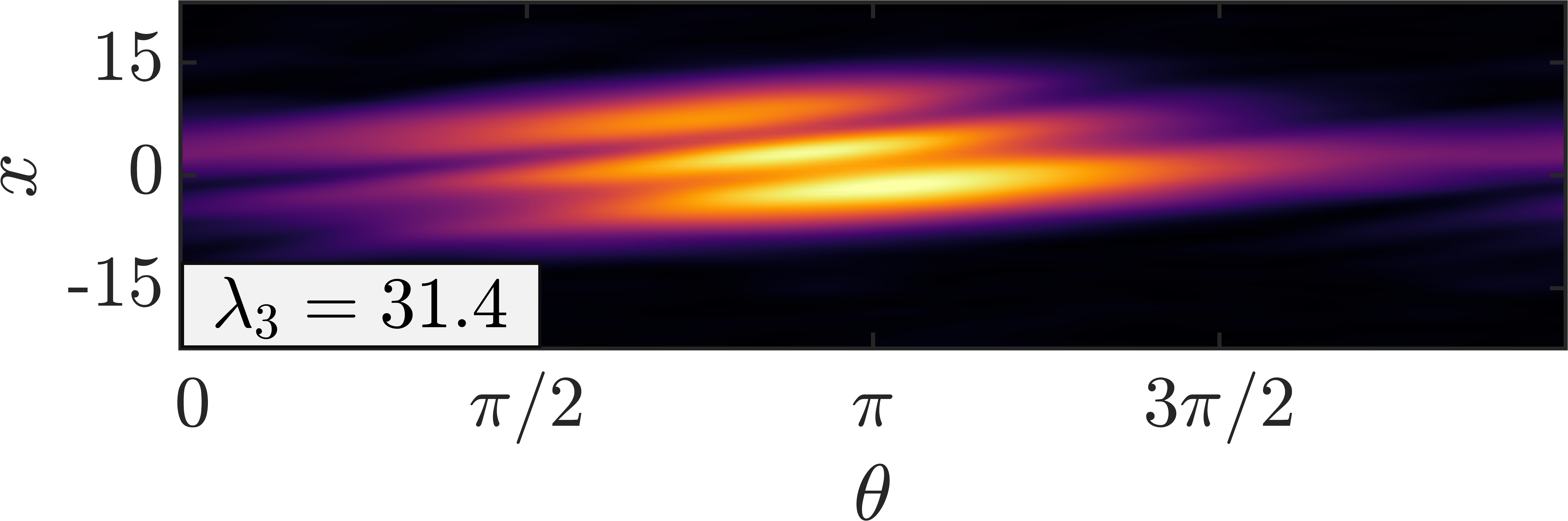}
\includegraphics[height=0.14003\textwidth]{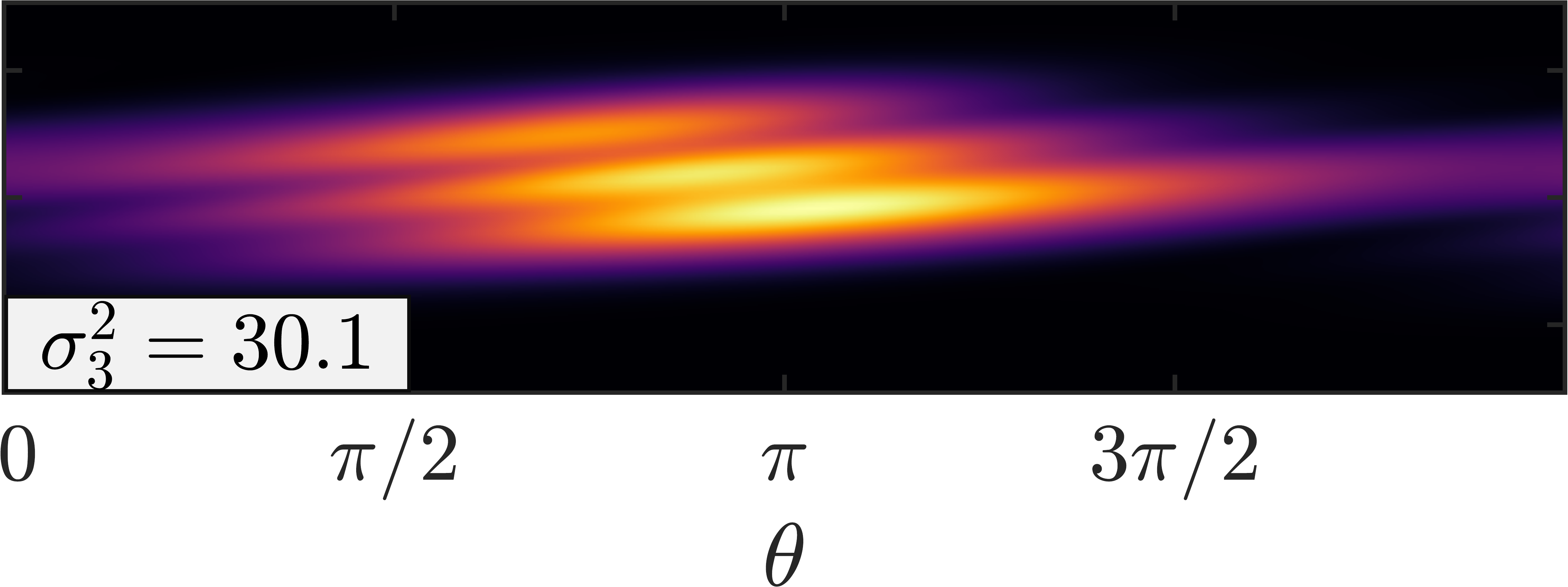} \hfill 
        \caption{Comparison of the magnitude of the three most energetic CS-SPOD (left) and harmonic resolvent (right) modes at $\fci = 0.05$ for the Ginzburg-Landau system at $A_\mu = 0.4$. The contour limits for the CS-SPOD modes are set equal to the corresponding harmonic resolvent modes $[0, |\psib_j(\x, t)|_\infty]$.}
    \label{fig:CSHARModes}
\end{figure}

Figure \ref{fig:CSHARModes} shows the magnitude of the time evolution of the three most energetic CS-SPOD and harmonic resolvent modes at $\fci = 0.05$, which we see are almost indistinguishable.  The similarity between the CS-SPOD and harmonic resolvent modes is quantified using the projection $\pcoeff_{jk}(f) = \langle \psib_j(\fci), \UCHb_k(\fci) \rangle_x$ and the harmonic-resolvent-mode expansion-coefficient CSD $S_{\beta_j \beta_k}(\fci)$ given by \eqref{eqn:sbetabeta}. To compute $S_{\beta_j \beta_k}(f)$, we take two inner products with respect to $\UCHb_j(\fci)$ and $\UCHb_k(\fci)$ and then divide by $\sigma_j(\fci)$ and $\sigma_k(\fci)$, obtaining
\begin{equation}
    S_{\beta_j \beta_k}(\fci) = \sum_{j = 1}^\infty \frac{\lambda_n(\fci)}{\sigma_j(\fci) \sigma_k(\fci)} \pcoeff_{nj}(\fci)\pcoeff^*_{nk}(\fci).
\end{equation}

\begin{figure}
\centering
  \begin{subfigure}[t]{0.4\textwidth}
  \vskip 0pt
  \centering
    \includegraphics[height=0.7\textwidth]{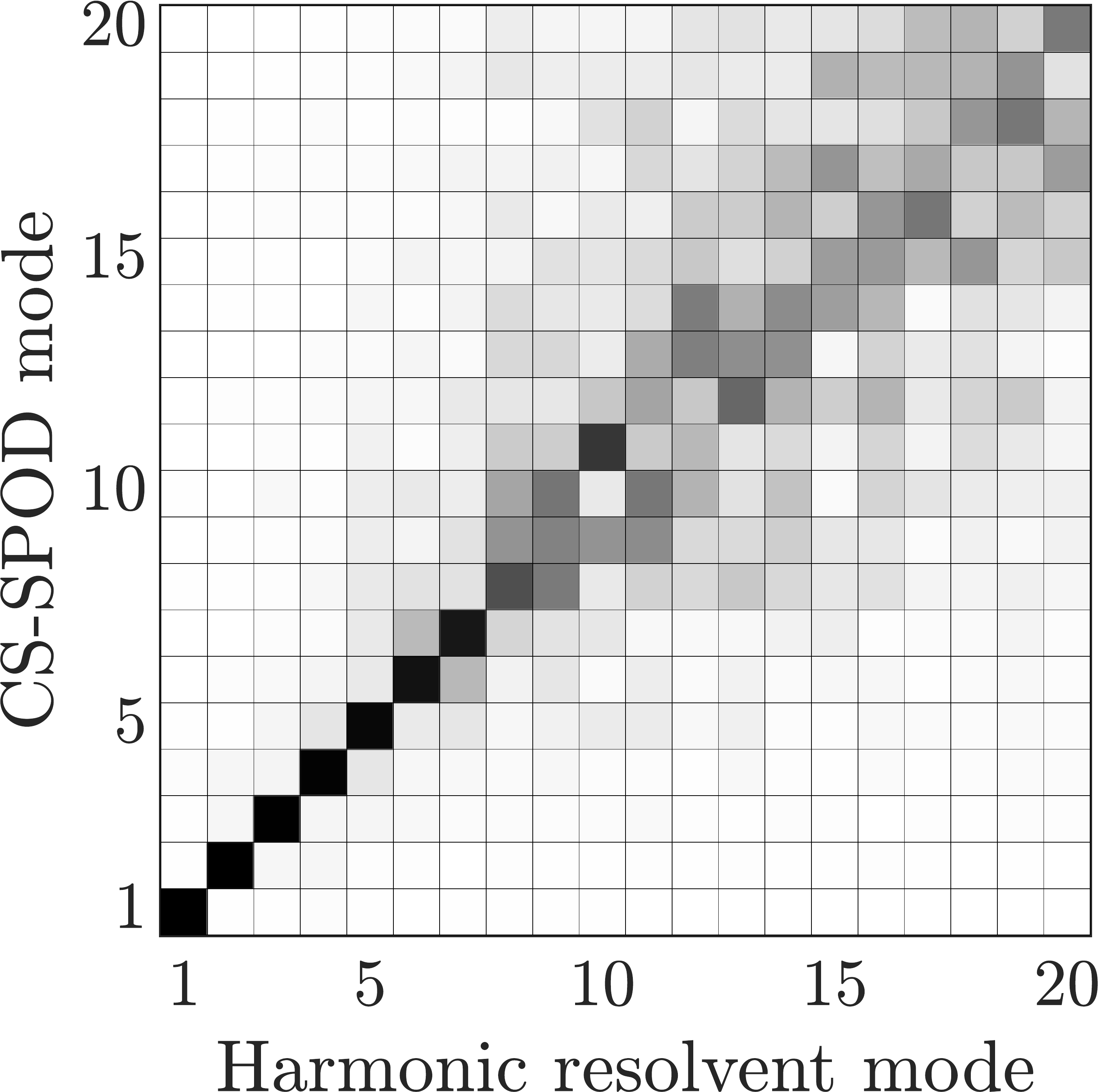}
    \caption{$\pcoeff_{jk}$ \vphantom{$\frac{|S_{\beta_j \beta_k}|}{|S_{\beta_j \beta_k}|_\infty}$} }
  \end{subfigure} 
  \begin{subfigure}[t]{0.4\textwidth}
  \vskip 0pt
  \centering
\includegraphics[height=0.7\textwidth]{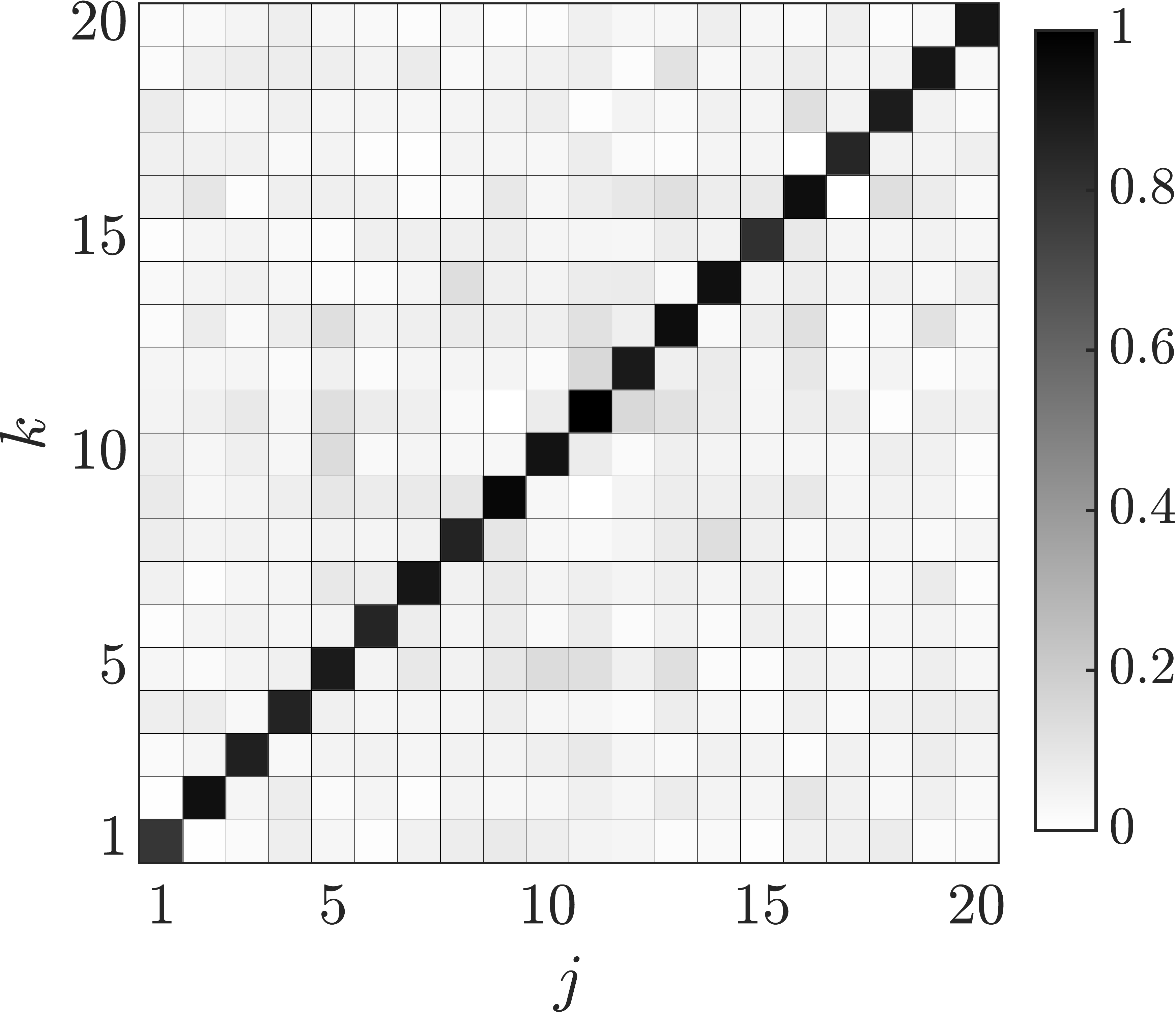}
    \caption{$\frac{|S_{\beta_j \beta_k}|}{|S_{\beta_j \beta_k}|_\infty}$}
  \end{subfigure}
\caption{CS-SPOD and harmonic resolvent analysis mode projection coefficient  (a) and magnitude of the normalized harmonic-resolvent-mode expansion-coefficient CSD (b) at $\fci = 0.05$ for the Ginzburg-Landau system with white noise forcing. }
\label{fig:VarMetrics_white}
\end{figure}

The projection $\pcoeff_{jk}$ and $\frac{|S_{\beta_j \beta_k}|}{|S_{\beta_j \beta_k}|_\infty}$ are shown in figure \ref{fig:VarMetrics_white} for $\fci = 0.05$.  $|S_{\beta_j \beta_k}|$ is, by construction, diagonal, and this should result in a diagonal $\pcoeff_{jk}$.  This is verified for the first eight modes, but for increasingly subdominant modes,  off-diagonal terms become increasingly apparent, which is owing to a lack of full statistical convergence. 

\begin{figure}
        \centering
        \includegraphics[width=0.9 \textwidth]{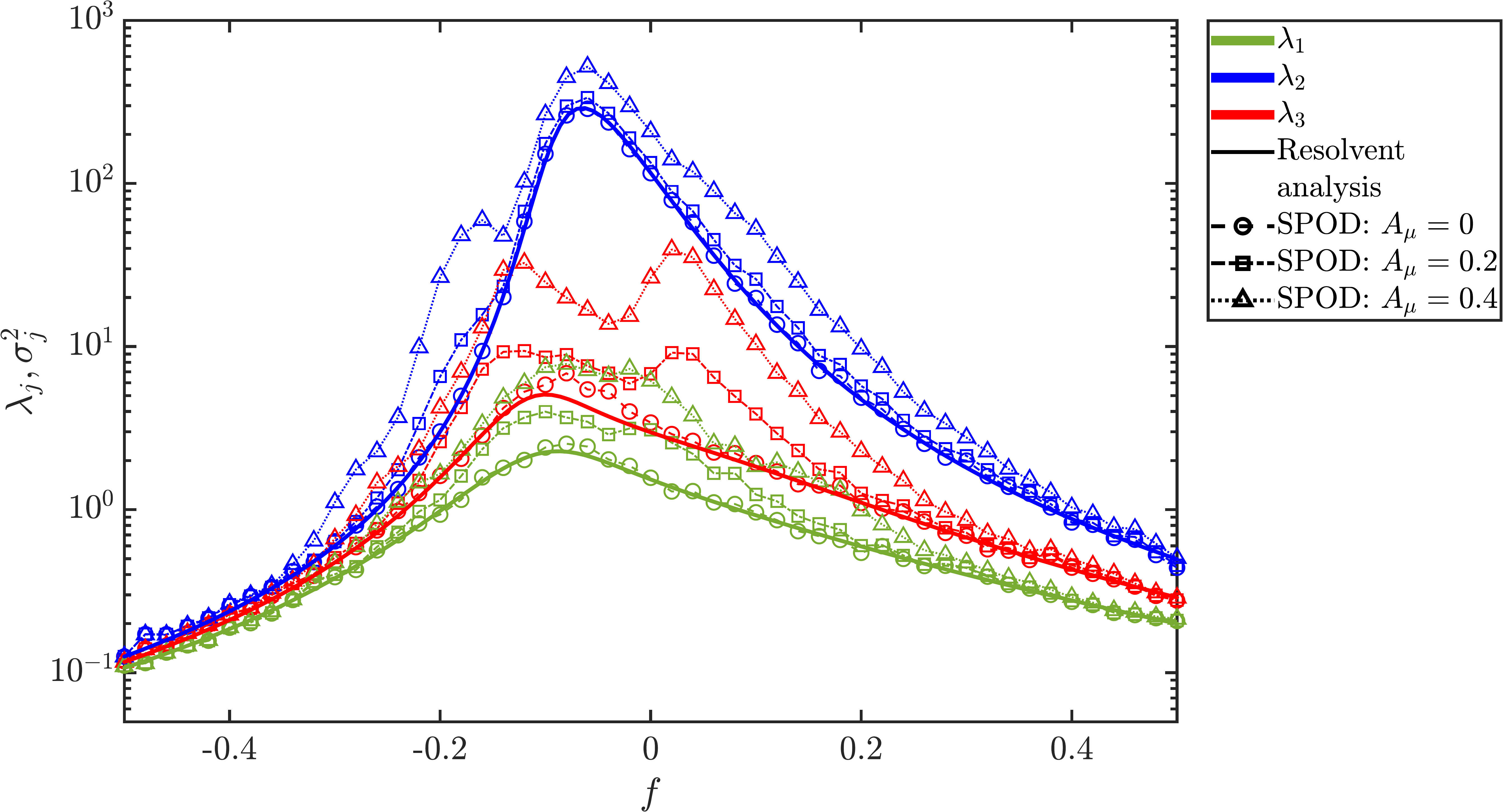}
        \caption{Comparison of the first three resolvent analysis gains $\sigma_j^2$ and SPOD eigenvalues $\lambda_j$ as a function of frequency $f$ for the white noise forced Ginzburg-Landau system at $A_{\mu} = 0.0, 0.2$, and $0.4$. For clarity, every second SPOD eigenvalue has been omitted.}
        \label{fig:SPODRescompare}
\end{figure}

Finally, to demonstrate the necessity of using harmonic resolvent and CS-POD to model and educe structures for time-periodic mean flows, we compare our results with a naive application of SPOD and standard resolvent to the time-periodic GL system.  Figure \ref{fig:SPODRescompare} compares the (standard) resolvent gains and SPOD eigenvalues for  $A_{\mu}=0$, 0.2, and 0.4. When $A_{\mu} = 0$, the system is stationary, and the resolvent gains and SPOD energies agree (as expected), but there are significant and growing discrepancies as $A_{\mu} \ne 0$ is increased and the base flow is increasingly oscillatory.  For systems with periodic statistics, CS-SPOD and harmonic resolvent analysis must be used to analyze these flows. 

\section{Low-frequency and high-frequency forcing limits}
In many flows, the frequency of the forcing may be either low or high with respect to the dynamics of interest. In both cases, simplifications can be made to the analysis. \par
For low-frequency forcing, CS-SPOD and HR tend towards systems that link all frequency components together, thereby making the analysis of the resulting system impractical. However, in many cases, we are interested in frequencies that are much larger than the forcing frequency. \citet{franceschini2022identification} showed that high-frequency structures evolving on a low-frequency periodic motion could be analyzed using a quasi-steady approach which they named phase-conditioned localized SPOD (PCL-SPOD) and quasi-steady (QS) resolvent analysis. These methods require $f >> f_0$ and that at each fixed time (or phase) $t$, the cross-correlation tensor, around that phase, only depends on the time lag $\tau$. At each phase, all standard SPOD and resolvent analysis properties are satisfied in PCL-SPOD and QS resolvent analysis, and we refer the reader to \citet{franceschini2022identification} for a detailed discussion. Although PCL-SPOD was developed without reference to cyclostationary theory and computational methods, by employing a similar derivation to \citet{franceschini2022identification}, PCL-SPOD can be written as
\begin{align}
\int_{\Omega} \WVb(\x, \x^\prime, f, t) \Wb(\x^\prime) \psib(\x^\prime, f, t)  \dx^\prime &= \lambda \psib(\x, f, t), 
\label{eqn:PCLSPOD}
\end{align}
where $\WVb(\x, \x^\prime, f, t)$ is the Wigner-Ville spectrum and $\psib(\x^\prime, f, t^\prime)$ are the PCL-SPOD eigenvectors that only contain a single frequency component $f$ and are independent over time. This is analytically identical to the PCL-SPOD shown in \citet{franceschini2022identification}, but is numerically determined using a different computational procedure. QS resolvent analysis is similarly written as 
\begin{equation}
    (i2\pi f\ICb - \Ab(t))\hat{\gb}_f = \hat{\bf{\etab}}_f, 
\end{equation} 
where $\Rb(t) = (i2\pi f\ICb - \Ab(t))$ is the QS resolvent operator, and the solution at each time-instance $t$ is independent of the solution at any other time-instance. We then seek to solve the forcing mode $\hat{\bf{\etab}}_f$ that results in the most energetic response $\hat{\gb}_f$, which is determined via the singular value decomposition of the weighted harmonic resolvent operator
\begin{equation}
\WCb_{g}^{1/2} \Rb(t) \WCb_{\eta}^{-1/2} = \widetilde{\UCb}_\dag \Sigmab_\dag \widetilde{\VCb}^{*}_\dag , 
\end{equation}
where $\WCb_{\eta}$ and $\WCb_{g}$ are the norms on the input and output space, respectively, and are defined similarly to equation \ref{eqn:normdef}. The diagonal matrix $\Sigmab_\dag = \text{diag}[\sigma_1^2, \sigma_2^2, \cdots]$ contains the ranked gains and the columns of $\VCHb_\dag = \WCb_{\FC}^{-1/2} \widetilde{\VCb}_\dag$ and  $\UCHb_\dag= \WCb_{\GC}^{1/2} \widetilde{\UCb}_\dag$ contain the forcing and response modes, respectively. \par

\begin{figure}
\centering
    \begin{subfigure}[t]{0.49\textwidth}
    \vskip 0pt 
\includegraphics[height=0.204226415\textwidth]{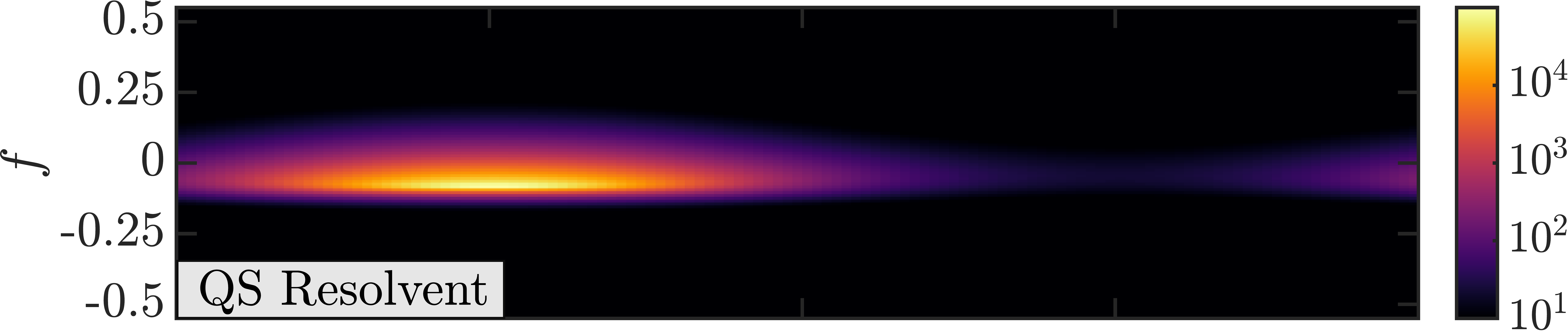} \hfill \vspace{0.45mm} \\
        \includegraphics[height=0.198\textwidth]{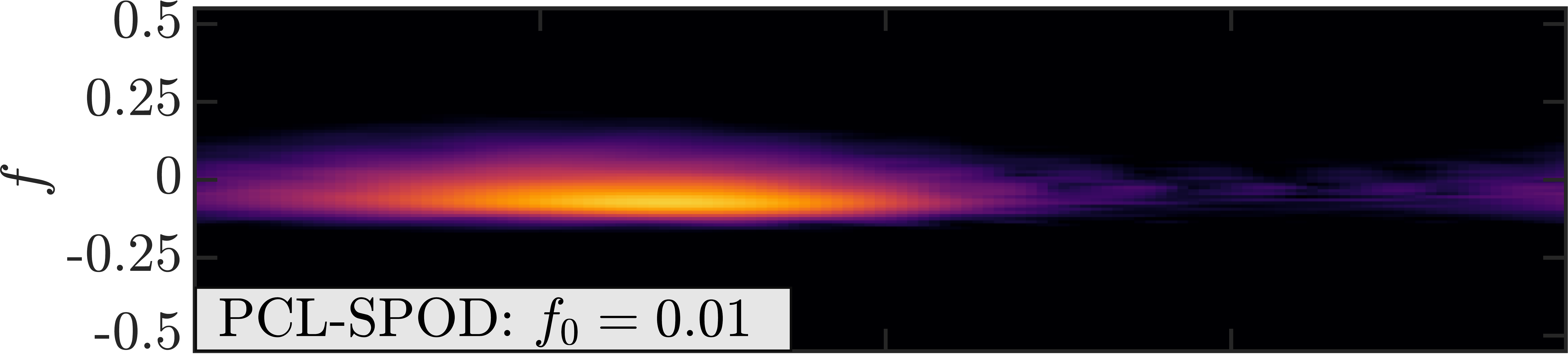} \hfill \vspace{0.7mm} \\
        \includegraphics[height=0.198\textwidth]{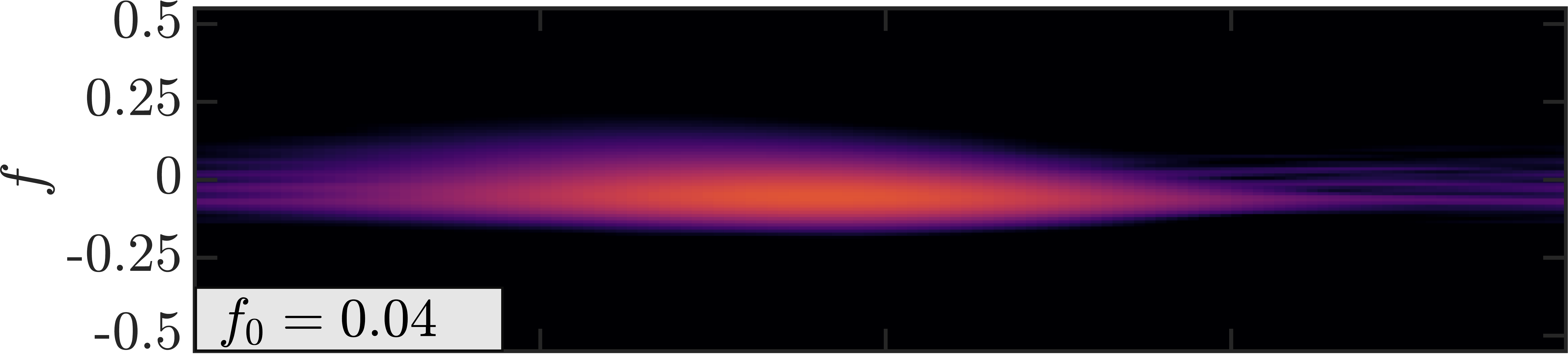} \hfill \vspace{0.7mm} \\
\includegraphics[height=0.270226415\textwidth]{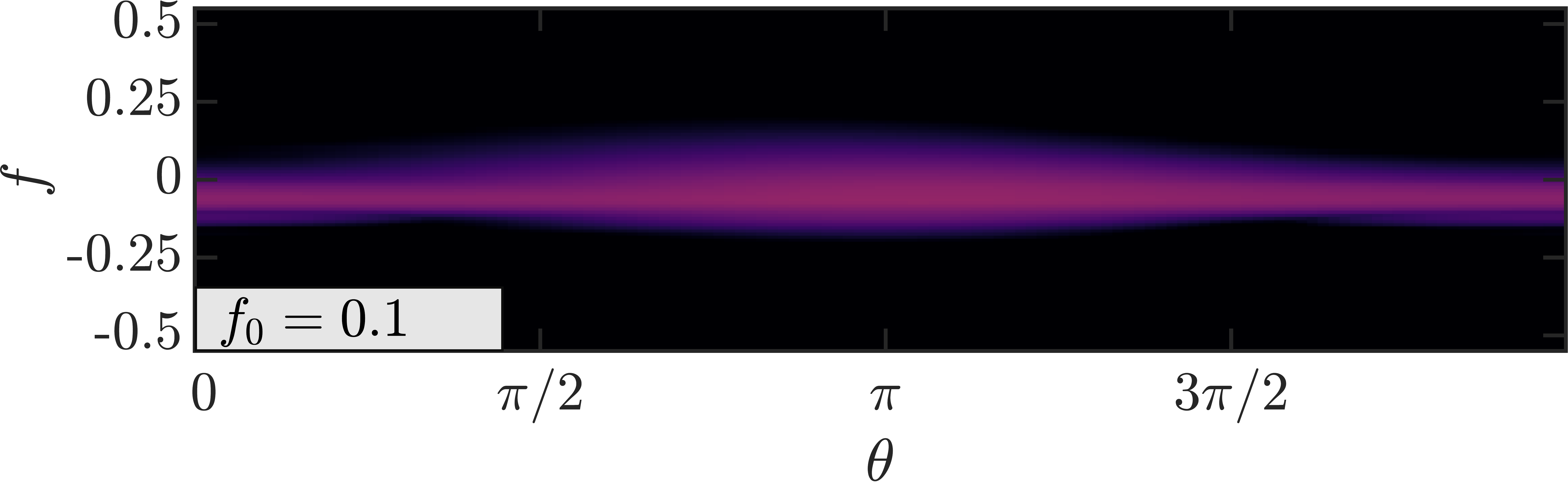} \hfill  \\ 
        \caption{Contours of QS resolvent gain ($\sigma_j^2$) and PCL-SPOD energy ($\lambda_j$) as a function of frequency $f$ and phase $\theta$.}
        \label{fig:GLLowHigh:Gain}
    \end{subfigure}\hfill 
    \begin{subfigure}[t]{0.49\textwidth}  
        \vskip 0pt  
\includegraphics[height=0.204226415\textwidth]{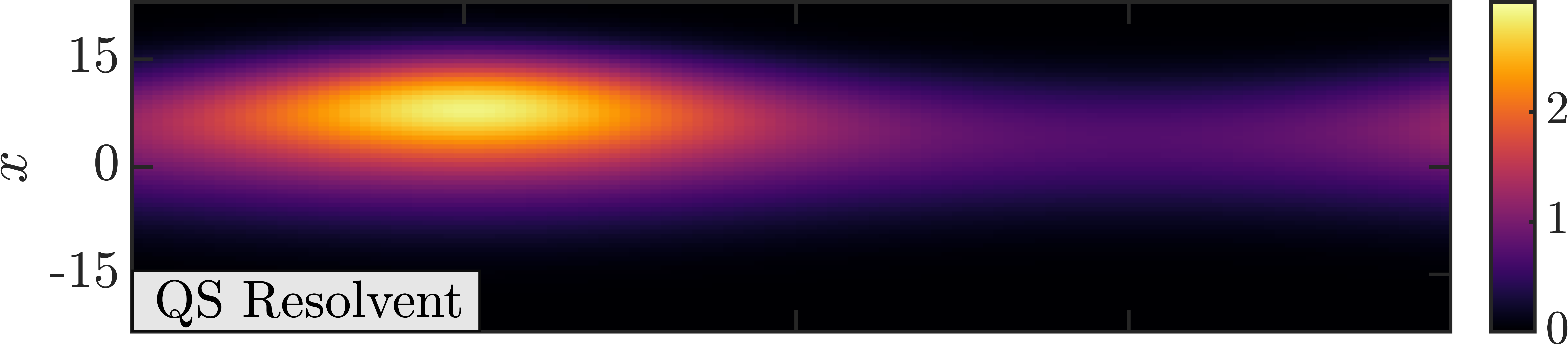} \hfill \vspace{0.5mm} \\
        \includegraphics[height=0.198\textwidth]{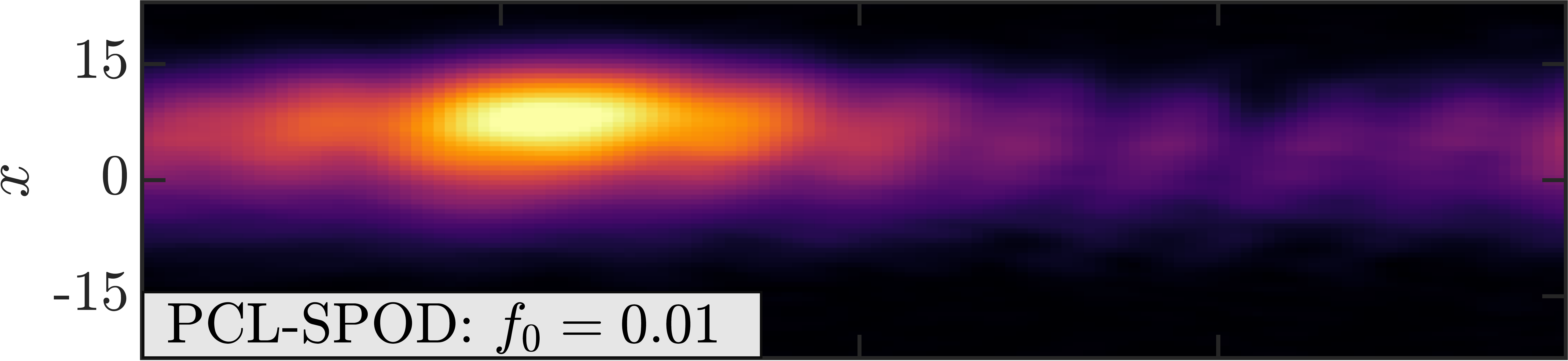} \hfill \vspace{0.8mm} \\
        \includegraphics[height=0.198\textwidth]{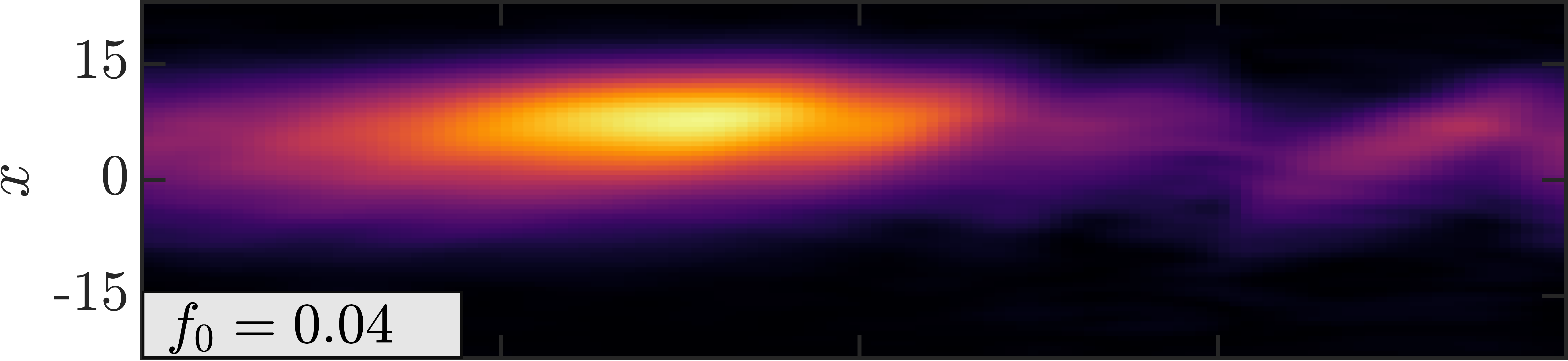} \hfill \vspace{0.8mm} \\
\includegraphics[height=0.270226415\textwidth]{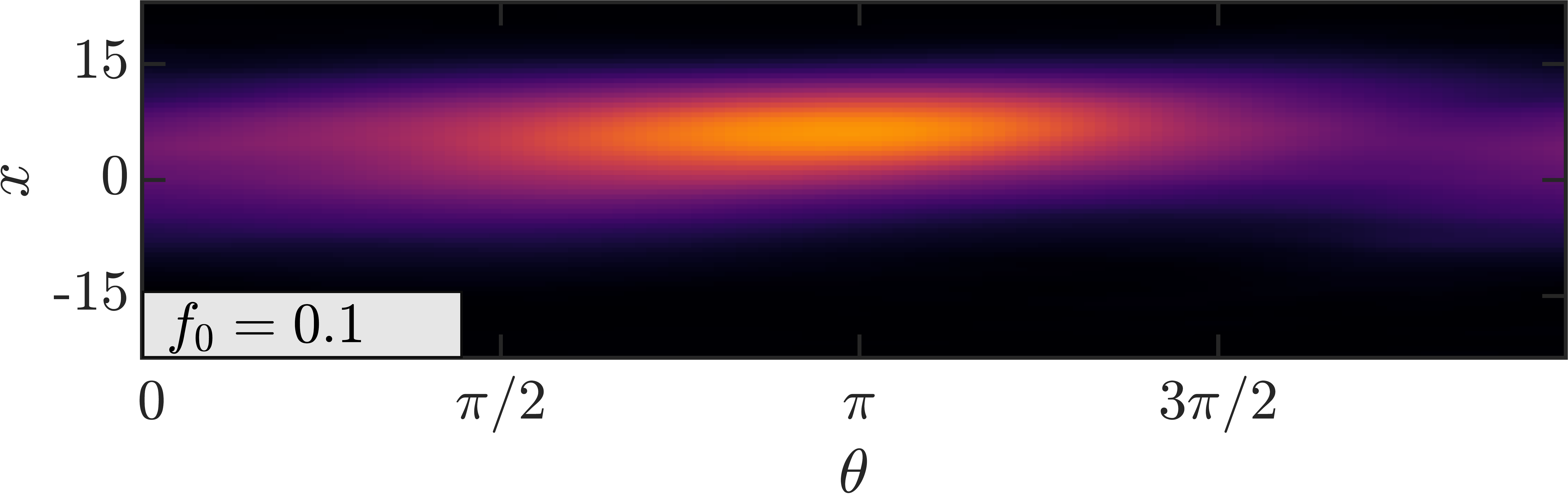} \hfill\\
       \caption{Weighted mode shapes in $\theta-x$ space of the dominant QS resolvent ($\sigma_j(f)|\hat{\bm{u}}_j(\x, f)|$) and PCL-SPOD ($\sqrt{\lambda_j}|\psib(\x^\prime, f, t)|$) mode at $f = 0.1$. }
       \label{fig:GLLowHigh:Mode}
    \end{subfigure}% 
    \caption{Contours of the gain and weight modes shapes of the white noise forced Ginzburg-Landau system at $A_\mu = 0.2$ and $f_0 = 0.01, 0.04,$ and $0.1$.}
    \label{fig:GLLowHighGainMode}
\end{figure}

Using equation \ref{eqn:wvdef}, algorithm \ref{alg:cap}, and a procedure similar to that of regular SPOD, we compute PCL-SPOD of the Ginzburg-Landau systems with white-noise forcing for several different forcing frequencies $f_0 = 0.01, 0.04,$ and $0.1$ at $A_{\mu} = 0.2$. Due to the substantially lower forcing frequency, $2\times 10^5$ snapshots are saved instead of $4\times 10^4$. We then compare the PCL-SPOD and QS resolvent results in figure \ref{fig:GLLowHighGainMode} where we see excellent agreement for small $f_0$. We see that as $f_0$ increases, the PCL-SPOD and QS resolvent results increasingly deviate as the two aforementioned assumptions are increasingly violated. \par
\begin{figure}
        \centering
        \includegraphics[width=1\textwidth]{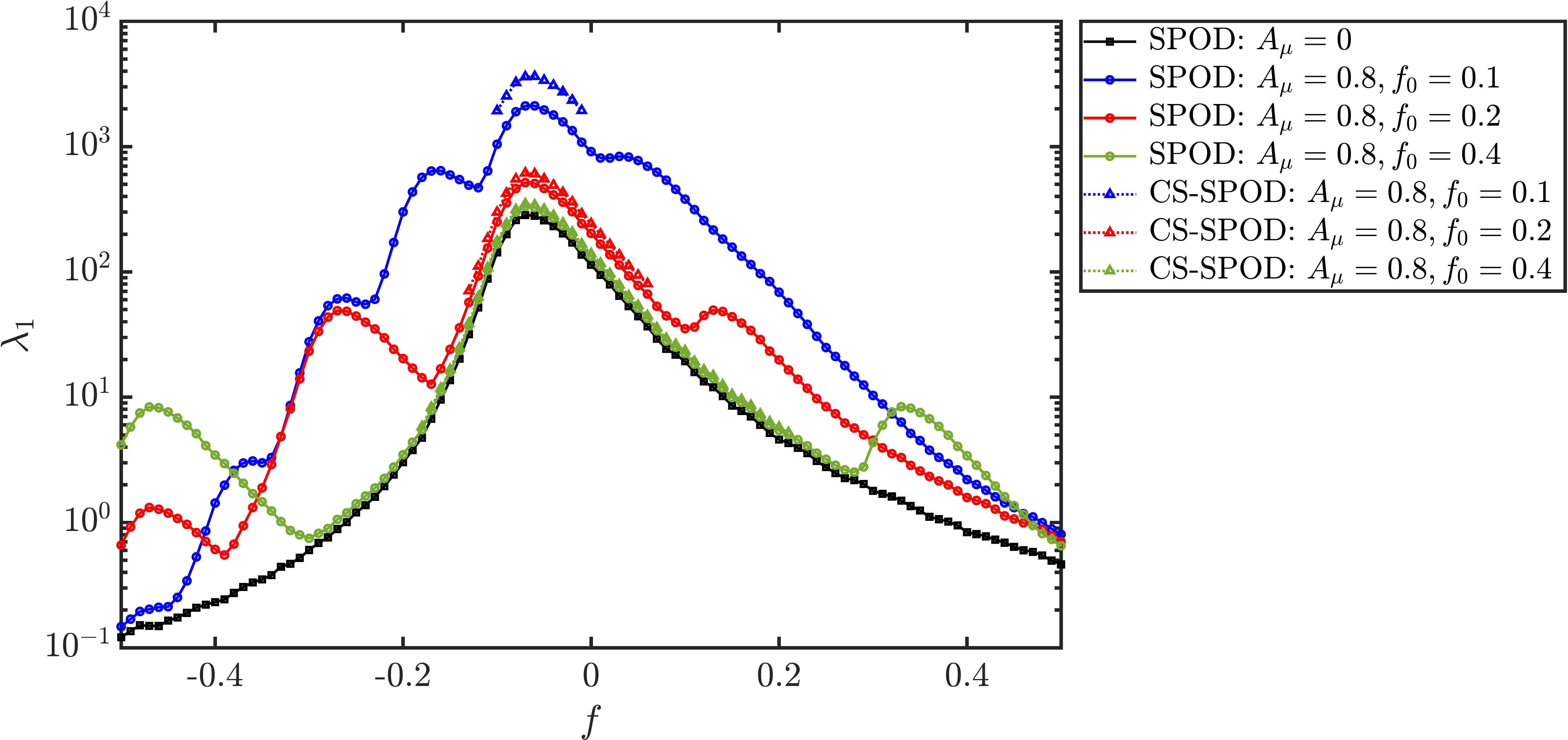}
        \caption{Comparison of the dominant SPOD and CS-SPOD eigenvalues $\lambda_1$ as a function of frequency $f$ for the white noise forced Ginzburg-Landau system at $A_{\mu} = 0.8$ for $f_0 = 0.1, 0.2, 0.4$. The SPOD eigenvalues for $A_{\mu} = 0$ are overlaid to show the impact of the forcing on the spectrum. }
        \label{fig:HighFreqCompare}
\end{figure}

Many physical systems exhibit some form of spectral peak. If the forcing frequency is sufficiently large, such that the energy contained at $f + k\alpha_0, k \in \mathbb{Z}\ \&\ k \ne 0$ is substantially lower than at $f$, one can see that the CS-SPOD and harmonic resolvent systems (given by equations \ref{eqn:finalcsspod}, \ref{eqn:HRsystem}, respectively) can be approximated by the block diagonal term that corresponds to $f$ (i.e. the most energetic component in $\Fset_{\fci}$). Furthermore, for many systems, the impact of a high-frequency forcing on the low-frequency dynamics is not direct, instead, the low-frequencies are modified as a result of nonlinear interaction that modifies mean flow. Thus, for a large forcing frequency, CS-SPOD and harmonic resolvent analysis approach SPOD and standard resolvent analysis, respectively. In figure \ref{fig:HighFreqCompare}, we show the SPOD and CS-SPOD eigenspectrum of the white-noise forced Ginzburg-Landau system at $A_{\mu} = 0.8$ for $f_0 = 0.1, 0.2, 0.4$. To assess the convergence of CS-SPOD to SPOD for large forcing frequencies, the CS-SPOD modes have been mapped to the SPOD mode of greatest alignment (computing over the same set of frequencies $\Fset_{\fci}$). This is similar to what was performed in \S \ref{sec:FGL} during the comparison between SPOD and CS-SPOD modes. We see that as the forcing frequency increases, the CS-SPOD and SPOD eigenvalues begin to converge in the region where the energy at $f + k\alpha_0 << f, k \in \mathbb{Z}\ \&\ k \ne 0$. 

\section{Conclusions} \label{sec:conclusion}
In this paper, we have proposed CS-SPOD for the extraction of the most energetic coherent structures from complex turbulent flows whose statistics vary time-periodically (i.e. flows that have cyclostationary statistics). This is achieved by an extension of the one-dimensional technique developed by \citet{kim1996eofs} to large high-dimensional data through the use of the method-of-snapshots to make the algorithm computationally feasible for large data. The orthogonality/optimality properties of the modes generated by CS-SPOD are shown, where, similar to SPOD analysis of stationary flows, CS-SPOD determines the set of orthogonal modes that optimally reconstruct the statistics of these flows in terms of the space-time norm. 

In contrast to SPOD, where the modes oscillate at a single frequency and have a constant amplitude in time, CS-SPOD modes oscillate at a set of frequencies separated by the fundamental cycle frequency (typically the frequency of modulation), have a periodic amplitude in time, and optimally reconstruct the second-order statistics. We show that CS-SPOD naturally becomes SPOD when analyzing a stationary process, allowing the CS-SPOD results to be interpreted in a familiar manner. Furthermore, we develop an efficient computational algorithm to compute CS-SPOD with a computational cost and memory requirement similar to SPOD, thus allowing CS-SPOD to be computed on a wide range of problems. Lastly, similar to the relationship that exists between SPOD and standard resolvent analysis \citep{towne2018spectral}, CS-SPOD modes are identical to harmonic resolvent modes in the case where the harmonic-resolvent-mode expansion coefficients are uncorrelated. We also discuss simplifications that can be made when forcing at a low or high frequency. \par

We applied the CS-SPOD algorithm to two datasets. The first is data from a modified linearized complex Ginzburg-Landau equation with time-periodic dynamics, which represents a simple model of a flow exhibiting non-modal growth. As the amplitude of the imposed time-periodicity is increased, CS-SPOD yields modes that are increasingly phase-dependent. We demonstrated the inability of SPOD to capture these dynamics, which is shown through both an analysis of the temporal evolution of the modes and by the ability of CS-SPOD to capture substantially more energy than SPOD. In addition, we show that when the system is forced with unit-variance white noise, the CS-SPOD modes from the data were identical (up to statistical convergence) with modes computed by harmonic resolvent analysis. For cyclostationary processes, we show that (standard) resolvent analysis cannot predict the time-averaged statistics even when the white-forcing conditions are met. This shows that CS-SPOD and harmonic resolvent analysis should be used to correctly analyze and/or model flows with cyclostationary statistics. \par

We next considered a forced, turbulent  high-Reynolds-number jet, demonstrating CS-SPOD on a turbulent flow for the first time. We identified coherent structures that differed in important ways from their SPOD-identified cousins in natural jets.  In particular, CS-SPOD clarifies how the dynamics of the coherent structures are altered by the forcing. For example, the axisymmetric CS-SPOD structure at a low Strouhal number featured finer-scale axisymmetric Kelvin-Helmholtz roll-up in the near-nozzle region that is absent in natural jets at a high Reynolds number.  This roll-up waxed and waned at those phases of the forcing cycle where the initial shear layer was thinned and thickened, respectively.  

Overall, our results show that CS-SPOD successfully extends SPOD to flows with cyclostationary statistics. This allows us to study a wide range of flows with time-periodic statistics such as turbomachinery, weather and climate, and flow control with harmonic actuation, and wake flows rendered cyclostationary through the (arbitrary) choice of a phase reference for the dominant shedding frequency. Although we focused on strictly cyclostationary processes, further generalizations are possible to almost periodic flows and flows forced with several non-commensurate frequencies.

\bigbreak
\footnotesize
\textbf{Acknowledgements.}
The authors gratefully acknowledge support from the United States Office of Naval Research under contract N00014-20-1-2311 with Dr. S. Martens as program manager and the Federal Aviation Administration under grant 13-C-AJFE-UI. This work was supported in part by high-performance computer time and resources from the DoD High Performance Computing Modernization Program. This work used Stampede2 at Texas Advanced Computing Center through allocation CTS120005 from the Advanced Cyberinfrastructure Coordination Ecosystem: Services \& Support (ACCESS) program, which is supported by National Science Foundation grants \#2138259, \#2138286, \#2138307, \#2137603, and \#2138296. \par \vspace{3mm}

\textbf{Declaration of Interests.} The authors report no conflict of interest.
\normalsize
\appendix
\section{}\label{appen:csspodderivation}
To derive the eigenvalue problem given by \eqref{eqn:finalcsspod}, we rewrite $\Rb(\x, \x^\prime, t, t^\prime) \rightarrow \Rb(\x, \x^\prime, t, \tau) \equiv E\{\qd(\x, t + \tau/2 )\qd^*(\x^\prime, t - \tau/2)\}$, where $\tau = t - t^\prime$. Recalling that for a cyclostationary process, the two-point space-time correlation density is a periodic function in time and can be expressed as a Fourier series
\begin{equation}
    \Rb(\x, \x^\prime, t, \tau) = \sum_{n \in \mathcal{A}_n} \widetilde{\Rb}_{n\alpha_0}(\x, \x^\prime, \tau) e^{i2\pi n \alpha_0  t}, 
\end{equation}
where $\widetilde{\Rb}_{n\alpha_0}(\x, \x^\prime, \tau)$ are the cyclic autocorrelation functions of $\Rb(\x, \x^\prime, t, \tau)$ at cycle frequency $n\alpha_0$ and $\mathcal{A}_n = \{\cdots, -1, 0, 1, \cdots\}$ is, in general, the infinite set of harmonics of the fundamental cycle frequency present in the flow. One can also decompose the two-point space-time correlation density as the following phase-shifted Fourier series
\begin{equation}
    \Rb(\x, \x^\prime, t, \tau) =  \sum_{n \in \mathcal{A}_n} \hat{\Rb}_{n\alpha_0}(\x, \x^\prime, \tau)e^{-i\pi n \alpha_0 \tau} e^{i2\pi n \alpha_0 t}, 
    \label{eqn:tpstctfourier}
\end{equation}
where the two Fourier coefficients are related by
\begin{equation}
    \widetilde{\Rb}_{n\alpha_0}(\x, \x^\prime, \tau)e^{i\pi n\alpha_0 \tau} = \hat{\Rb}_{n\alpha_0}(\x, \x^\prime, \tau).
\end{equation}
Although somewhat unusual, this simply applies a phase shift to the resulting Fourier series coefficients that, after Fourier transforming, shifts the center frequency of the CCSD. This is identical to the phase shift that relates the symmetric and asymmetric definitions of the cyclic cross-correlation functions and CCSD. Due to this, one can derive CS-SPOD using the symmetric definitions and a phase shift or using the asymmetric definition. We choose the former as it results in a simpler derivation later. This phase shift is required to ensure the resulting eigensystem is Hermitian and positive semi-definite.  Substituting the cyclic Wiener-Khinchin relation from \eqref{eqn:cwkt} into \eqref{eqn:tpstctfourier} and then into the Fredholm eigenvalue problem \eqref{eqn:KLEquation} results in 
\begin{align}
\int_{-\infty}^\infty \int_{\Omega} \int_{-\infty}^{\infty}  \sum_{n \in \mathcal{A}_n} \Sb_{n\alpha_0}(\x, \x^\prime, f )  e^{i 2\pi n\alpha_0 t} e^{i 2\pi (f - \frac{1}{2} n\alpha_0)  \tau}\Wb(\x^\prime)&\phib(\x^\prime, t^\prime) \df \dx^\prime \dt^\prime \nonumber\\ 
& = \lambda \phib(\x, t).
\label{eqn:ctos3}
\end{align}
Since $\tau = t - t^\prime$, this leads to the following simplifications,
\begin{align}
\int_{-\infty}^\infty \int_{\Omega}  \sum_{n \in \mathcal{A}_n} \Sb_{n\alpha_0}(\x, \x^\prime, f )  e^{i 2\pi n\alpha_0 t} e^{i 2\pi (f - \frac{1}{2} n\alpha_0)  t}    \Wb(\x^\prime)& \nonumber\\ 
\int_{-\infty}^{\infty}  \left[\phib(\x^\prime, t^\prime) e^{-i 2\pi (f - \frac{1}{2}n \alpha_0)  t^\prime} \dt^\prime \right] \df dx^\prime  &= \lambda \phib(\x, t), \\
    \int_{-\infty}^\infty \int_{\Omega}  \sum_{n \in \mathcal{A}_n}  \Sb_{n\alpha_0}(\x, \x^\prime, f )   e^{i 2\pi (f + \frac{1}{2} n \alpha_0)t}    \Wb(\x^\prime)   \hat{\phib}(\x^\prime, f - \frac{1}{2}n\alpha_0) &\df \dx^\prime\hspace{-0.5mm} =\hspace{-0.5mm}\lambda \phib(\x, t),
    \label{eqn:beforesub}
\end{align}
where $\hat{\phib}(\x^\prime, f)$ is the temporal Fourier transform of $\phib(\x^\prime, t^\prime)$. Similar to SPOD, we must choose a solution ansatz. In SPOD, we can solve a single frequency at a time as there is no correlation between different frequency components. However, since cyclostationary processes have spectral components that are correlated, we are unable to solve for each frequency component separately. Instead, we solve multiple coupled frequencies together by choosing our solution ansatz as 
\begin{equation}
    \phib(\x, t) = \sum_{m \in \mathcal{A}_m} \psib(\x, \fci + m\alpha_0) e^{i2\pi (\fci + m\alpha_0)t},
\end{equation} giving, 
\begin{equation}
    \hat{\phib}(\x, f) = \sum_{m \in \mathcal{A}_m} \psib(\x, \fci + m\alpha_0) \delta(f - (\fci + m\alpha_0)), 
\end{equation}
 where $\mathcal{A}_m = \{\cdots, -1, 0, 1, \cdots\}$ gives, in general, the infinite set of frequencies present in the solution (all separated by $\alpha_0$). The frequency-shifted version of $\hat{\phib}(\x, f)$ is given by 
\begin{equation}
    \hat{\phib}(\x, f - \frac{1}{2} n \alpha_0) = \sum_{m \in \mathcal{A}_m} \psib(\x, \fci + m\alpha_0) \delta(f  - (\fci + (m + \frac{1}{2}n)\alpha_0)).
\end{equation}
Substituting these expressions into \eqref{eqn:beforesub} and integrating with respect to $f$ results in 
\begin{align}
    &\int_{\Omega}  \sum_{n \in \mathcal{A}_n}  \sum_{m^\prime \in \mathcal{A}_m} \Sb_{n\alpha_0}(\x, \x^\prime, \fci + (m^\prime +\frac{1}{2}n)\alpha_0 )   e^{i 2\pi (\fci + (n+m^\prime) \alpha_0)  t}    \Wb(\x^\prime)  \nonumber\\
    &\psib(\x, \fci + m^\prime\alpha_0) \dx^\prime   = \lambda \sum_{m \in \mathcal{A}_m} \psib(\x, \fci + m\alpha_0) e^{i2\pi (\fci + m\alpha_0)t}.
\end{align}
For this equation to hold over all time, we perform a harmonic balance where each frequency component must hold separately. This gives $\fci + (m^\prime + n)\alpha_0  = \fci + m\alpha_0 \rightarrow m^\prime + n = m$. An equation for each frequency component of our ansatz is formed as
\begin{align}
     \int_{\Omega}  \sum_{n \in \mathcal{A}_n}  \sum_{m^\prime \in \mathcal{A}_m} \Sb_{n\alpha_0}(\x, \x^\prime, \fci + (m^\prime +\frac{1}{2}n)\alpha_0 )    \Wb(\x^\prime)  & \psib(\x, \fci + m^\prime\alpha_0) \dx^\prime \delta_{n + m^\prime, m} \nonumber\\
    & =  \lambda \psib(\x, \fci + m\alpha_0).
\end{align}
Substituting $n = m - m^\prime$, this expression simplifies to
\begin{align}
    \int_{\Omega}  \sum_{m^\prime \in \mathcal{A}_m} \Sb_{(m - m^\prime)\alpha_0}(\x, \x^\prime, \fci + \frac{1}{2}(m + m^\prime)\alpha_0)    \Wb(\x^\prime)   & \psib(\x, \fci + m^\prime\alpha_0) \dx^\prime\nonumber\\
    & =  \lambda \psib(\x, \fci + m\alpha_0), 
    \label{eqn:finalCSPOD}
\end{align}
where we ignore $m - m^\prime \notin \mathcal{A}_n$. Expanding \eqref{eqn:finalCSPOD} gives the final CS-SPOD eigenvalue problem \eqref{eqn:finalcsspod}.

\bibliographystyle{jfm}
\bibliography{jfm-instructions}

\end{document}